\documentclass{raa}            % referee version: for submission

\usepackage{graphicx,times}             %for PS/EPS graphics inclusion, new
\usepackage{natbib}
\usepackage{amssymb,amsmath}
\bibpunct{(}{)}{;}{a}{}{,}

\usepackage{hyperref}
\hypersetup{colorlinks = true, linkcolor = purple, anchorcolor = red, citecolor = blue, filecolor = red, pagecolor = red, urlcolor = red}

\begin{document}

   \title{Spectral energy distribution similarity of the local galaxies and the 3.6$\mu${\MakeLowercase m} selected galaxies from the Spitzer Extended Deep Survey
%\footnotetext{$*$ Supported by the National Natural Science Foundation of China.}
}
%   \subtitle{I. Place Your Subtitle Here}

   \volnopage{Vol.0 (20xx) No.0, 000--000}      %%preserved for Editor. DOn't remove!
   \setcounter{page}{1}          %%starting page, preserved for Editor. DOn't remove!
   \author{Cheng Cheng\inst{1,2},
   Jia-Sheng Huang\inst{1,2},
   Hai Xu\inst{1,2},
   Gaoxiang Jin\inst{1,2}
   Chuan He\inst{1,2},
   Tianwen Cao\inst{1,2},
   Zijian Li\inst{1,2},
   Shumei Wu\inst{1,2},
   Piaoran Liang\inst{1,2},
   Yaru Shi\inst{1,2},
   Xu Shao\inst{1,2},
   Y. Sophia Dai\inst{1,2},
   Cong Kevin Xu\inst{1,2},
   Marat Musin\inst{1,2}
   }

   \institute{Chinese Academy of Sciences South America Center for Astronomy, National Astronomical Observatories, CAS, Beijing 100101, China, Email: chengcheng@nao.cas.cn\\
        \and
		CAS Key Laboratory of Optical Astronomy, National Astronomical Observatories, Chinese Academy of Sciences, Beijing 100101, China\\
\vs\no
   {\small Received~~20xx month day; accepted~~20xx~~month day}}

\abstract{
The Spitzer Extended Deep Survey (SEDS) as a deep and wide mid-infrared (MIR) survey project provides a sample of 500000+ sources spreading 1.46 square degree and a depth of 26 AB mag (3$\sigma$). Combining with the previous available data, we build a PSF-matched multi-wavelength photometry catalog from u band to 8$\mu$m. We fit the SEDS galaxies spectral energy distributions by the local galaxy templates. The results show that the SEDS galaxy can be fitted well, indicating the high redshift galaxy ($z \sim 1$) shares the same templates with the local galaxies. This study would facilitate the further study of the galaxy luminosity and high redshift mass function.
}

   \authorrunning{Cheng, C., Huang, J.-S. et al. }            %author_head in even pages
   \titlerunning{Multi-wavelength catalog of SEDS}  % title_head in odd pages

   \maketitle
%% The author head (on even pages) and the title head (on odd pages) will be
%% automatically extracted from \author{} and \title{}. Whenever the title is too long,
%% you will be asked to supply a shorter one by inserting either \authorrunning{} or
%% \titlerunning{} before \maketitle. Anyway, you can specify your own heads.
%%
%%
%% Note: In the following text body of your manuscript, please note several differences from
%%       other major journals:
%% (1) \subsection{Please Capitalize the First Letter of Each Notional Word in Subsection Title}
%% (2) Please Capitalize the First Letter of Each Notional Word in all tables' captions

%
%________________________________________________ sections below
%
\section{introduction}
The mid-infrared (MIR) survey permits an unbiased census of red and blue galaxies. The {\it Spitzer space telescope} Infrared Array Camera \citep[IRAC, ][]{Fazio2004} can efficiently detect the faint galaxies at redshift about 1 with much shorter exposure time because the galaxy spectral energy distribution (SED) shape peaks in the rest frame near infrared band. The galaxy rest-frame Near-infrared (NIR) band flux is dominated by the old stellar population, which traces the underlying stellar mass \citep{Cowie1996, Huang1997}. Thus the rest frame NIR selected galaxy sample is equivalent to a mass selected sample \citep{Huang2013} completed for both the blue and red population. Therefore, a deep IRAC survey would provide us a unique chance to understand the galaxy properties as the stellar population, galaxy color evolution and so on with highly completeness and much wider redshift range.

Galaxy populations are bimodally distributed in color, morphology, metallicity and so on which indicates a divergence galaxy evolution path. The mechanism that governs these observation results is still under debate. This depart phenomenon can be illustrated more clearly in color-magnitude diagram \citep{Baldry2004, Faber2007, Schawinski2014}, where the galaxies locate in blue cloud (BC) or red sequence (RS). The blue cloud galaxies are still star-forming while the red sequence galaxies are always full of quiescent galaxies.
	
Previous IRAC source population was mainly studied to the depth about 23 AB mag or less. The shallow survey as SWIRE \citep{Rowan2005} and Bootes fields \citep{Eisenhardt2004} have shown the color properties \citep{Rowan2005}, redshift distribution \citep{Brodwin2006, Rowan2008, Rowan2013}, galaxy population \citep{Davoodi2006} and the clustering of the IRAC galaxy \citep[][and ref. therein]{Zeimann2013} with wide area. The deep IRAC survey project as the EGS \citep{Barmby2008}, COSMOS \citep{Sanders2007}, HUDF and GOODS-South \citep{Labbe2005, Labbe2015, Damen2011}, SPLASH \citep{Capak2012} revealed us the rest frame K band luminosity function, stellar mass density evolution \citep{Huang2013}, the star formation rate estimation \citep{Barro2011a, Barro2011b}, star formation history \citep{Steinhardt2014}, high redshift red galaxies \citep{Papovich2006}, lyman break galaxies \citep{Huang2005, Rigopoulou2006}; optical/NIR dropout galaxies \citep{Yan2005, Huang2011, Caputi2012} and high redshift dusty galaxies \citep{Wang2016, 2019Natur.572..211W}. IRAC survey deeper than 24 AB mag projects, such as the Spitzer Extended Deep Survey (SEDS) are just beginning to be explored \citep{Ashby2013, Ashby2015}. In this paper, we aim to study the SEDS source population by comprising the galaxy spectral energy distribution (SED) with the local well-observed galaxies. 

The SEDS survey covered in total 1.46 square degree with the 3$\sigma$ depth about 26 AB mag in $3.6\mu \rm m$ and $4.5\mu \rm m$ bands. The SEDS fields include the Extended Groth Strip (EGS), The Cosmic Evolution Survey (COSMOS), the UKIDSS Ultra-Deep Survey (UDS), the Extended Chandra Deep Field South (ECDFS) and the Hubble Deep Field North (HDFN) region. All the five survey fields have also been covered by many previous optical/NIR survey project. Moreover, Cosmic Assembly Near-infrared Deep Extragalactic Legacy Survey \citep[CANDELS;][PIs: S. Faber, H. Ferguson]{Grogin2011, Koekemoer2011} project performed a NIR survey within the SEDS region and obtained F120W and F160W band {\it Hubble Space Telescope (HST)} WFC3 image up to 27 AB mag with high completeness and spatial resolution. Thus we can derived the SED of the SEDS source from all the available archive data, and investigate the IRAC detected source properties.

To understand the multi-wavelength properties, we also need galaxy templates for the SED fitting. Templates such as the BC03\citep{2003MNRAS.344.1000B} have already shown a great power in describing nearly all the galaxy SEDs \citep{2009ApJ...700..221K, 2008MNRAS.388.1595D}. However, BC03 only have the stellar spectra with no emission line from ionized gas, and lack of the dust components in MIR\citep{Brodwin2006, Huang2013, 2021ApJ...912..161H}, and thus do not suitable for fitting our SEDs from u to 8$\mu$m. \citet{Brown2014} provides 129 galaxy templates that either directly extracted from the observed spectrum or interpolated from Multi-wavelength Analysis of Galaxy Physical Properties code \citep[MAGPHYS,][]{Cunha2008} based on the multi-wavelength SED. The templates show a good coverage of the real galaxy in the color-color diagram like the R-I, B-R diagram and [5.8] - [8.0], [4.5] - [8.0] diagram, which would be more reliable in reflecting and represent the local galaxies SEDs. 

The observed 3.6$\mu$m bright galaxies would include galaxies at a large redshift range, however, high and low redshift galaxies share the same templates is still an assumption. Therefore, in this work, we fit the 3.6$\mu$m selected sample with local galaxy templates to study whether the high-z galaxy SED have similar templates. The structure of this paper is arranged as follows: Section 2 is data introduction and reduction; in Section 3, we show the main results and then briefly summary in Section 4.

\section{Observation and data reduction}
\subsection{Data for the SEDS region}

\citet{Ashby2013}has introduced the SEDS data mosaic and the source detection process. All the $3.6\mu \rm m$ and $4.5\mu \rm m$ mosaic image and coverage map can be obtained from \url{https://www.cfa.harvard.edu/SEDS/}.

\begin{figure}[ht!]
\centering
\includegraphics[width=0.9\textwidth]{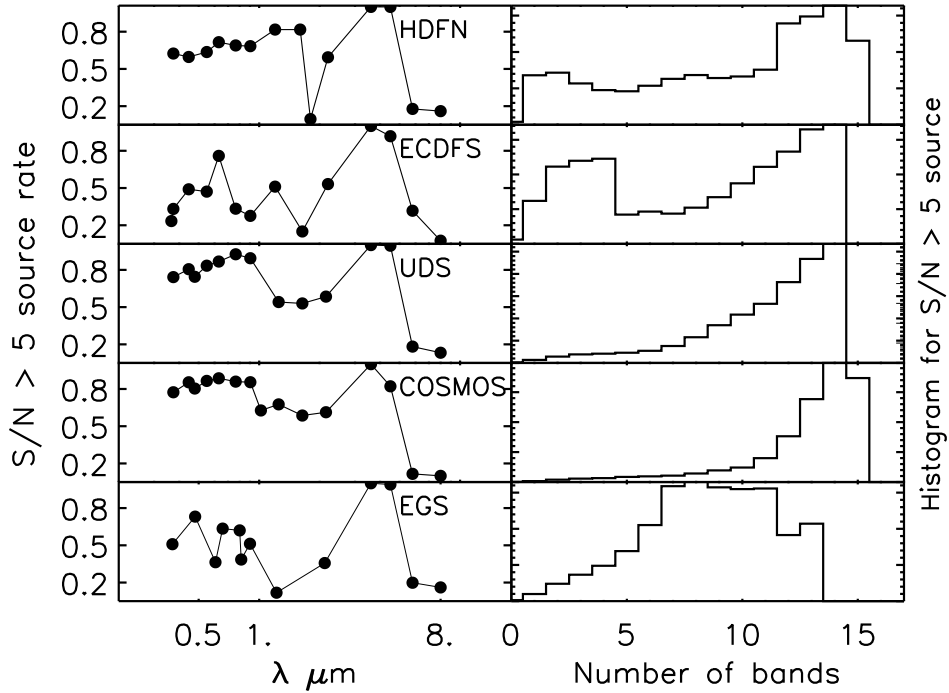}
\caption{Distribution of the broad band $S/N > 5$ source fraction and histogram of the $[3.6] < 24$ source for each SEDS region. The left panel is the $S/N > 5$ source fraction as the function of the band wavelength. The optical bands have been covered by the previous survey projects and 80\% of the SEDS source have high S/N photometry. For the NIR band, high S/N source are not as many as the optical band data which is caused by the NIR band image area and depth. The right panel shows the histogram of high S/N band number for the SEDS source. Most of the sources are covered by more than 10 bands. We only add up the broad band photometry results.
}\label{detection_rate}
\end{figure}

We collect the available optical, NIR and the $5.8\mu \rm m$ and $8.0\mu \rm m$ IRAC data to proceed the multi-wavelength photometry. Figure \ref{detection_rate} shows the detection rate distribution along the broad band wavelength and the histogram of high S/N source detection band. For the left panel, higher fraction means more complete sample for the multi-band catalog and we can find the shortage always comes in the $5.8\mu \rm m$ and $8.0\mu \rm m$ band, which is limited by the noise and instruments. The weakness also comes from the NIR band, implying the further requirement of NIR observations. The medium and narrow band photometry in the COSMOS and ECDFS fields would help a lot in populating the galaxy properties  \citep{Cardamone2010}. The medium or narrow band data is not take into account in Figure \ref{detection_rate}. We briefly introduce the data we used for the five SEDS regions as follows.

\textbf{SEDS-EGS:} The image data of SEDS-EGS region we obtained including the u, g, i, z band by the MMT and R band image by the Subaru telescope \citep{Zhao2009} which cover the whole EGS field. The image data by the MMT and the Subaru telescope have been mosaiced into four separate mosaic images, we perform the photometry on each image and then median the flux for each panel. The HST image is available in HST/ACS F606W, F814W bands\footnote{\url{http://www.stsci.edu/~koekemoe/egs/}}, covered about half area of SEDS-EGS. For the near infrared band, we obtained the Ks band image by Subaru MOIRCS(PIs: Fukugita, Yamada) and the J, H, Ks band image by the WIRCAM Deep Survey (WIRDS) \footnote{\url{http://terapix.iap.fr/rubrique.php?id_rubrique=261}}\citep{Bielby2012}. Additionally, we use the Subaru telescope Y band data covering the whole EGS map in four pointings (PIs: Newman, Ashby, 2010). Exposure time for the four pointings are 3 minutes, 3 minutes, 3 hours and 9 minutes. The 3 $\sigma$ limit magnitudes for 1'' aperture are about 24., 23.9, 26.3, 24.7, individually. This Y band data is deducted following the Suprime-Cam pipeline {\it sdfred2}. The raw data of Subaru Y band image can be downloaded from the SMOKA database\footnote{\url{http://smoka.nao.ac.jp}}.

In the MIR bands, the $5.8\mu \rm m$, $8.0\mu \rm m$ band are fully covered by IRAC EGS survey \citep{Barmby2008}.

Archive catalogs we can make use of in this field include the u, B, g, r, i, z band from CFHTLS, J, Ks or H band from CAHA, Palomar and HST catalog. Figure \ref{detection_rate} show the detected source number histogram. $83\%$ sources have been detected at least 5 bands and $50\%$ sources have been detected at least 9 bands. More detail introduction about the EGS image data can be found in \citet{Barro2011a, Barro2011b, Huang2013}.

\textbf{SEDS-COSMOS:}
The SEDS-COSMOS region lies within the 0.3 degree$^{2}$ of the whole COSMOS field. As a well observed region \citep{Capak2007, Ilbert2009, Muzzin2013, Laigle2016}, COSMOS have been covered by many telescopes in almost all wavelength. We use the u band image by CFHT \citep{McCracken2010}, B, g, V, r, i, z, as well as 12 optical medium bands (IA427, IA464, IA484, IA505, IA527, IA574, IA624, IA679, IA709, IA738, IA767, and IA827) and two narrow bands (NB711, NB816) image by Subaru telescope \citep{Taniguchi2007, Taniguchi2015}. For the NIR band, we use the image from UltraVista observation in Y, J, H, Ks bands \citep{McCracken2012}. For the MIR band, we use the $5.8\mu \rm m$, $8.0\mu \rm m$ data from the S-COSMOS project  \citep{Sanders2007}. All the COSMOS image data can be 
download at {\url{http://irsa.ipac.caltech.edu/data/COSMOS/images/}} or SEDS website. 

Detection band histogram of the SEDS-COSMOS field can be found in Figure \ref{detection_rate}. The $4.5\mu \rm m$ band source rate is about 85\% which is caused by the $4.5\mu \rm m$ band image overlapping the $3.6\mu \rm m$ image about 85\%.

\textbf{SEDS-UDS:}
The SEDS-UDS region locates in center part of the UDS field \citep{Lawrence2007, Cirasuolo2007} with the area about 0.3 degree$^2$. We use the u band image from the CFHT Large Area U-band Deep Survey \citep[CLAUDS][]{2019MNRAS.489.5202S} project, the B, V, R, i, z bands from the Subaru telescope SXDS project\footnote{\url{http://soaps.nao.ac.jp/SXDS/Public/DR1/index_dr1.html}}  \citep{Furusawa2008} and the J, H, Ks bands from UKIDSS \footnote{\url{http://www.nottingham.ac.uk/astronomy/UDS/data/dr3.html}}\citep{Lawrence2007}. The $5.8\mu \rm m$, $8.0\mu \rm m$ band IRAC image data can be found in the Spitzer Public Legacy Survey of the UKIDSS Ultra Deep Survey (SpUDS)\footnote{\url{http://irsa.ipac.caltech.edu/data/SPITZER/SpUDS/images/irac/}}. 

\textbf{SEDS-ECDFS:}
In the SEDS-ECDFS fields region, we use the optical data by ESO U38, U, B, V, R, I band image and 14 Subaru medium bands (IA427, IA445, IA464, IA484, IA505, IA505, IA527, IA550, IA574, IA598, IA624, IA651, IA679, IA709, IA738, IA767, IA797, IA827, IA856) image by MUSYC project\footnote{\url{http://www.astro.yale.edu/MUSYC/}} \citep{Cardamone2010}. The NIR band image data we used are the J, Ks band by CTIO and H band by ESO NTT. We also use the J, Ks band image data by the Taiwan ECDFS Near-Infrared Survey \footnote{\url{http://www.asiaa.sinica.edu.tw/~bchsieh/TENIS/About.html}} \citep[TENIS;][]{Hsieh2012}. The ECDFS field have been covered by the {\it Spitzer} IRAC/MUSYC Public Legacy Survey in the Extended Chandra Deep Field South \cite[SIMPLE;][]{Damen2011} in $5.8\mu \rm m$ $8.0\mu \rm m$ band \footnote{\url{http://irsa.ipac.caltech.edu/data/SPITZER/SIMPLE/}}.
	
The ECDFS detection band histogram in Figure \ref{detection_rate} shows a peak when detection band $ = 4$, which is caused by the optical and NIR region only covers about 70\% of the SEDS-ECDFS field and thus there are 30\% source have only been covered by the IRAC.

\textbf{SEDS-HDFN:}
The optical bands of SEDS-HDFN have been observed by KPNO in the U band, and by Subaru telescope Superime-Cam in the B, V, R, i, z band. The NIR band image data we used are the HK' band by UH 2.2m telescope \footnote{\url{http://www.astro.caltech.edu/~capak/hdf/}}
\citep{Capak2004} and the Ks band by CFHT WIRCam \citep{Wang2010}. We also take the public catalog by \citet{Yang2014} into account. The $5.8\mu \rm m$ $8.0\mu \rm m$ band image can be found in the Spitzer Heritage Archive \footnote{\url{http://sha.ipac.caltech.edu/applications/Spitzer/SHA/}}.

Optical and NIR band image of SEDS-HDFN field covered about 90\%. Therefore, similar as SEDS-ECDFS regions, the histogram of detected band number also shows a little peak for the low band number end in Figure \ref{detection_rate}.

\subsection{PSF-matched photometry}

\citet{Ashby2013} provides us a well defined $3.6\mu \rm m$ and $4.5 \mu \rm m$ selected catalog individually. As the $3.6\mu \rm m$ and $4.5 \mu \rm m$ band source are too crowd relative to the image spatial resolution, The source detection method is fitting every source beyond 3 $\sigma$ of the image noise by the image PSF, and then iterate the detection with the bright-source-subtracted image. Finally, all the three detection catalogs are combined as the final coordinate catalog. To reduce the contamination of the nearby source in the crowd field, for each IRAC target, all the nearby sources are subtracted with the flux measured by PSF fitting, leaving the only source that we would aperture photometried. The background within the aperture is estimated by the mod pixel value of the annulus. We match the $4.5 \mu \rm m$ band catalog with the $3.6\mu \rm m$ band catalog position. The 2.4'' diameter aperture photometry results of  $3.6\mu \rm m$ and $4.5\mu \rm m$ band source are corrected by the PSF growth curve to approach the total flux. Details of the SEDS catalog can be found in \citep{Ashby2013}.

We perform the PSF-matched photometry with the target coordinates from SEDS catalog to the multi-wavelength images. We build the image PSF by stacking the point source image. Then we create and convolve the PSF kernel from each optical and NIR band image to the $3.6\mu \rm m$ band PSF by assuming a gaussian shaped PSF in each band. Relation between the $\sigma$ of the kernel and PSFs is, $\sigma_{\rm opt}^{2} + \sigma_{\rm kernel}^{2} = \sigma_{3.6\mu \rm m}^{2},$ where $\sigma_{\rm kernel}$ is the kernel we used to convolve the optical and NIR image to the same resolution as $3.6\mu \rm m$ band image. We perform the 2.4'' diameter aperture photometry on the convolved stamp image, and correct the aperture flux to total flux from the PSF growth curve. Then we calibrate the aperture flux with available catalog, which will give us the photometry results following the same manner as the $3.6\mu \rm m$ band.

The PSF width of the $5.8\mu \rm m$, $8.0\mu \rm m$ band image is similar to the $3.6\mu \rm m$ and $4.5\mu \rm m$ image, so we simply perform the aperture photometry without convolution and also corrected by PSF growth curve individually to obtain the total flux.

Uncertainty of photometry can be derived from the standard aperture photometry formula \footnote{\url{https://wise2.ipac.caltech.edu/staff/fmasci/ApPhotUncert.pdf}} which contains the CCD gains, pixel exposure time, etc. However, as most of the image data we used are the mosaic image and the information as CCD gains, background counts and coverage map for each pixel and so on are not always provided, and hence unable to derive the Poisson error follow the photometry formula. To estimate our photometry accuracy, we randomly insert 100 fake sources into the image with a known magnitude and then perform the aperture photometry exactly as the way we have done on the real source. The histogram of the (input magnitude - measured magnitude) shows the accuracy and bias of our photometry. Then we interpolate the photometry results with the simulated magnitude v.s. magnitude error relation.

\begin{figure}[ht!]
\centering
\includegraphics[width=0.95\textwidth]{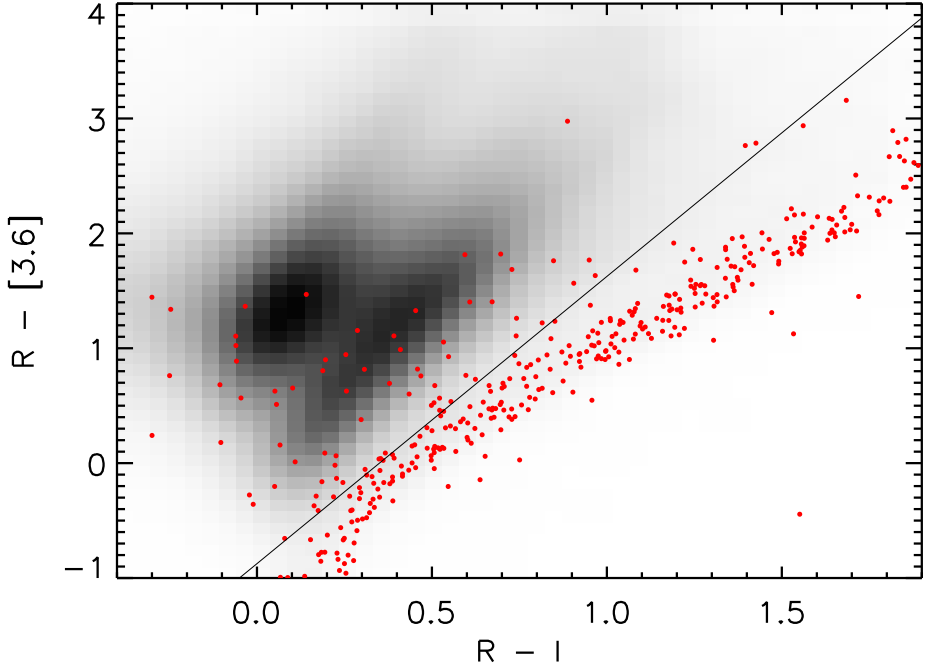}
\caption{R - I vs. R - [3.6] color-color diagram for the source $[3.6] < 24.5$. The black shadows are the SEDS source and the red dots are the stars identified by CANDELS catalog. Although the R and I band filters in each SEDS fields are slightly different, we still can find a clear stellar branch which contains mainly metal poor stars and the two distinct clouds of galaxies which contain the low and high redshift galaxies. Typical error bar is smaller than the scatter.}\label{star_galaxy_separation}
\end{figure}

\subsection{Star/galaxy separation}

There are many color color plot can be used for the Star/Galaxy separation\citep{Huang1997, Daddi2004, Barmby2008, Huang2013}. For the SEDS catalog, the optical band image is deeper and provide more coverage than the NIR band, and thus we plot R - I V.S. R - [3.6] color diagram in Figure \ref{star_galaxy_separation} with all the five SEDS region catalog to perform the star galaxy separation. Despite the difference for each optical image filter and depth, we can still see the clear separations as two islands and one branch. The two islands mainly consist low and high redshift galaxies separately \citep{Huang2013} while the branch are formed by the halo K or M star with low metallicity \citep{Guo2013}. We also match the SEDS catalog with the CANDELS catalog and plot the CANDELS STAR\_FLAG $>$ 0.98 with the red color dots in Figure \ref{star_galaxy_separation}. In this paper, we classify a source as galaxy with $R-[3.6] > ( R - I - 0.35)/0.4$, which is shown in solid lines in Figure \ref{star_galaxy_separation}.

\subsection{$3.6\mu \rm m$ selected catalog completeness: compare with HST CANDELS sample}

\begin{figure}[ht!]
\centering
\includegraphics[width=0.52\textwidth]{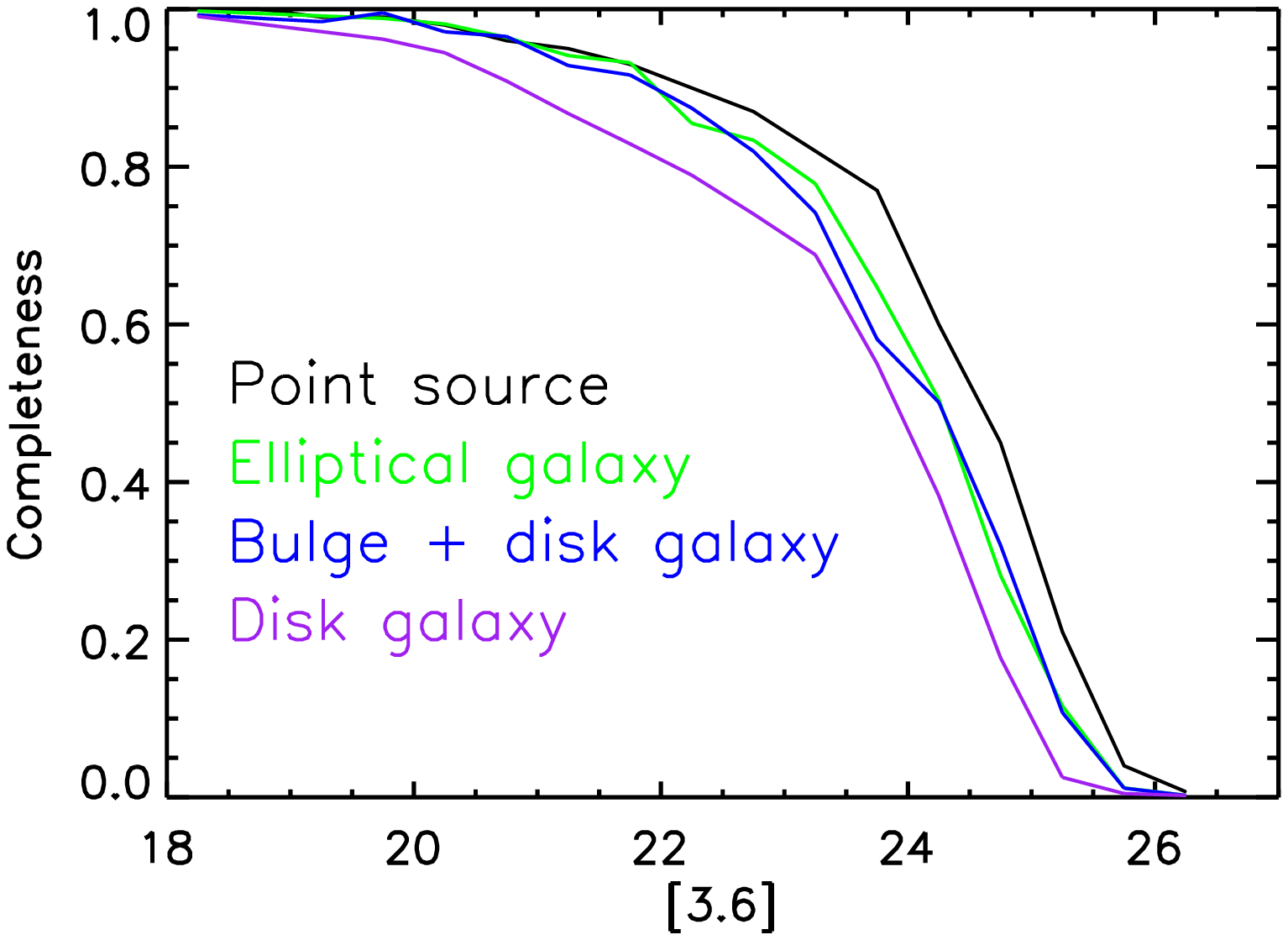}
\includegraphics[width=0.47\textwidth]{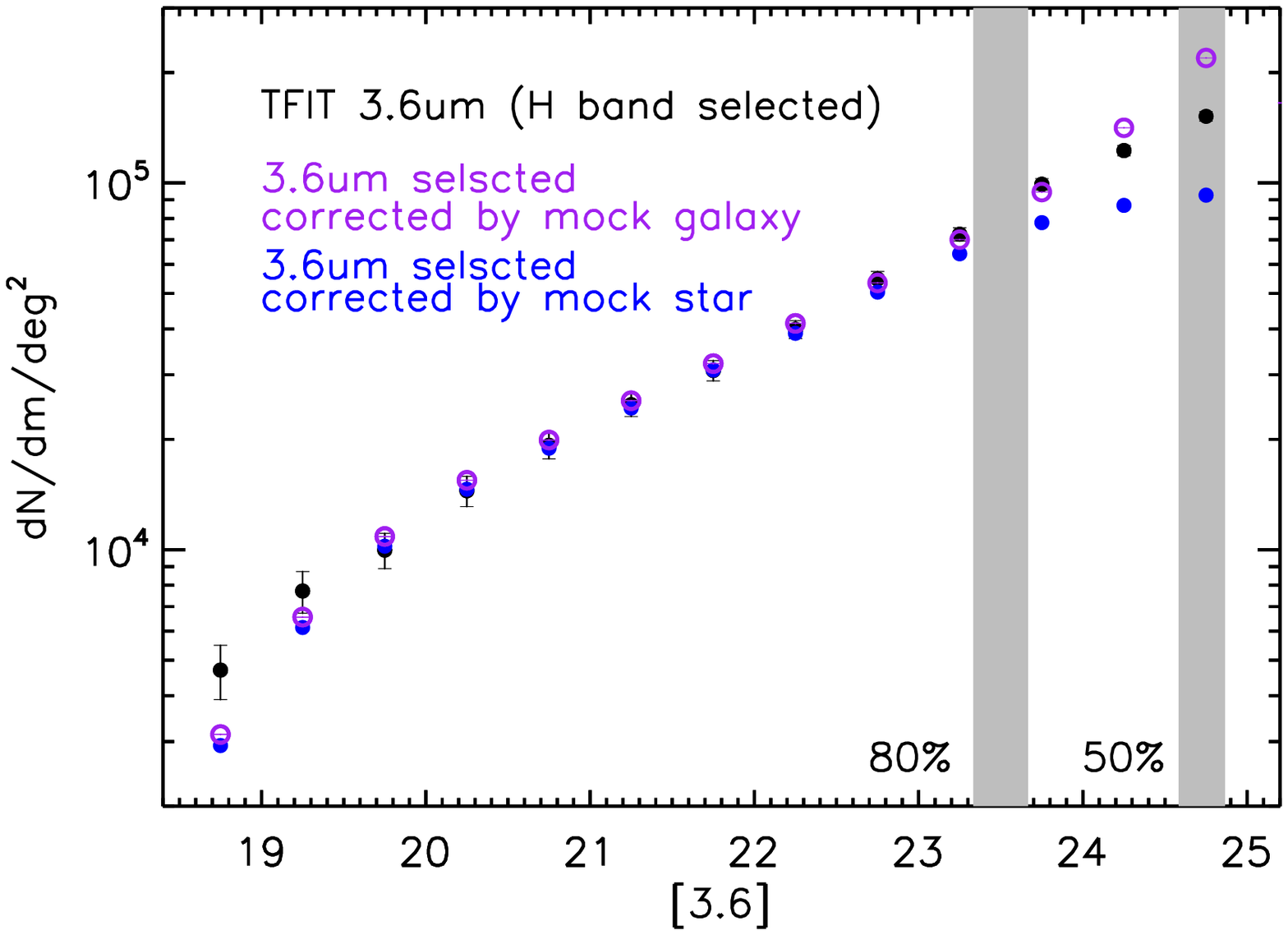}
\caption{{\bf Left panel:}Completeness curve of our source detection method for the point source and extend source. We simulate the source detection with four kinds of the source with given magnitude: point source, elliptical galaxy, disk galaxy and disk galaxy with a classical bulge. 
{\bf right panel:}Number counts density of the SEDS catalog and the CANDELS TFIT catalog. The blue and black dots are the galaxy number counts density with [3.6] for the SEDS and CANDELS TFIT catalogs. The purple open circles are the SEDS number counts density corrected by the disk+classical bulge completeness curve. The grey shadows are 80\% and 50\% completeness magnitude and we can see the number counts are consistency between the SEDS and CANDELS TFIT catalog up to 25 AB mag. 
}\label{completeness}
\end{figure}

Each SEDS region has been covered by CANDELS project in the center 200 arcmin$^2$ in F120W, F160W band \citep{2013ApJS..206...10G, 2013ApJS..207...24G, 2017ApJS..228....7N, 2017ApJS..229...32S, 2019ApJS..243...22B}. Facilitating for the {\it HST} spatial resolution, CANDELS provide a TFIT catalog \citep{2007PASP..119.1325L} by fitting the low resolution band image based on the prior HST source catalog to derive the source flux of multi-wavelength. This method would trace an accuracy photometry for the low resolution band as well as hold the spatial information.	
We compare the number density for the CANDELS five region and SEDS in Figure \ref{completeness} right panel with solid points. Assuming the CANDELS catalog is complete down to $[3.6] = 24$ AB mag, We would find that after recovering the number counts by completeness curve, up to 80\% completeness, the number counts density offset between SEDS and CANDELS is less than 10\%. We also plot the 50\% complete magnitude of $3.6\mu \rm m$ and not surprisingly, the number density division turns larger for the lower completeness. 	

Nevertheless, we should also notice that the number counts given by \citet{Ashby2013} is recovered from the point source simulation. If the completeness is simulated by an extended source, the detection rate would be lower than the point source \citep{Barmby2008}. Hence we re-run the completeness simulation with not only the point source but also the extend source as shown by the Figure \ref{completeness}. We perform the standard ``artificial object'' method with three kinds of galaxy: elliptical galaxy with ``de Vaucouleurs'' profile, compose spiral galaxy with half ``de Vaucouleurs'' profile and half exponential disk (1:1 flux ratio), spiral galaxy with pure exponential disk profile \citep{1997PhDT........19H}. All those mocked artificial galaxies are set as axial ratios = 1, effective radii $r_e = 1$ pixel and being convolved with image PSF. The artificial sources are normalized to the AB mag range from 18 to 26 and inserted in the mosaic image, then detected and photometried with the same manner as the real source. In order not to affect the noise of the mosaic image, we input less then 500 artificial source for each run. We consider the object as recovered when the position is within a radius of 1'' and the source measured flux is within 80\% to 120\% of the source input flux. The source detection method we employed here is PSF fitting, hence the artificial point source can be recognised easier, so we can see the source recover rate is lower for the extent source than the point source. For those artificial galaxies, the surface brightness profile of the elliptical galaxy is sharper than the exponential disk profile in the center region, therefore the disk galaxy recover rate is the lowest while the elliptical galaxy is slightly larger than that of the compose spiral galaxy from 22 to 24 AB mag. Although these detection simulations cannot cover all galaxy morphology, we can still expect the real galaxy detection rate would below that of the point source.

Considering the completeness curve of galaxies, we re-plot the number counts density of [3.6]. As the open circles in the right panel of Figure \ref{completeness}, which stand for the source number counts corrected by the extent source completeness curve, we can see the galaxy number counts density is more consistency with that of HST image up to 24 AB mag.

\section{Similarility of the local and SEDS galaxy sample: SED fitting}

\begin{figure*}[ht!]
\centering
\includegraphics[width=0.45\textwidth]{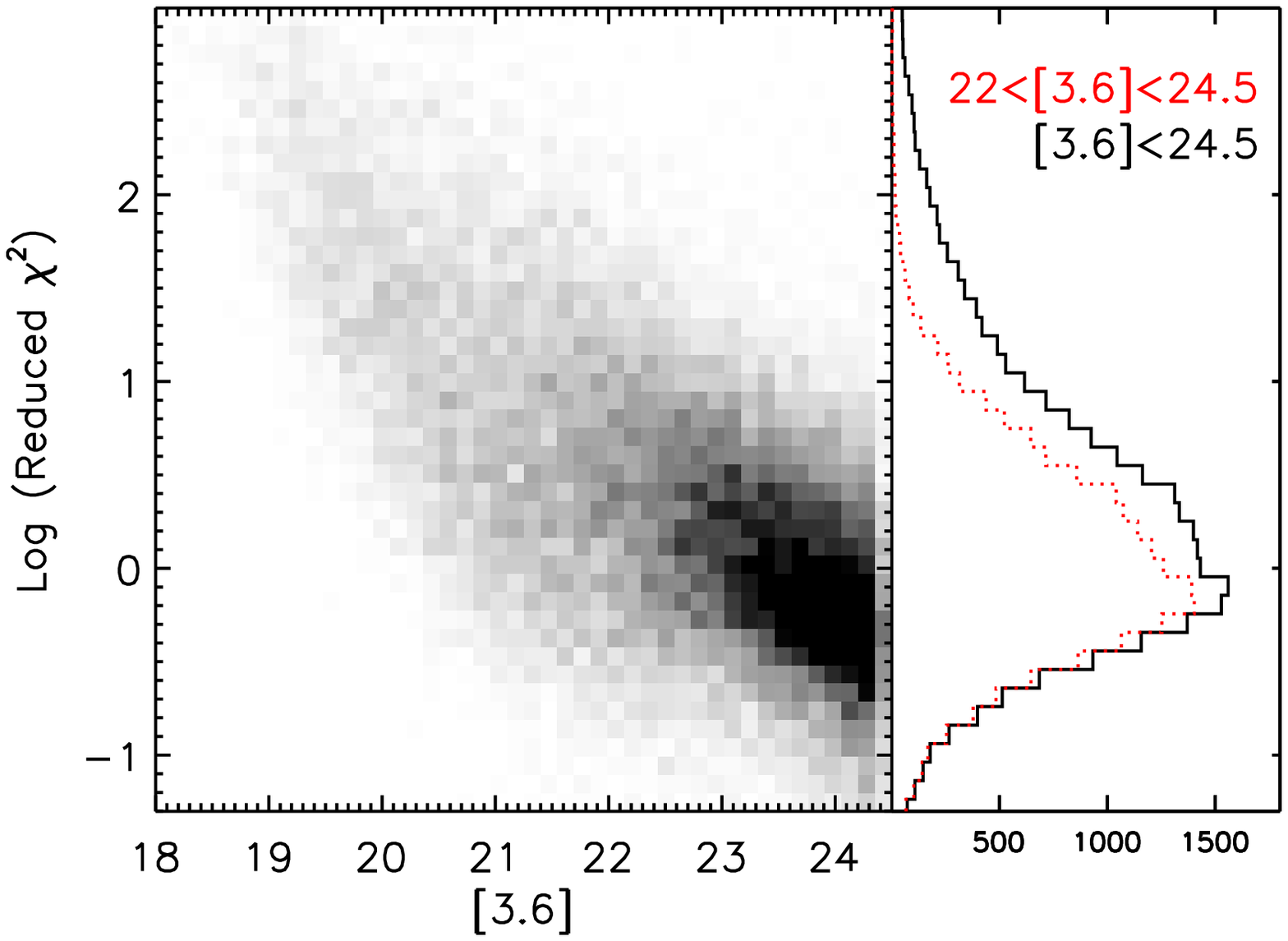}
\includegraphics[width=0.45\textwidth]{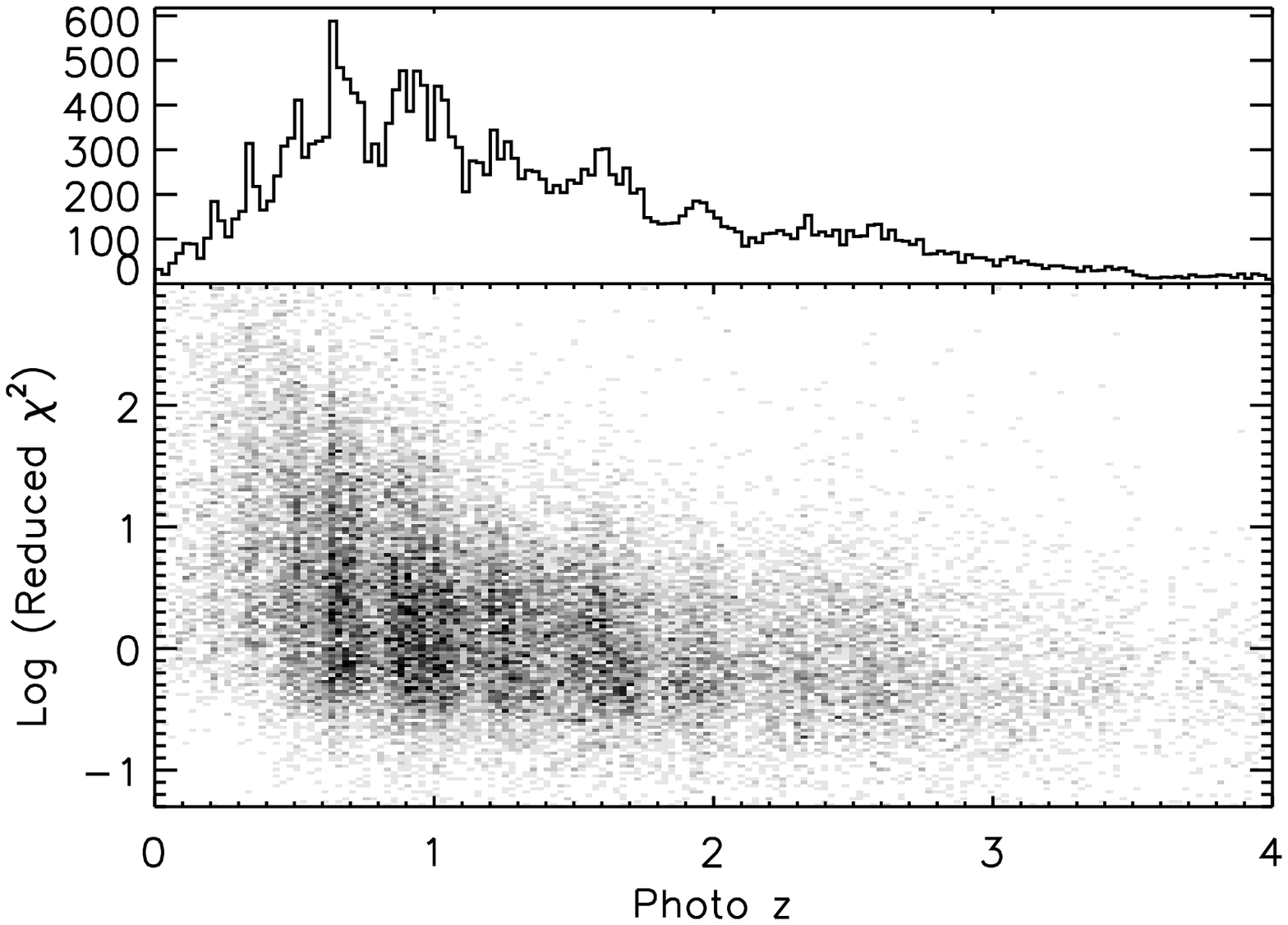}
\caption{{\bf left panel:}Reduced $\chi^2$ as function of the photometric redshift and [3.6]. The left panel shows the histogram of the $\chi^2$. We can see the $\chi^2$ is approaching 1 for the source fainter than [3.6] $ = 22$ (red dot line in the histogram). {\bf right panel:} The $\chi^2$ as a function of the photometric redshift. Most of the SEDS source locates at about redshift 1. Some of the $z < 0.5$ source are not fitting well by the model because of the small flux error and for the $z > 0.5$ source, $\chi^2$ is about 1 with large scatter. }\label{Hist_Chi2}
\end{figure*}

\subsection{SED fitting with local galaxy templates}
The CANDELS catalog provide the photo-z based on the TFIT photometry result by the template fitting method \citep{Dahlen2013, Grazian2015, Song2016}. We match the SEDS galaxy with the CANDELS target to get the phot-z of our SEDS galaxy, and fit the CANDELS-matched SEDS source by the local galaxy templates assuming the CANDELS photo-z can stand for the SEDS target. Since the SEDS catalog, CANDELS phot-z and the local galaxy templates are obtained independently, the fitting results would be an consistent check of the CANDELS phot-z accuracy and the local galaxy templates. We fit the SED from the u band to Ks band. Each SEDS galaxy can be found a galaxy template that gives minimal $\chi^2$:
\begin{equation}
	\chi^2 = \frac{1}{\rm N_{\rm DOF}}\sum_{\rm X}\frac{(\rm mag_{\rm _X} - \rm mag^{\rm t}_{\rm _X} - \rm  NF)^2}{\rm magerr_{\rm _X}^2},
\end{equation}
where the $\rm mag_{_X}$, $\rm magerr_{_X}$ are the observed mag and error of the band X and the $\rm mag^{t}_{_X}$ is the galaxy template mag measured from the template spectrum. The $\rm N_{\rm DOF}$ is the degree of freedom (number of fitting bands). The normalize factor, NF, that minimal the $\chi^2$ is
\begin{equation}
	\rm NF = {\sum_{\rm X}\left( \frac{\rm mag_{_X} - mag^{t}_{_X}}{\rm magerr_{_X}^2} \right) }/{ \sum_{\rm X}\left( \frac{1}{\rm magerr_{\rm _X}} \right)^{2} }.
\end{equation}

Figure \ref{Hist_Chi2} shows the reduced $\chi^2$ of the template fitting results. The left panel of the Figure \ref{Hist_Chi2} shows the $\chi^2$-[3.6] relation. We also show the $\chi^2$ histogram for the targets with $22<[3.6]<24.5$ in red dot line in Fig. 4 left panel. Most of the targets with $\chi^2>10$ are the [3.6] brighter than 22 AB mag targets. 

Right panel of Figure \ref{Hist_Chi2} shows the $\chi^2$ distribution along the photometric redshift. Several filaments of the redshift shows the cosmology structure. The low redshift sources are also not fitted well with large $\chi^2$. Most of the SEDS source locates at redshift about 1 and for the galaxy $z > 0.5$, the $\chi^2$ is in the order of 1. Therefore, the local galaxy templates can represent most of the SEDS source. Moreover, the local galaxy templates can represent the SEDS source properties for the redshift larger than 0.5. The large $\chi^2$ value for the 3.6$\mu$m bright source indicates that the 129 local galaxy templates are not enough to describe the SEDs. Based on the $\chi^2$ distribution in Fig. \ref{Hist_Chi2}, we set $\chi^2 < 10$ as `good fit' in this work, which means our SED can be represented by one local galaxy SED.  

The criterion of $\chi^2 < 10$ is based on the histogram in the left panel of Fig. \ref{Hist_Chi2}. A similar SED fitting work by \citet{2021ApJ...912..161H} fitted the 16$\rm \mu$m selected galaxy SED by the templates given by Brown+ 2014, and treat $\chi^2 < 10$ targets as outlier. we follow the same criterion as \citet{2021ApJ...912..161H}.

However, the similarity of the SED between our targets and the local galaxies may not imply the similarity between the galaxy other properties, such as morphology, stellar mass etc \citep{2021ApJ...912..161H}. The connection between the SED similarity and other galaxy properties will be investigated in an upcoming work.

\begin{figure}[ht!]
\centering
\includegraphics[width=0.45\textwidth,angle=0.]{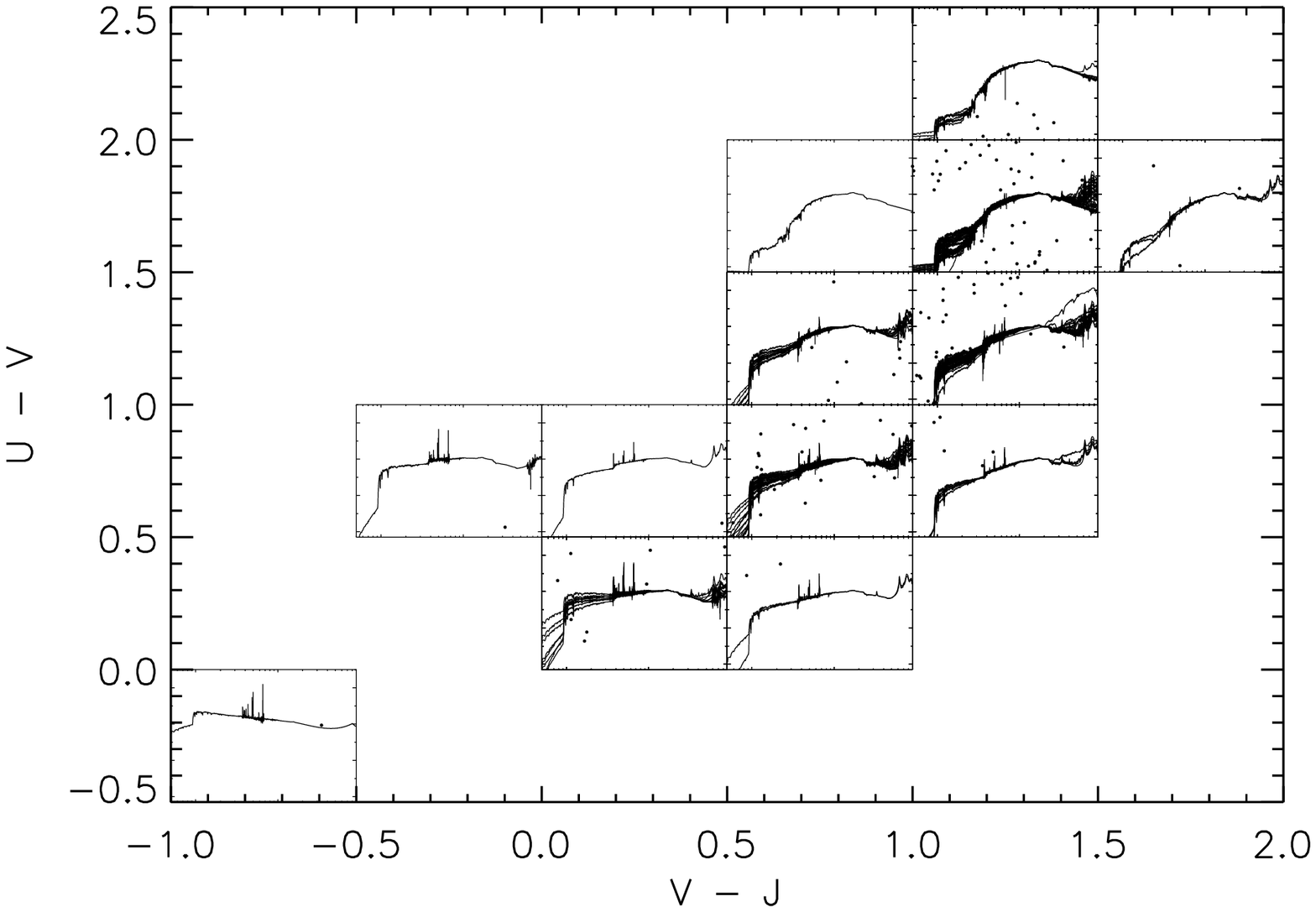}
\includegraphics[width=0.48\textwidth]{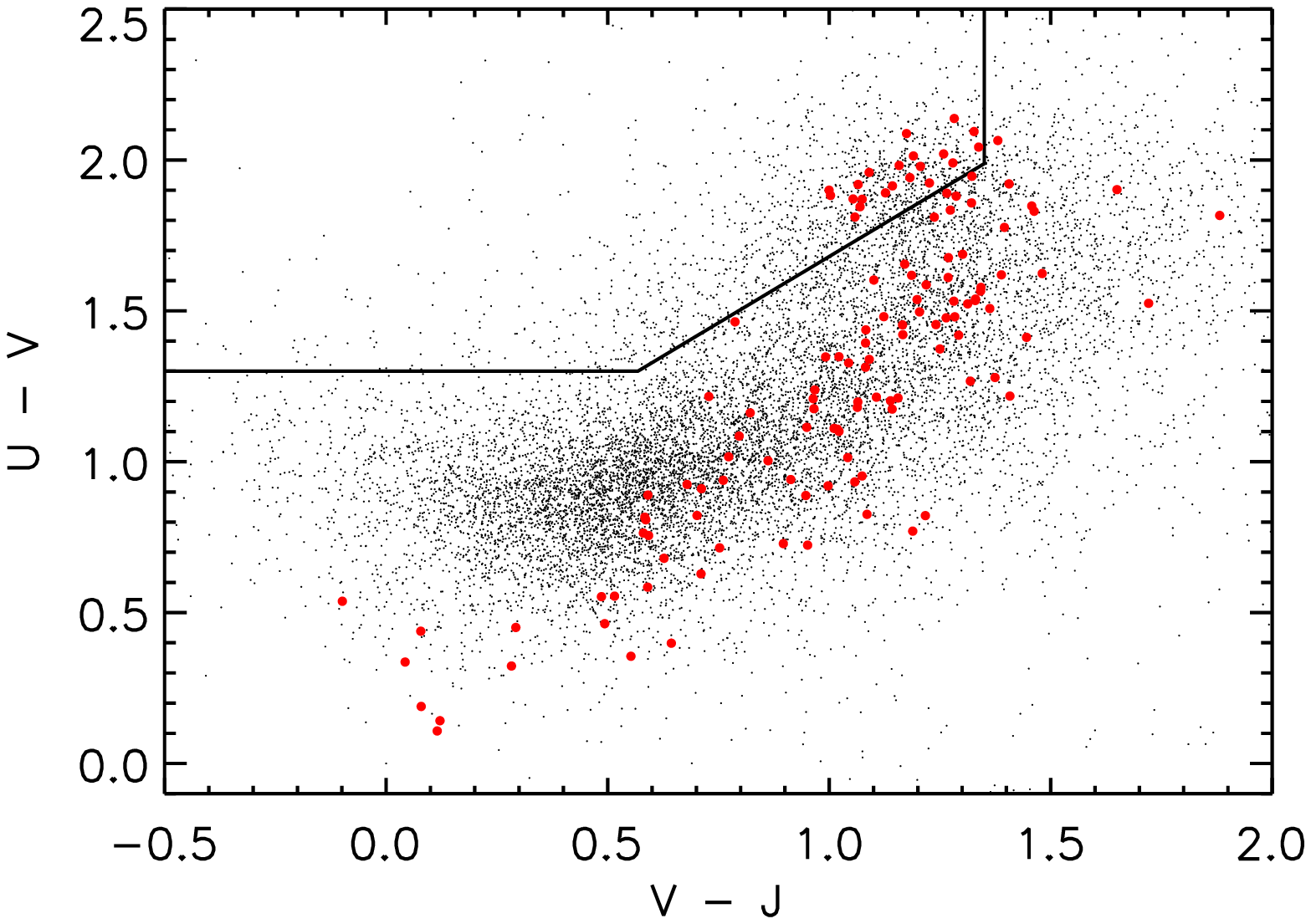}
\caption{{\bf left panel:}Coverage of the galaxy templates on the UVJ color-color diagram for the CANDELS source. The black dots show the galaxy template color while the SEDs in each color grid is the galaxy template \citep{Brown2014}. We normalize the galaxy templates in 2$\mu \rm m$. The template SEDs have been corrected the extinction of the milky way. The lower left conner lies a dwarf irregular galaxy UGCA 166. {\bf right panel:} UVJ diagram of SEDS source with matched CANDELS phot-z. Typical error bar is smaller than the scatter here. The K-correction are proceed based on the \citet{Brown2014} galaxy templates. The emission lines would affect the broad band color and yield a scatter in the blue end \citep{Speagle2015, 2019A&A...631A.123Y}. Emission lines in blue galaxies are very common and we can see a bump around the [OIII] doublet in Figure \ref{SED_fitting}.
}\label{galaxy_template}
\end{figure}

\subsection{SED fitting results in the UVJ diagram}
UVJ plot can effectively separate the red, blue and dusty galaxy\citep{Williams2009, Patel2012} because the dust extinction and old stellar population are evolving in orthogonal direction. We show the UVJ plot of the templates in Fig. \ref{galaxy_template} left panel and divide the color-color diagram into grids to show the galaxy population by the location of galaxy in this diagram. Template colors are obtained by performing the photometry with Bessell-U, Bessell-V, Palomar-J band filter on template SEDs. 

We derive the absolute mag by the nearest $S/N > 5$ band flux and the K-correction deduced from the best fitting templates and filters.  Formulas for the K-correction and absolute mag are given by \citet{Hogg2002}. The results is shown in Figure \ref{galaxy_template} right panel. The red dots in Figure \ref{galaxy_template} are the U-V, V-J colors measured from the templates.

\begin{figure*}[ht!]
\centering
\includegraphics[width=0.8\textwidth]{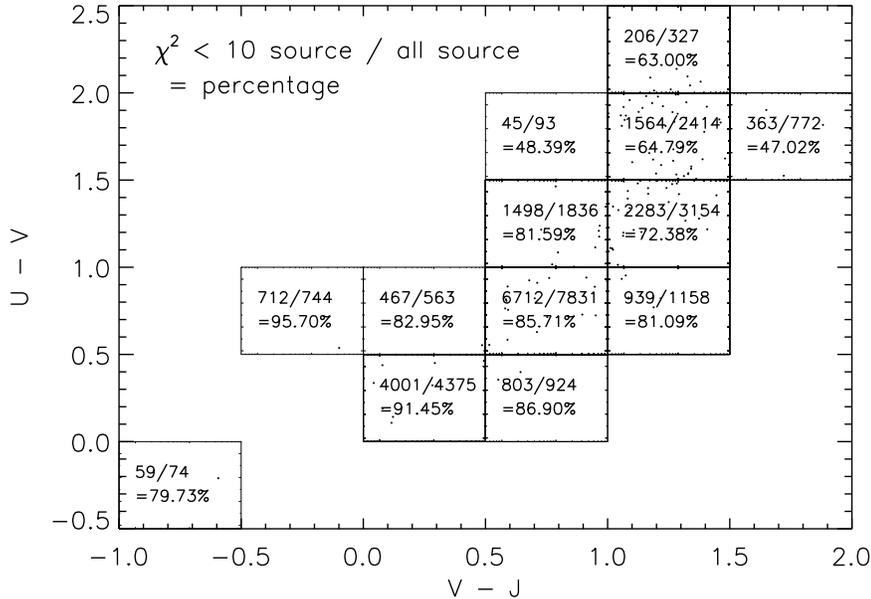}
\caption{Template fitting results for each color grid template. We fit every source from u to Ks band with every template and choose the lowest $\chi^2$. The ratio between the $\chi^2 < 10$ source number and the total source number is denoted in each grid. Nearly 80\% source can be fitted by the galaxy templates with $\chi^2 < 10$. Example of the fitting results can be found in Figure \ref{SED_fitting}.}\label{template_fitting_spec}
\end{figure*}

The fitting $\chi^2$ results in each UVJ grid is shown in Figure \ref{template_fitting_spec}. The percentage in each color grid shows the fraction of the $\chi^2 < 10$ source fitting by the templates in the grid. There are in total about 80\% SEDS source can be fitted by the local galaxy templates ($\chi^2 < 10$). This percentage shows the consistency between the local galaxy SED and SEDS multi-wavelength photometry. Nearly 90\% blue galaxies and 60\% red galaxies can be represented by the local galaxies template, which indicates no difference for the rest frame UV and optical band. The IRAC band data also shows consistency with the rest frame NIR band.

\subsection{Examples of the SED fitting}

\begin{figure*}[ht]
\centering
\includegraphics[width=0.45\textwidth]{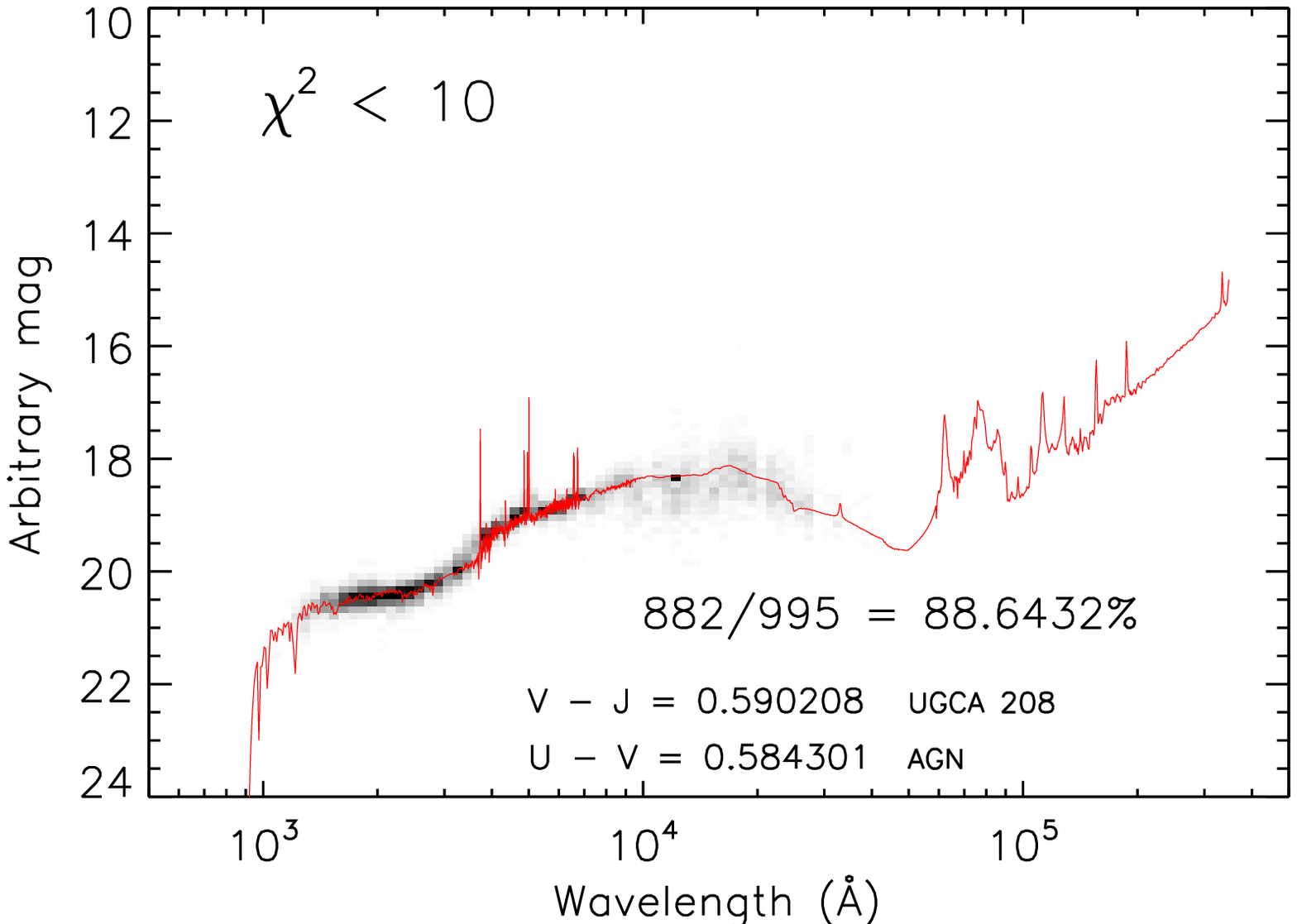}
\includegraphics[width=0.45\textwidth]{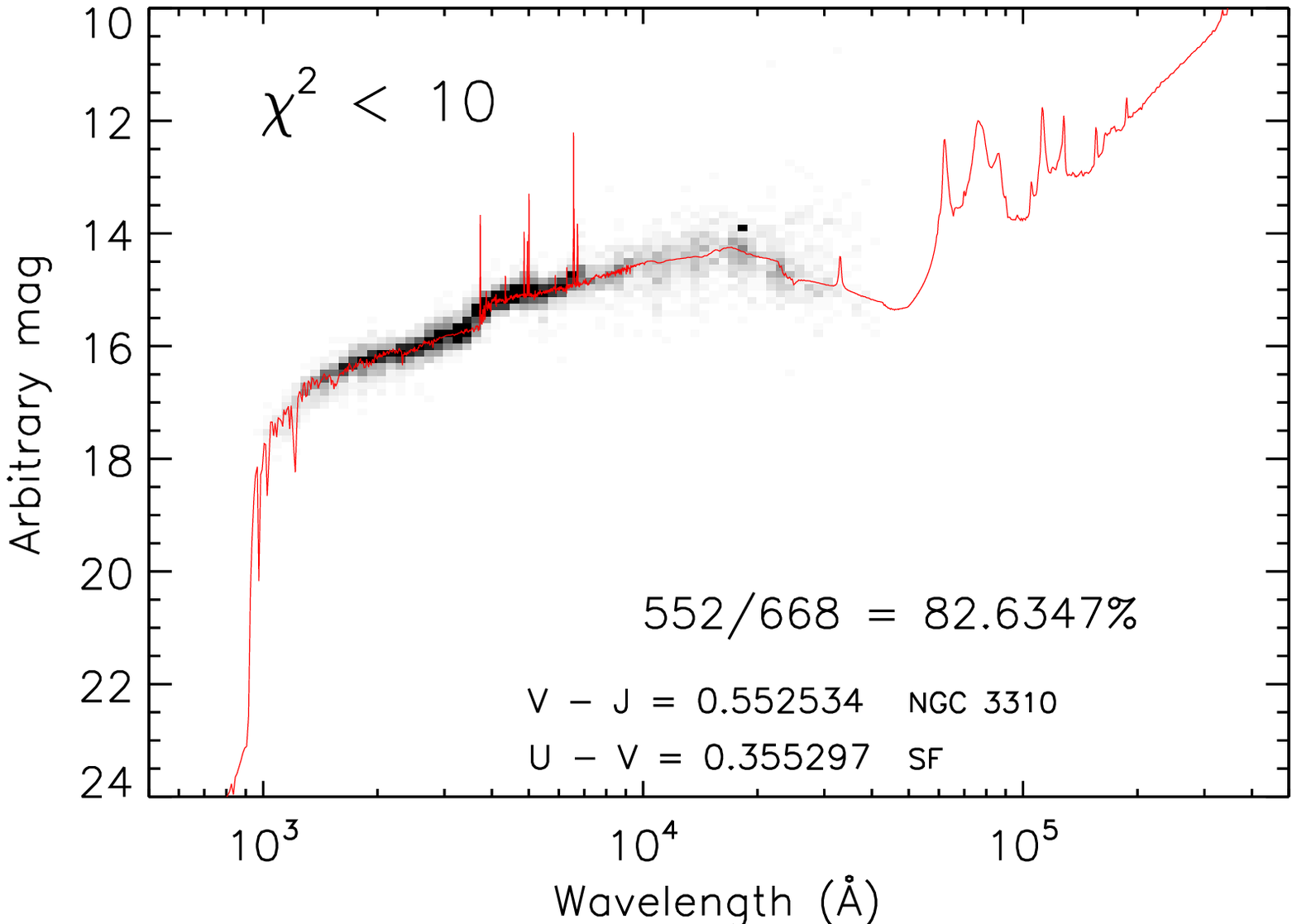}
\includegraphics[width=0.45\textwidth]{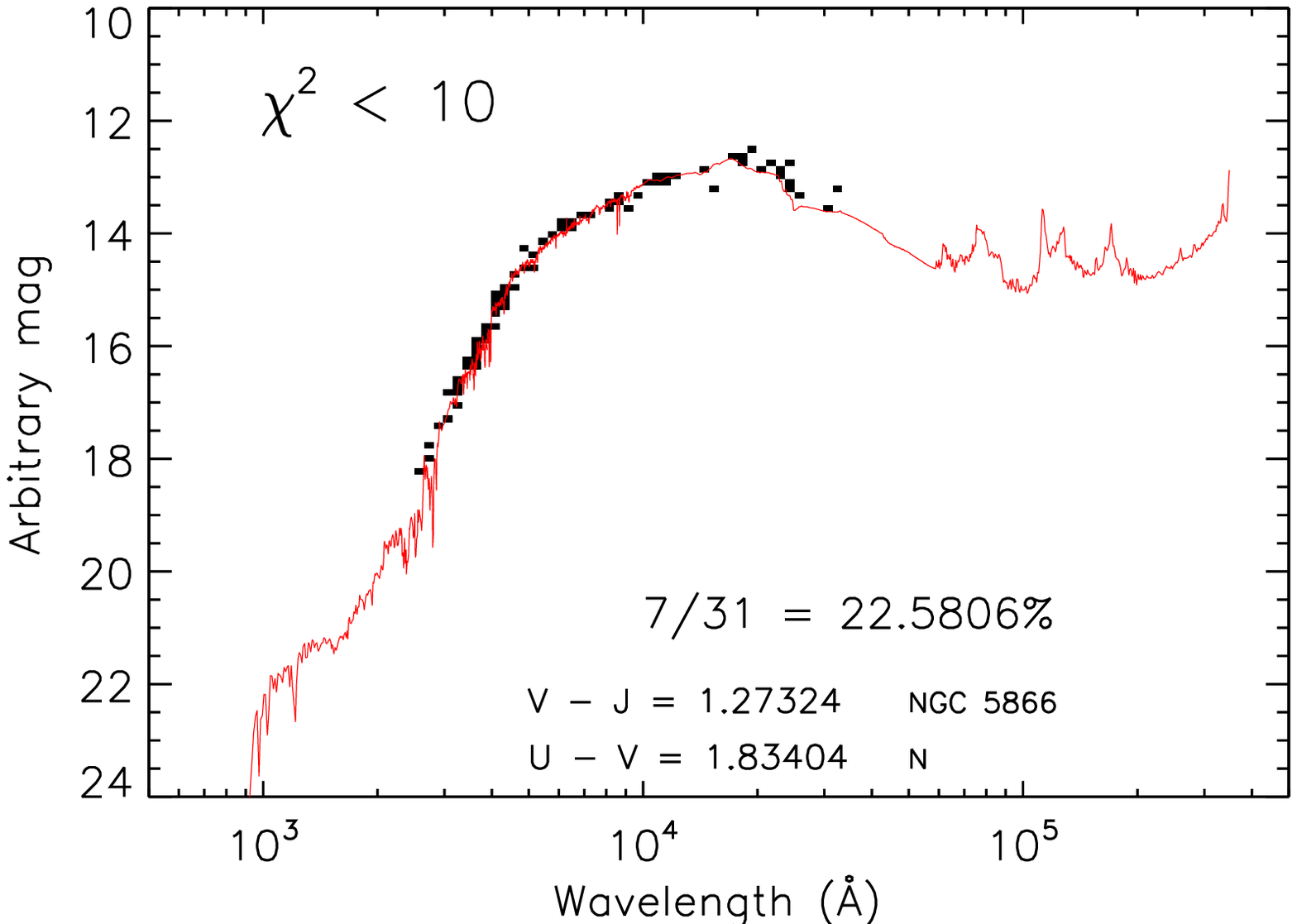}
\includegraphics[width=0.45\textwidth]{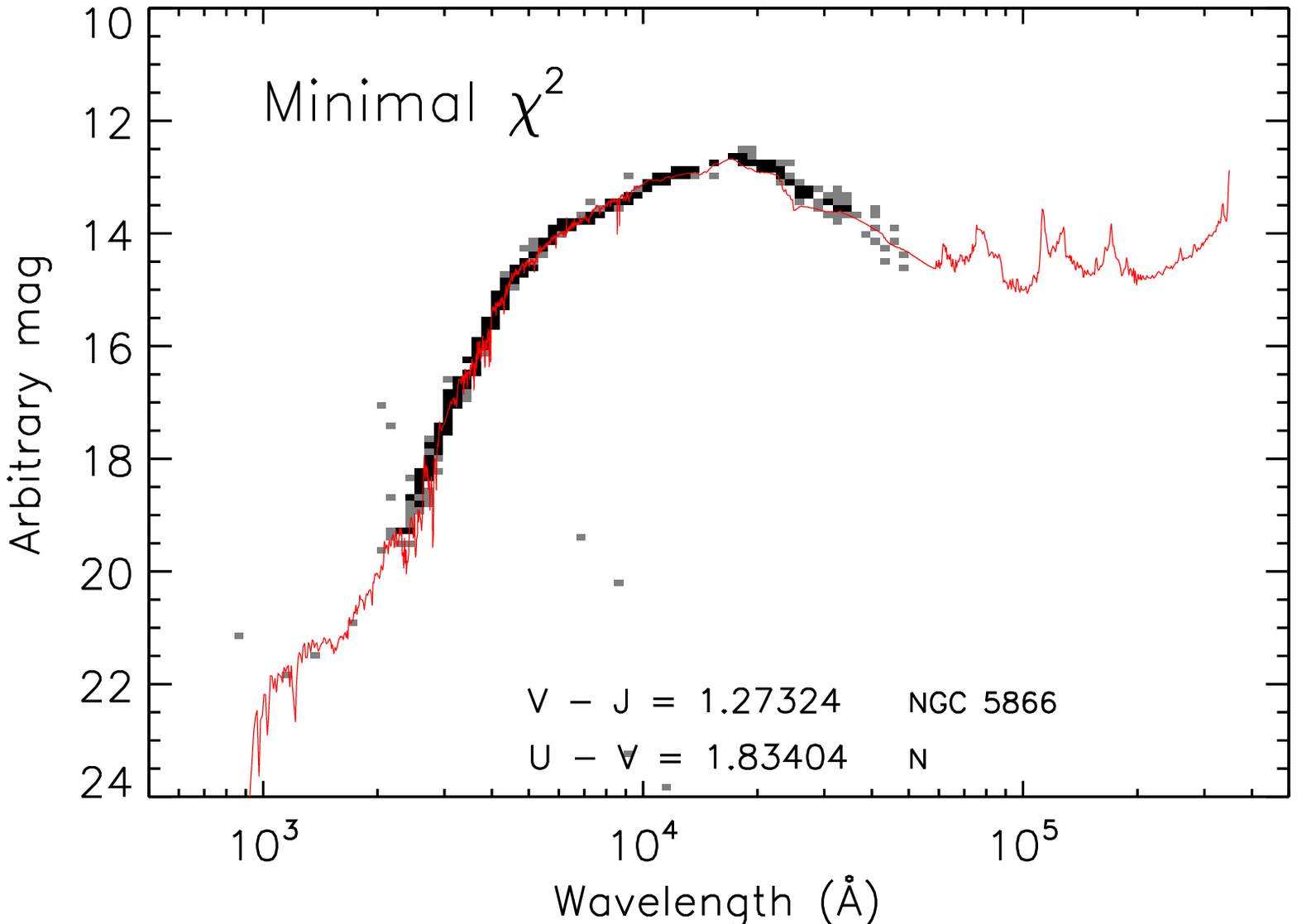}
\includegraphics[width=0.45\textwidth]{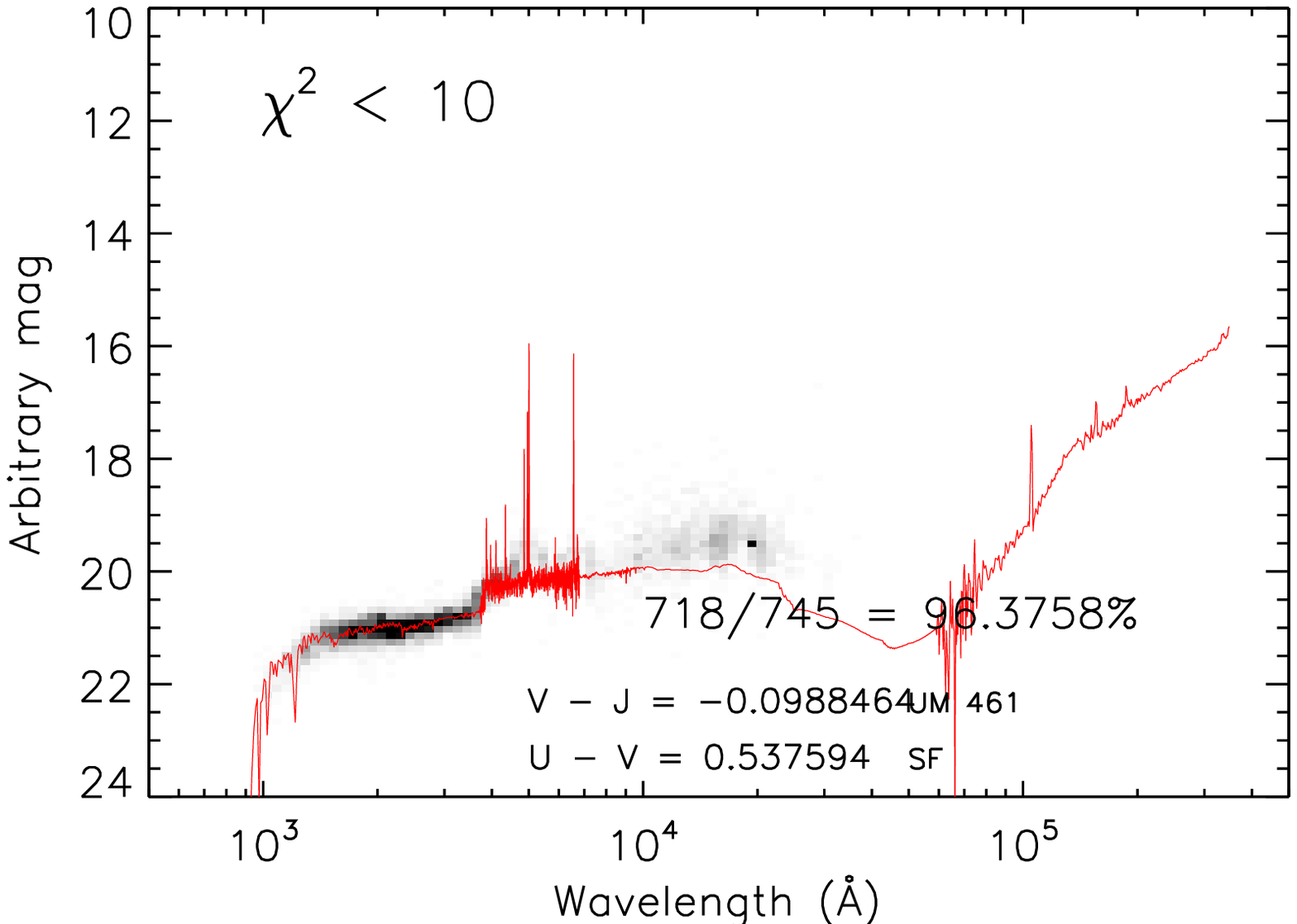}	
\includegraphics[width=0.45\textwidth]{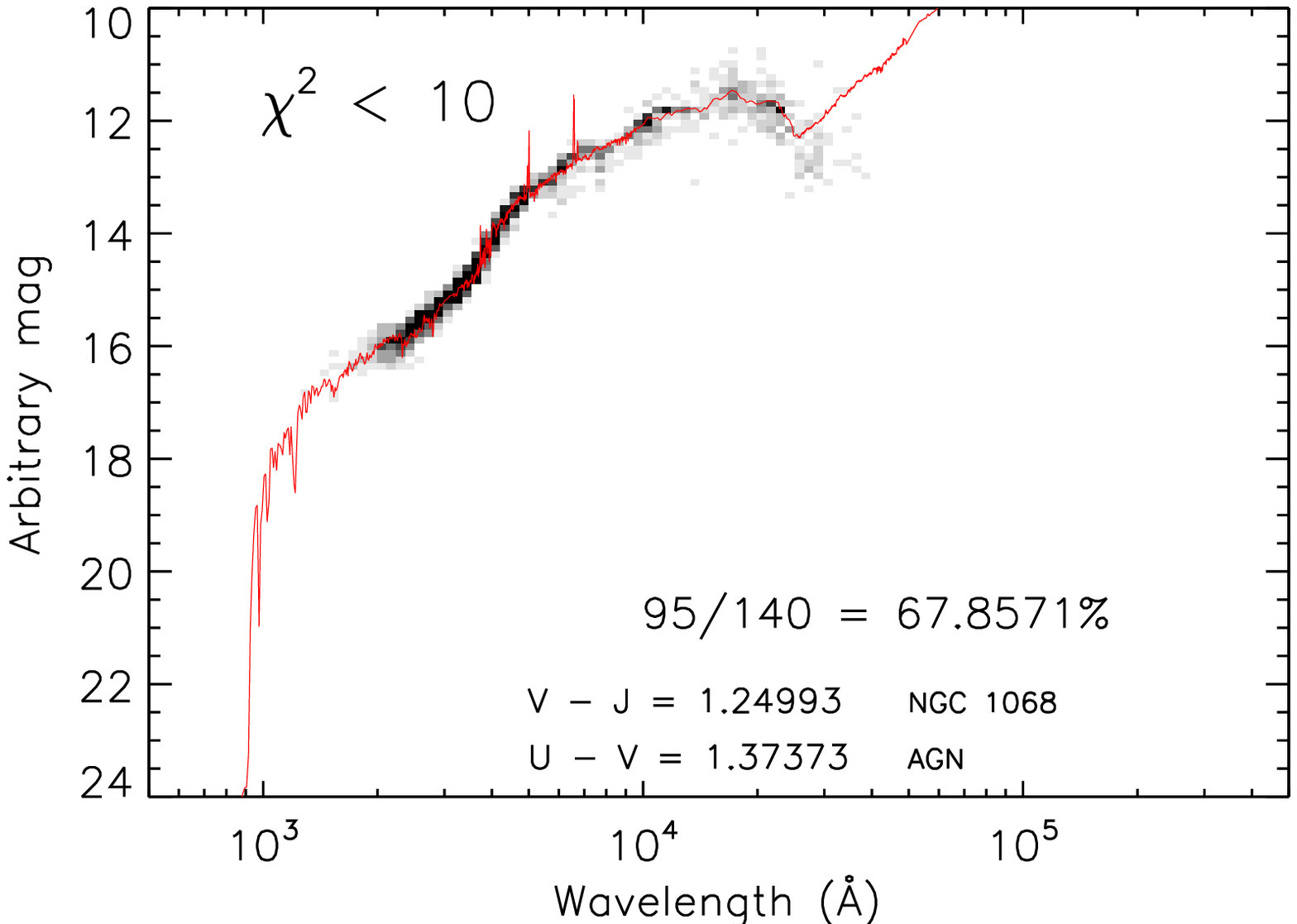}
\caption{Examples of the template fitting results. We plot every target SED onto the best fitting template with the $\chi^{2}$ cutting below 10, or all the target SED that minimized to the template, as denoted in each panel. The lower right notation is the flag of BPT class in \citet{Brown2014}.
}\label{SED_fitting}
\end{figure*}

The reduced $\chi^2$ as a summary of the deviation for every fitting band only reflect the integrated comparison between the data and model. To find the fitting goodness for each band, we overplot the source SED on the best fitting templates with the photometry data. Figure \ref{SED_fitting} shows some examples of the fitting results. We can see a bump near the [OIII] lines which means the emission lines are commonly existed in the blue galaxies, suggesting the necessity of templates with emission lines\citep[e.g., ][]{2017ApJ...837..170L}. The emission lines also caused the scatter for the blue galaxy in UVJ diagram \citep{Speagle2015, 2019A&A...631A.123Y}. 

For the blue galaxy example (upper left panel in Fig. \ref{SED_fitting}), UGCA 208, there are in total 995 SEDS sources that minimize the $\chi^2$ on this template. Most of the SEDS galaxies with $\chi^2<10$ locate in redshift 1 (Fig. \ref{Hist_Chi2}), thus the IRAC band data would be about rest frame NIR. We can find the IRAC data are also recovered by the blue galaxy UGCA 208 template. As we denoted in the upper left panel of the Figure \ref{SED_fitting}, 882 / 995 = 88.6\% source can be fitted within $\chi^2 < 10$. This result shows that the galaxies in redshift 1 are very similar to the local galaxies. The UGCA 208 is an AGN identified from the BPT diagram. From the fitting results, we conclude the AGN continuum is sinking inside the galaxy spectrum. For another local galaxy NGC 3310 (upper right panel in Fig. \ref{SED_fitting}), which have a similar U-V, V-J color as the UGCA 208, is star-forming galaxy due to the BPT criterion. This template can fit 83\% source with a $\chi^2 < 10$ with a consistency IRAC radiation.

%
%               one-column-spanning table
%________________________________________ Table 2: Use_of_the routines
\begin{table}
\begin{center}
\caption[]{ SEDS $U-V$ blue and red galaxy fraction\label{blue_red_fraction_tab}}

%%Please Capitalize the First Letter of Each Notional Word in table's caption

 \begin{tabular}{ccc}
  \hline\noalign{\smallskip}
    $[3.6]$ &  $U-V<1.5$&  $U-V>1.5$\\
  \hline\noalign{\smallskip}
	   	18.75  &  0.236111  &  0.763889\\
	   	19.75  &  0.355082  &  0.644918\\
	   	20.75  &  0.625544  &  0.374456\\
		21.75  &  0.803023  &  0.196977\\
		22.75  &  0.924637  & 0.0753633\\
		\noalign{\smallskip}\hline
\end{tabular}
\end{center}
\end{table}

In the case of the red galaxy NGC 5866 templates, we plot the SEDS source with $\chi^2 < 10$ and all the SEDS sources with minimal fitting $\chi^2$ on the NGC 5866 templates in the middle panel of Figure \ref{SED_fitting}. Two dots near 7000\AA are examples of bad fitting, which may lead to the $\chi^2 > 10$. Besides this outlier dots, there is no visually difference between the two results. The galaxies with $\chi^2>10$ are still consistent with the template.

For the example of NGC 1068, which is an AGN and shows power-law in MIR bands, which is the emission from AGN heated dust (lower right panel in Fig. \ref{SED_fitting}). The NGC 1068 template can also fit 68\% SEDS source from the u to Ks band, which means the MIR power-law galaxies SEDs are indistinguishable with a normal galaxies in optical and NIR band. Meanwhile, the SEDS sources well fitted by NGC 1068 template do not follow a similar power-law in MIR, indicating the MIR power-law galaxies are still normal in optical bands \citep{Alonso-Herrero2006}.

For some blue galaxy templates, we find the IRAC flux is brighter than the templates for about 0.5 mag (lower left panel of Fig. \ref{SED_fitting} and left panel of Fig. \ref{UM461_IRAC}). We discuss this feature in the next Section.  All the templates and the $\chi^2<10$ SEDs are shown in the appendix. 

\subsection{3.6$\mu$m flux excess in blue galaxies}
\begin{figure}[ht!]
\centering
\includegraphics[width=0.48\textwidth]{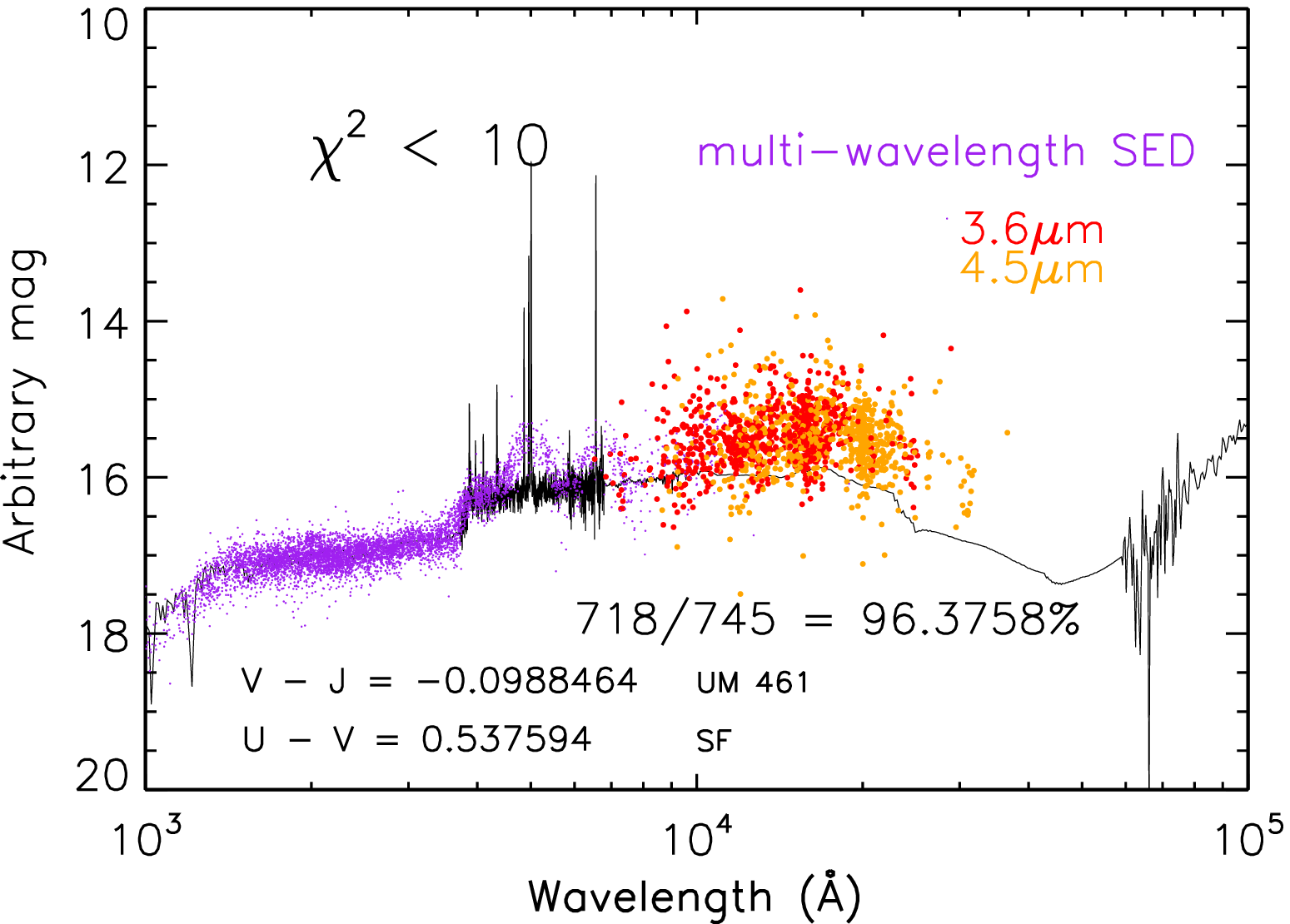}
\includegraphics[width=0.48\textwidth]{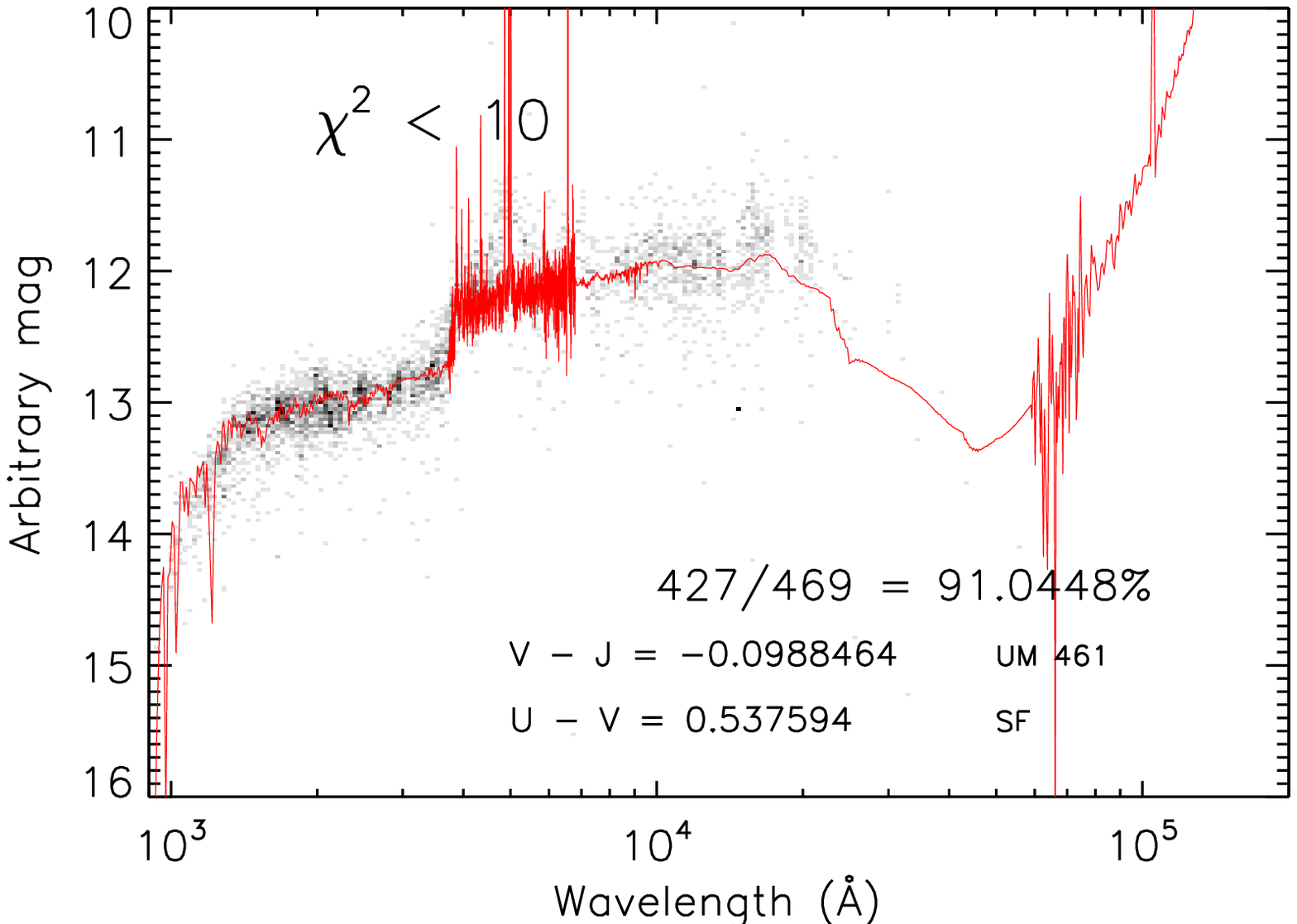}
\caption{
{\bf Left panel:} UM 461 template fitting result. The color dots are the SEDs can be fit with $\chi^2<10$. We highlight the IRAC 3.6$\mu$m and 4.5$\mu$m data with red and orange color, and the rest SEDs with purple color. {\bf Right panel: }
UM 461 template fitting result from u to $4.5\mu \rm m$ band. More than 90\% SEDS source can be Fitted from u to 4.5$\mu \rm m$ band with $\chi^2 < 10$. There is still a clear bump feature around the emission lines. We can see there is no excess in the IRAC band because the photometry data we used should fit the template within $\chi^2 < 10$.}\label{UM461_IRAC}
\end{figure}

For the extreme V-J blue galaxies, such as UM 461 in Fig. \ref{SED_fitting} the IRAC flux, which is about rest-frame J band, is excessing for about 0.5 mag (also see left panel of Fig. \ref{UM461_IRAC}). Similar excess also shows in the templates UGCA 166, UGCA 219, Mrk 0930, UGC 06850, Mrk 1450, Mrk 0475, all of which are very blue for $U - V \lessapprox 0.5$ and $V - J < 0.2$. Referring to the UVJ diagram for other surveys \citep{Whitaker2011, Muzzin2013}, there are very few galaxies bluer than $U-V \simeq 0.5$. These galaxies are young, low mass and very active in star formation. Their colors are blue from UV to about 5$\mu \rm m$, then turns red at longer wavelength.

A similar excess phenomenon can be found in fitting the GAMA SEDs data \citep{Taylor2011}. There is a systematically excess in rest frame NIR band flux if only u g r i z band is fitted by stellar population library (SPL) \citep[e.g. Fig. 7 \& A1 in ][]{Taylor2011} and by a series of analyse, \citet{Taylor2011} conclude that the NIR excess might caused by the NIR data, simple stellar population(SSP) model or the SPL. In our case, as the local galaxy templates can fit the optical and NIR data with consistency IRAC flux except for the six extreme blue galaxies, the offset is more likely not caused by the data. Otherwise there should be an offset for most of the templates.

This rest-frame NIR excess might be caused by the stellar population that exist in blue galaxies. \citet{Maraston2005} stress the importance of the thermally pulsating asymptotic giant branch (TP-AGB) star, which might contribute significantly to the rest-frame NIR luminosity in young galaxies. The follow up work  \citep{Maraston2006} also shows the consistency fitting results of the TP-AGB population, which might also help in alleviating the Age/metallicity degeneracy in the optical bands \citep{Worthey1994}. Due to the position of UM 461 in UVJ diagram, the age of the UM 461 well fitted SEDS sources are about 1 Gyr  \citep[e.g. Figure 1 in][]{Patel2012}. The bands we used for the template fitting are about the rest frame UV to the optical band and thus the minimal $\chi^2$ method might only catch the main features in optical and miss the accuracy in rest frame NIR band. However, the uncertainty of the stellar population synthesis model is controversy \citep{Maraston2006, Kriek2010}, and the template we used should include the TP-AGB stars. Maybe the local galaxy templates we used here are still not complete for the blue galaxies. We also fit the SEDS source from u to $4.5\mu \rm m$ band and the results are plotted in the Figure \ref{UM461_IRAC}. Target number with minimal $\chi^2$ at the template UM 461 decrease from 745 to 469. We can see no rest frame NIR excess in the template fitting from u to $4.5\mu$m. 

\section{Summary}
In this paper, we investigate the SEDS galaxy population by comparing the SEDS multi-wavelength SED with the local galaxy template. We build multi-wavelength catalog for the SEDS targets. After carefully simulate the completeness, we find that the IRAC number counts is consistent with the CANDELS F160W band selected sample. Our SEDS photometry, the CANDELS phot-z and the local galaxy templates show a consistency with each other. The SEDS galaxies mainly locate at redshift 1, which can be fitted well by the templates from the local galaxies. Consequently, the main evolution is the relative galaxy number counts, not the stellar population. Moreover, this result also confirms the CANDELS photometric redshift for the redshift 1 galaxy is consistent. 

As we only fit the galaxy SEDs from u to Ks band, we find a high consistency of MIR radiation between the local galaxy template SEDS galaxy IRAC flux. We also find the low mass galaxy templates do not fit the IRAC flux well and have about 0.5 mag excess. This excess may caused by the poor completeness of the low mass galaxy templates.

In the further work, we will focus on the other properties similarity of the SEDS galaxy SEDs and local galaxy templates. The local templates can also facilitate the further study of the SEDS galaxy photo-z, as well as the luminosity and mass function at high redshift.

\begin{acknowledgements}
The authors are very grateful to the anonymous referee for the helpful report.
This work is supported by the National Key R\&D Program of China grant 2017YFA0402704 and the National Natural Science Foundation of China, No. 11803044, 11933003. We acknowledge the science research grants from the China Manned Space Project with NO. CMS-CSST-2021-A05. This study makes use of data from AEGIS, a multiwavelength sky survey conducted with the Chandra, GALEX, Hubble, Keck, CFHT, MMT, Subaru, Palomar, Spitzer, VLA, and other telescopes and supported in part by the NSF, NASA, and the STFC. Based in part on data products produced by TERAPIX and the Cambridge Astronomy Survey Unit on behalf of the UltraVISTA consortium.The UKIDSS project is defined in Lawrence,  et al (2007). Further details on the UDS can be found in Almaini,  et al. (in prep). UKIDSS uses the UKIRT Wide Field Camera (WFCAM; Casali,  et al, 2007). The photometric system is described in Hewett,  et al (2006), and the calibration is described in Hodgkin,  et al. (2009). The pipeline processing and science archive are described in Irwin,  et al (in prep) and Hambly,  et al (2008). 
\end{acknowledgements}

\bibliographystyle{raa} % style aa.bst
\bibliography{raa.bib}{} % your references Yourfile.bib
%-------------------------------------------------------------------

\appendix
\begin{figure}
    \centering
\includegraphics[width=0.32\textwidth]{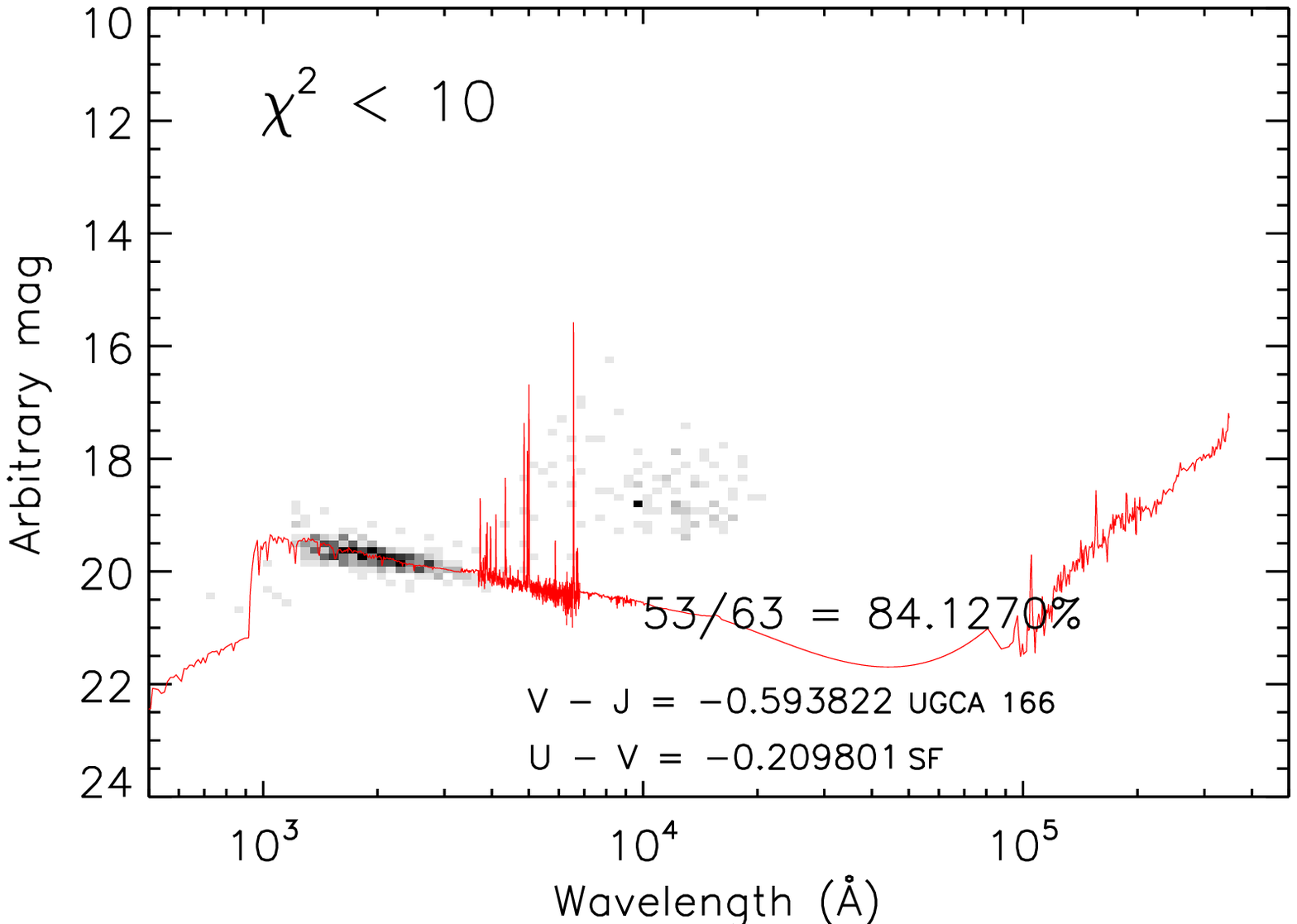}
\includegraphics[width=0.32\textwidth]{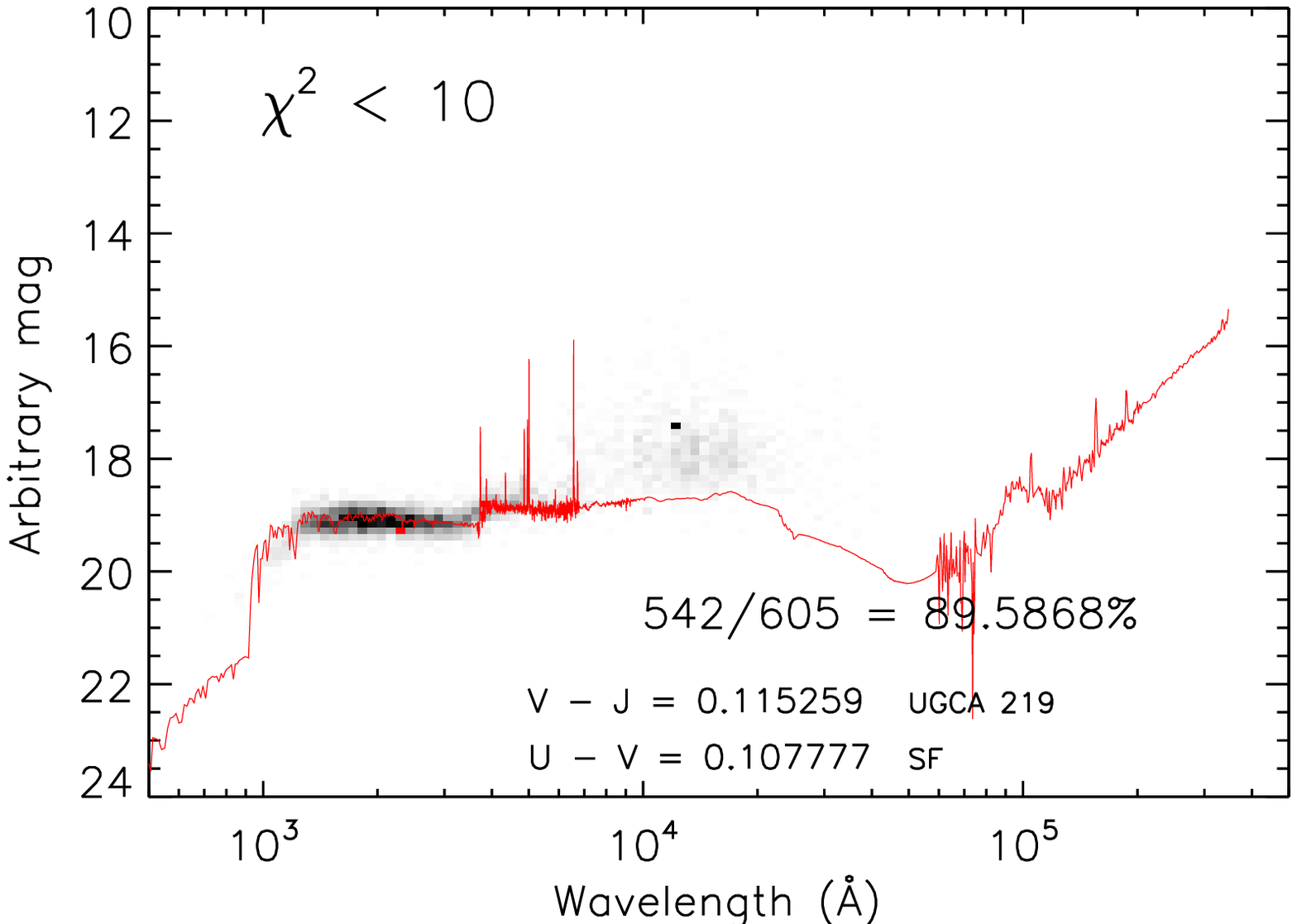}
\includegraphics[width=0.32\textwidth]{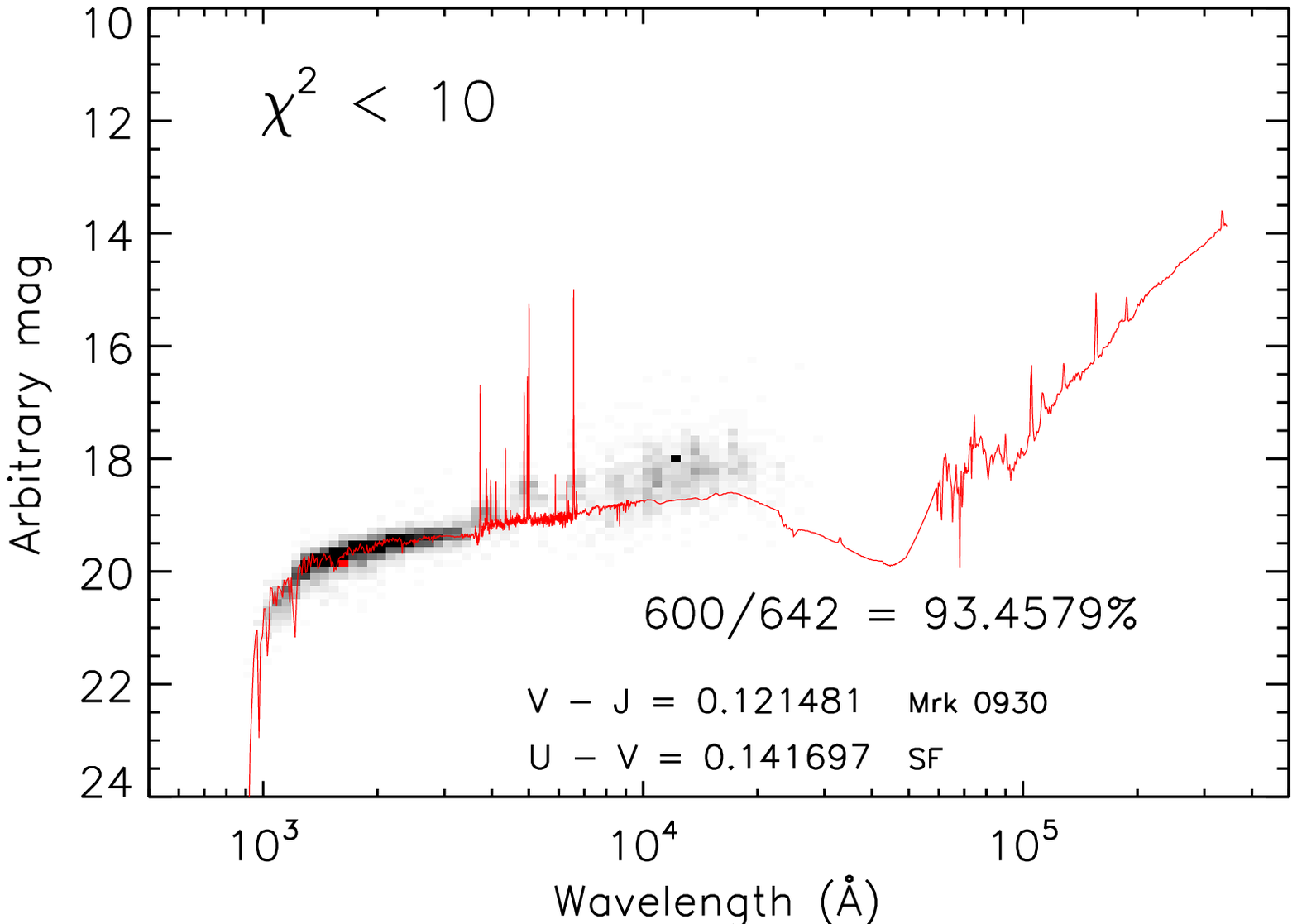}
\includegraphics[width=0.32\textwidth]{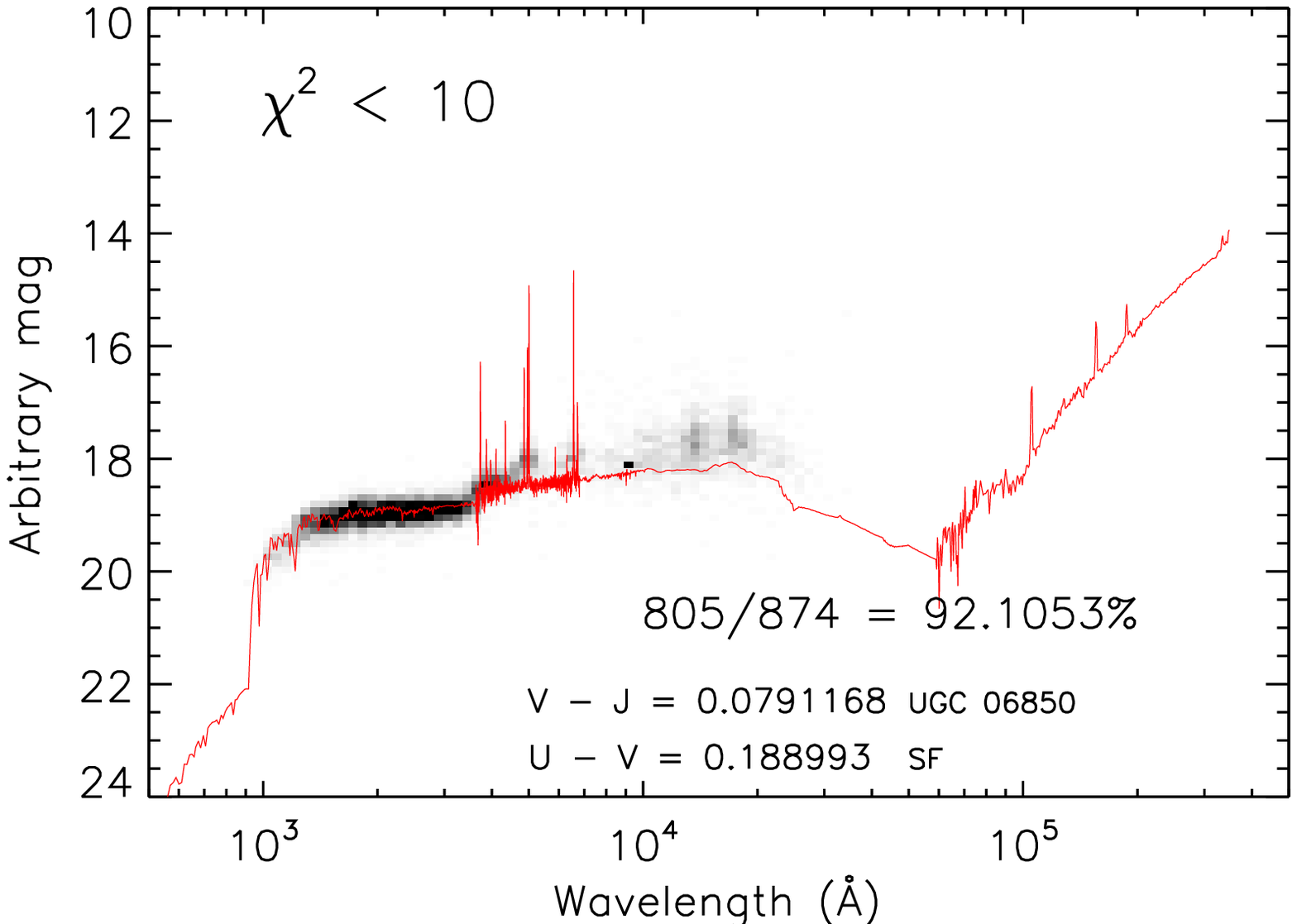}
\includegraphics[width=0.32\textwidth]{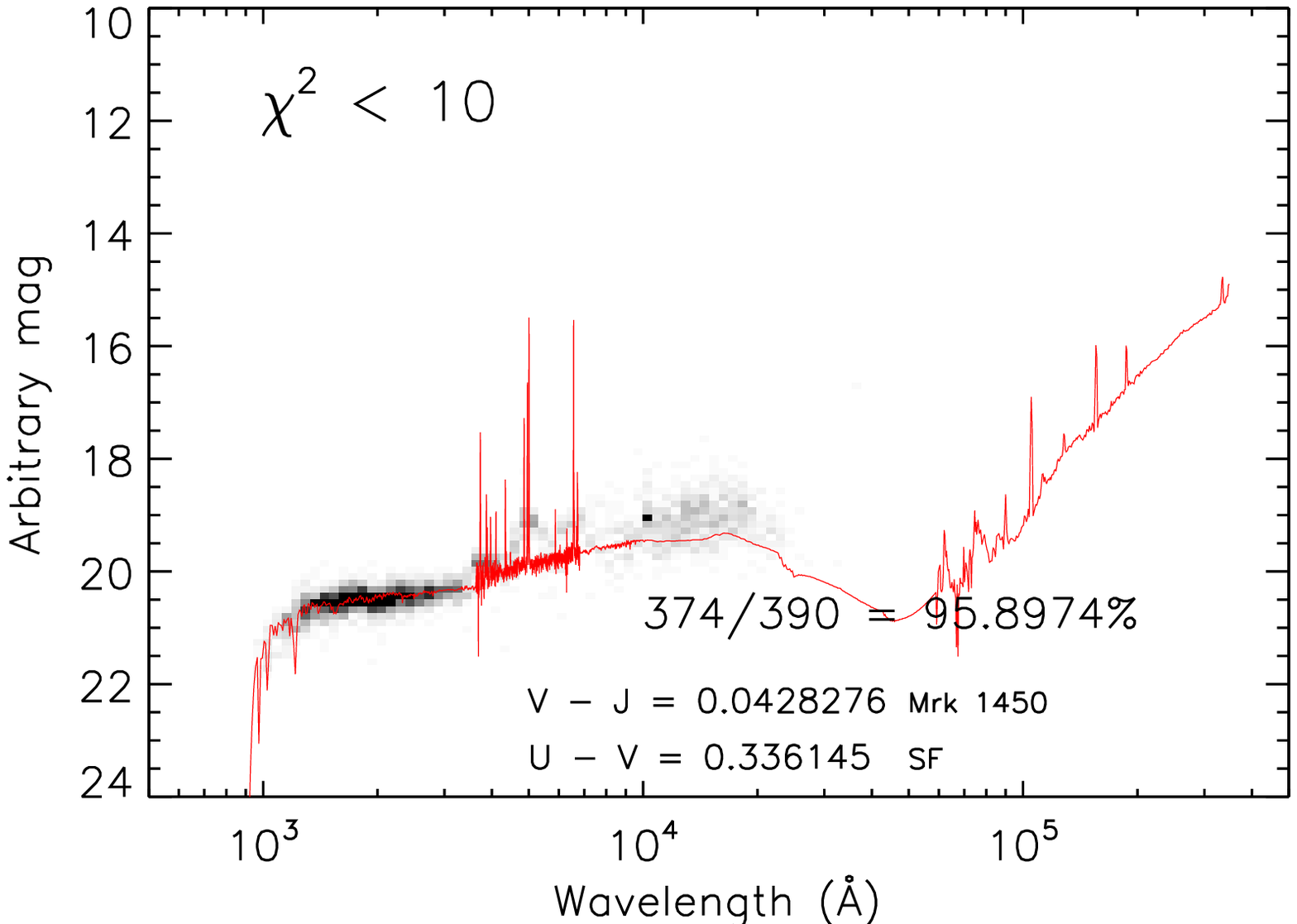}
\includegraphics[width=0.32\textwidth]{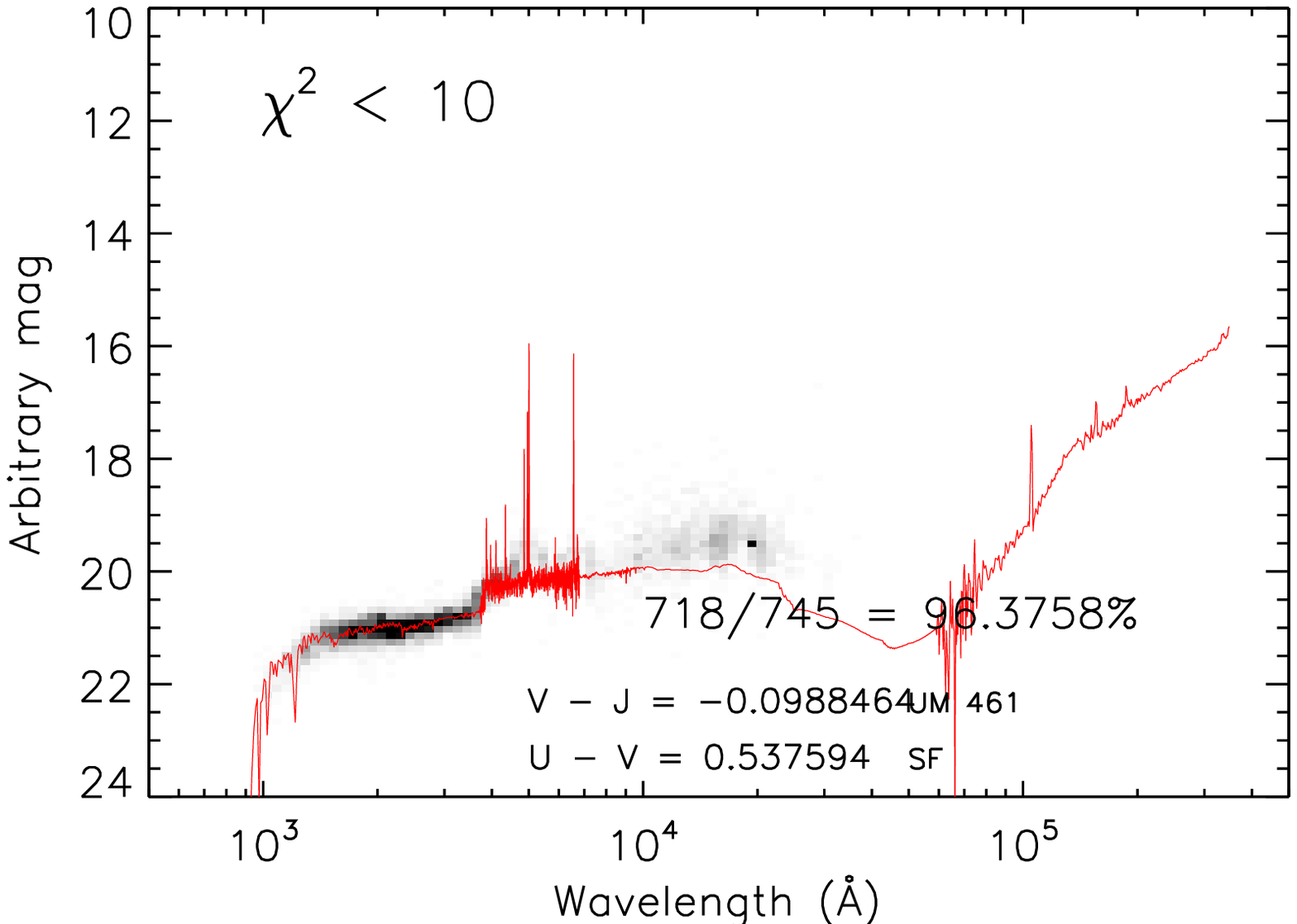}
\includegraphics[width=0.32\textwidth]{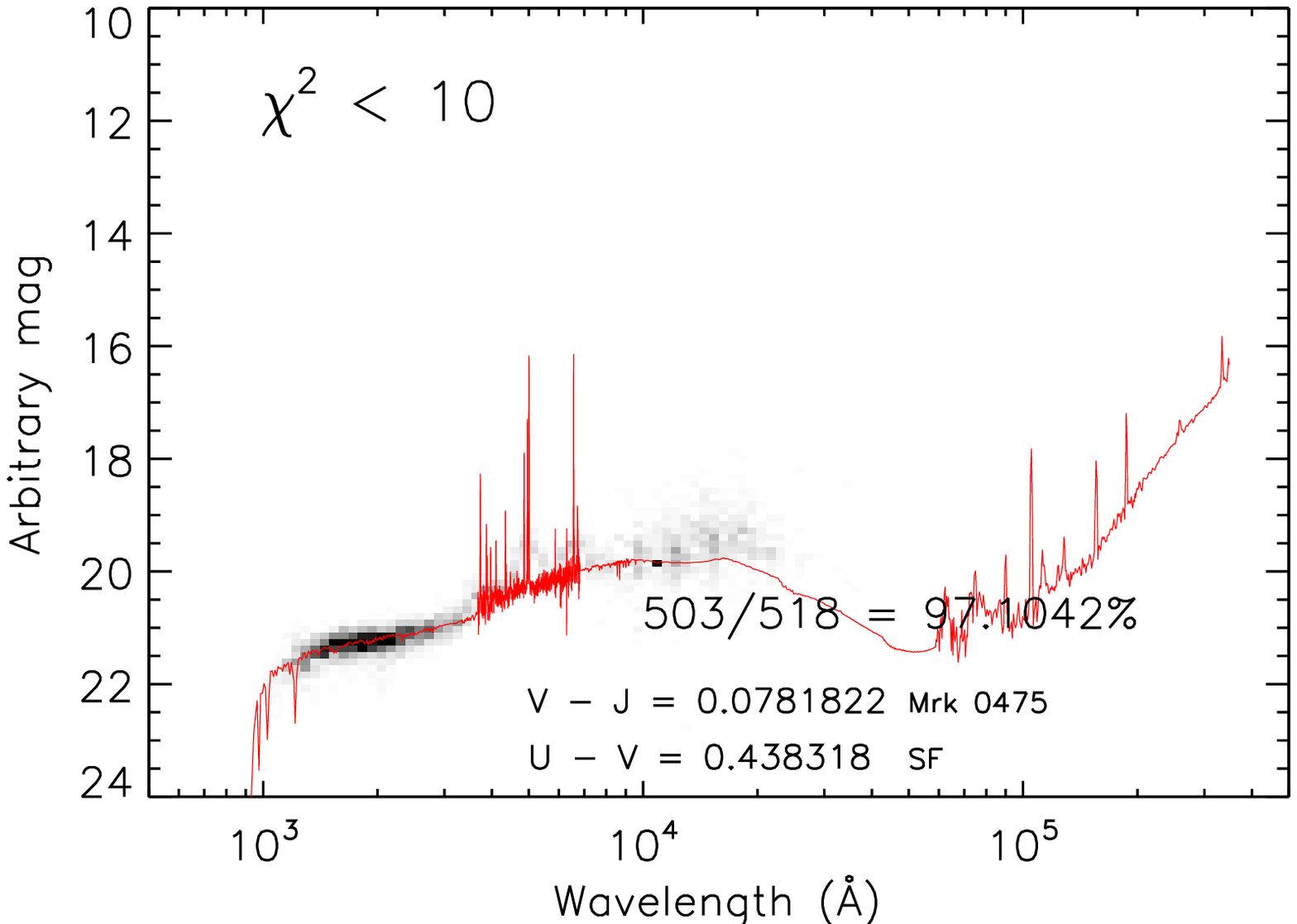}
\includegraphics[width=0.32\textwidth]{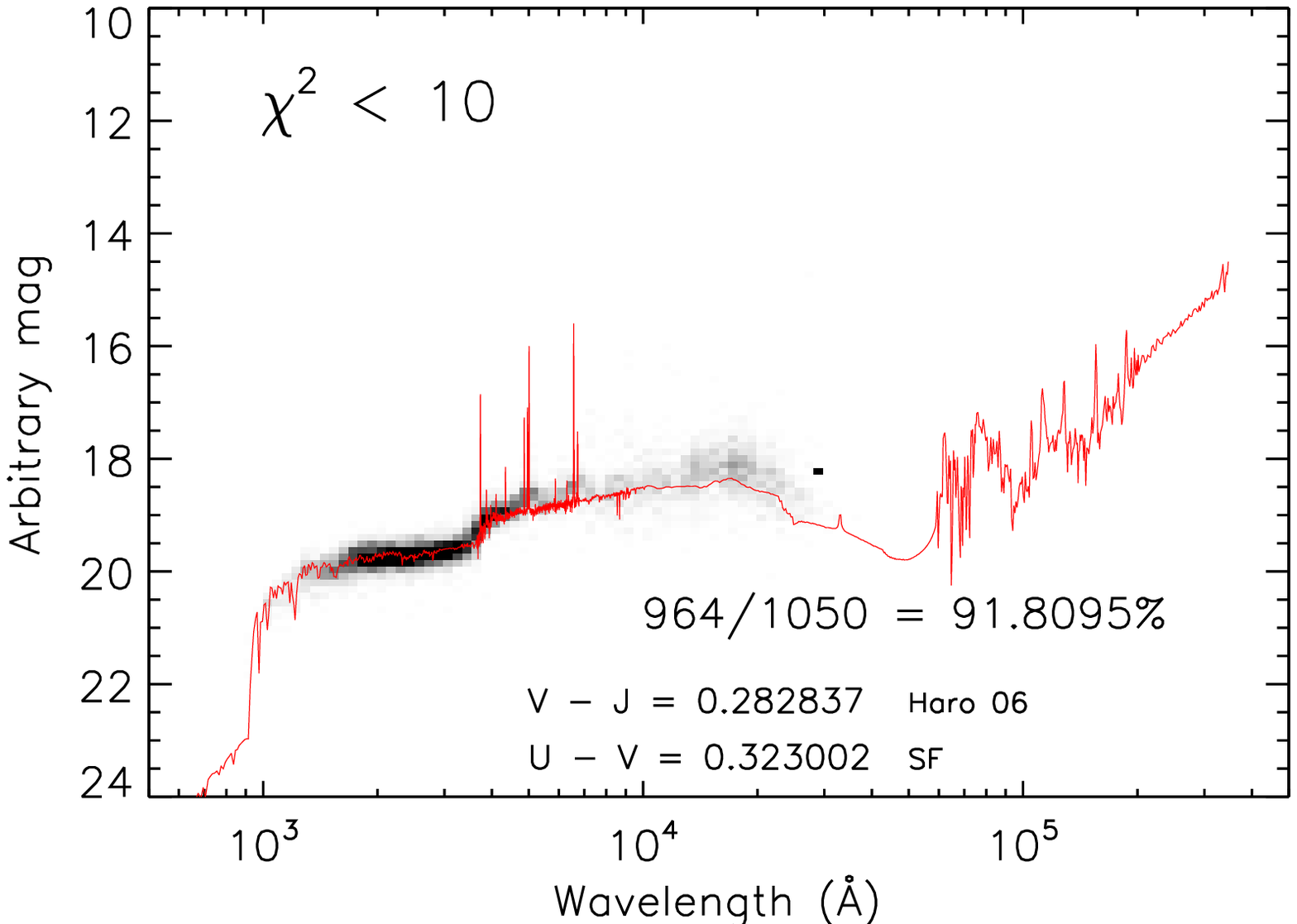}
\includegraphics[width=0.32\textwidth]{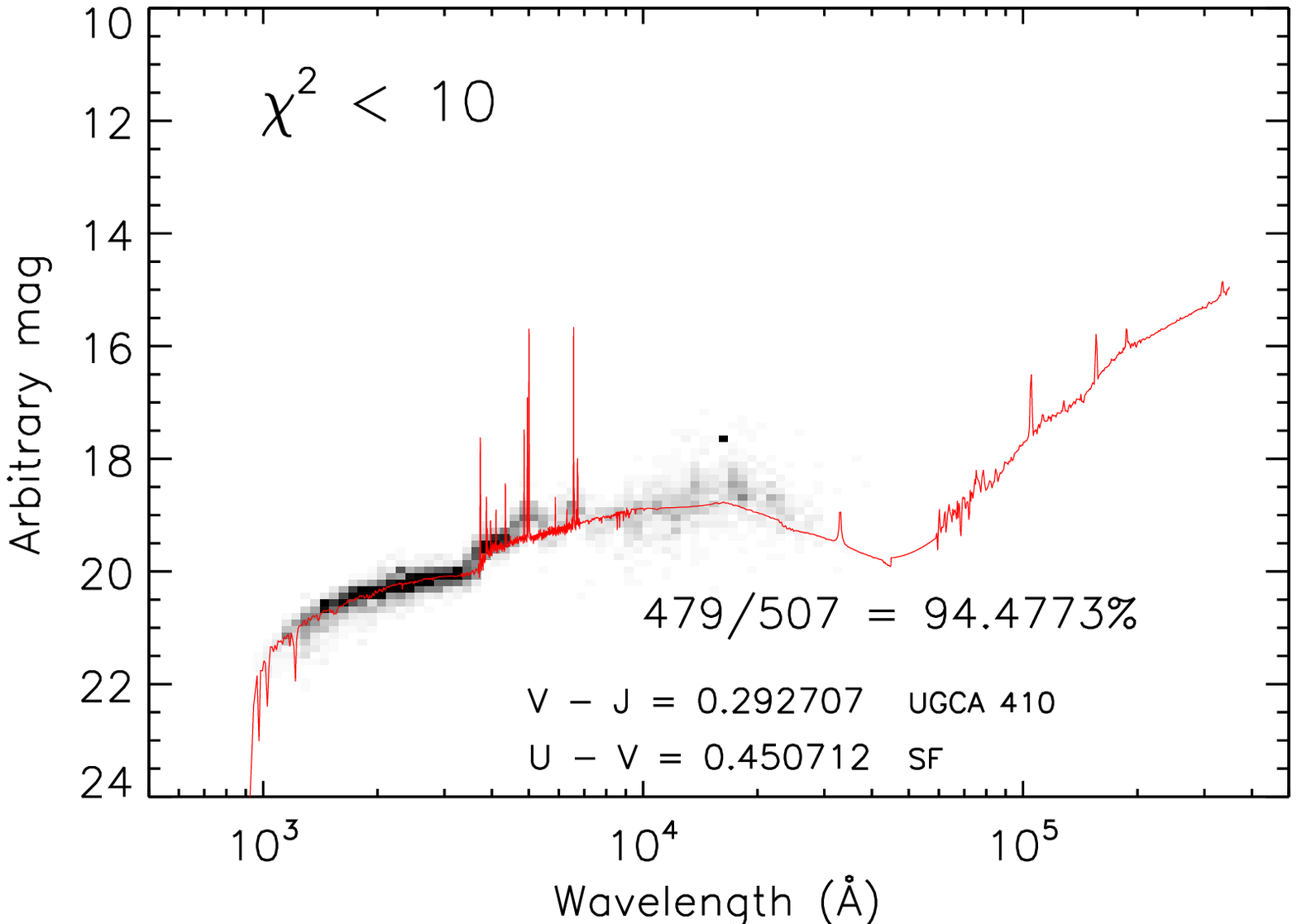}
\includegraphics[width=0.32\textwidth]{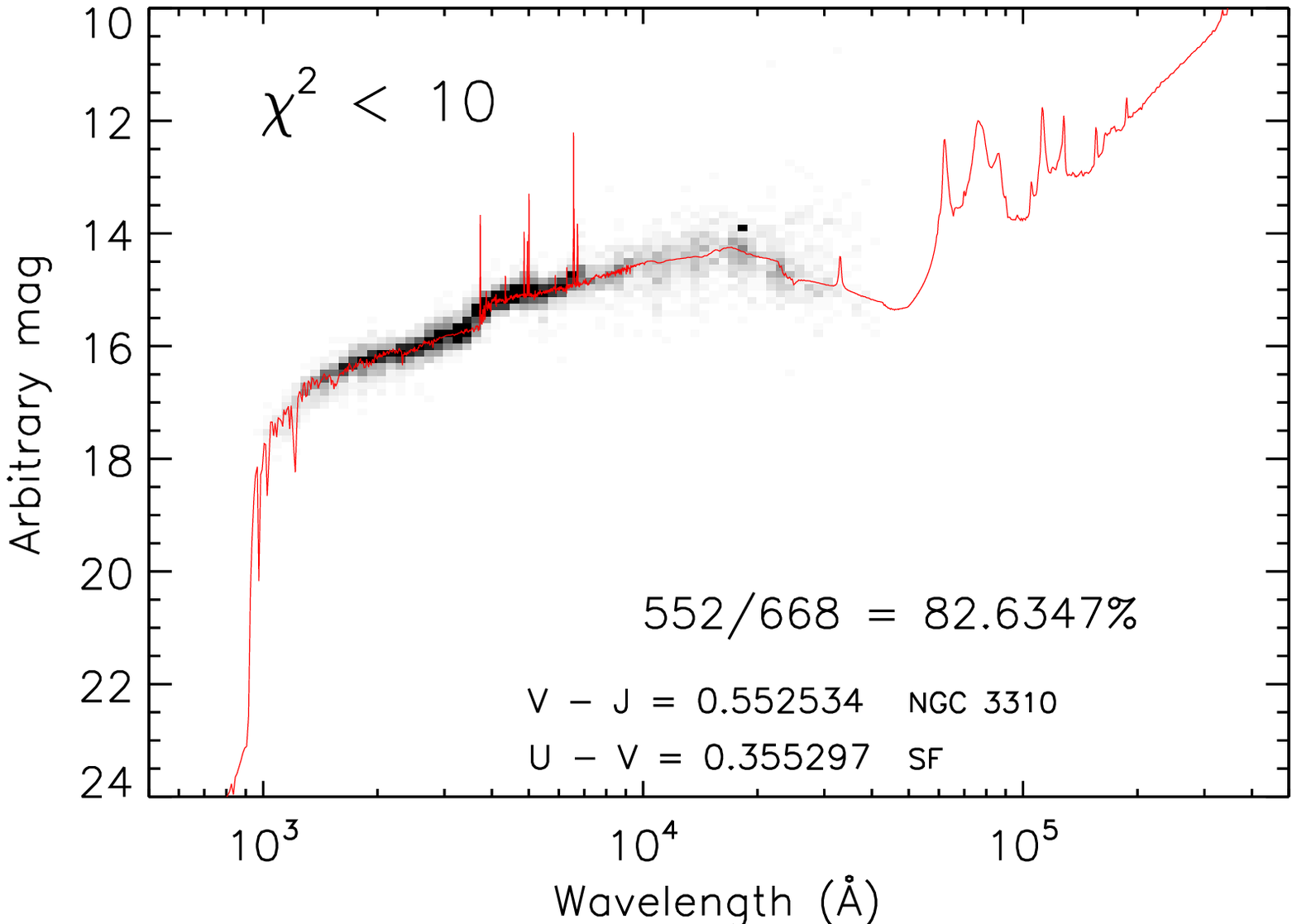}
\includegraphics[width=0.32\textwidth]{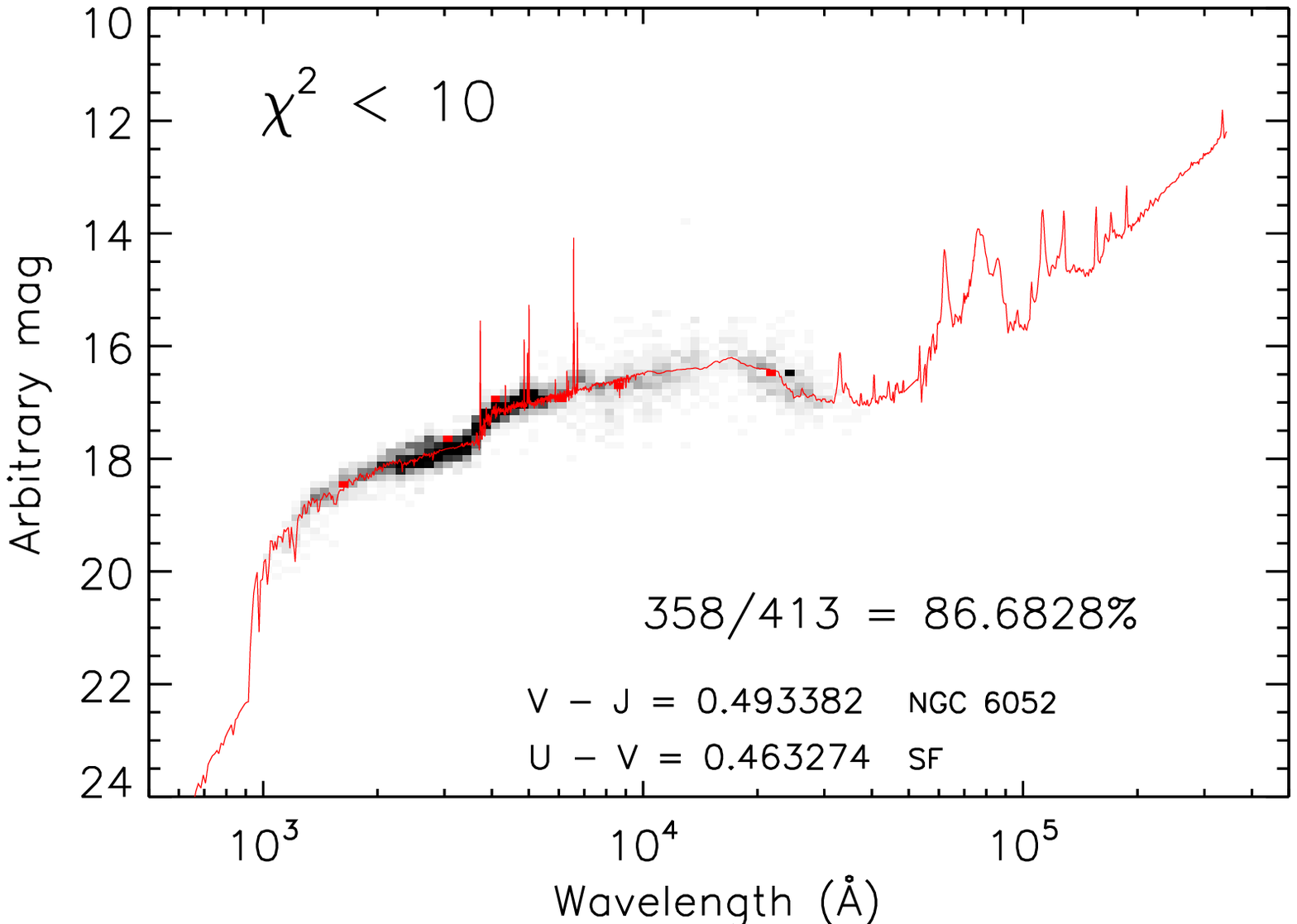}
\includegraphics[width=0.32\textwidth]{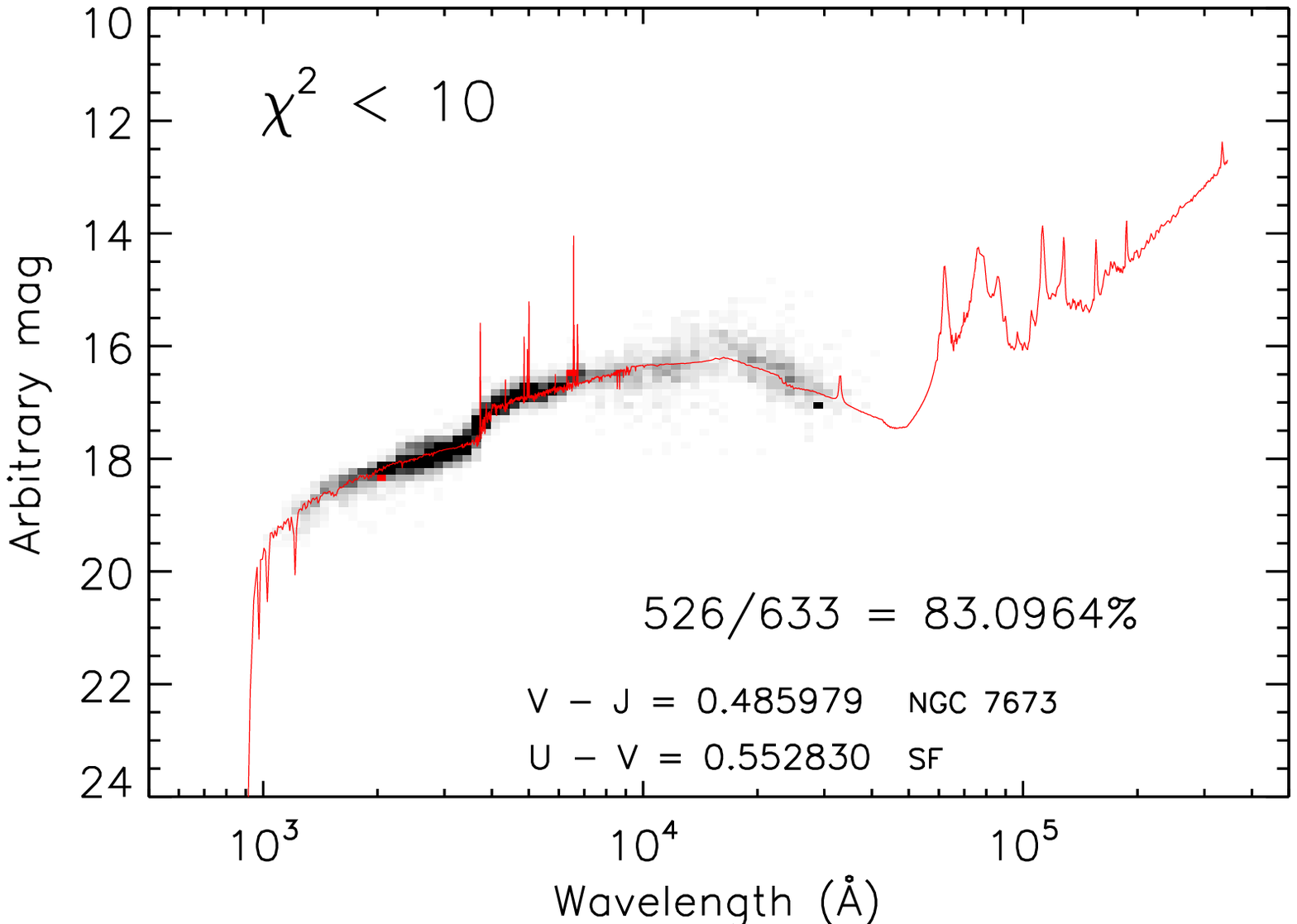}
\includegraphics[width=0.32\textwidth]{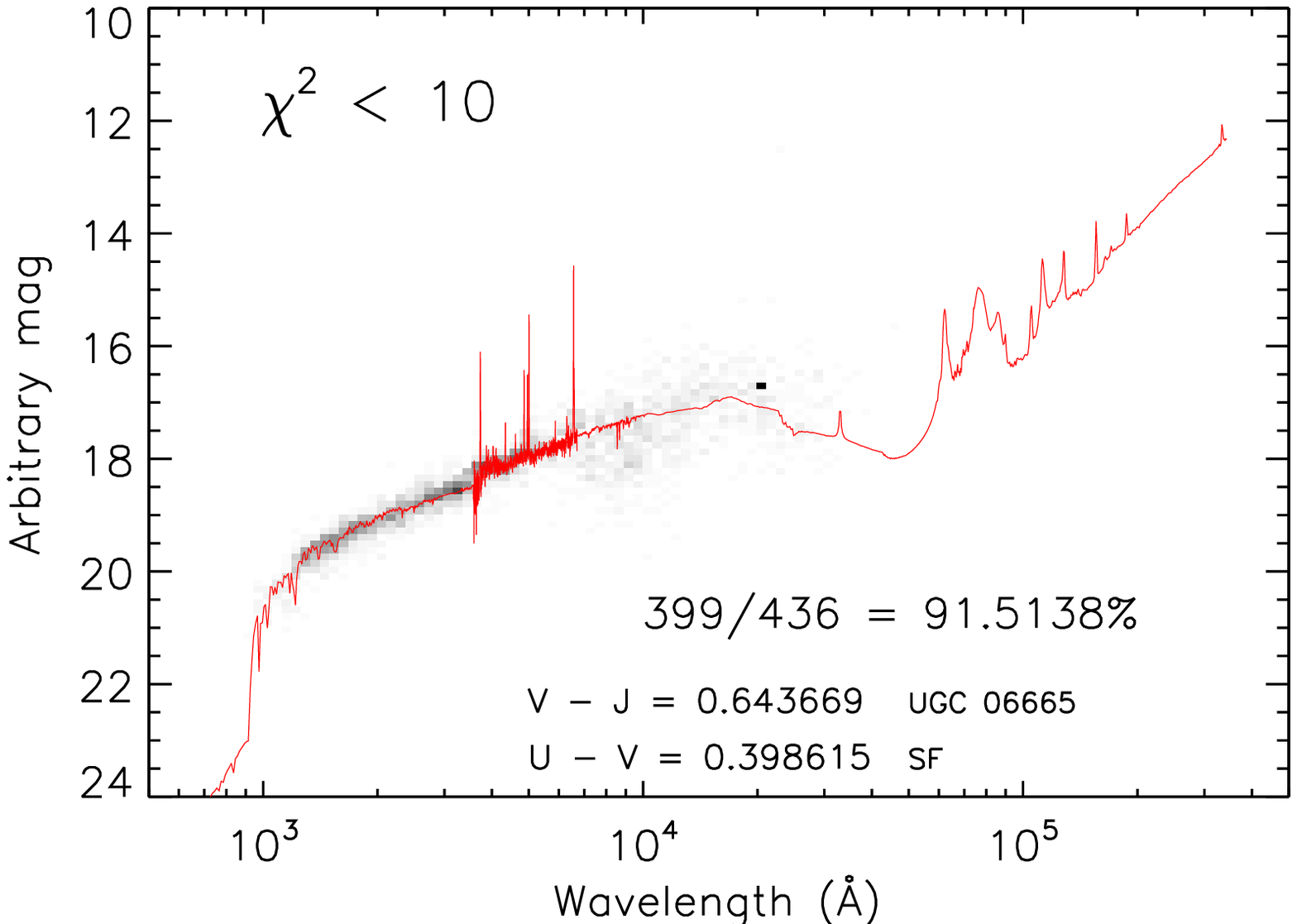}
\includegraphics[width=0.32\textwidth]{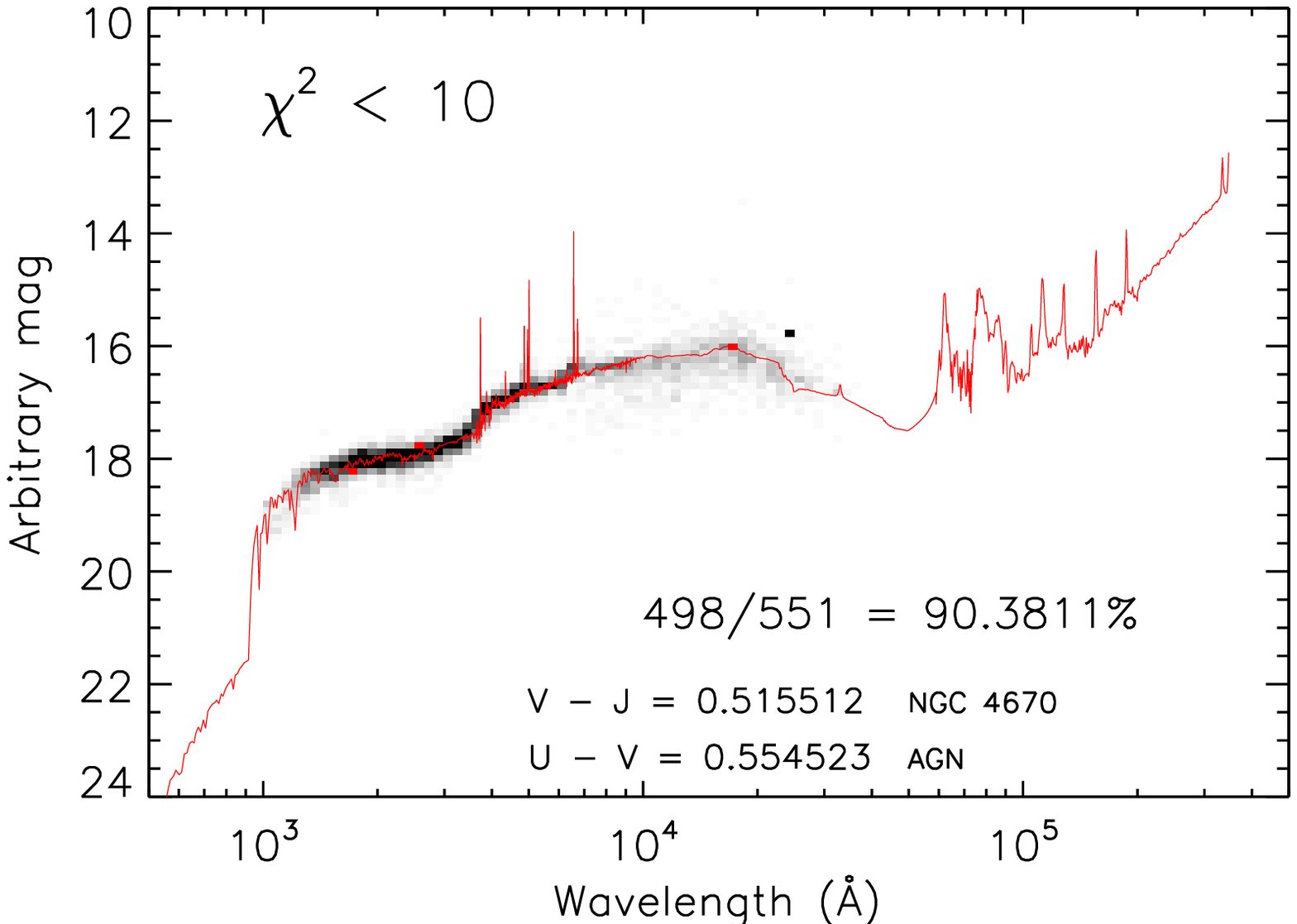}
\includegraphics[width=0.32\textwidth]{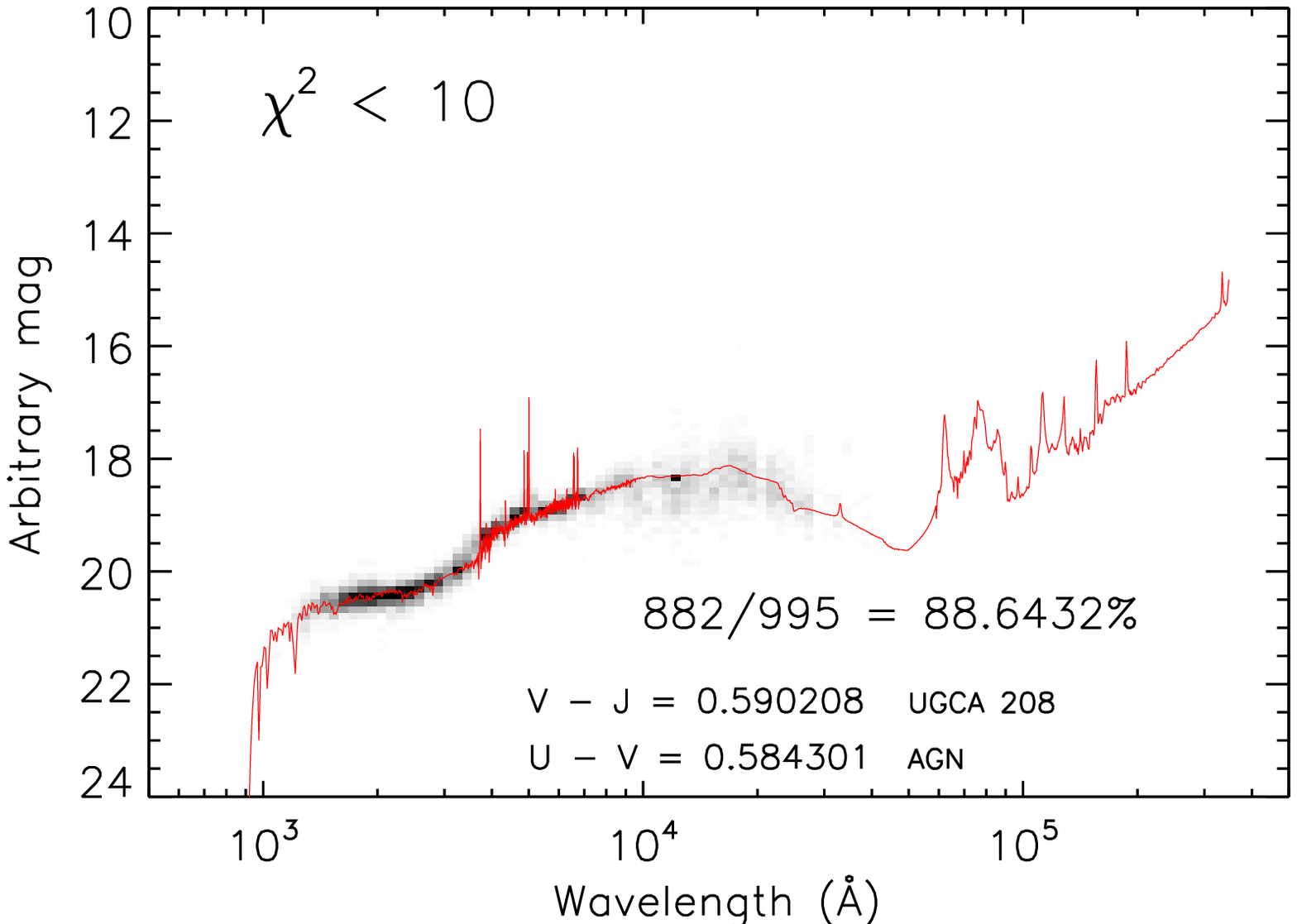}
\includegraphics[width=0.32\textwidth]{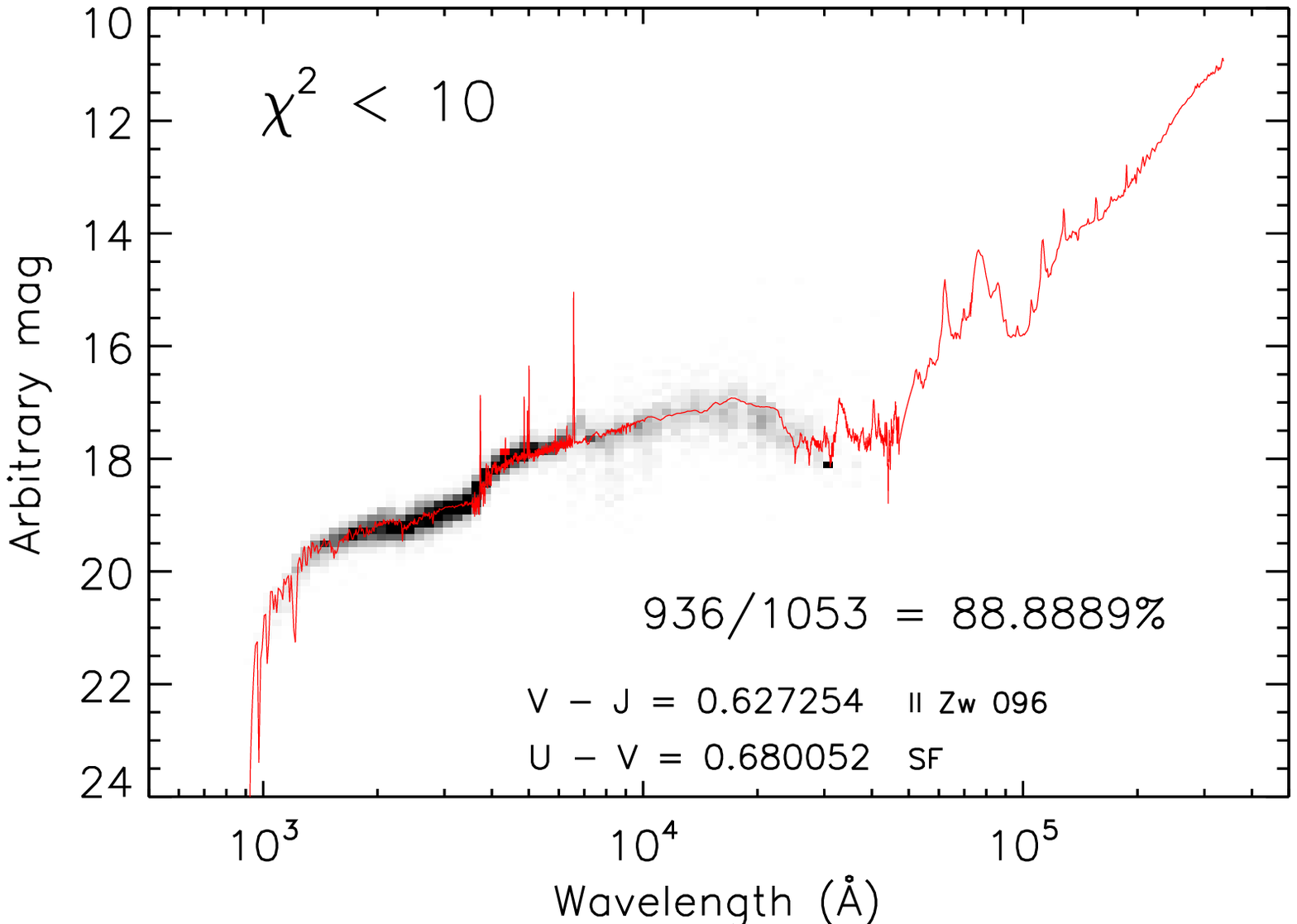}
\includegraphics[width=0.32\textwidth]{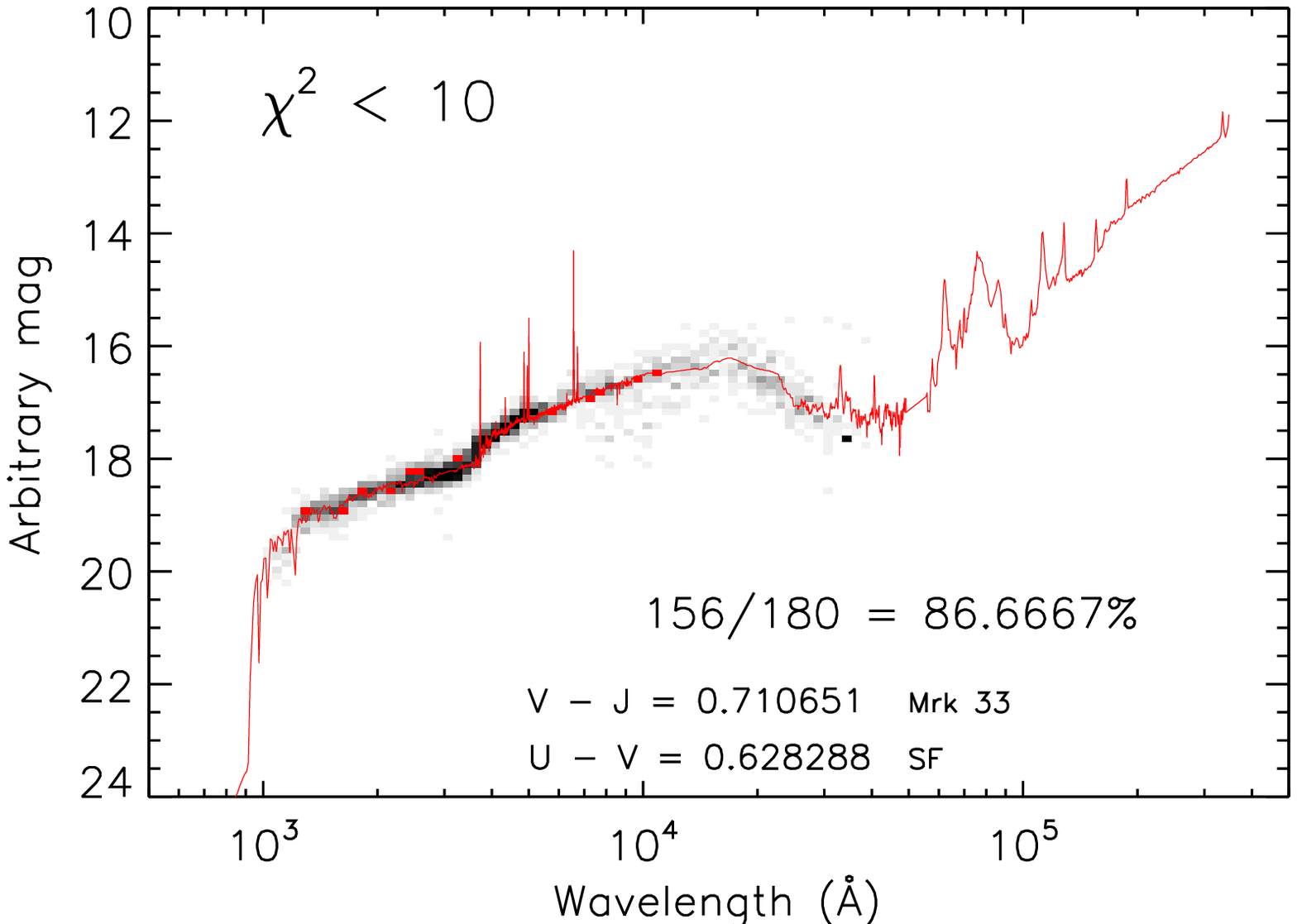}
\includegraphics[width=0.32\textwidth]{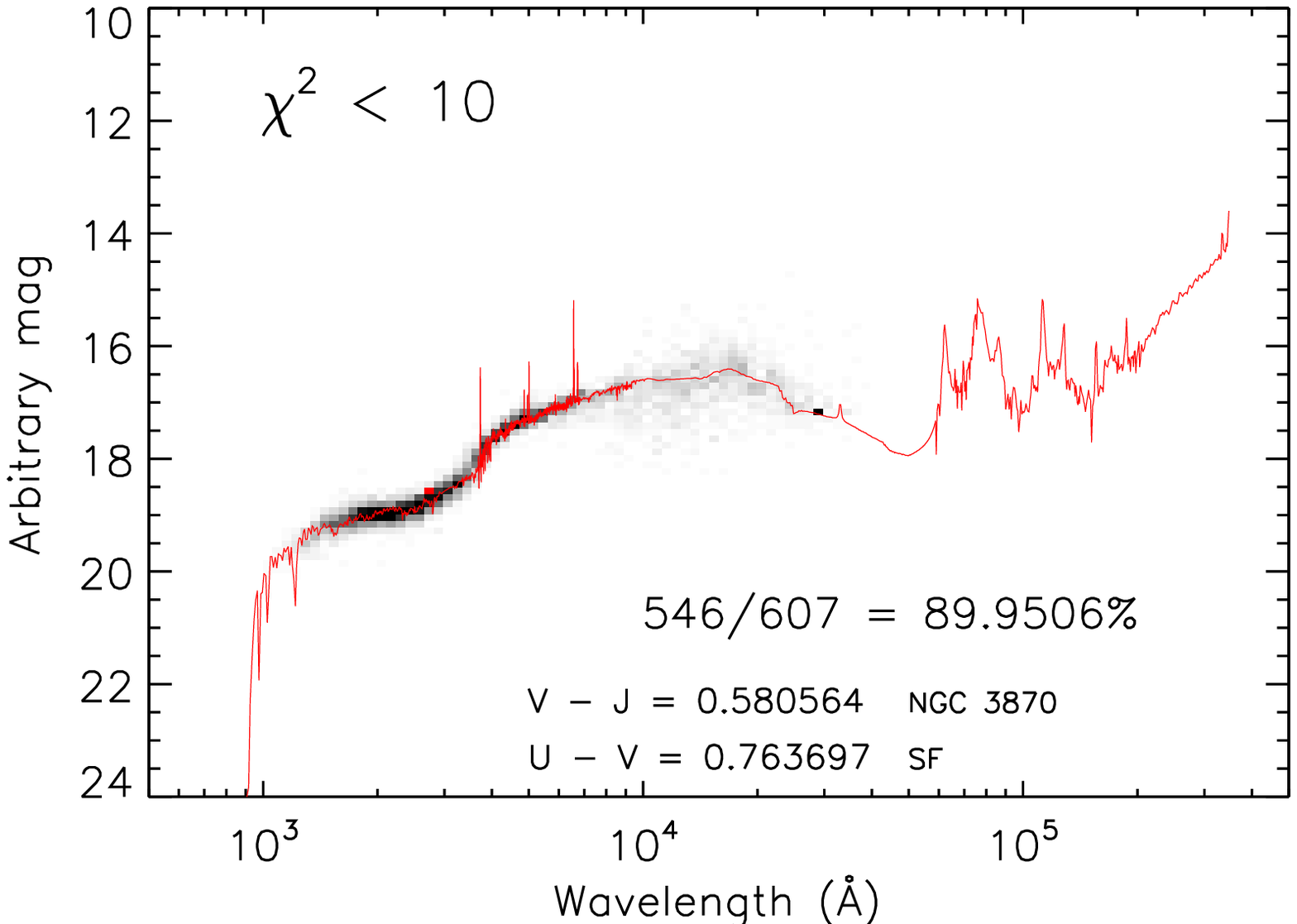}
\includegraphics[width=0.32\textwidth]{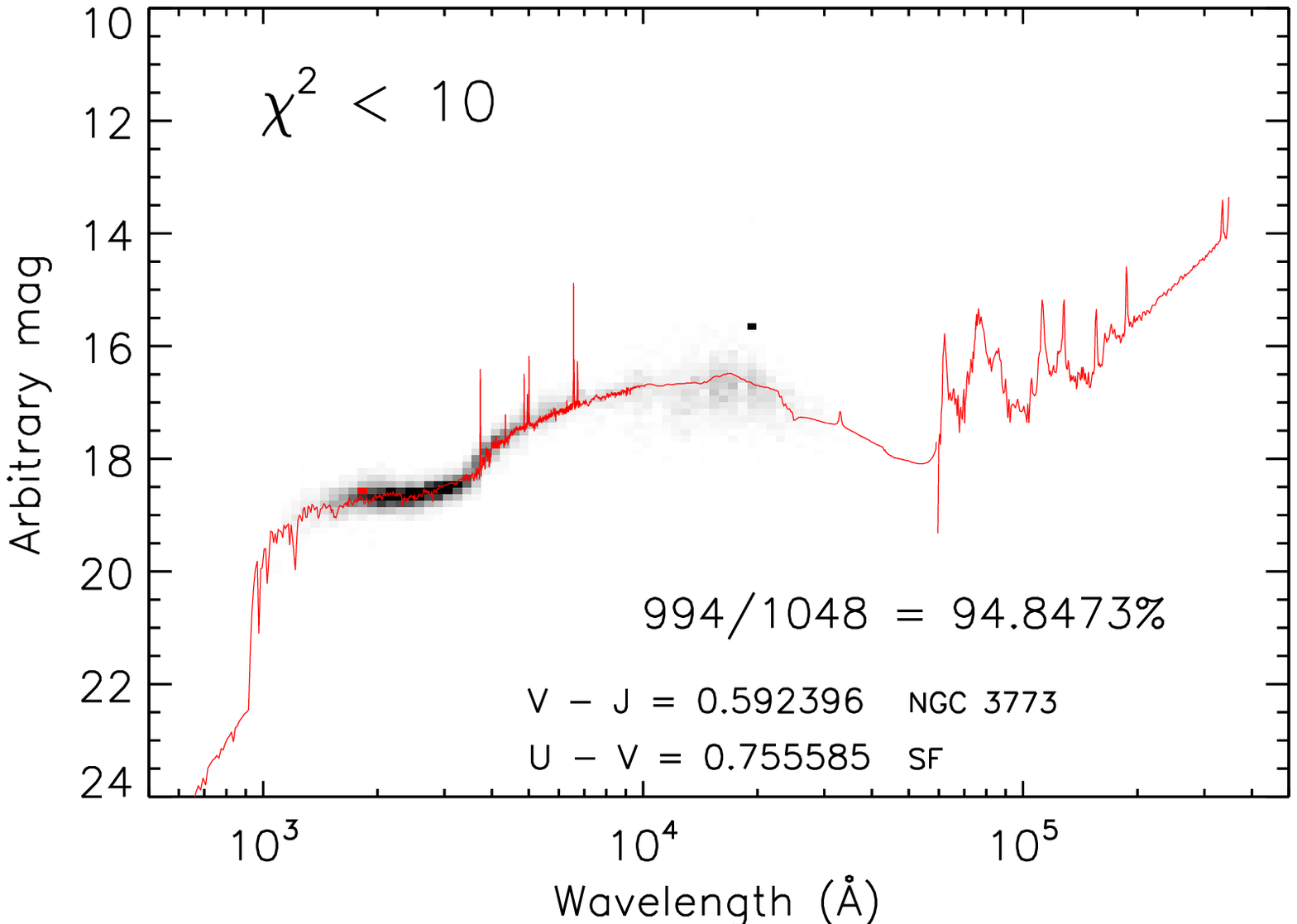}
\includegraphics[width=0.32\textwidth]{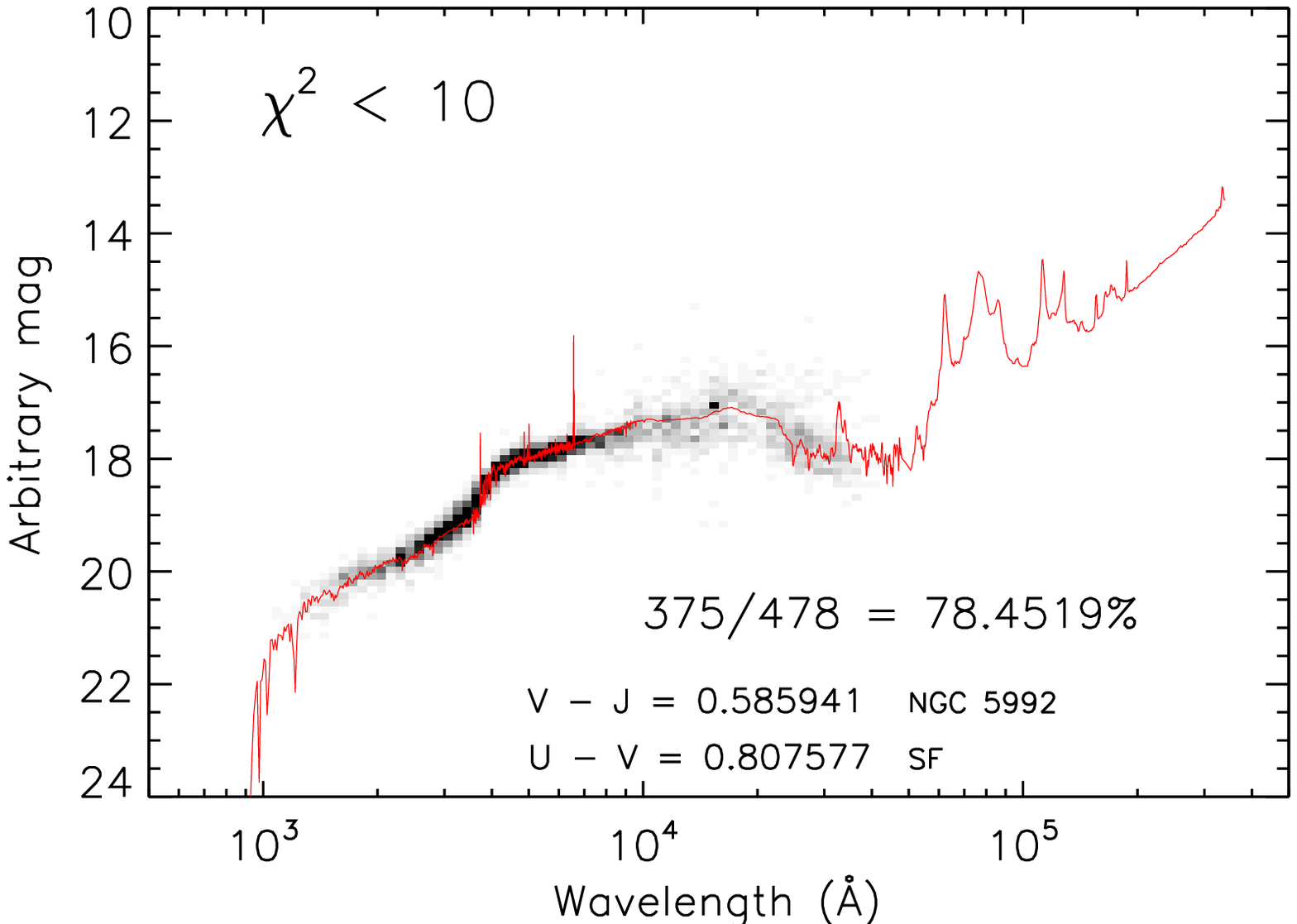}
\includegraphics[width=0.32\textwidth]{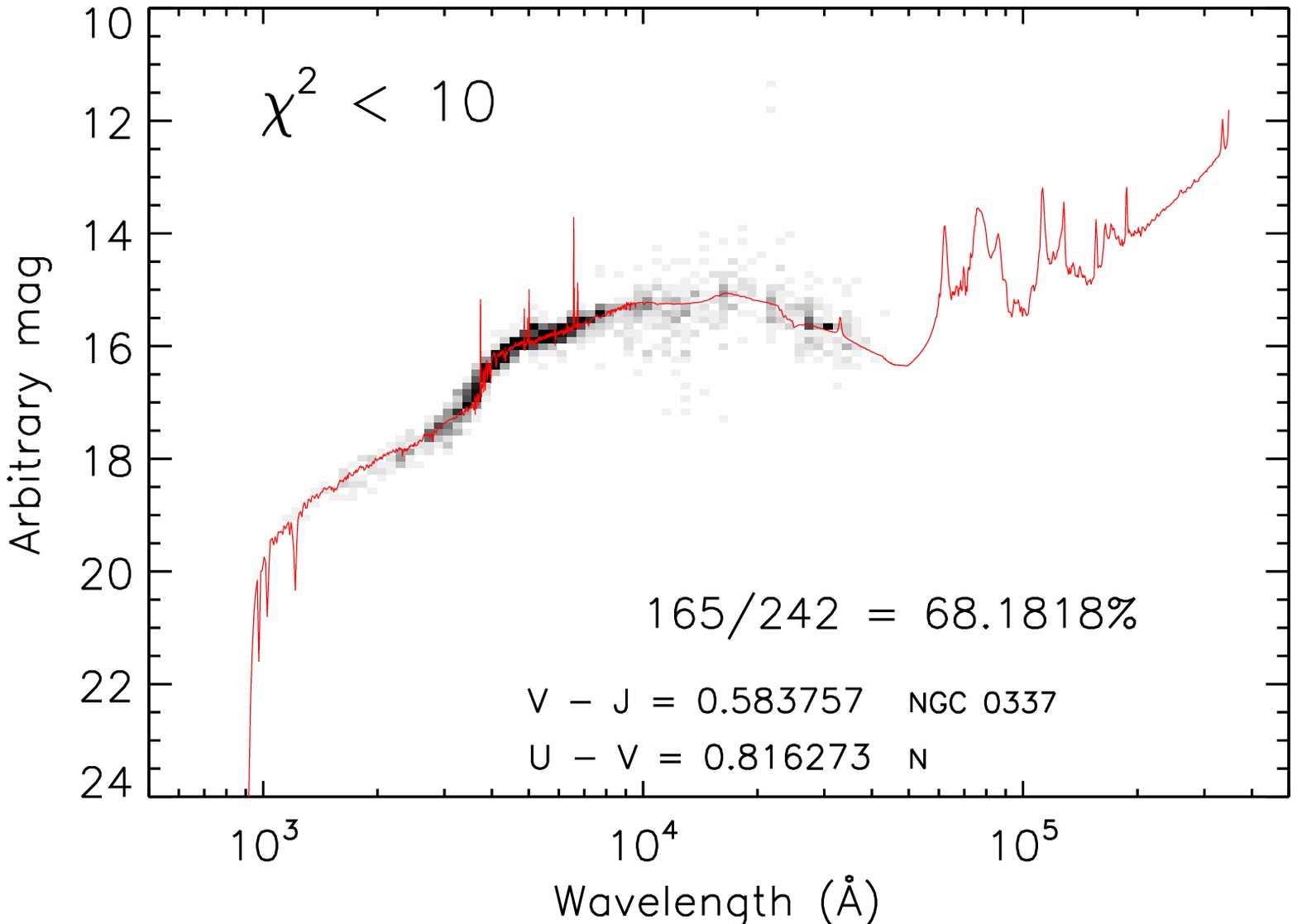}
    \caption{The local galaxy templates and the SEDs with $\chi^2 < 10$.}
    \label{fig:my_label}
\end{figure}

\begin{figure}
    \centering
\includegraphics[width=0.32\textwidth]{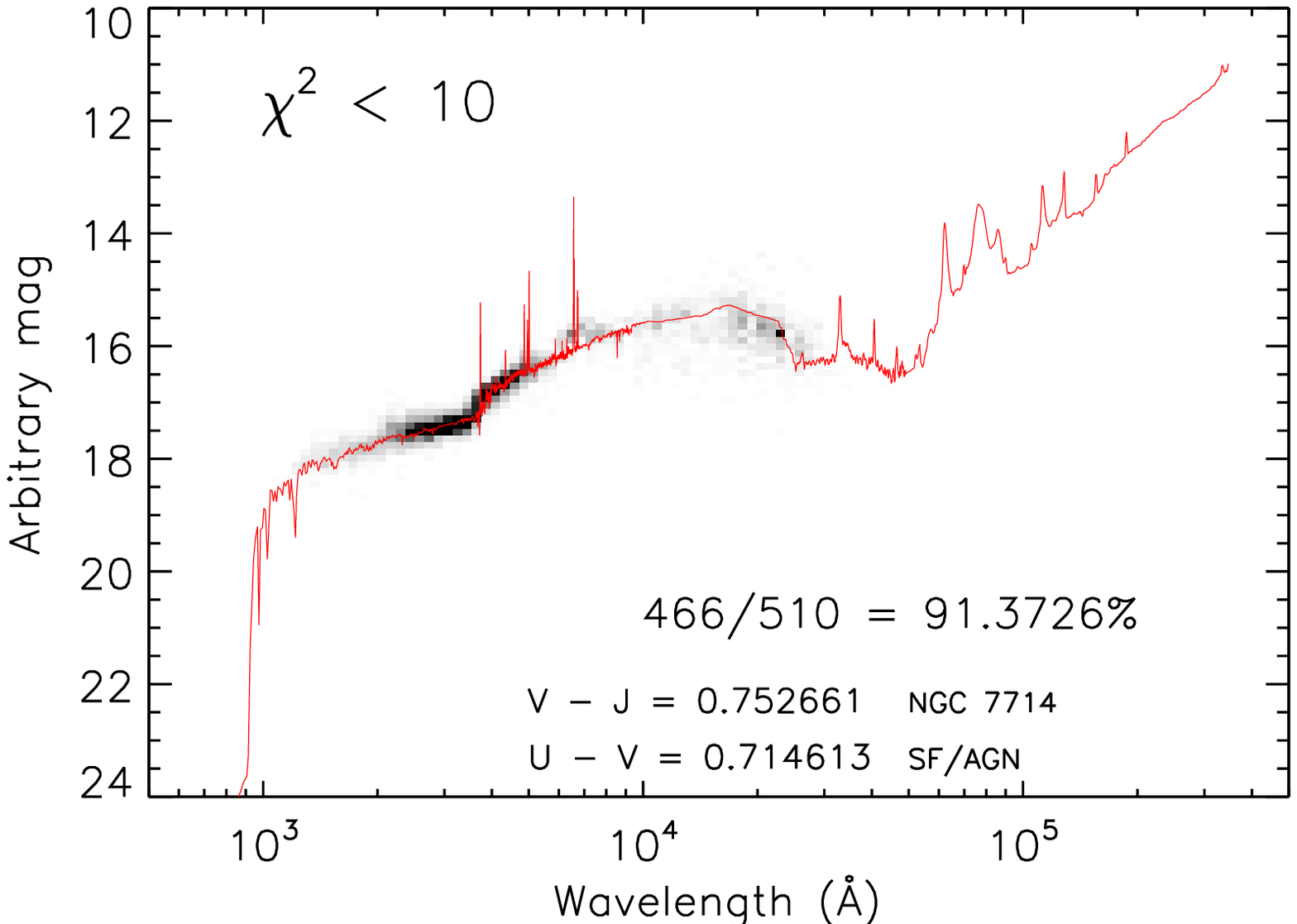}
\includegraphics[width=0.32\textwidth]{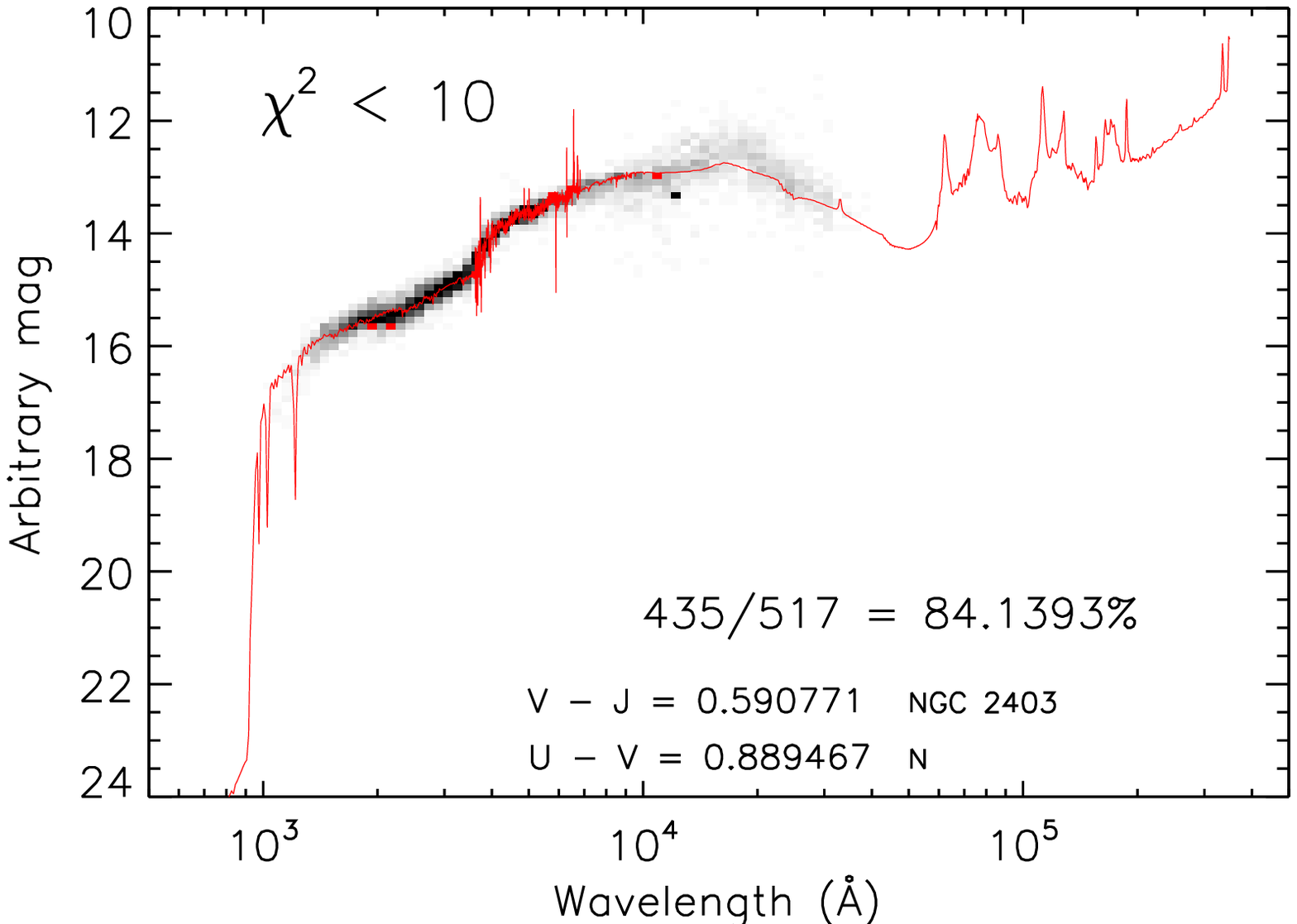}
\includegraphics[width=0.32\textwidth]{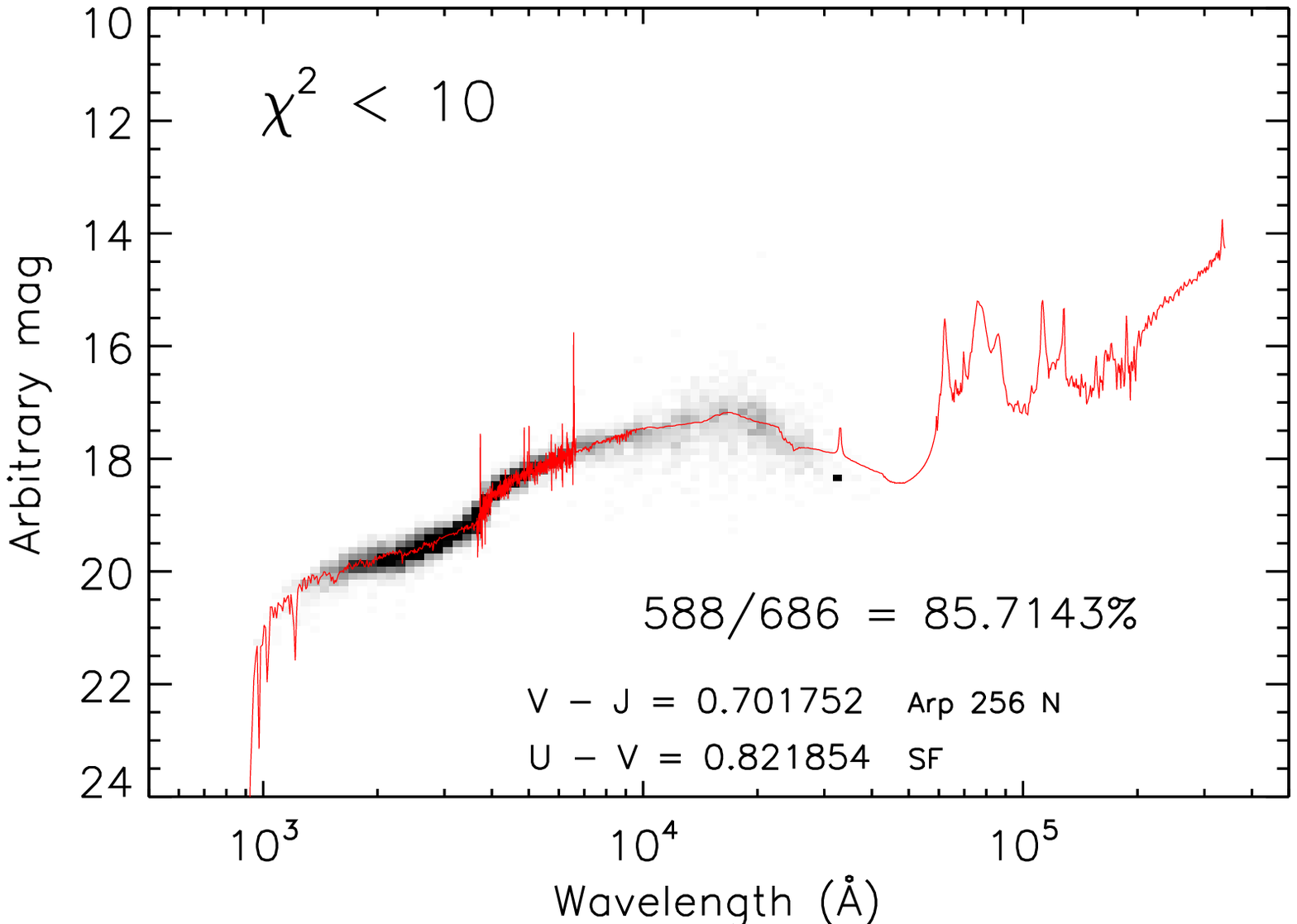}
\includegraphics[width=0.32\textwidth]{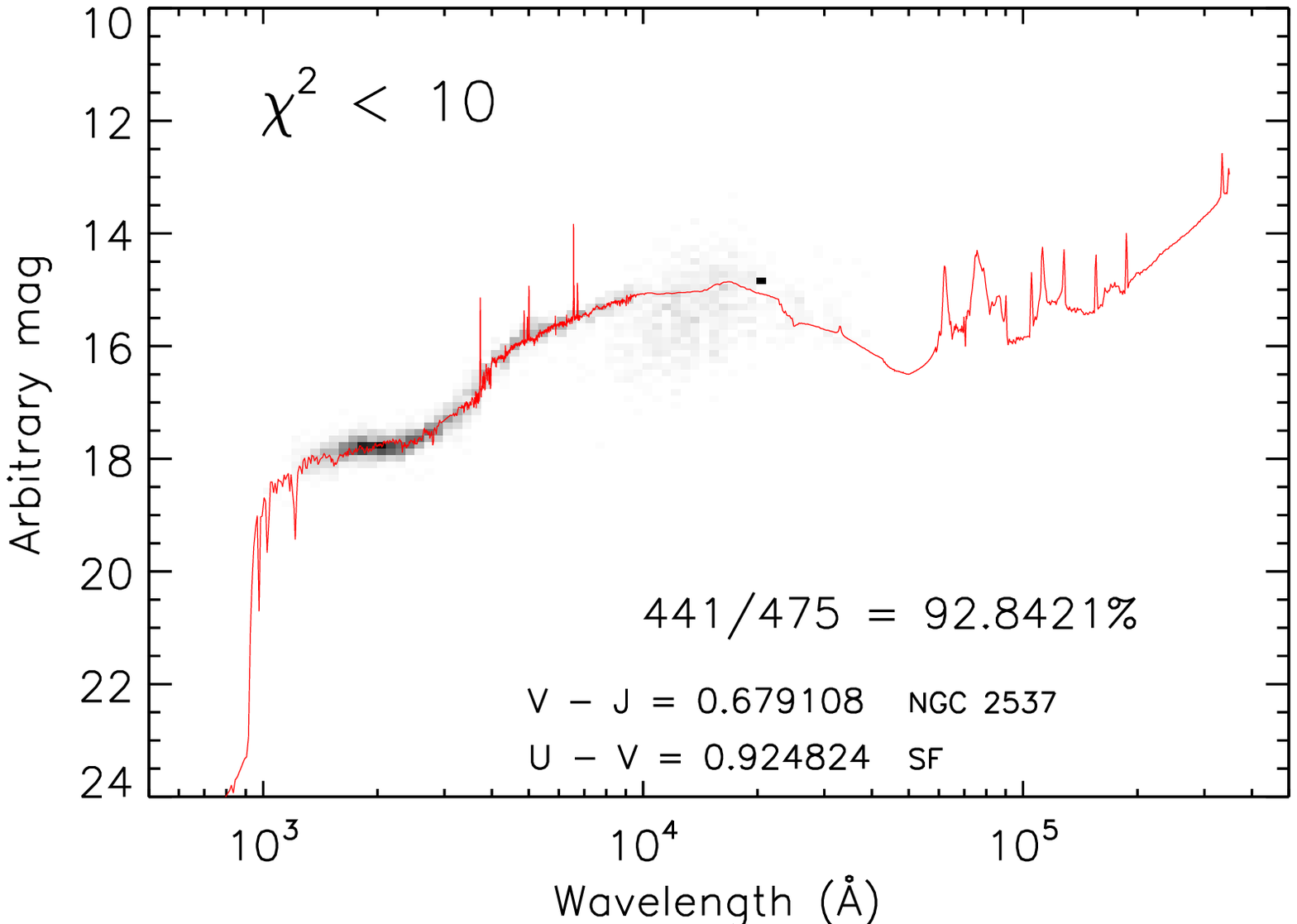}
\includegraphics[width=0.32\textwidth]{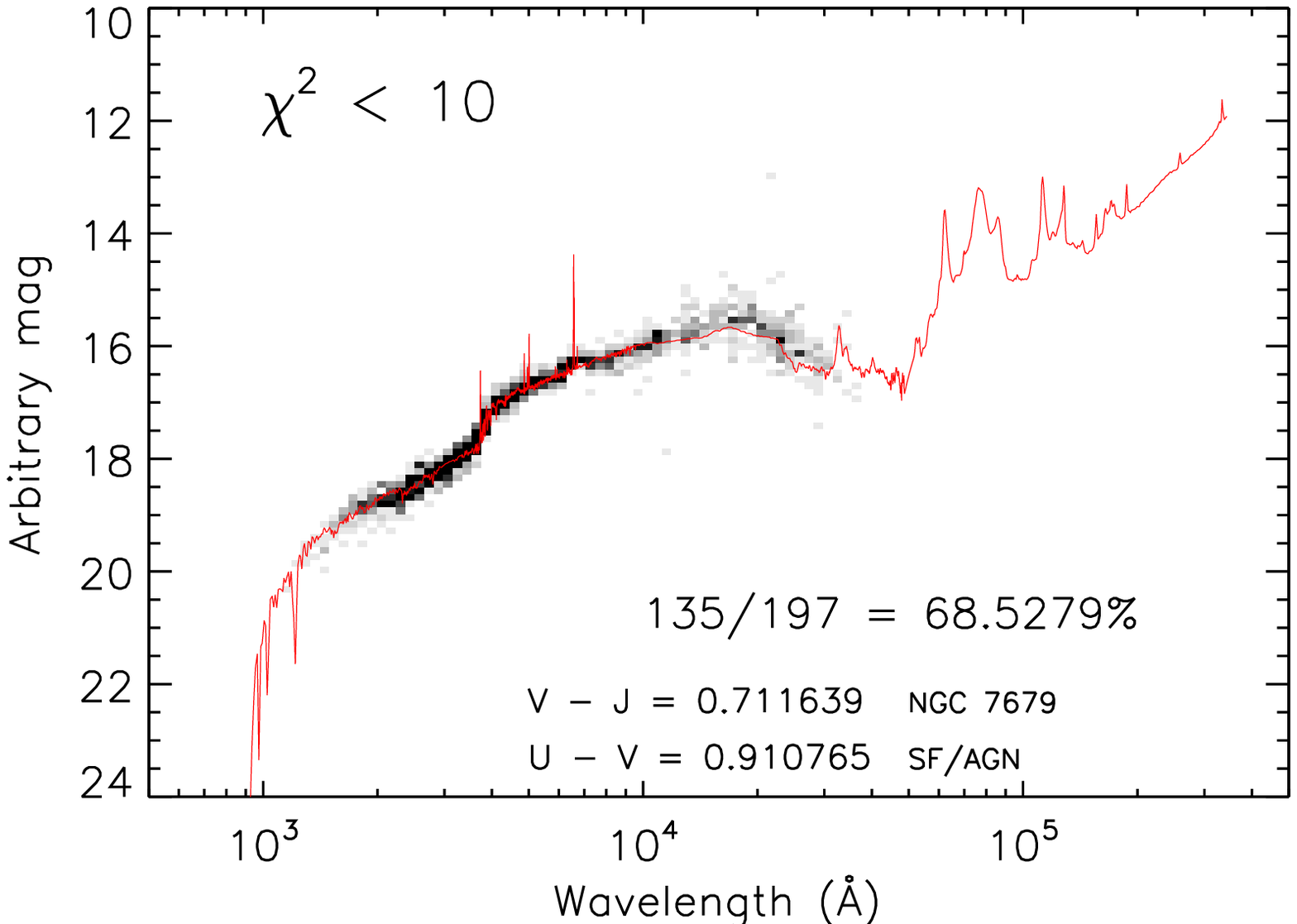}
\includegraphics[width=0.32\textwidth]{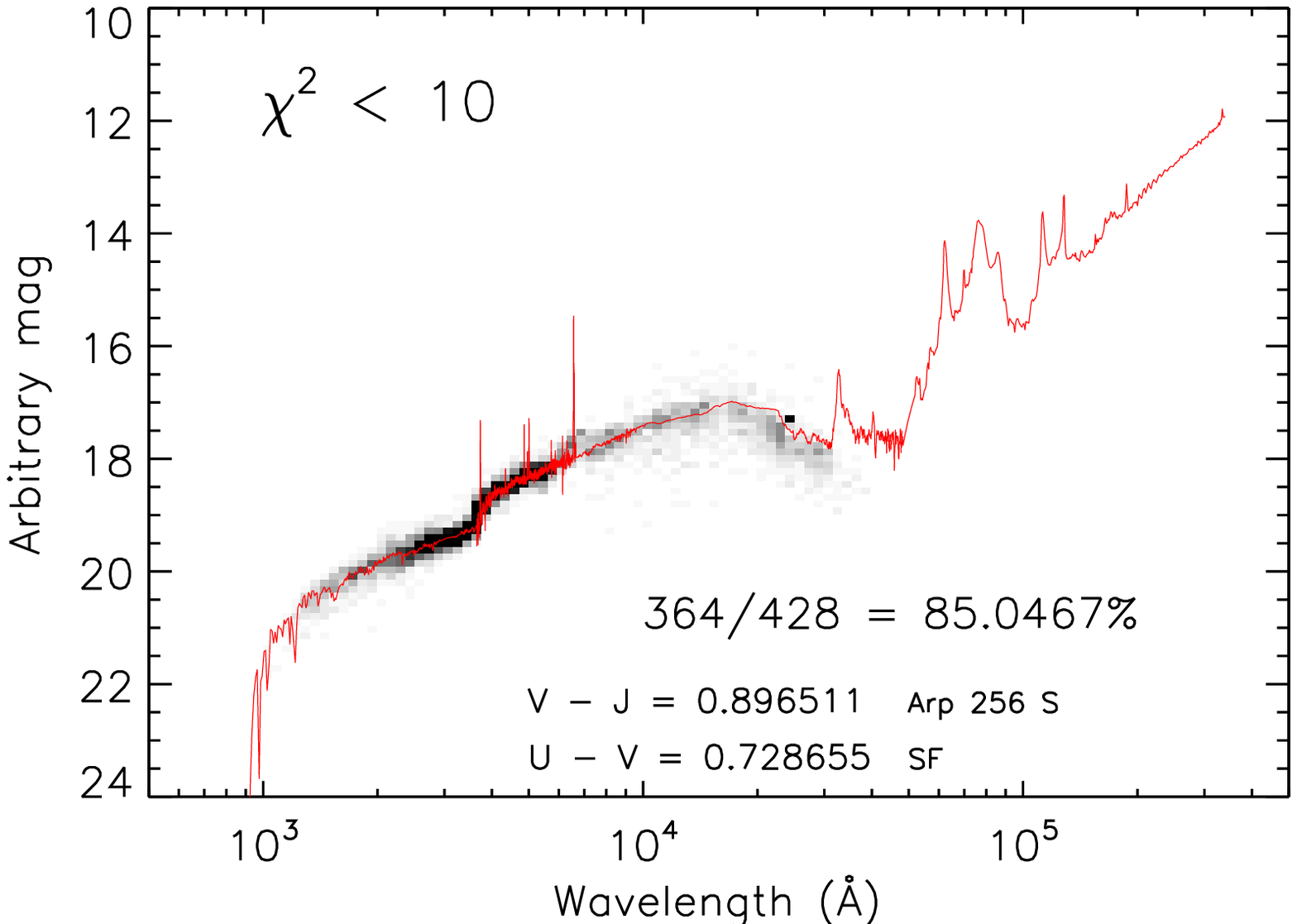}
\includegraphics[width=0.32\textwidth]{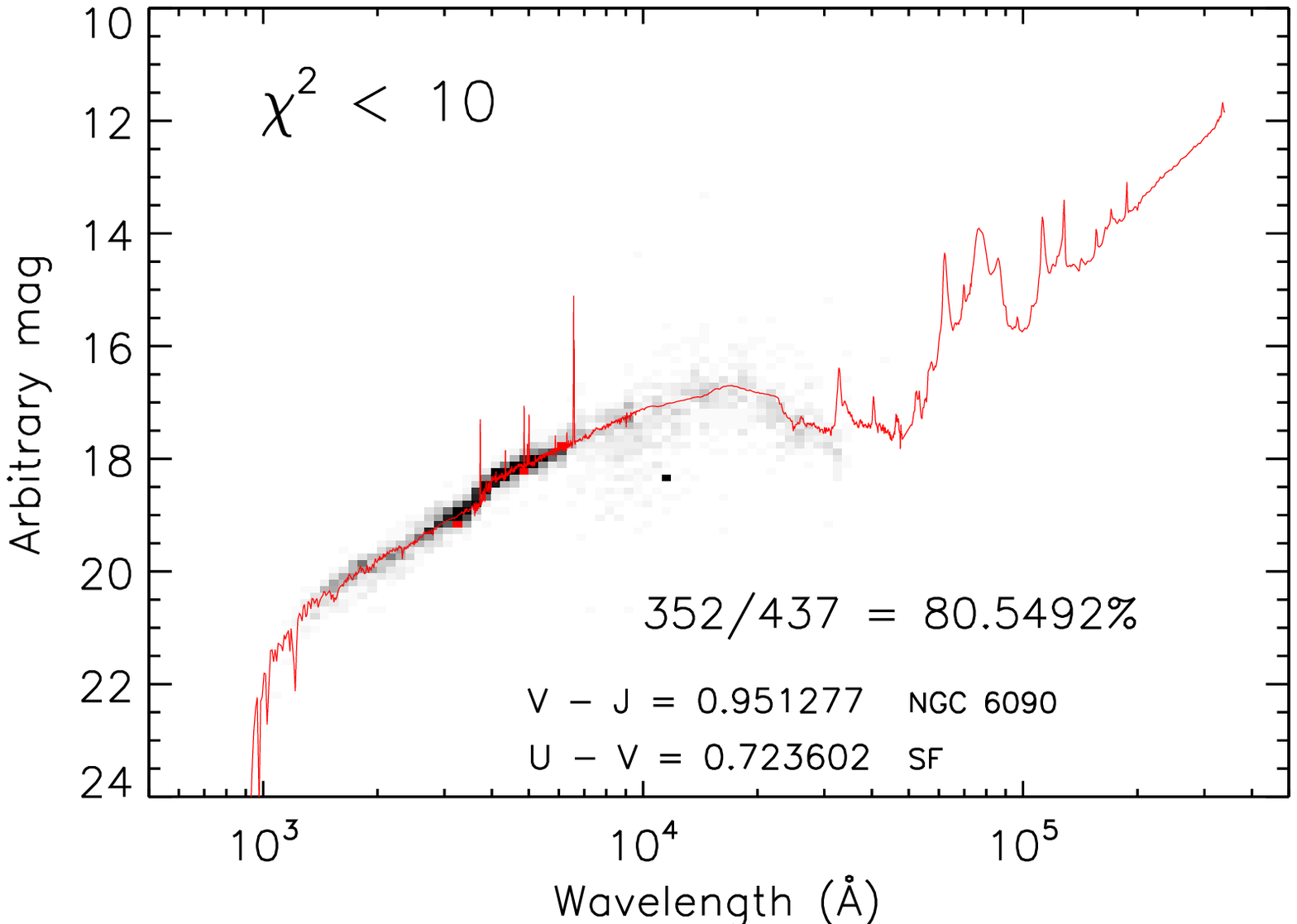}
\includegraphics[width=0.32\textwidth]{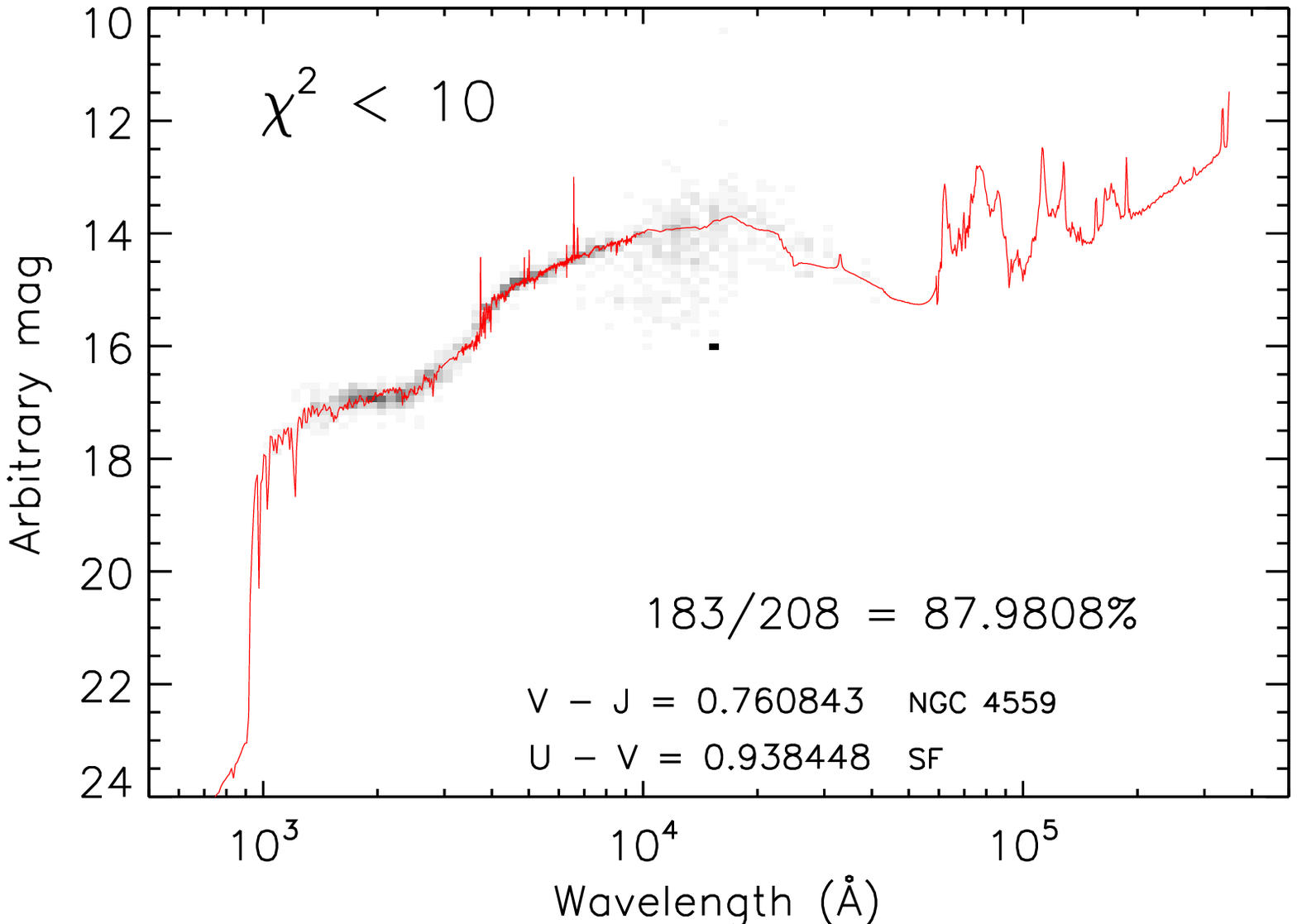}
\includegraphics[width=0.32\textwidth]{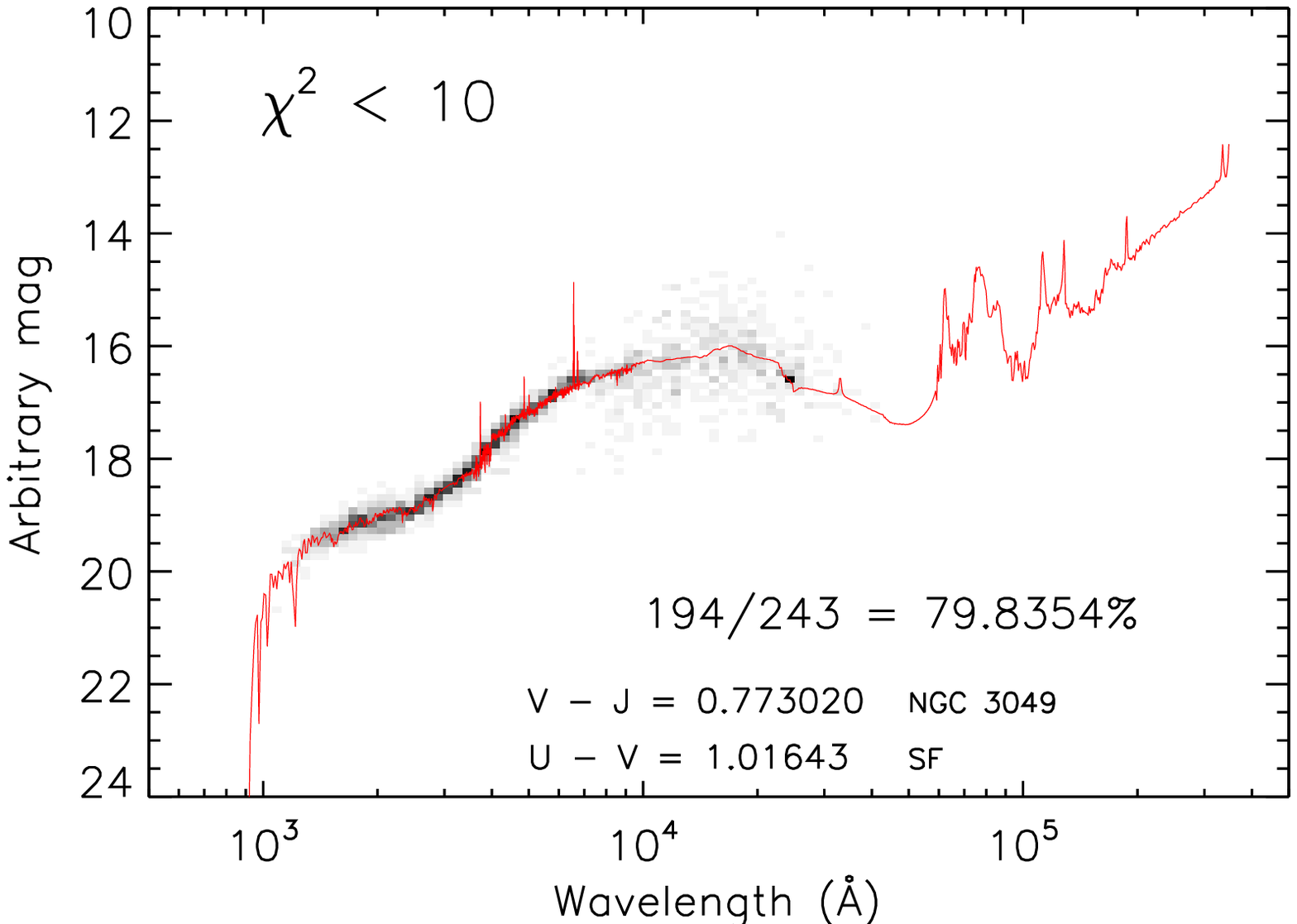}
\includegraphics[width=0.32\textwidth]{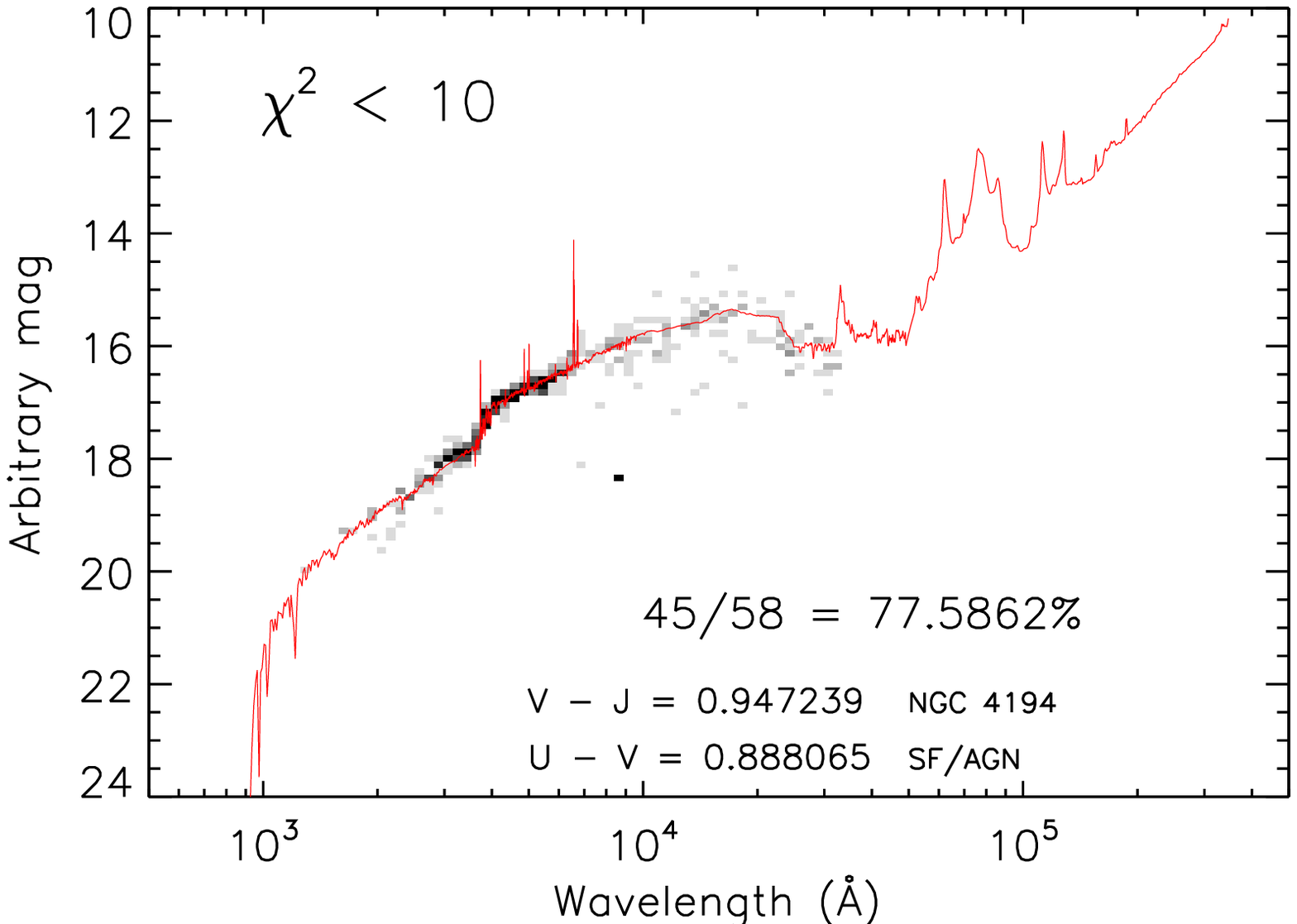}
\includegraphics[width=0.32\textwidth]{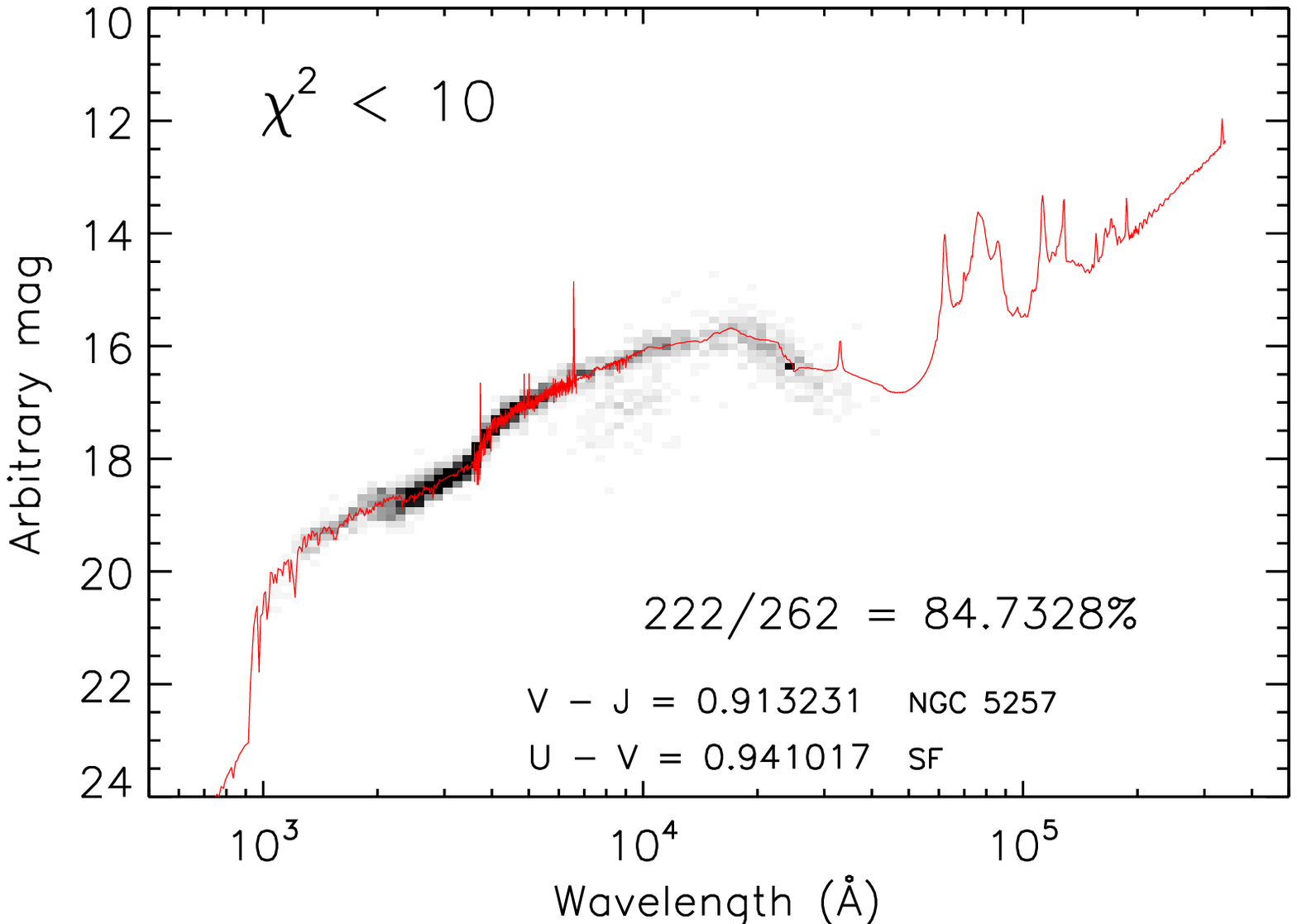}
\includegraphics[width=0.32\textwidth]{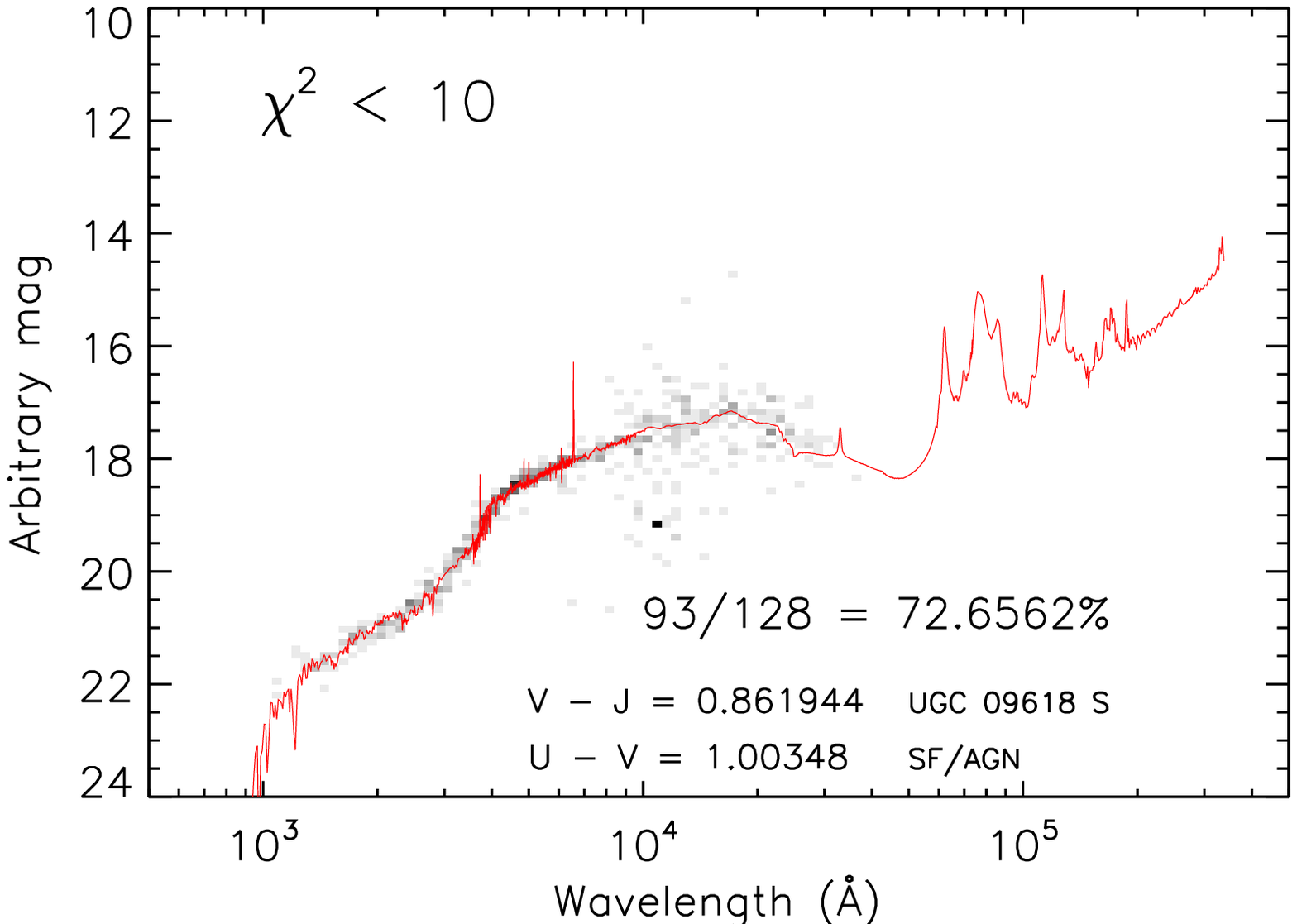}
\includegraphics[width=0.32\textwidth]{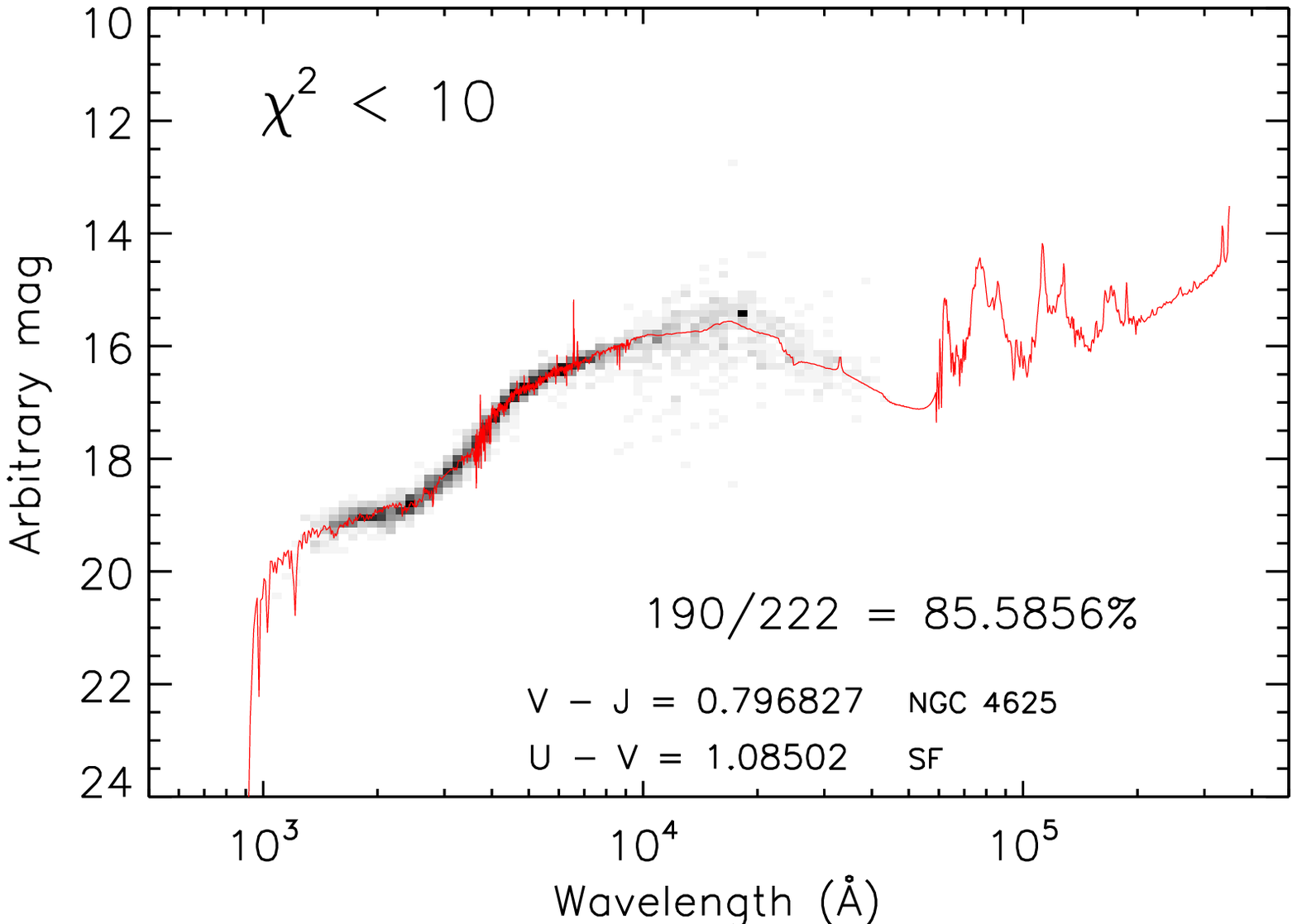}
\includegraphics[width=0.32\textwidth]{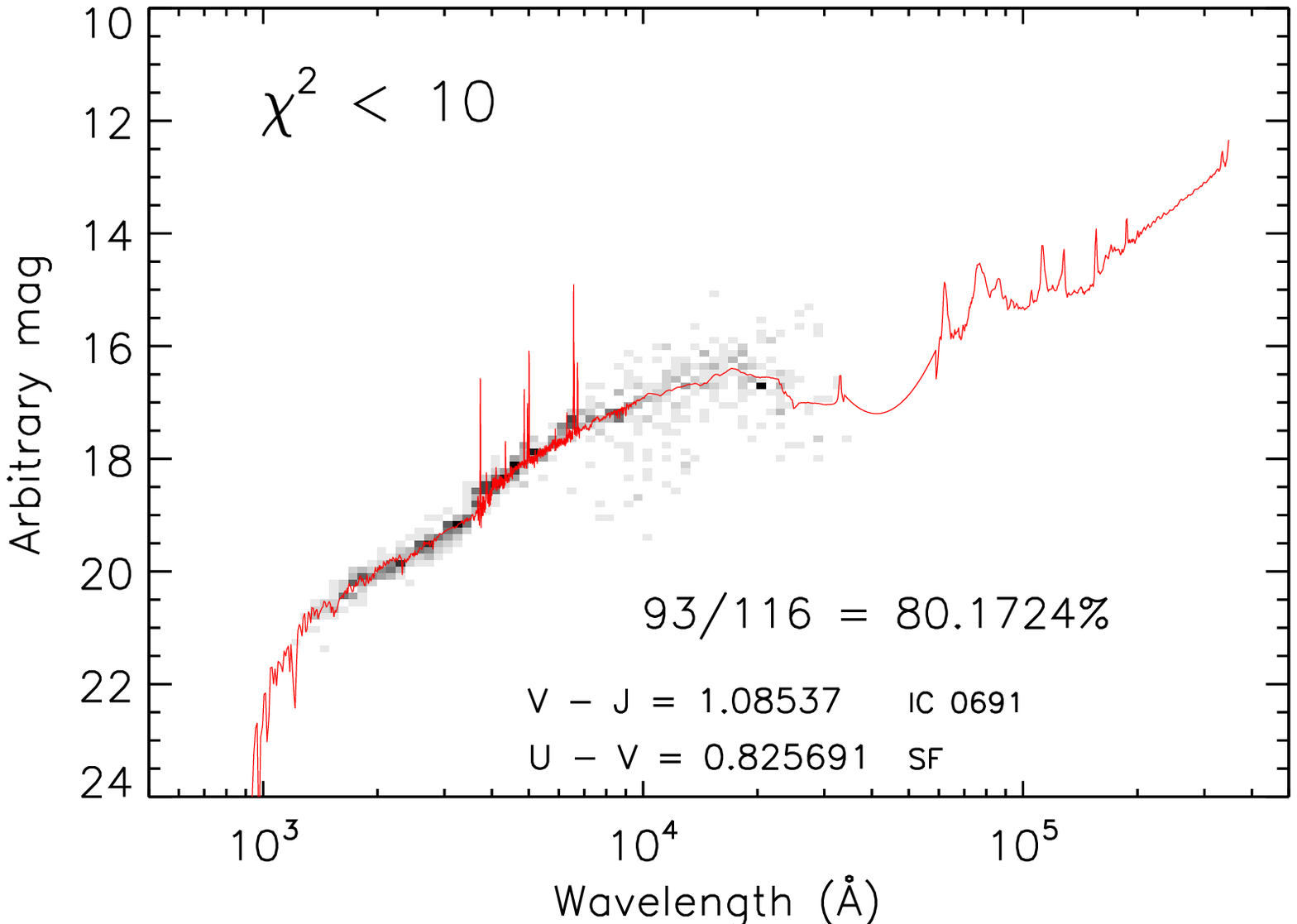}
\includegraphics[width=0.32\textwidth]{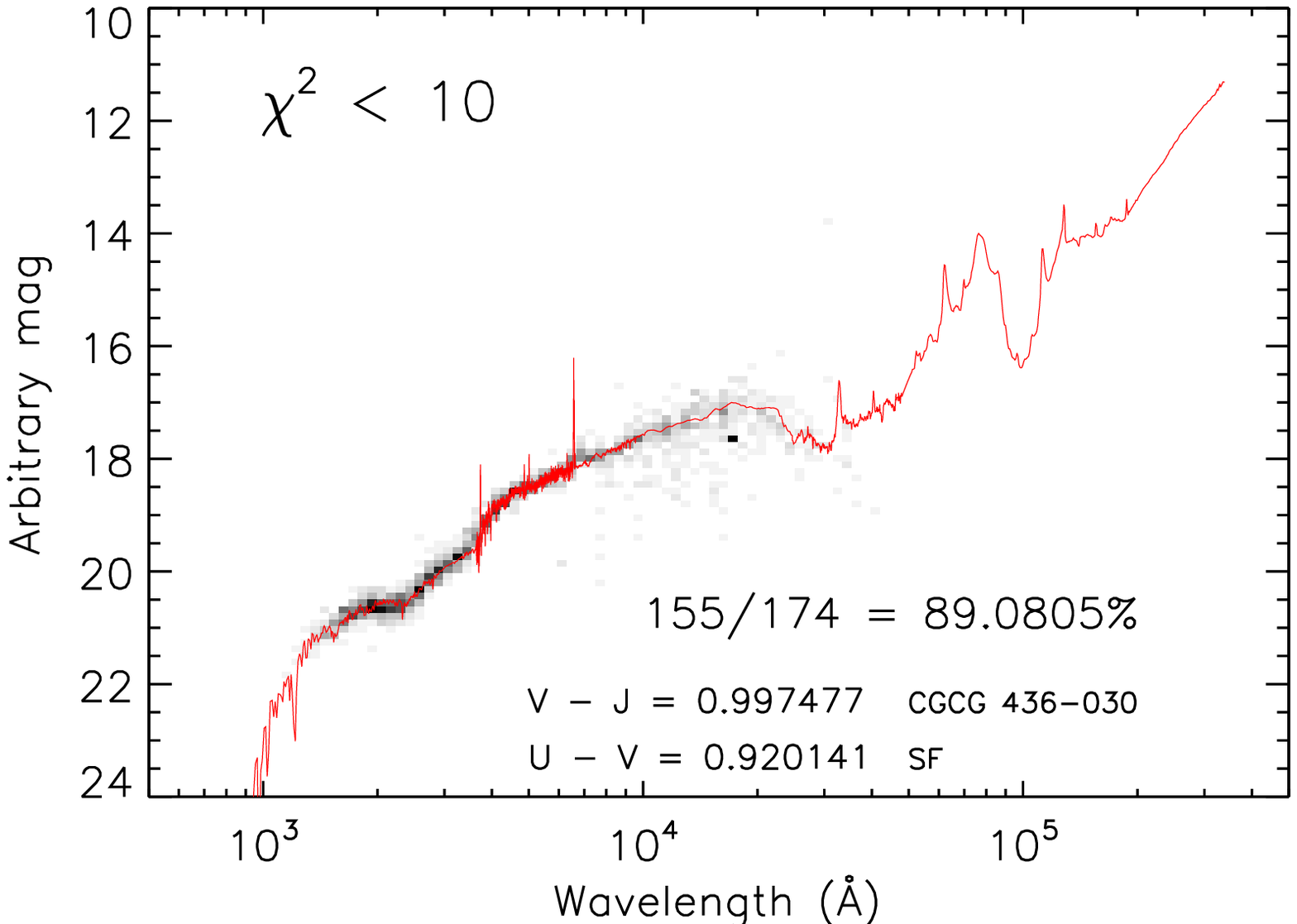}
\includegraphics[width=0.32\textwidth]{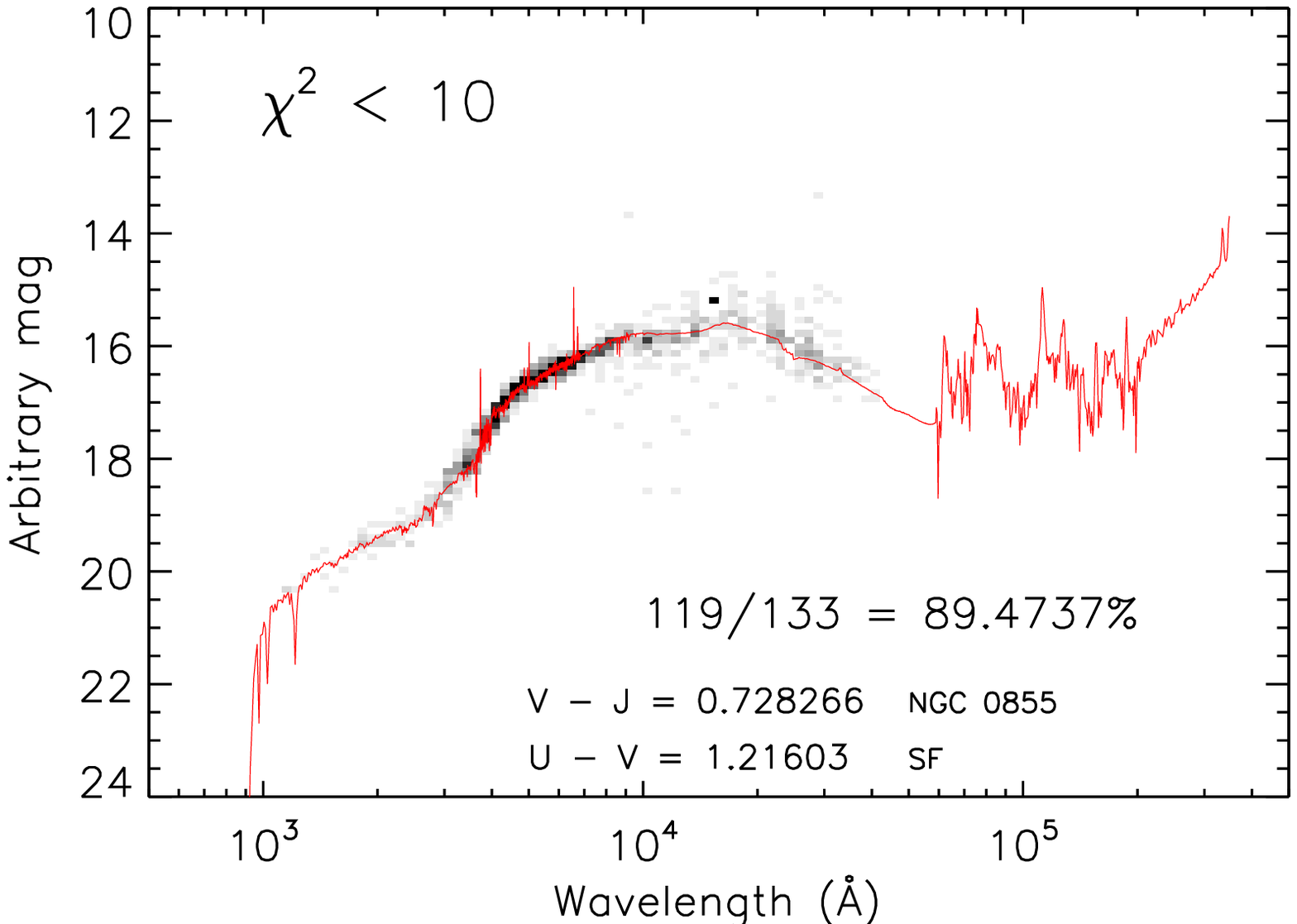}
\includegraphics[width=0.32\textwidth]{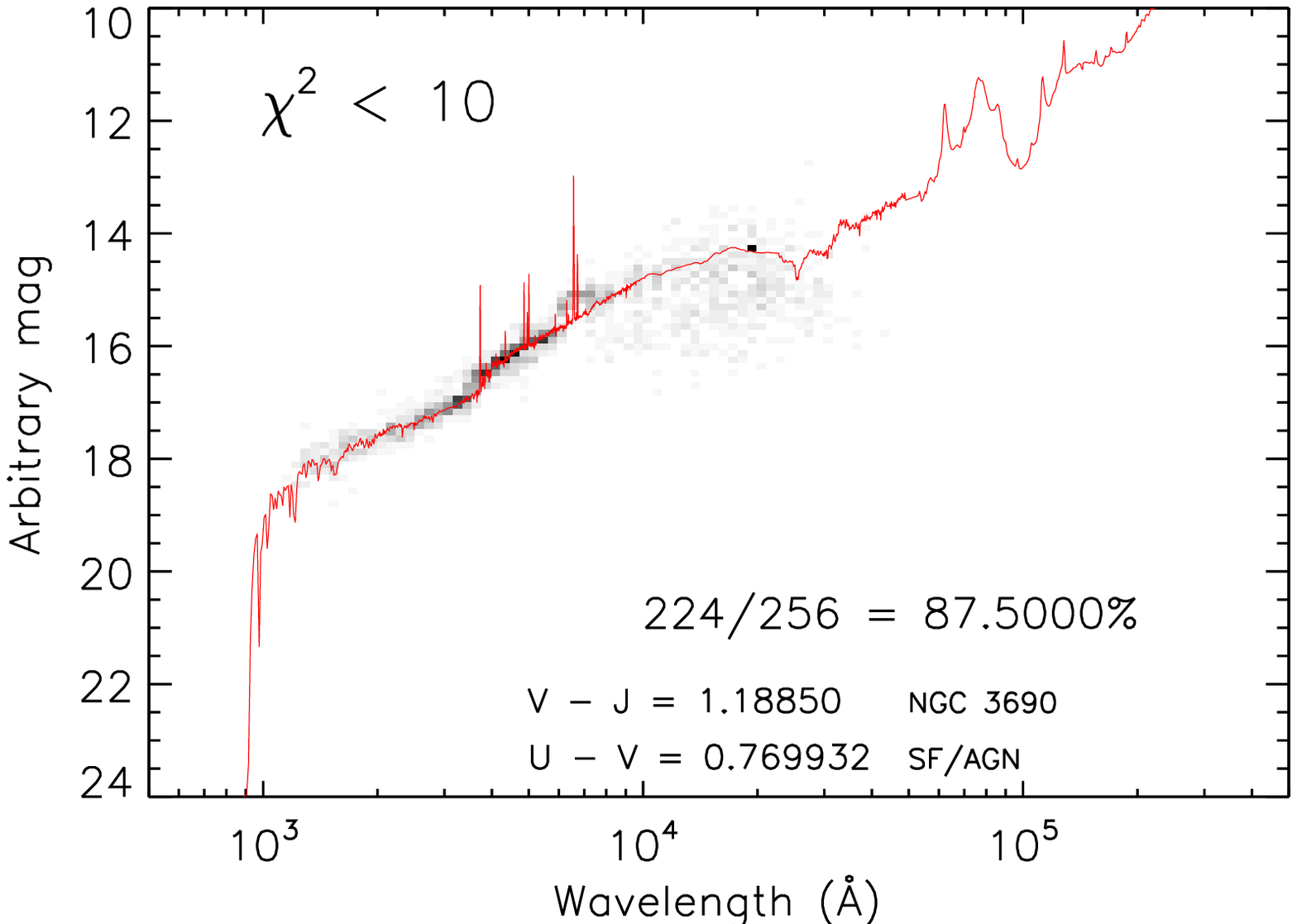}
\includegraphics[width=0.32\textwidth]{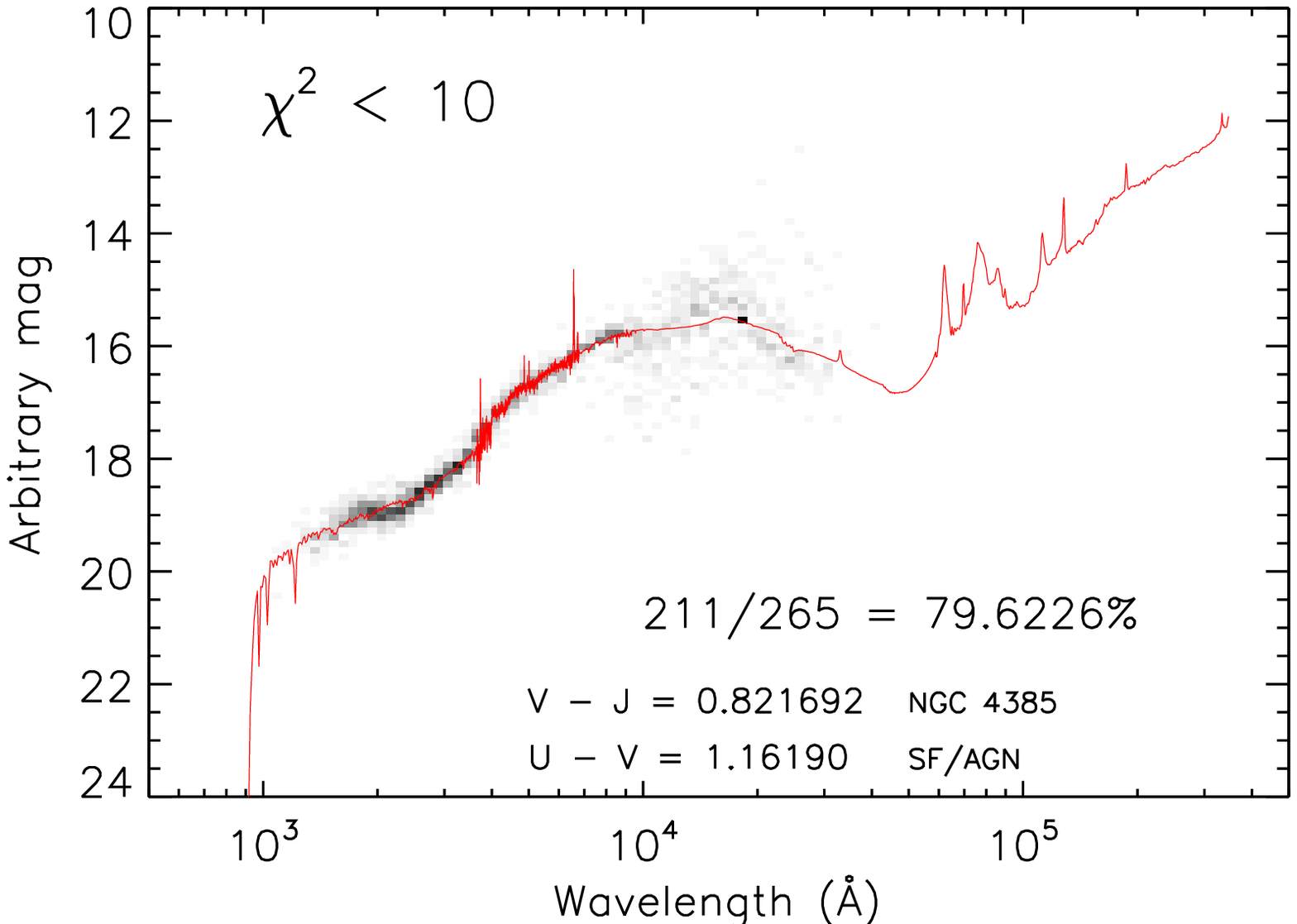}
\includegraphics[width=0.32\textwidth]{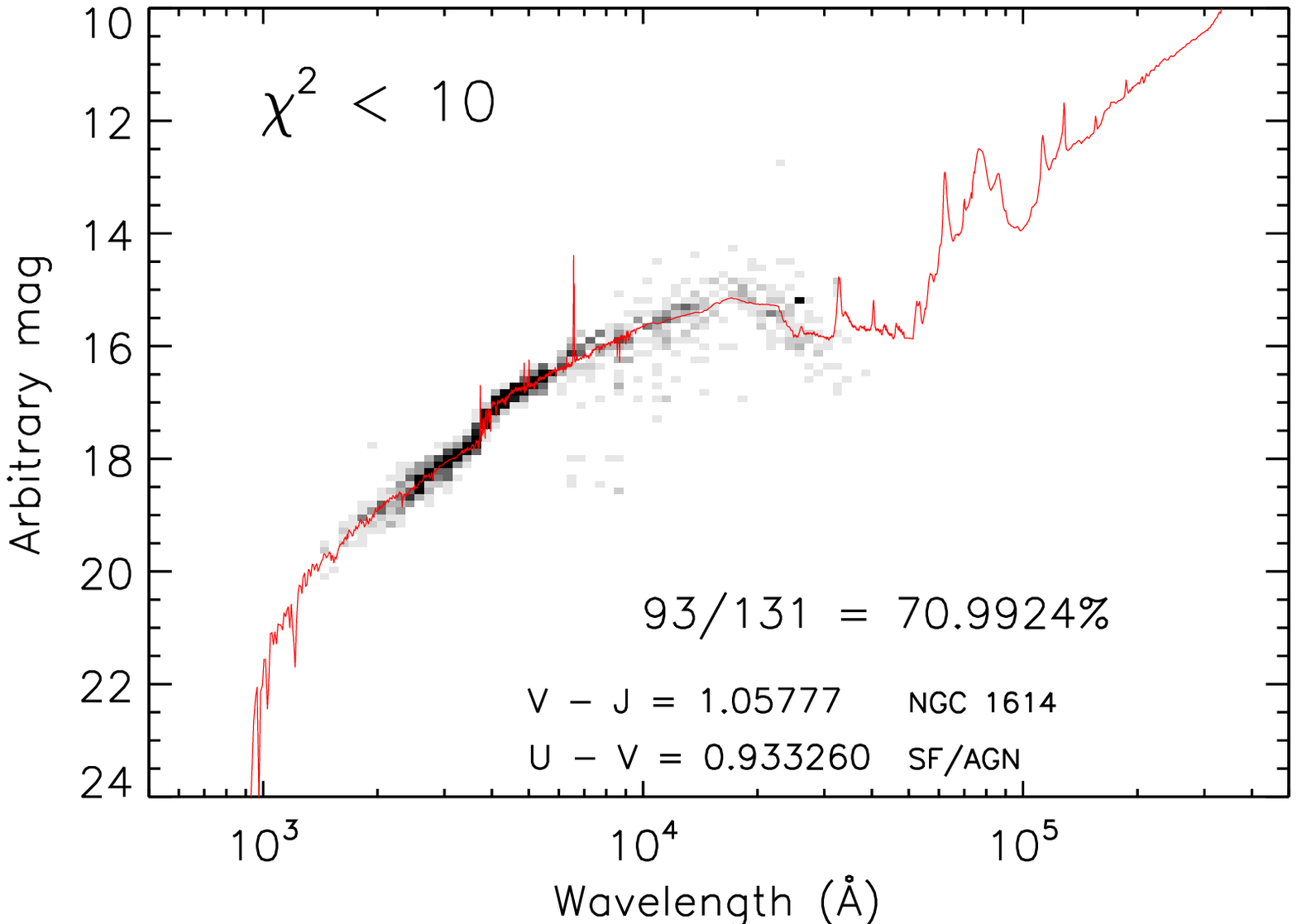}
\includegraphics[width=0.32\textwidth]{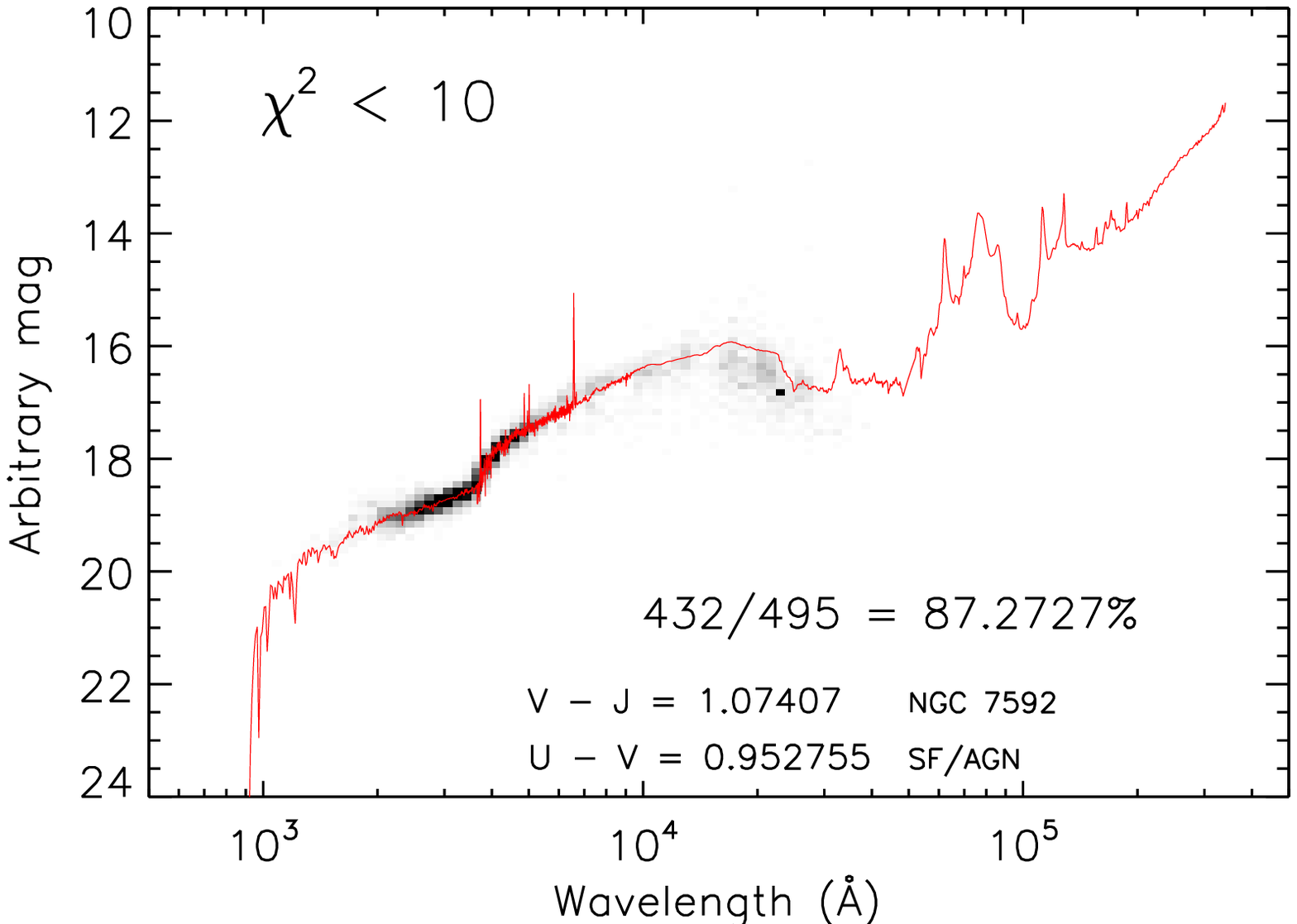}
\includegraphics[width=0.32\textwidth]{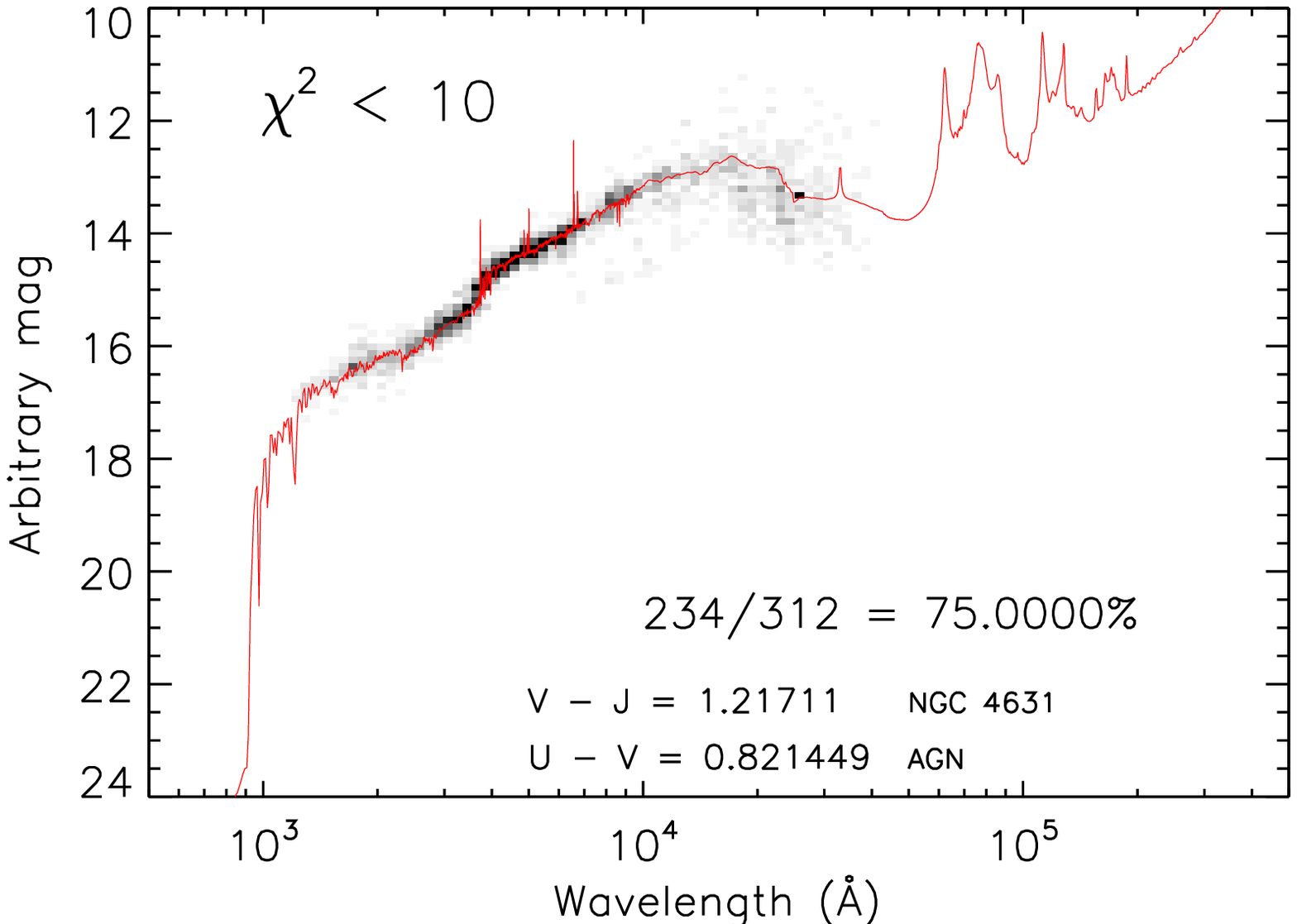}
    \caption{Continued.}
    \label{fig:my_label}
\end{figure}

\begin{figure}
    \centering
\includegraphics[width=0.32\textwidth]{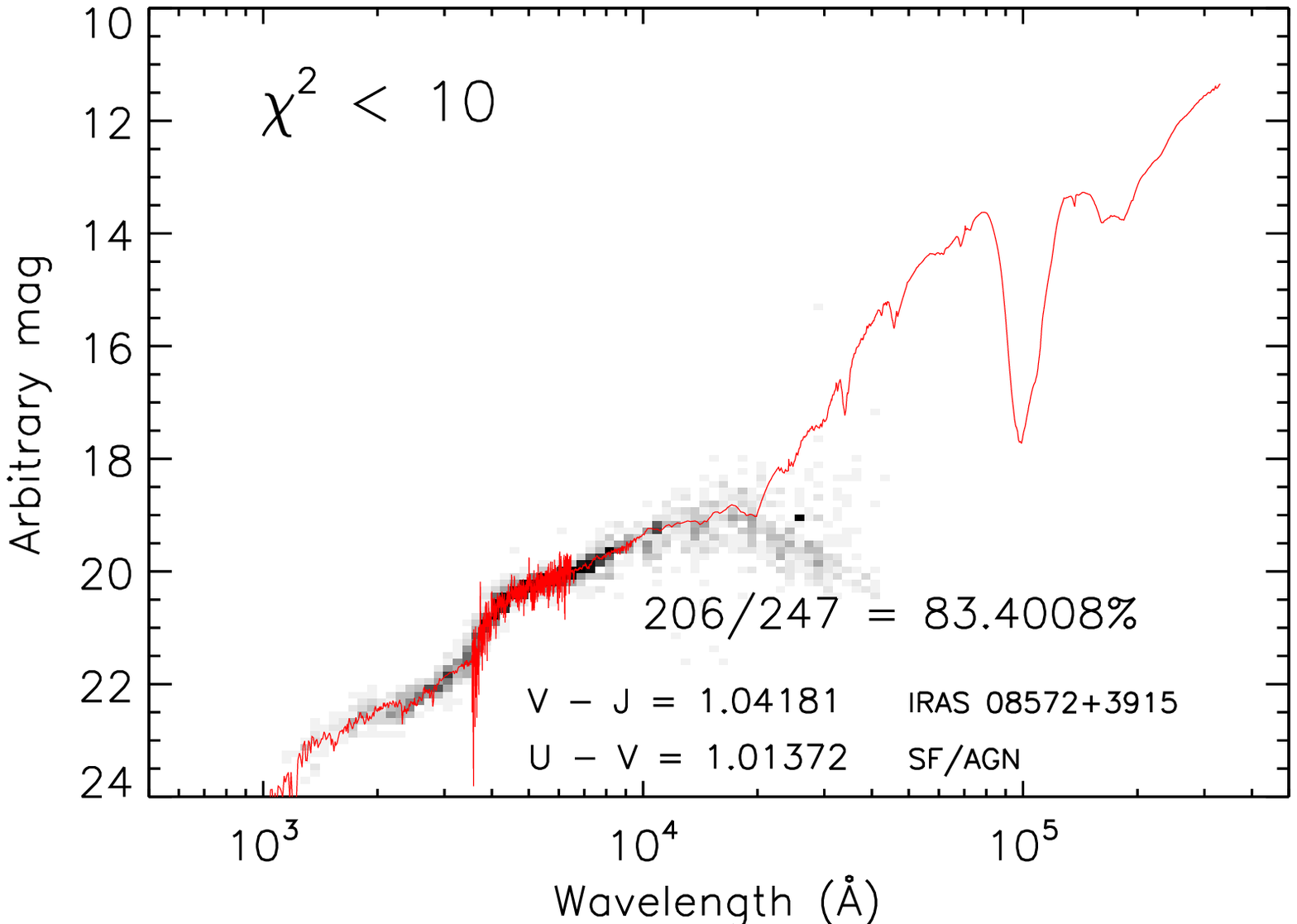}
\includegraphics[width=0.32\textwidth]{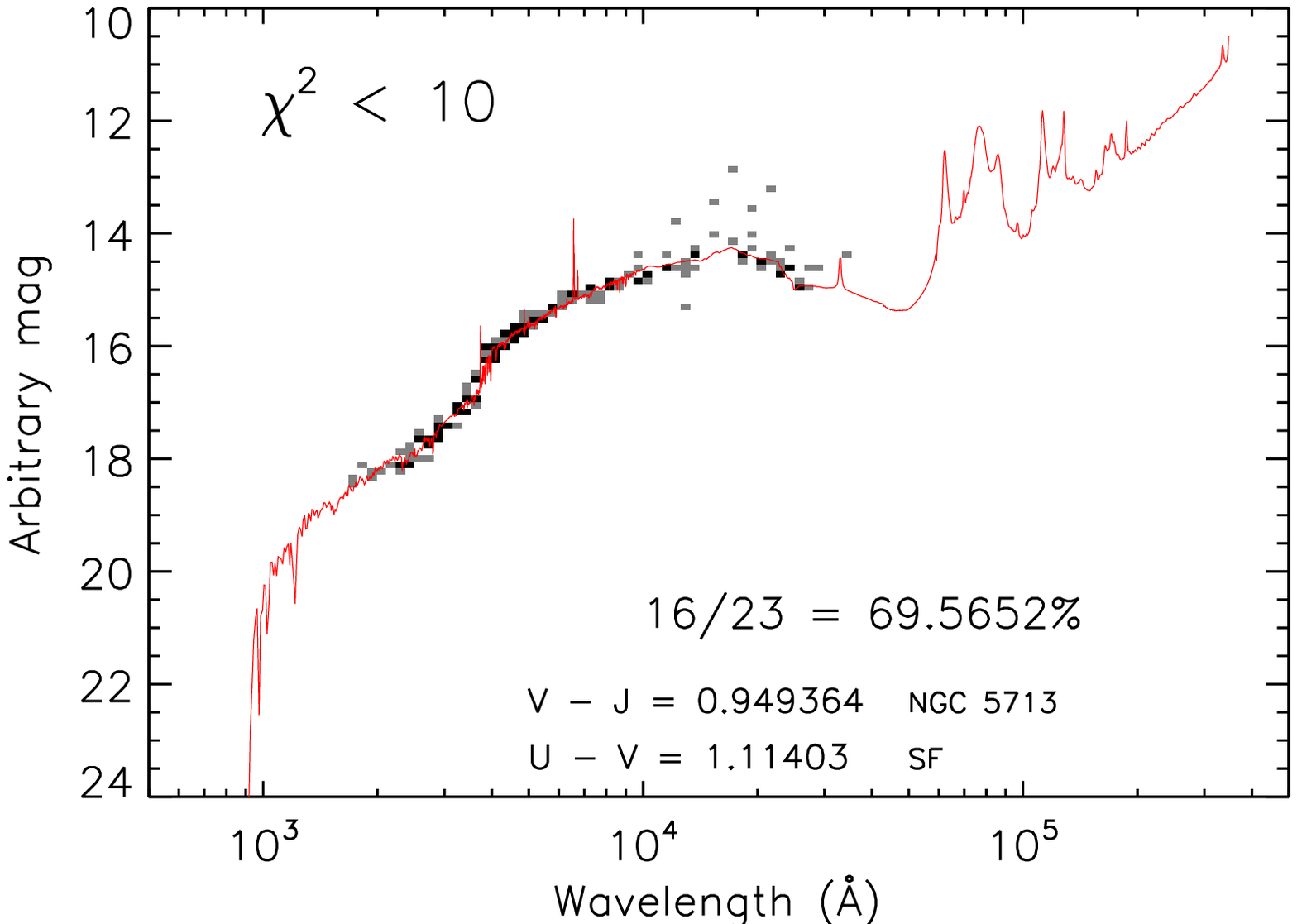}
\includegraphics[width=0.32\textwidth]{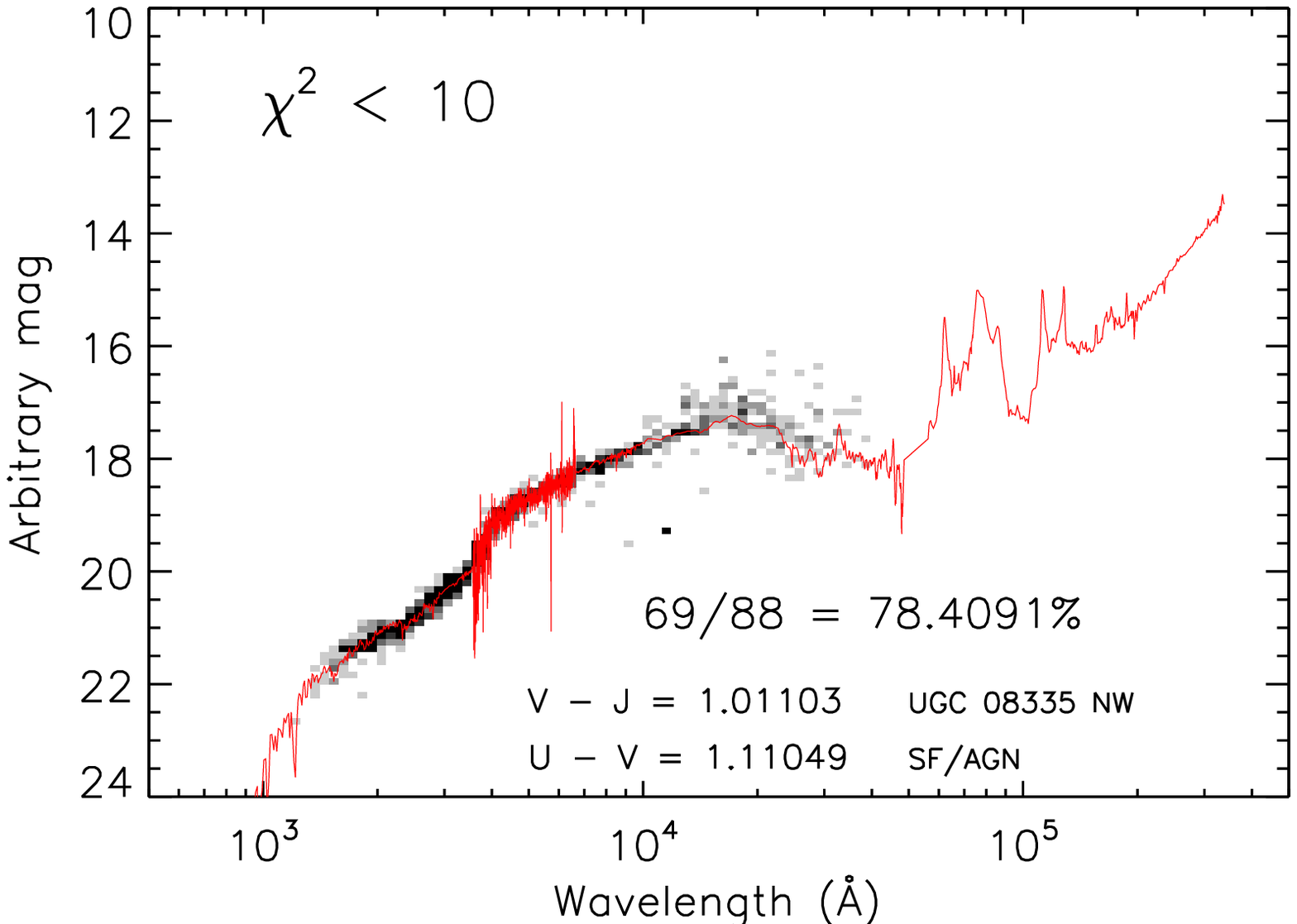}
\includegraphics[width=0.32\textwidth]{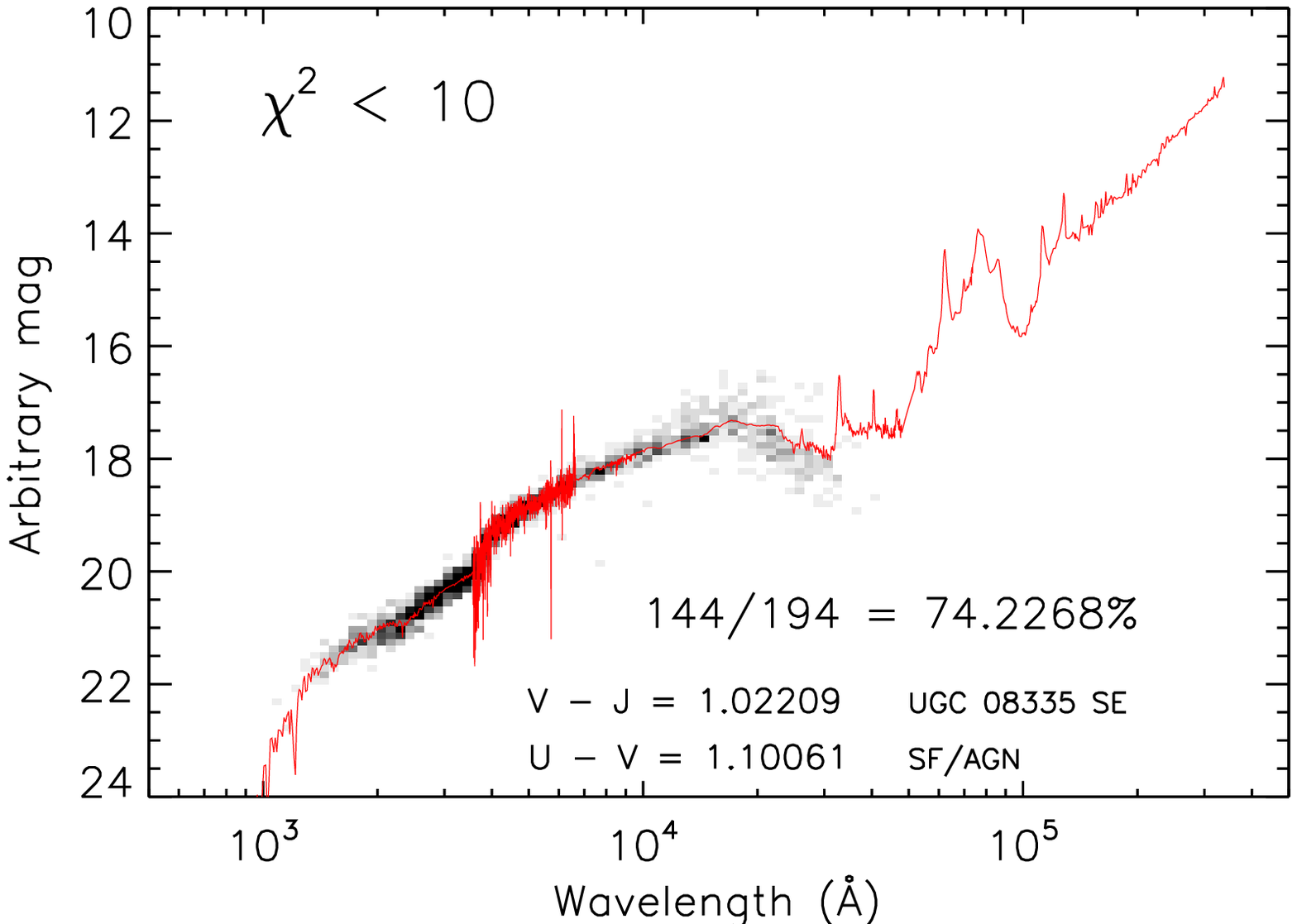}
\includegraphics[width=0.32\textwidth]{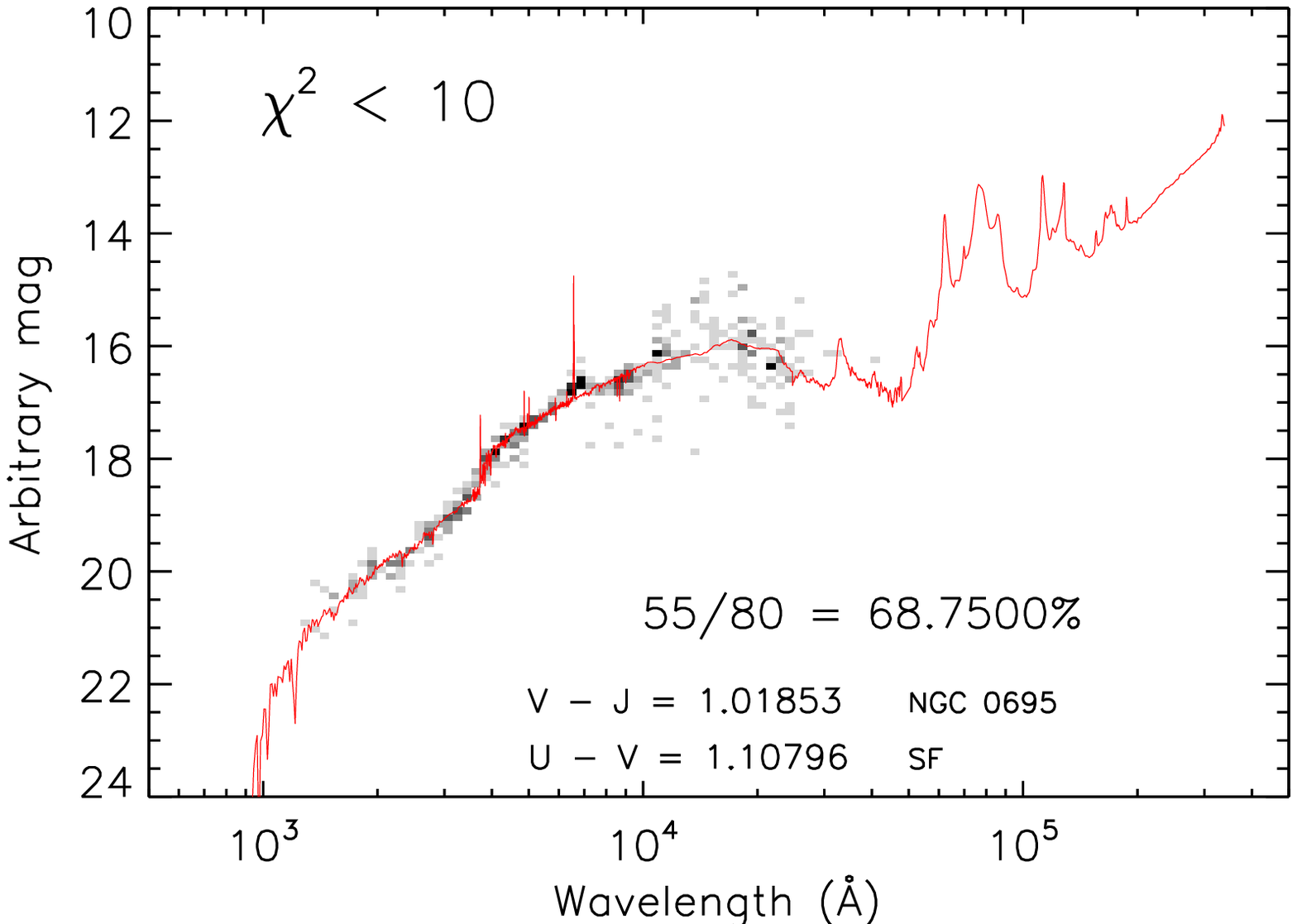}
\includegraphics[width=0.32\textwidth]{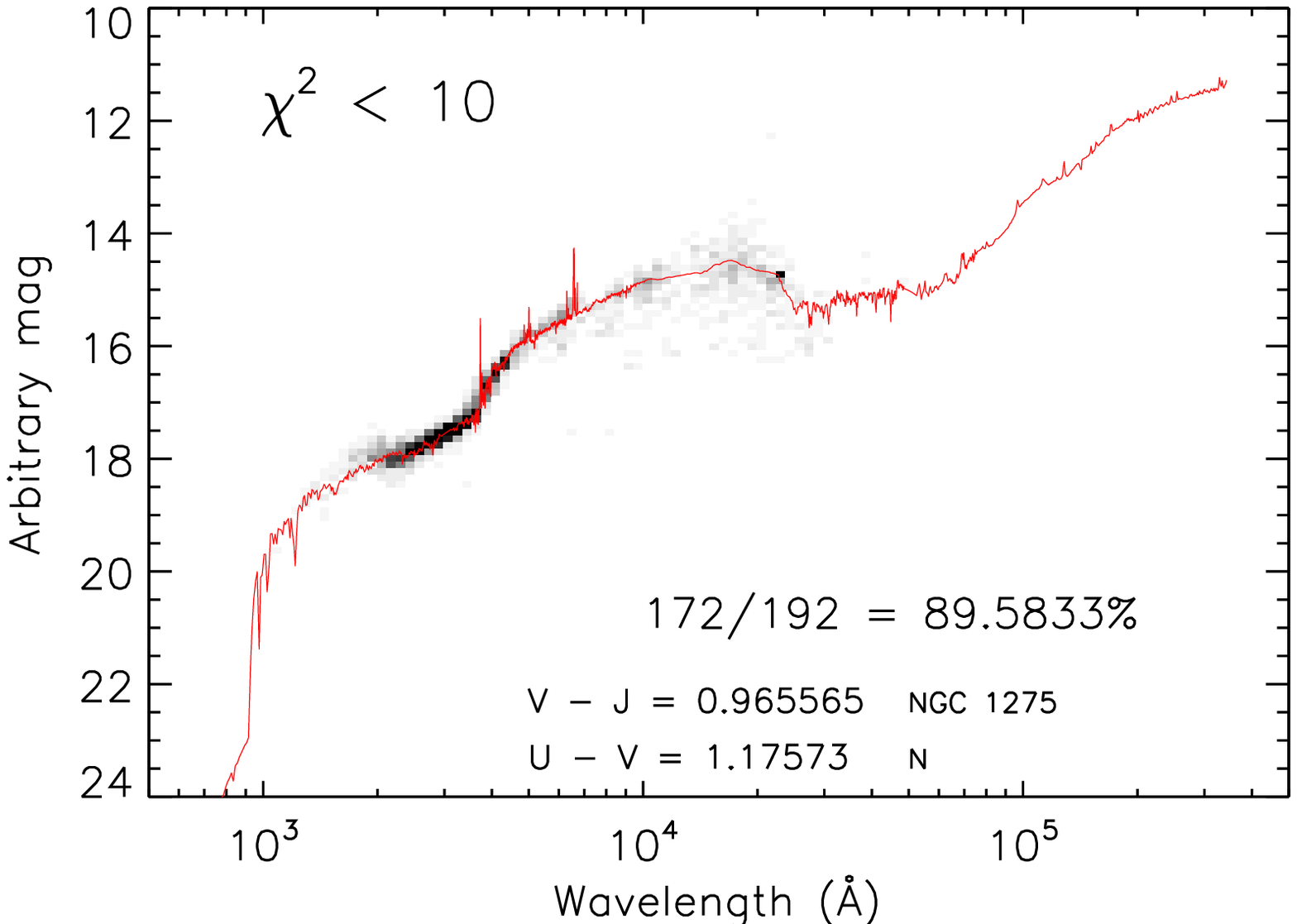}
\includegraphics[width=0.32\textwidth]{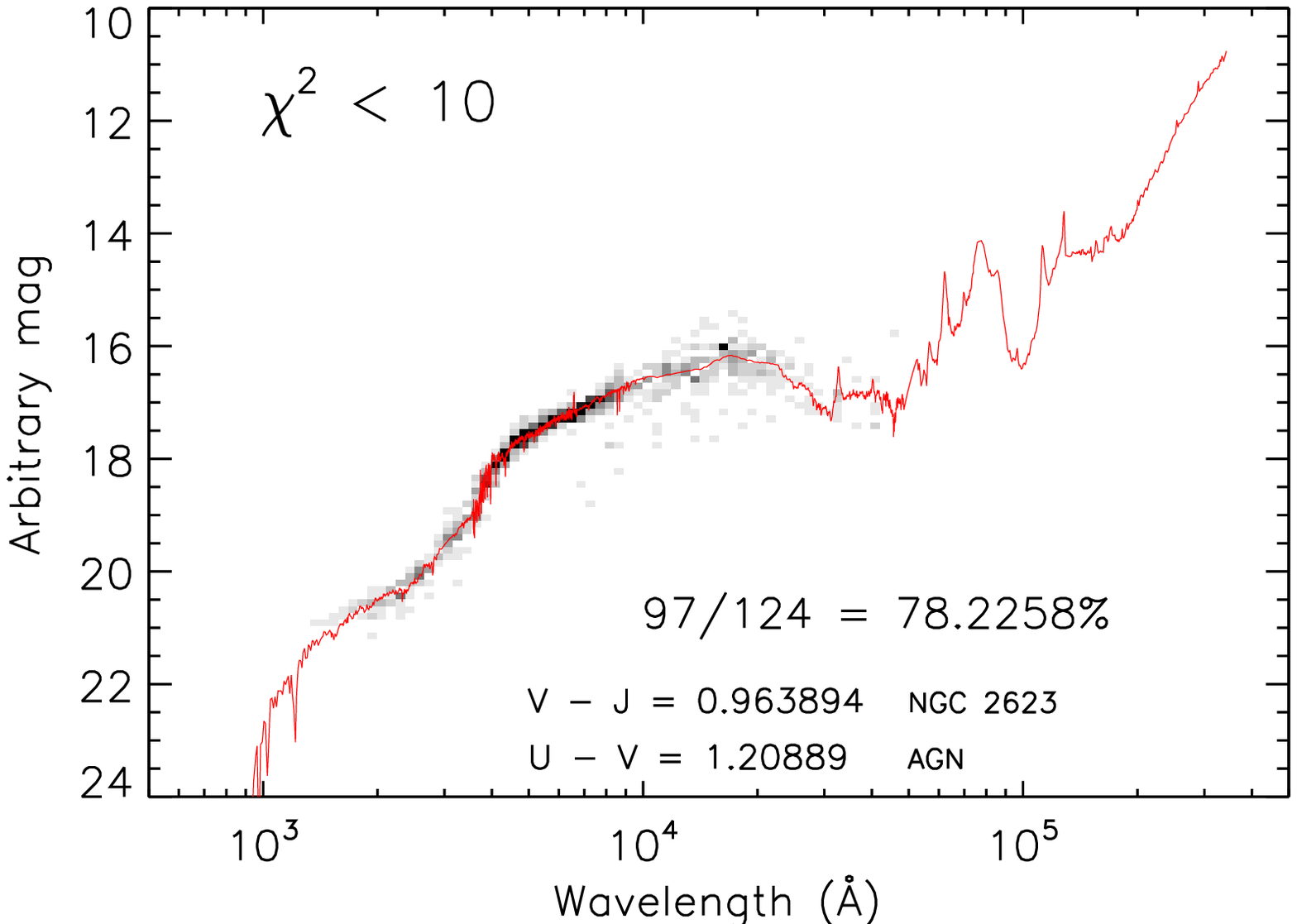}
\includegraphics[width=0.32\textwidth]{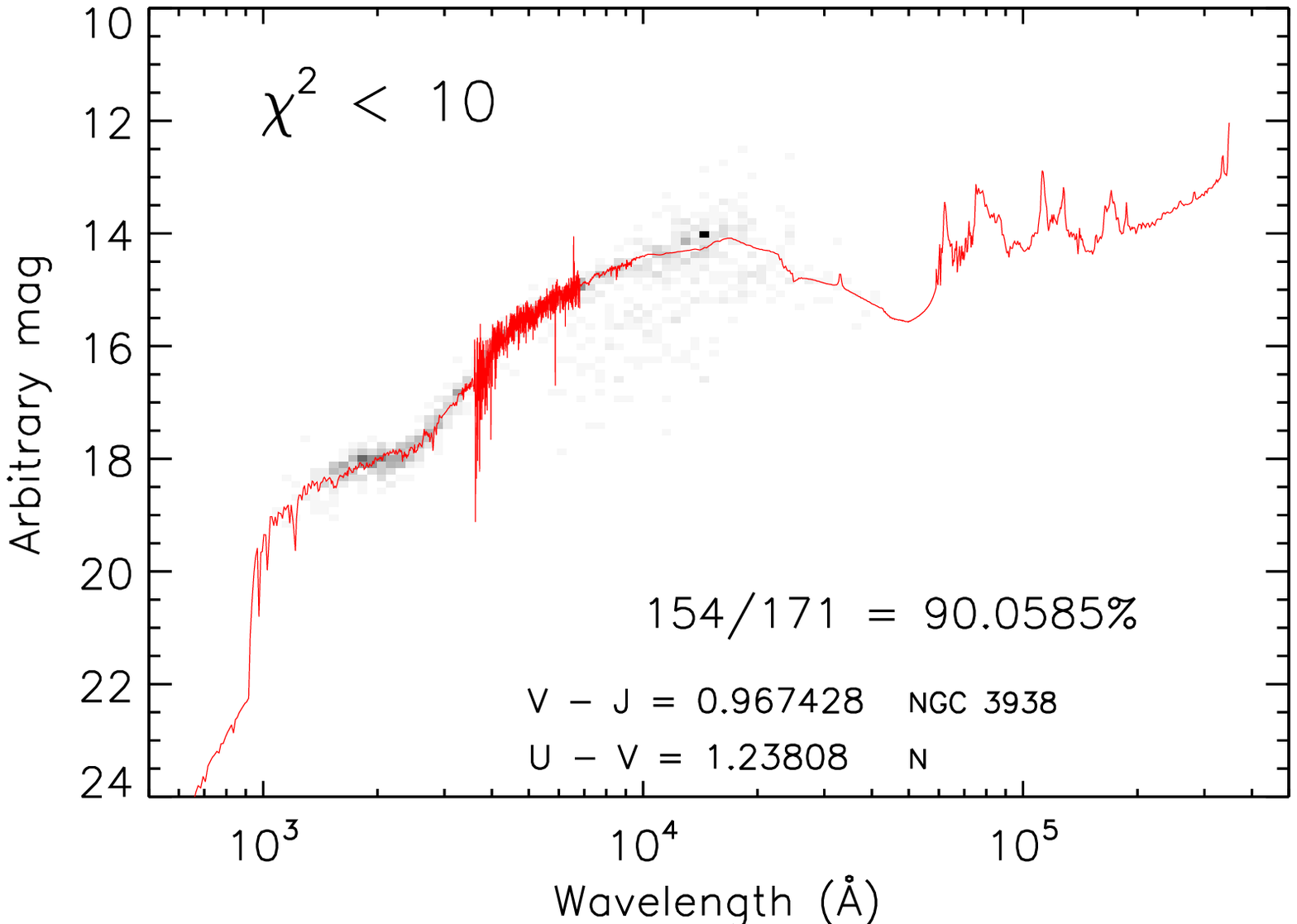}
\includegraphics[width=0.32\textwidth]{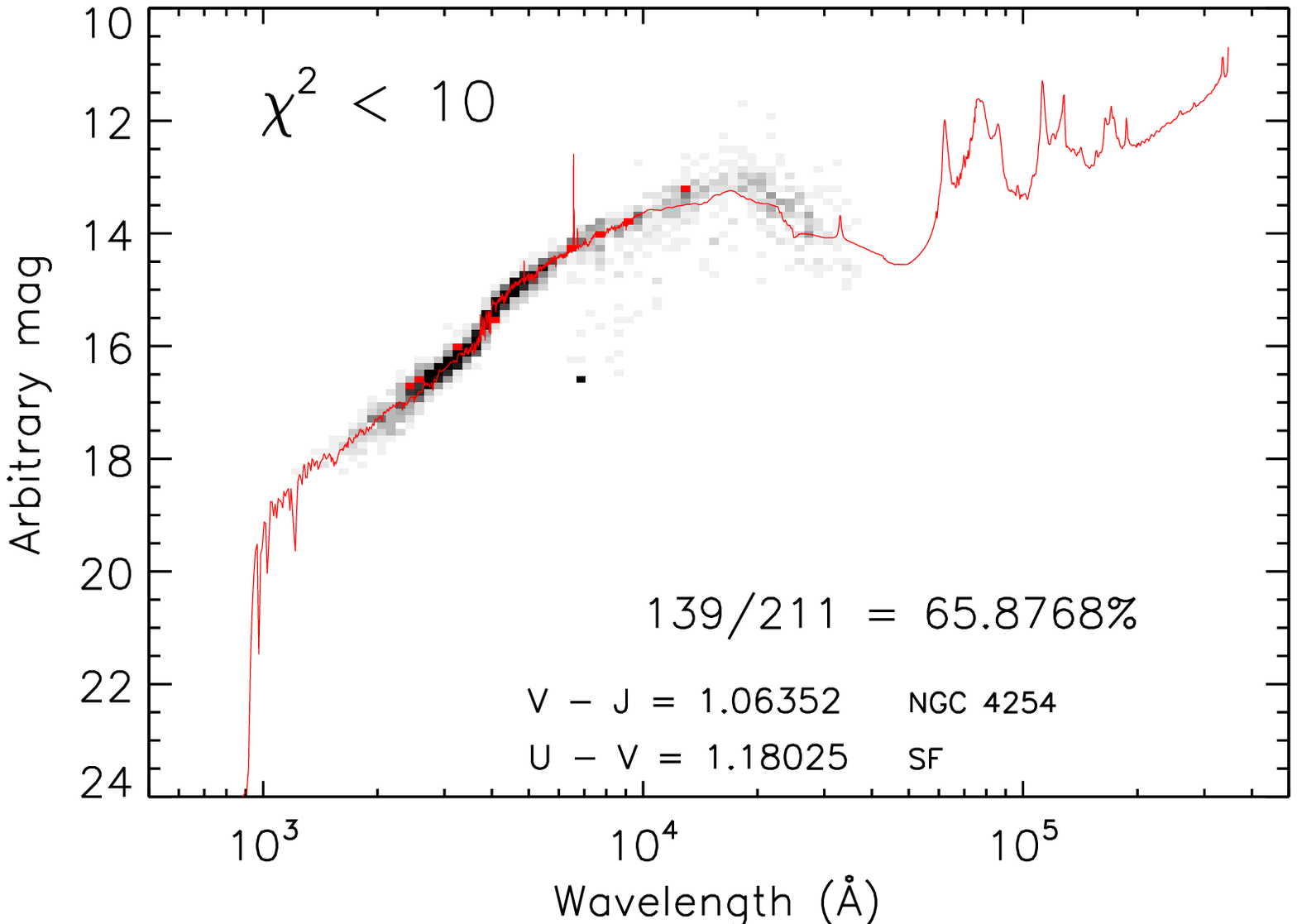}
\includegraphics[width=0.32\textwidth]{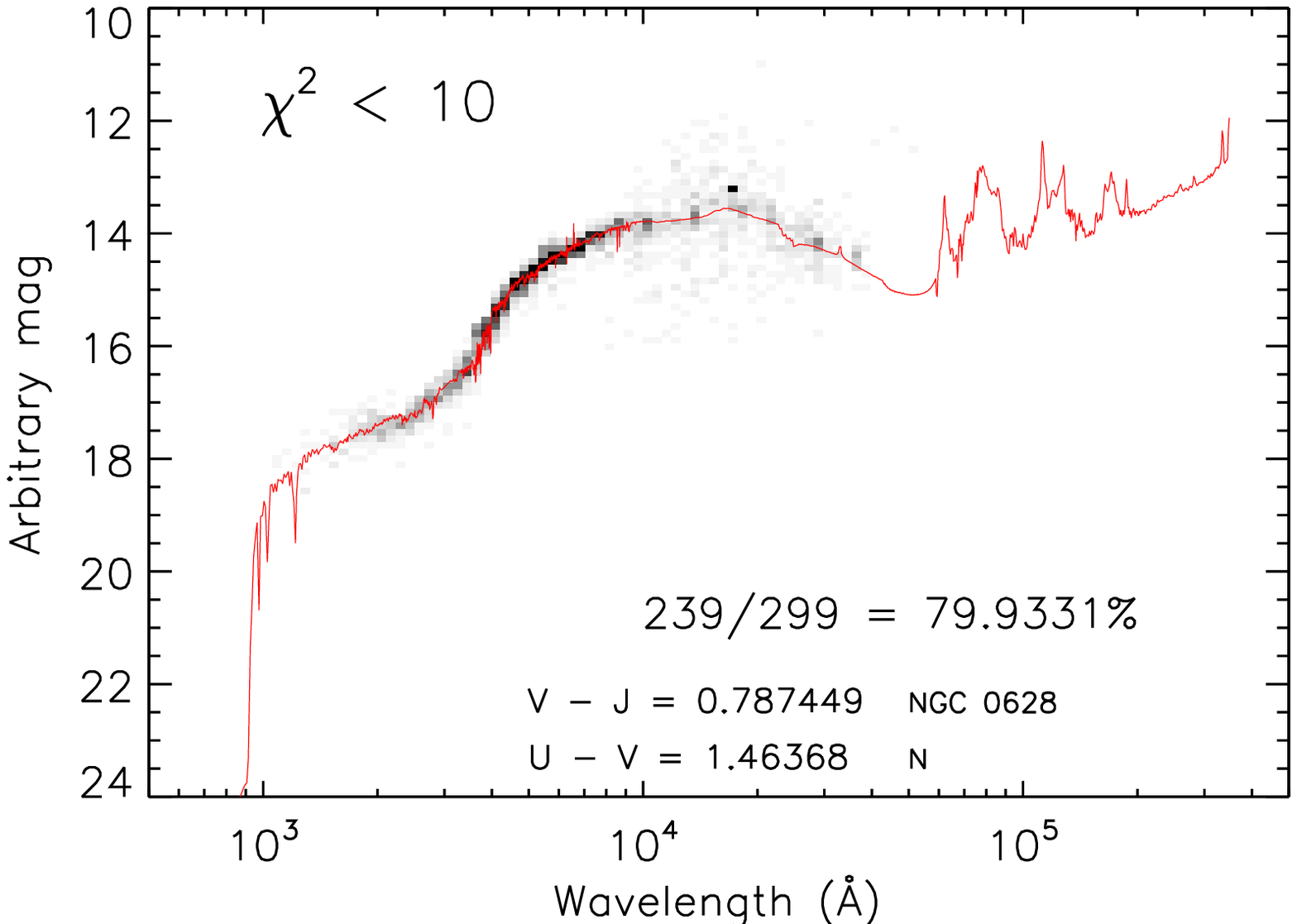}
\includegraphics[width=0.32\textwidth]{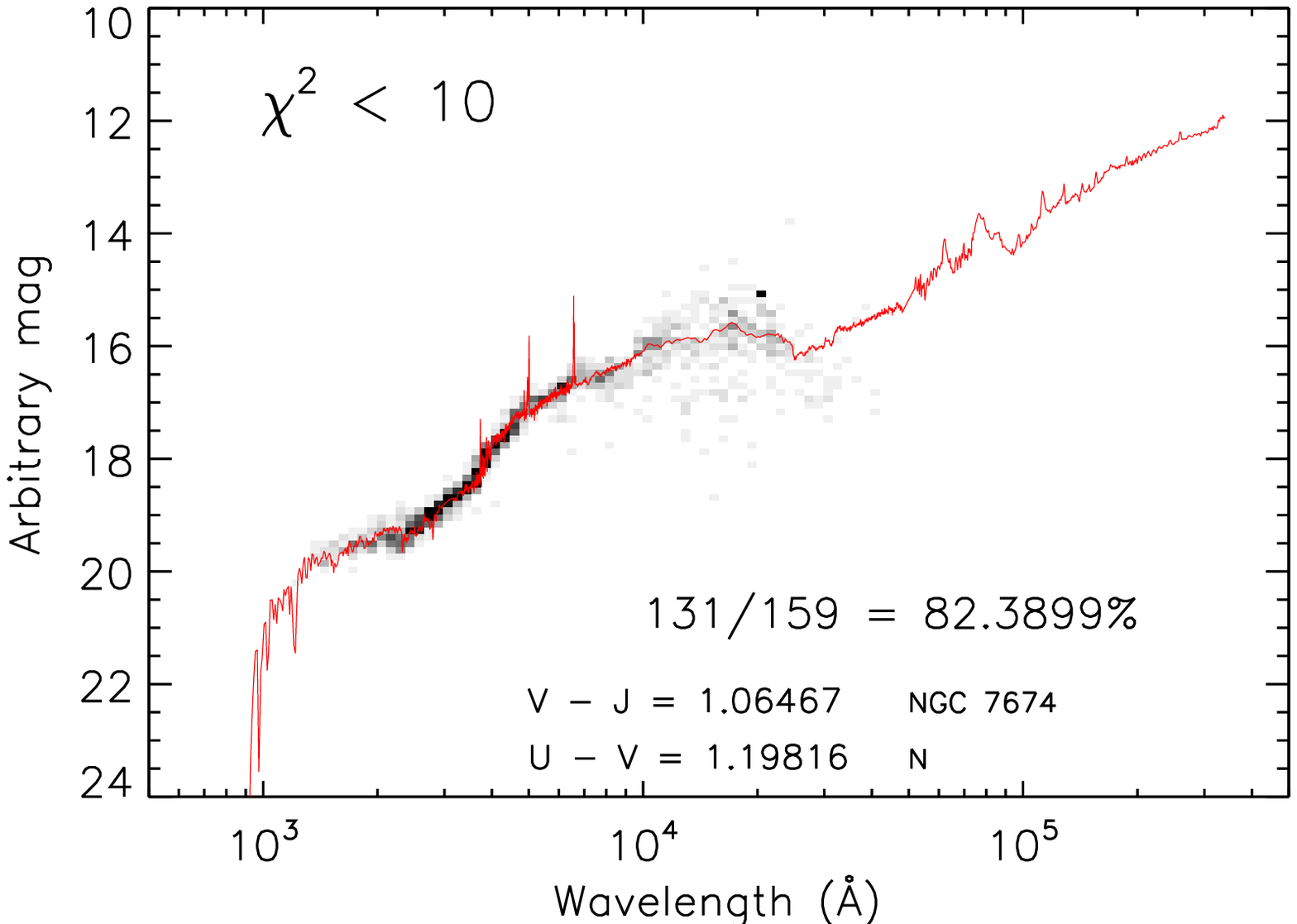}
\includegraphics[width=0.32\textwidth]{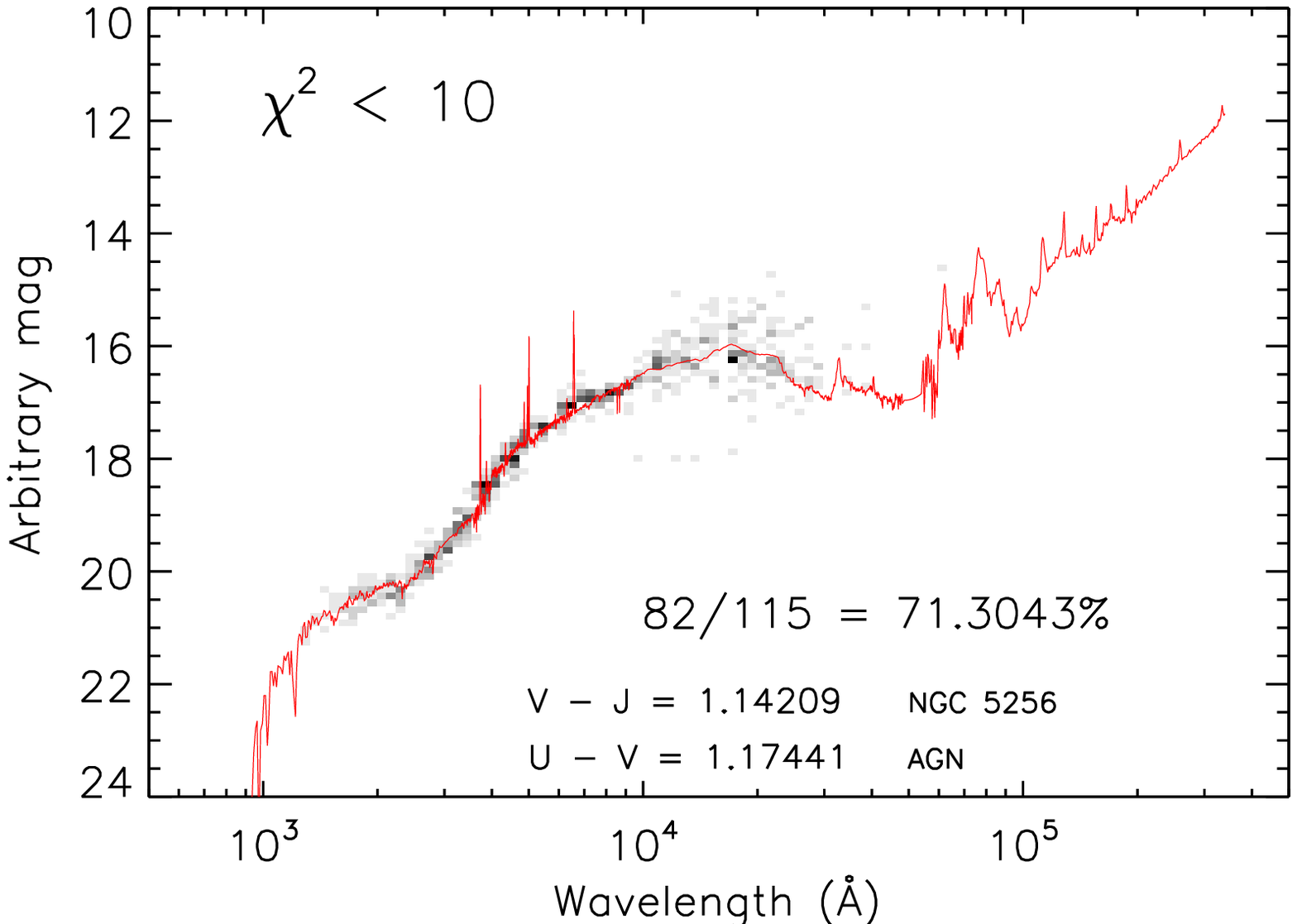}
\includegraphics[width=0.32\textwidth]{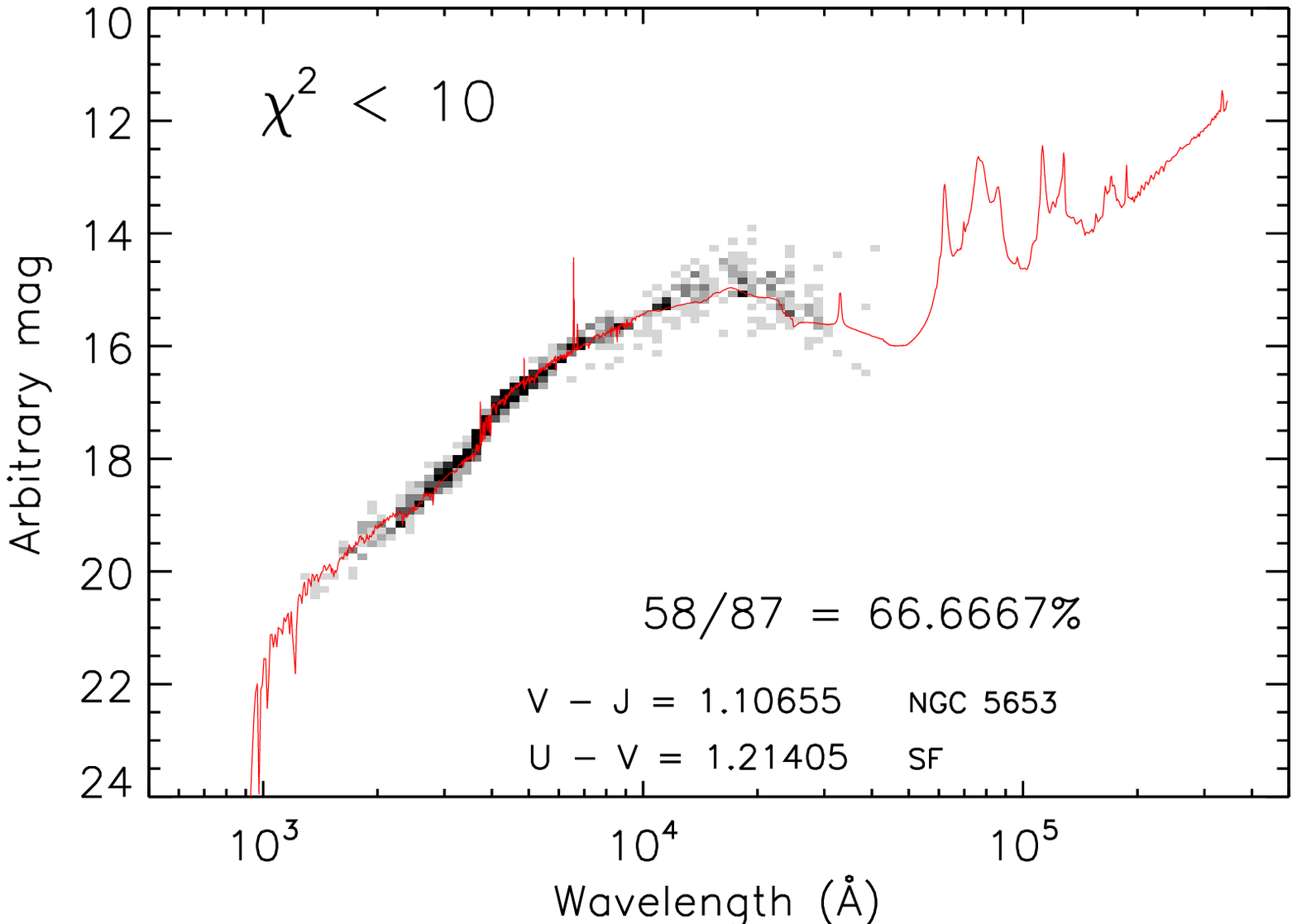}
\includegraphics[width=0.32\textwidth]{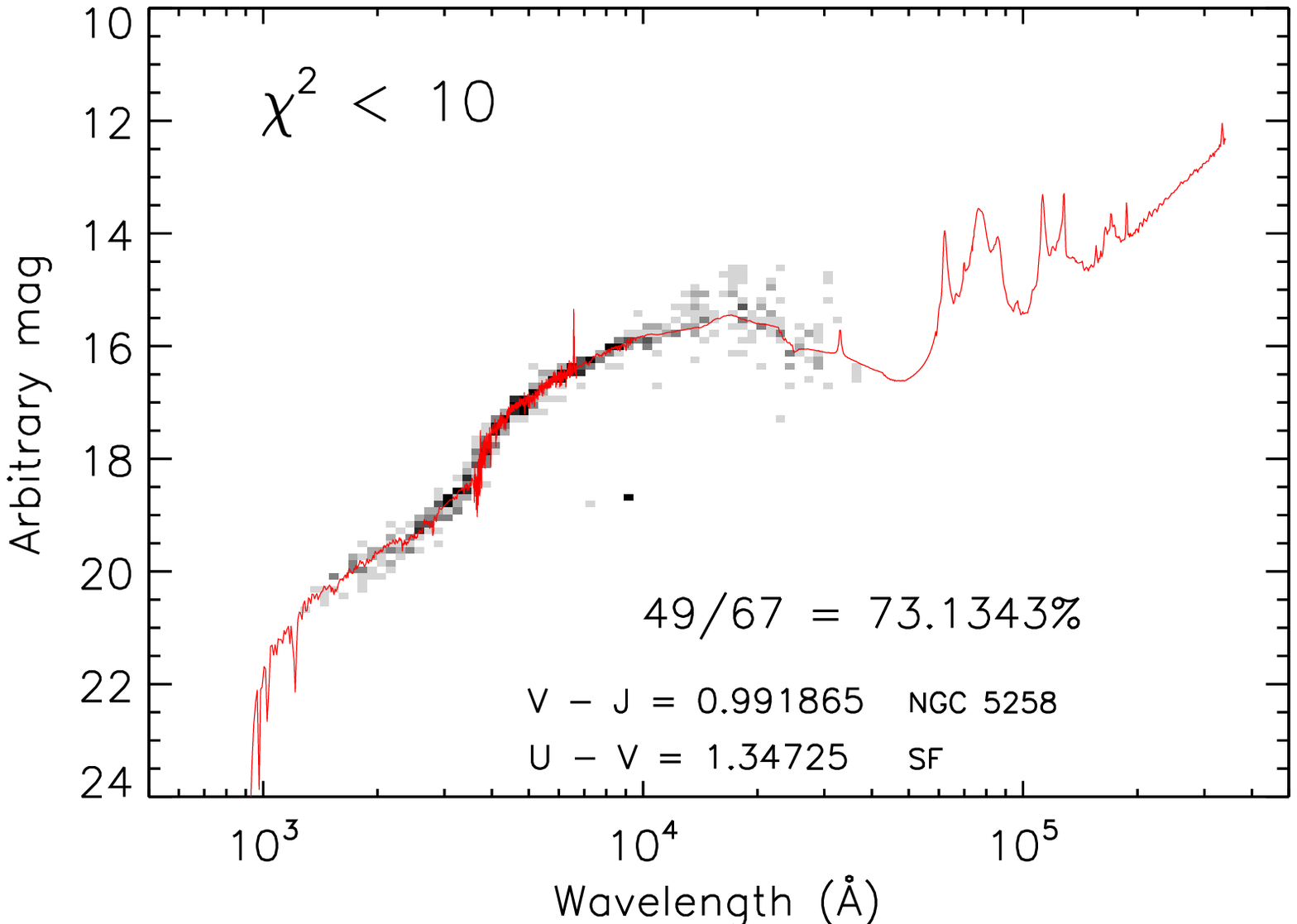}
\includegraphics[width=0.32\textwidth]{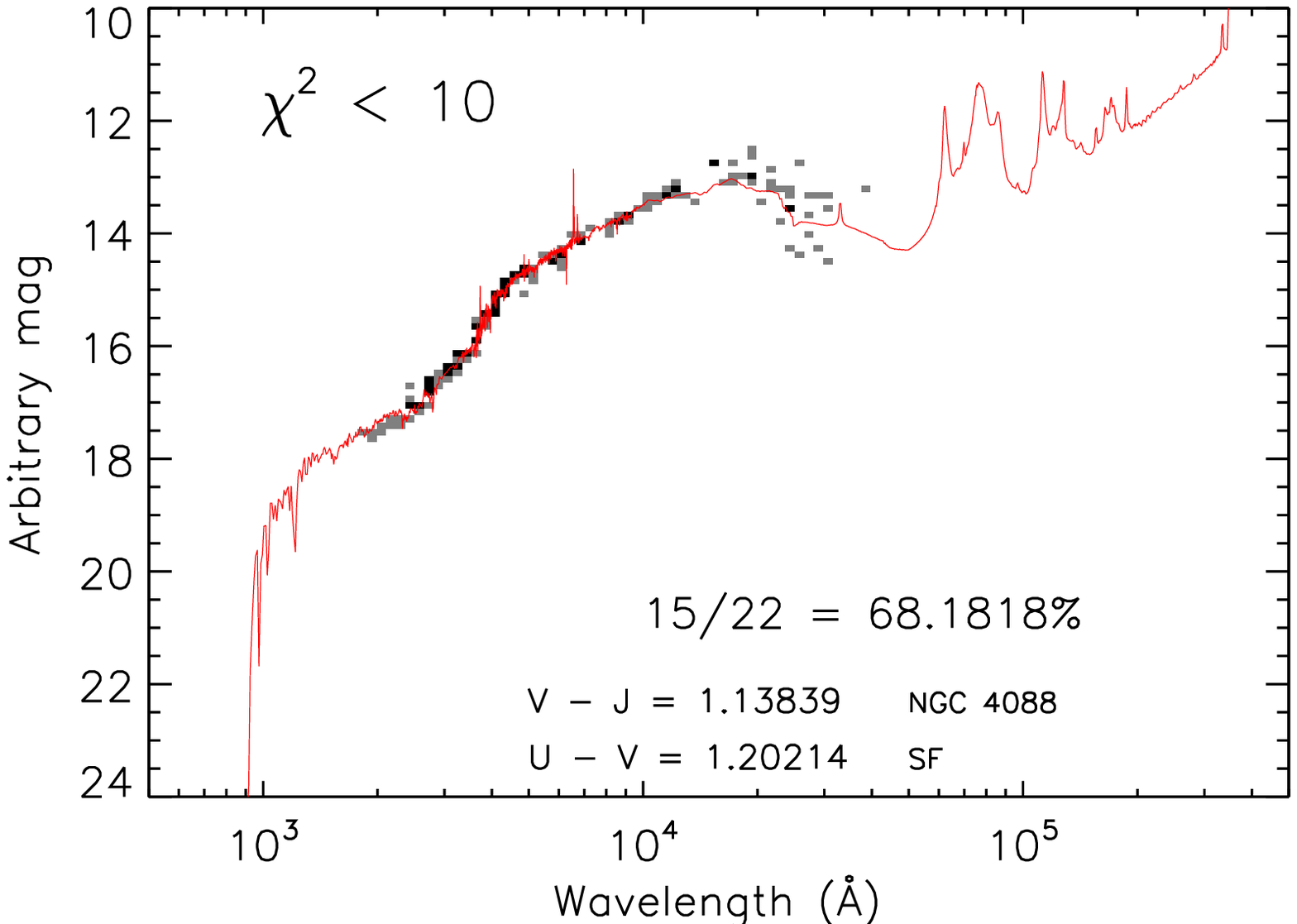}
\includegraphics[width=0.32\textwidth]{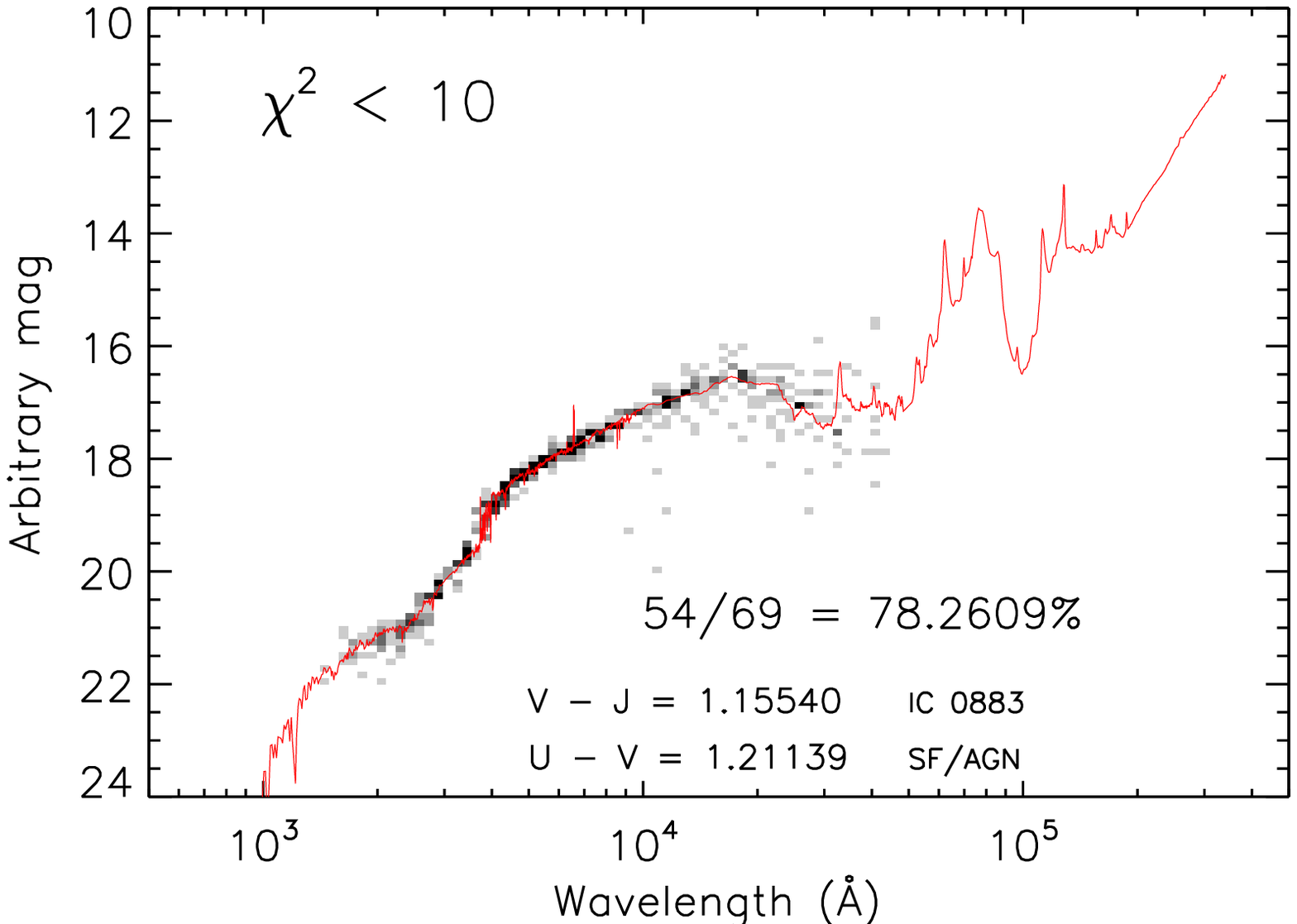}
\includegraphics[width=0.32\textwidth]{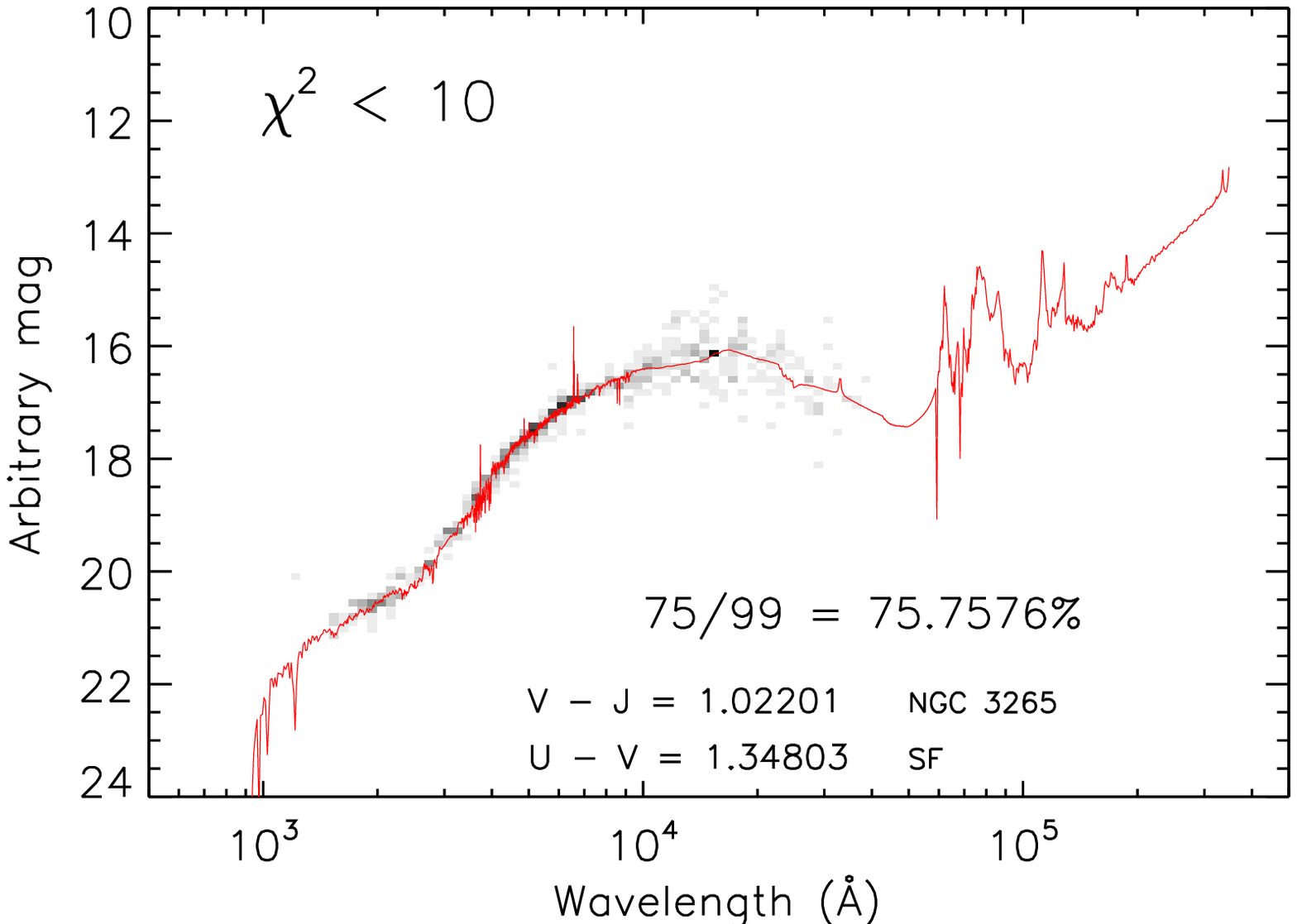}
\includegraphics[width=0.32\textwidth]{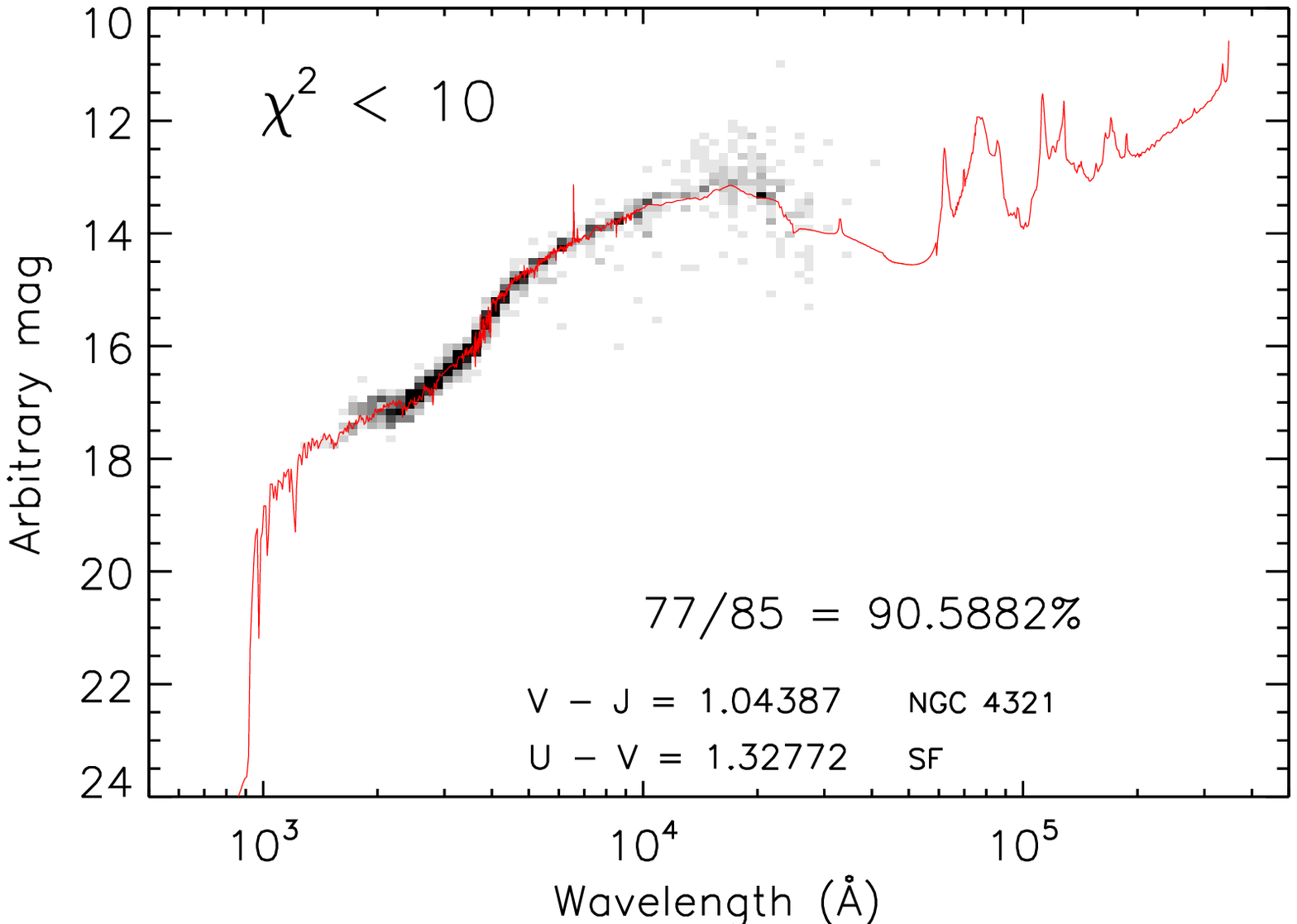}
\includegraphics[width=0.32\textwidth]{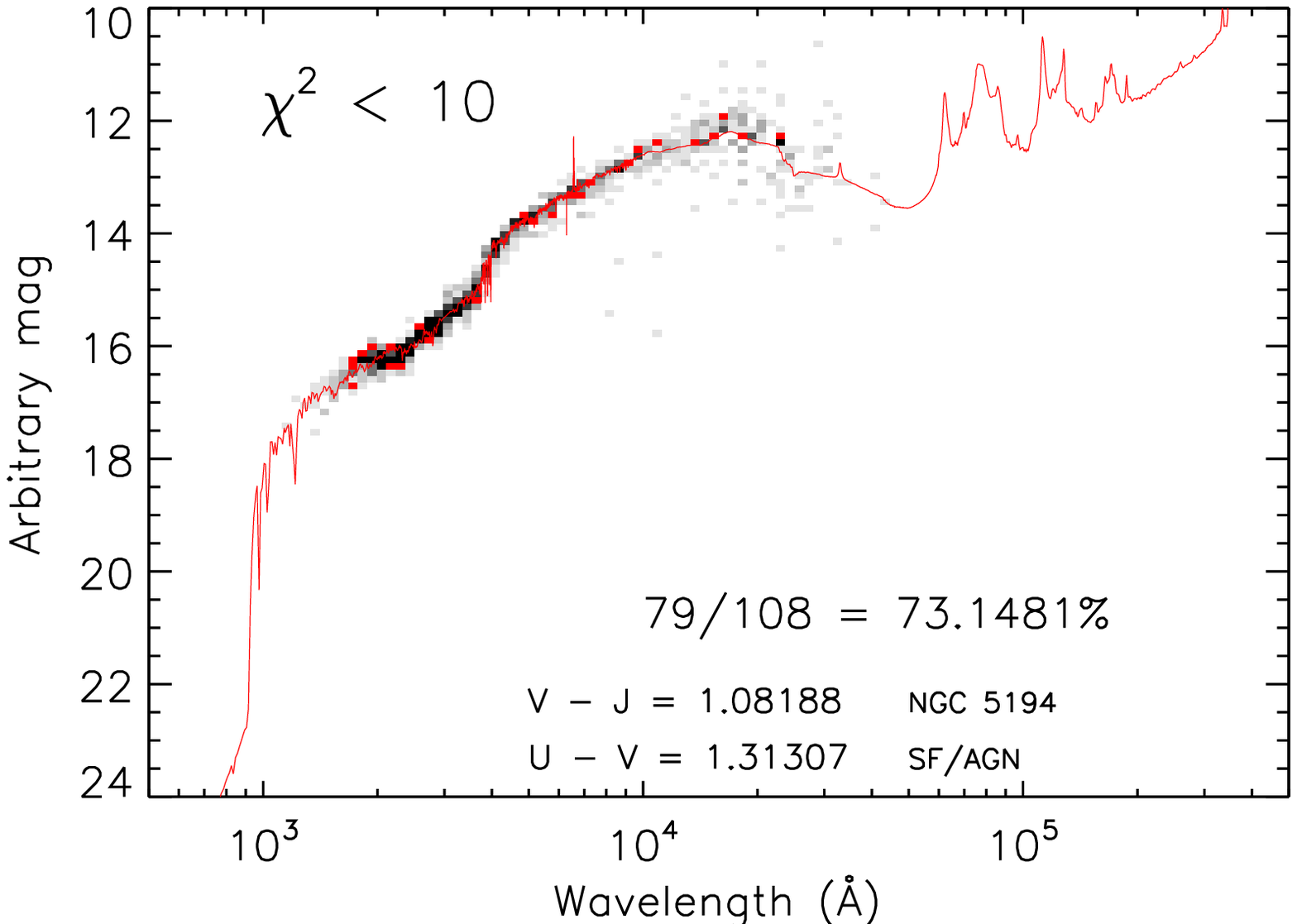}
\includegraphics[width=0.32\textwidth]{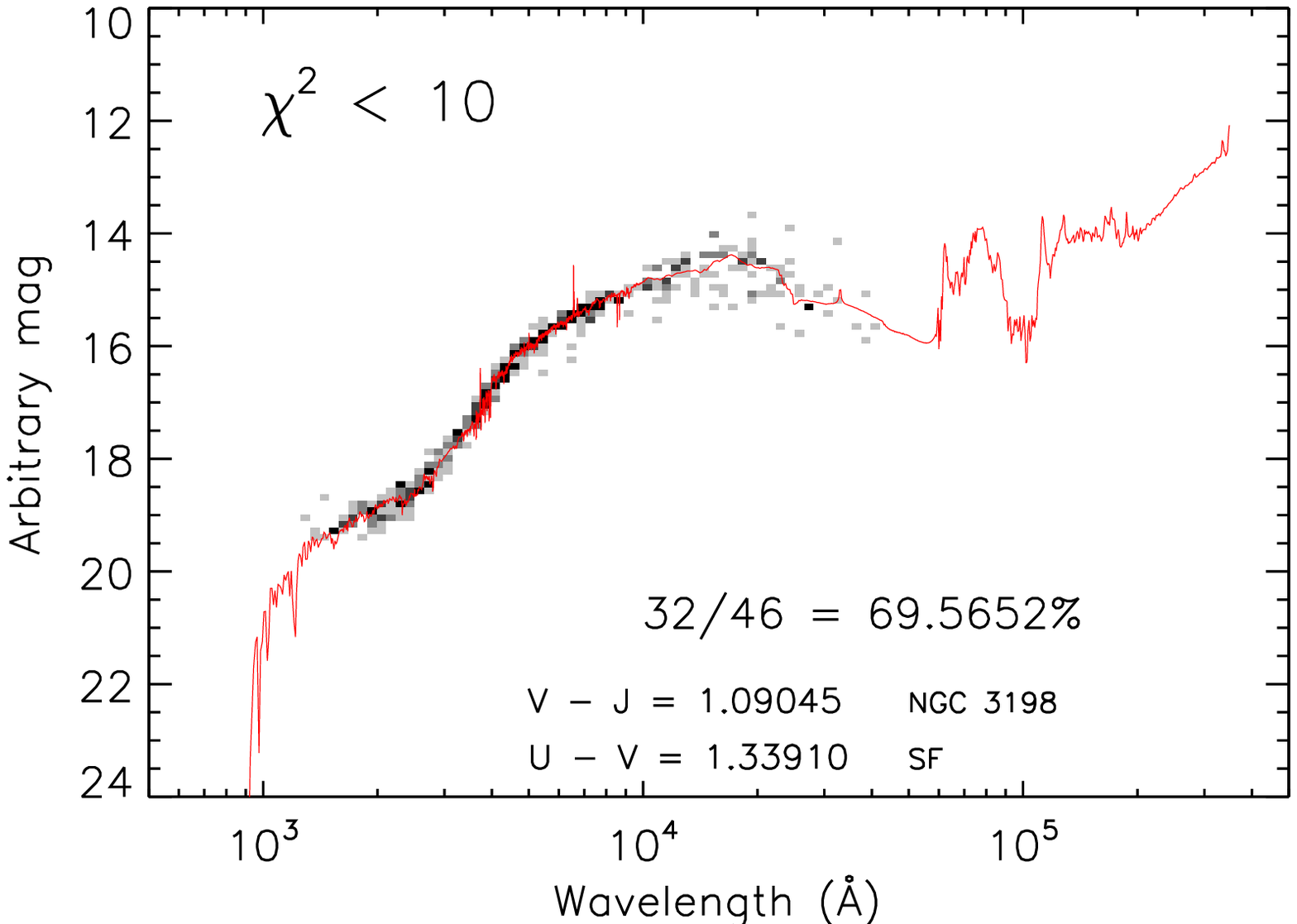}
\includegraphics[width=0.32\textwidth]{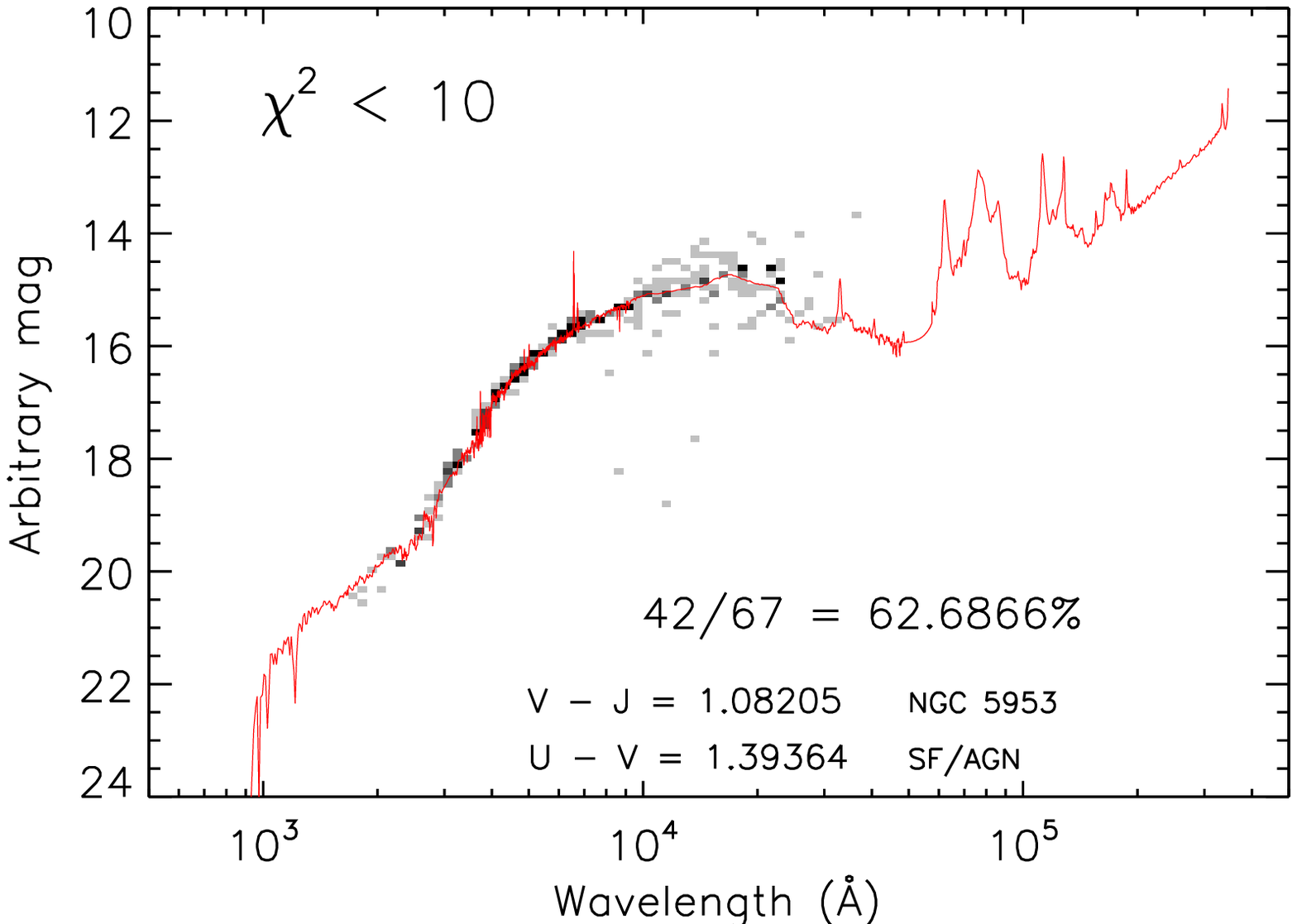}
    \caption{Continued.}
    \label{fig:my_label}
\end{figure}

\begin{figure}
    \centering
\includegraphics[width=0.32\textwidth]{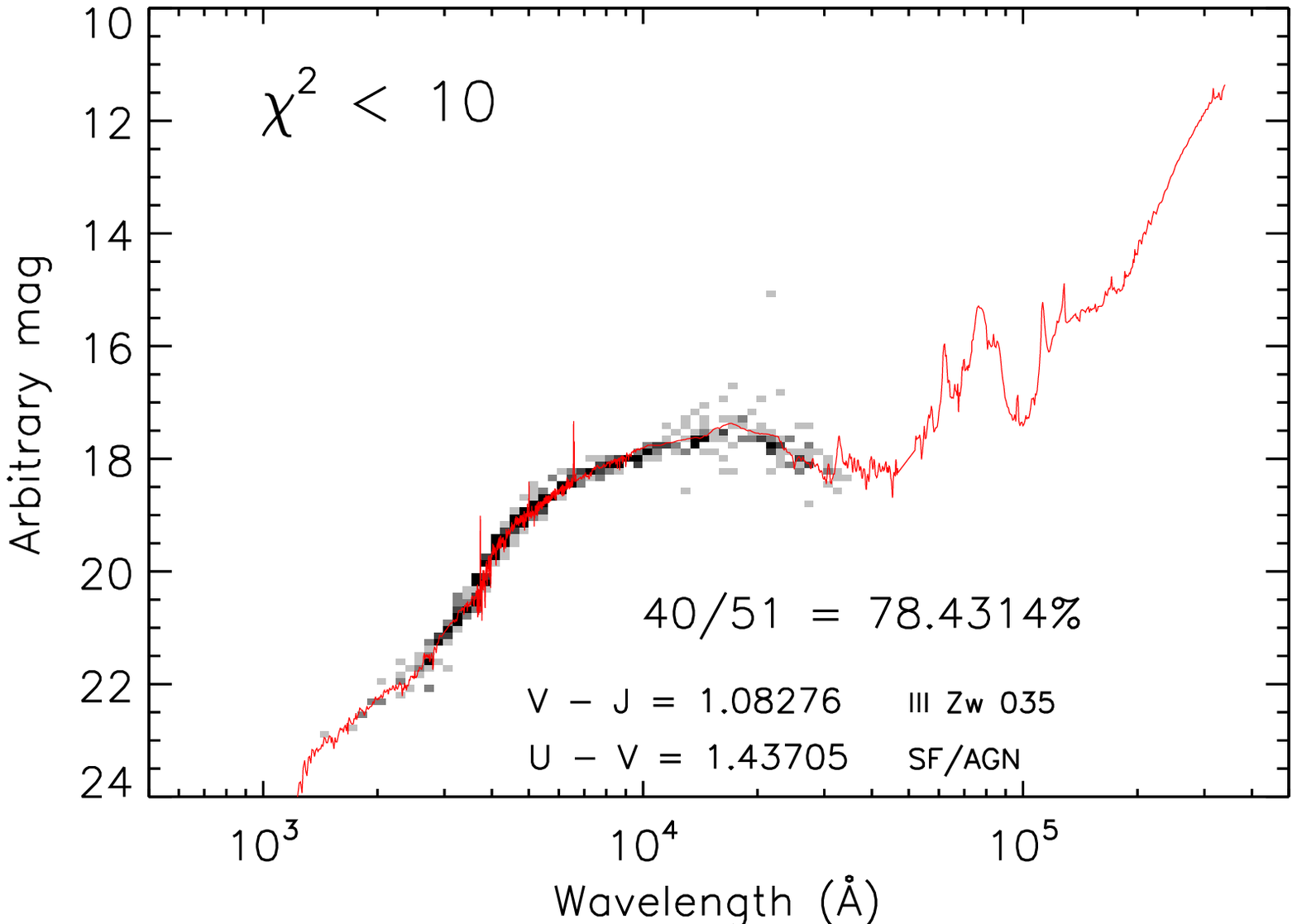}
\includegraphics[width=0.32\textwidth]{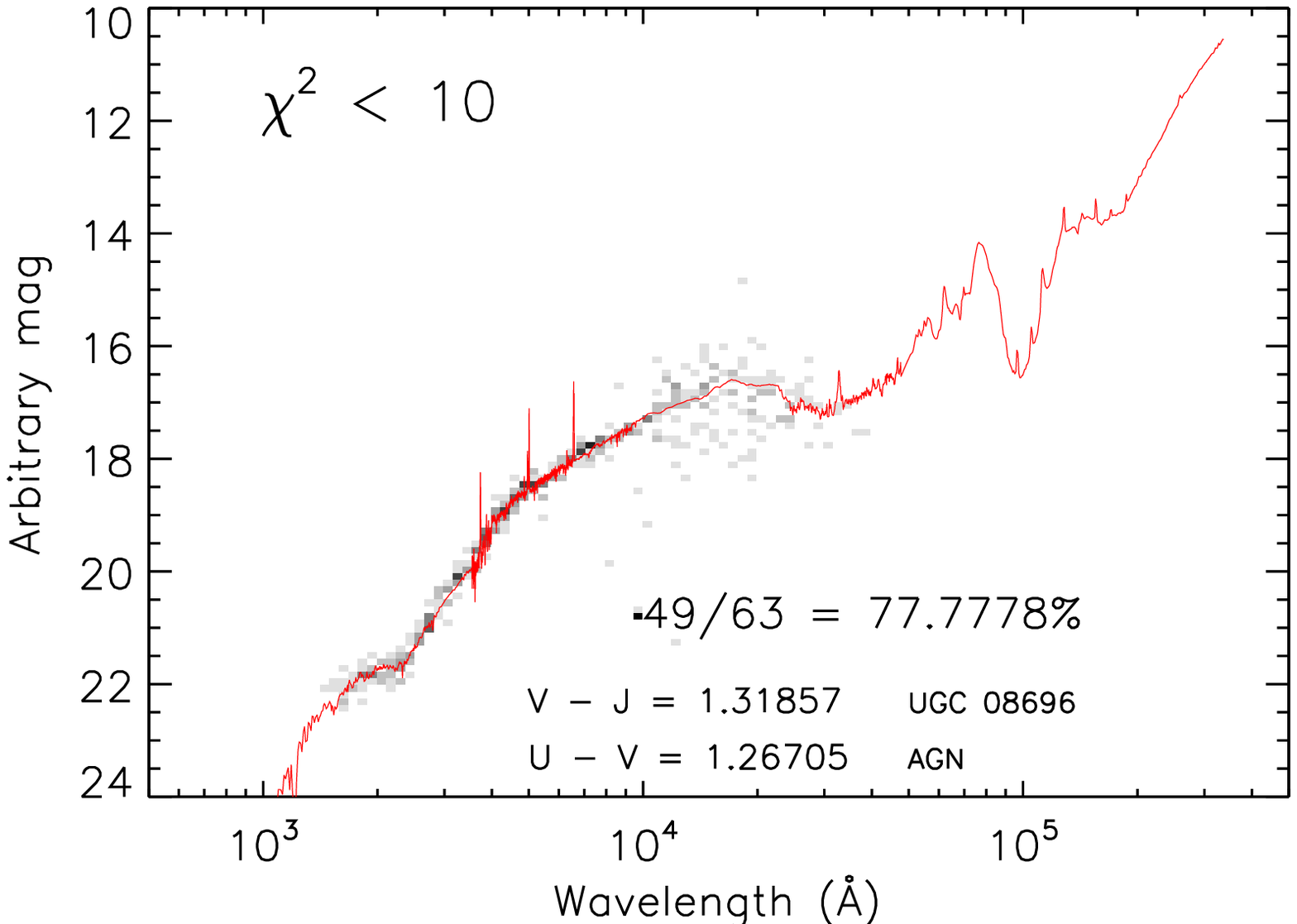}
\includegraphics[width=0.32\textwidth]{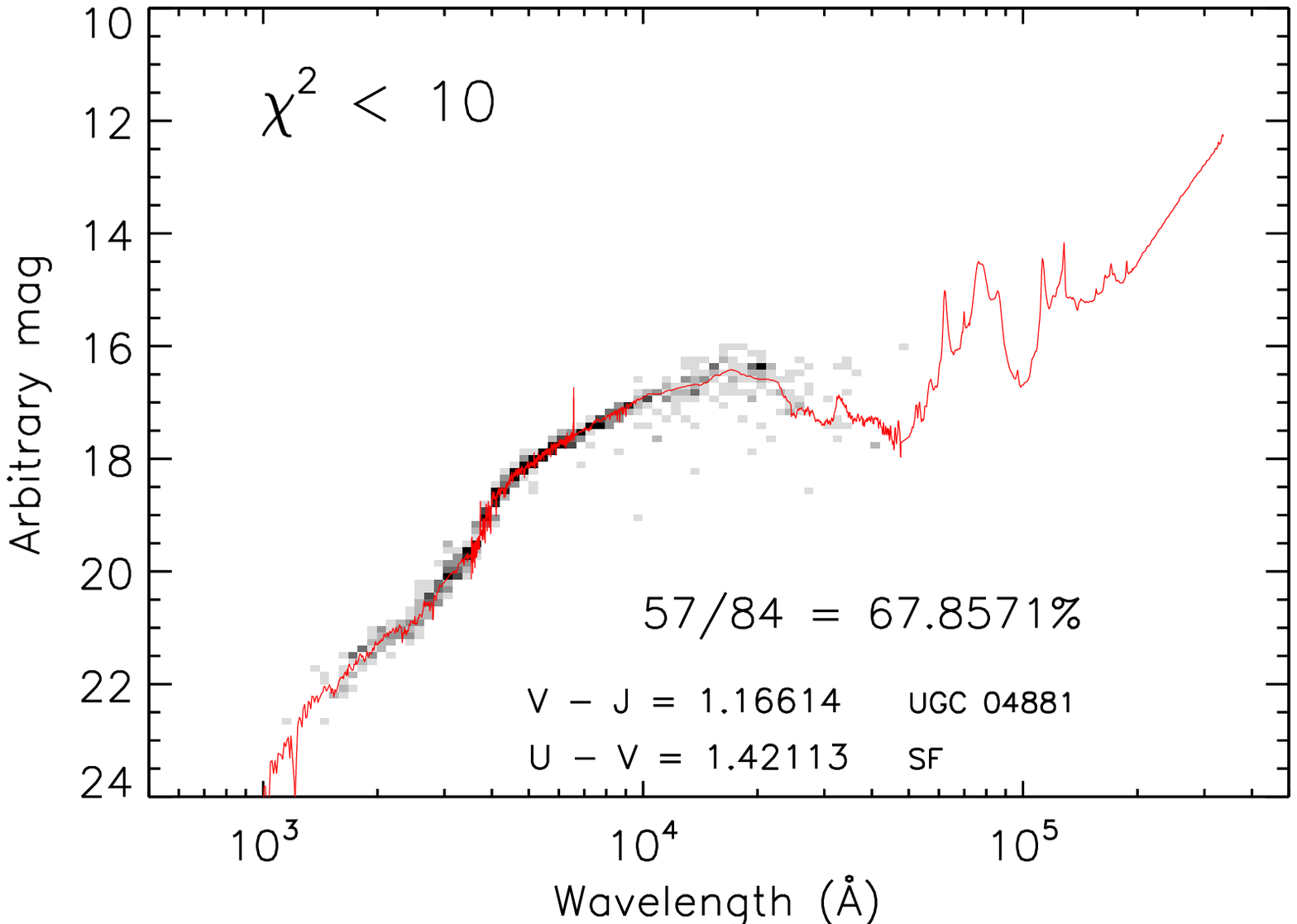}
\includegraphics[width=0.32\textwidth]{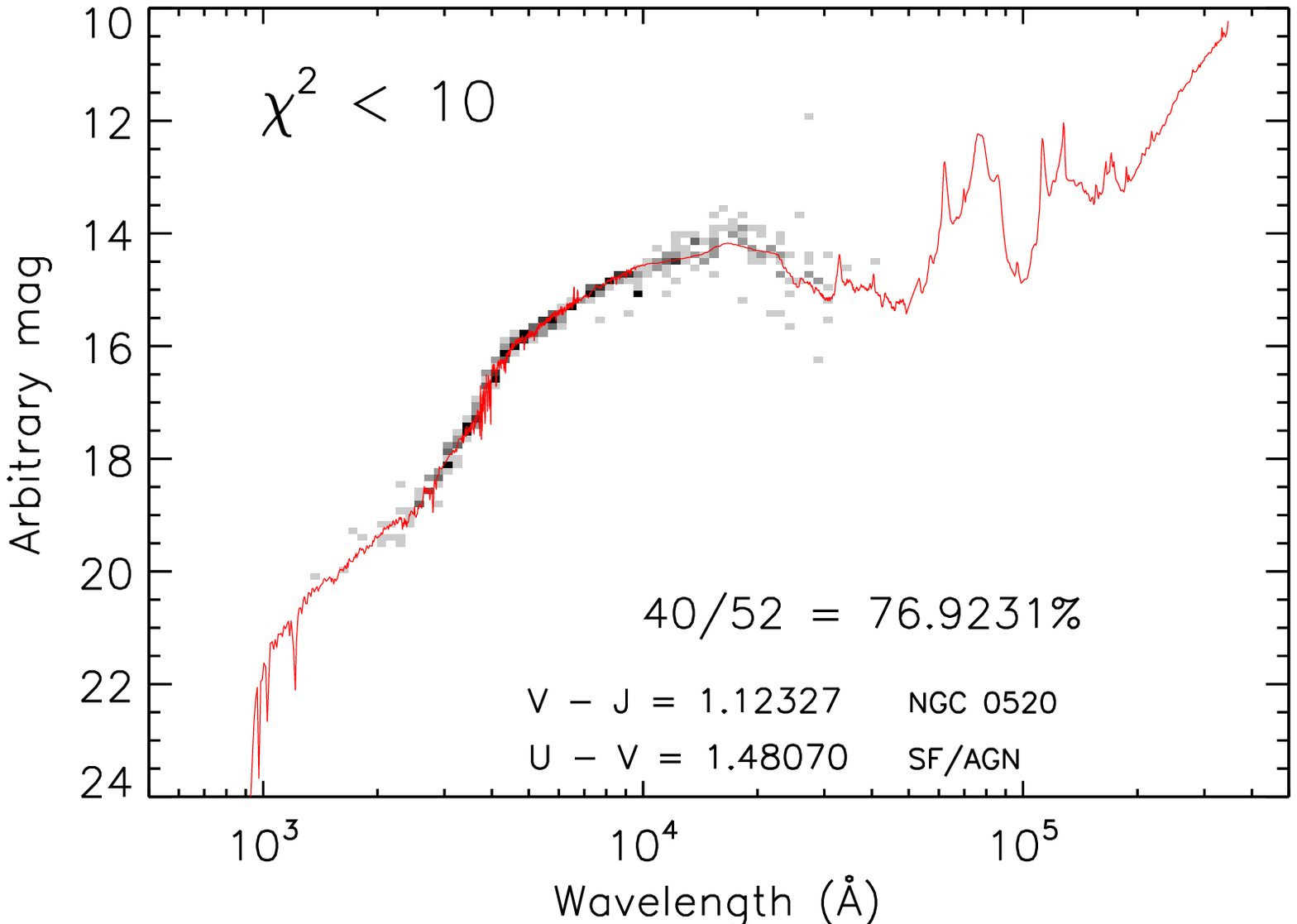}
\includegraphics[width=0.32\textwidth]{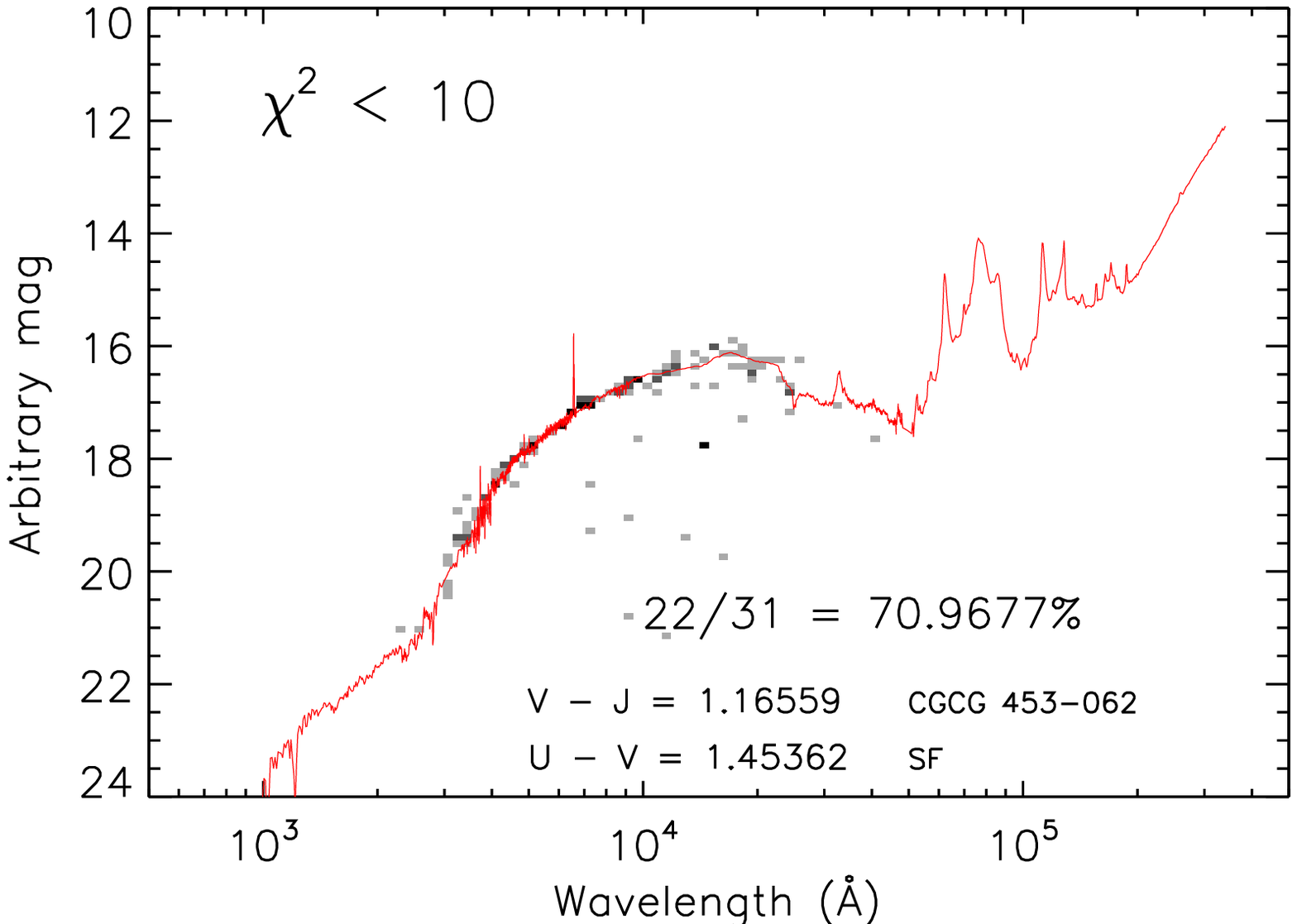}
\includegraphics[width=0.32\textwidth]{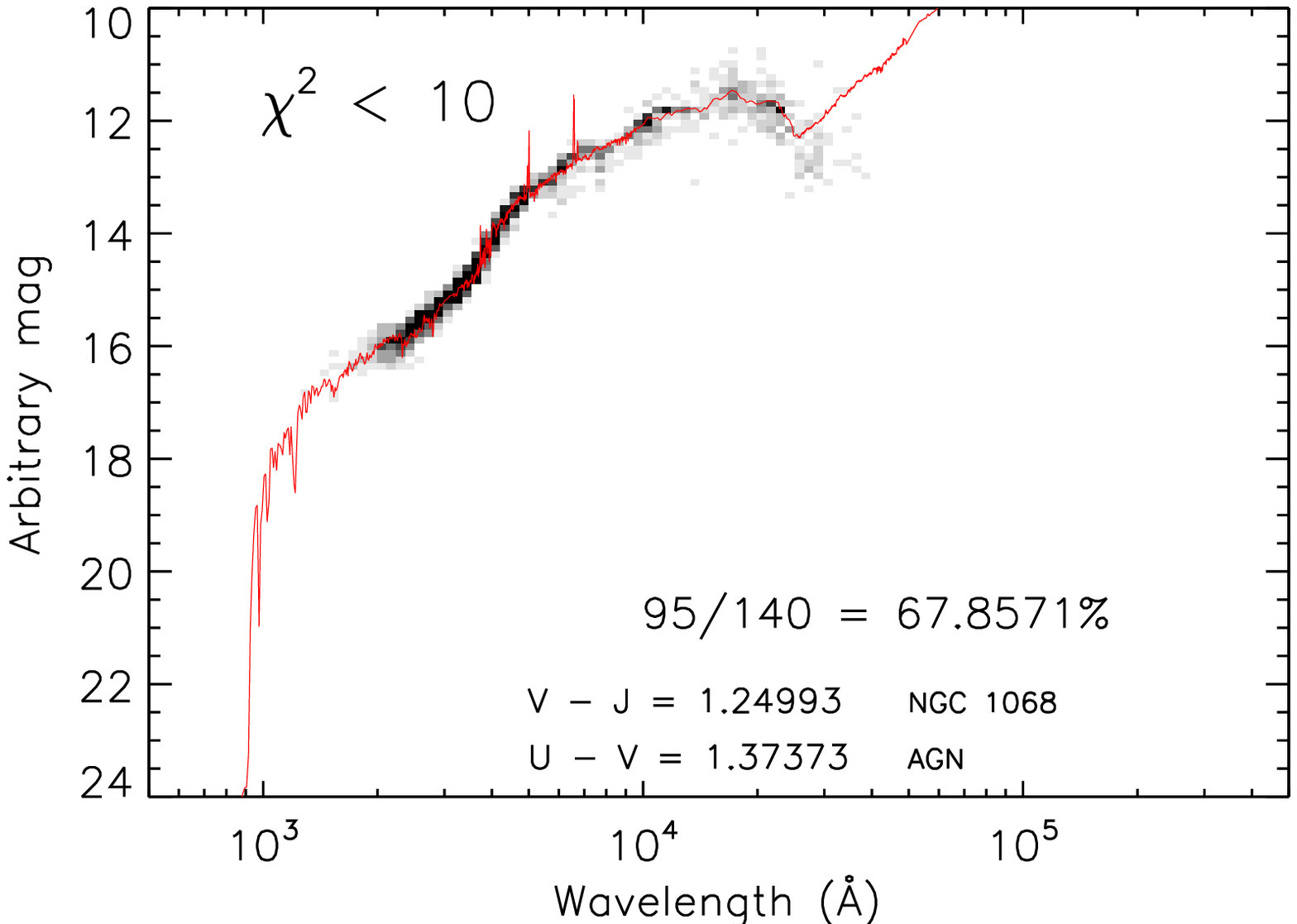}
\includegraphics[width=0.32\textwidth]{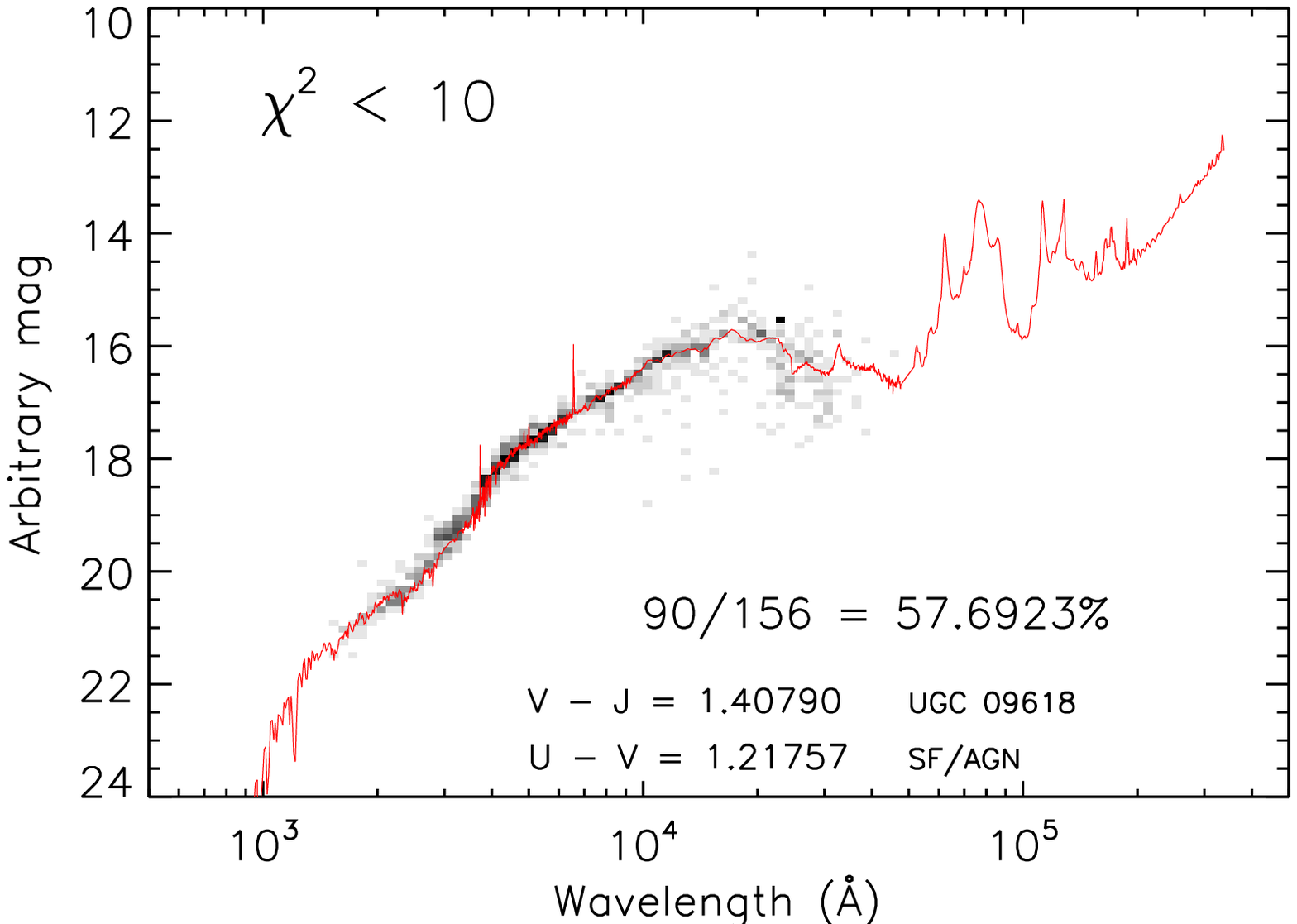}
\includegraphics[width=0.32\textwidth]{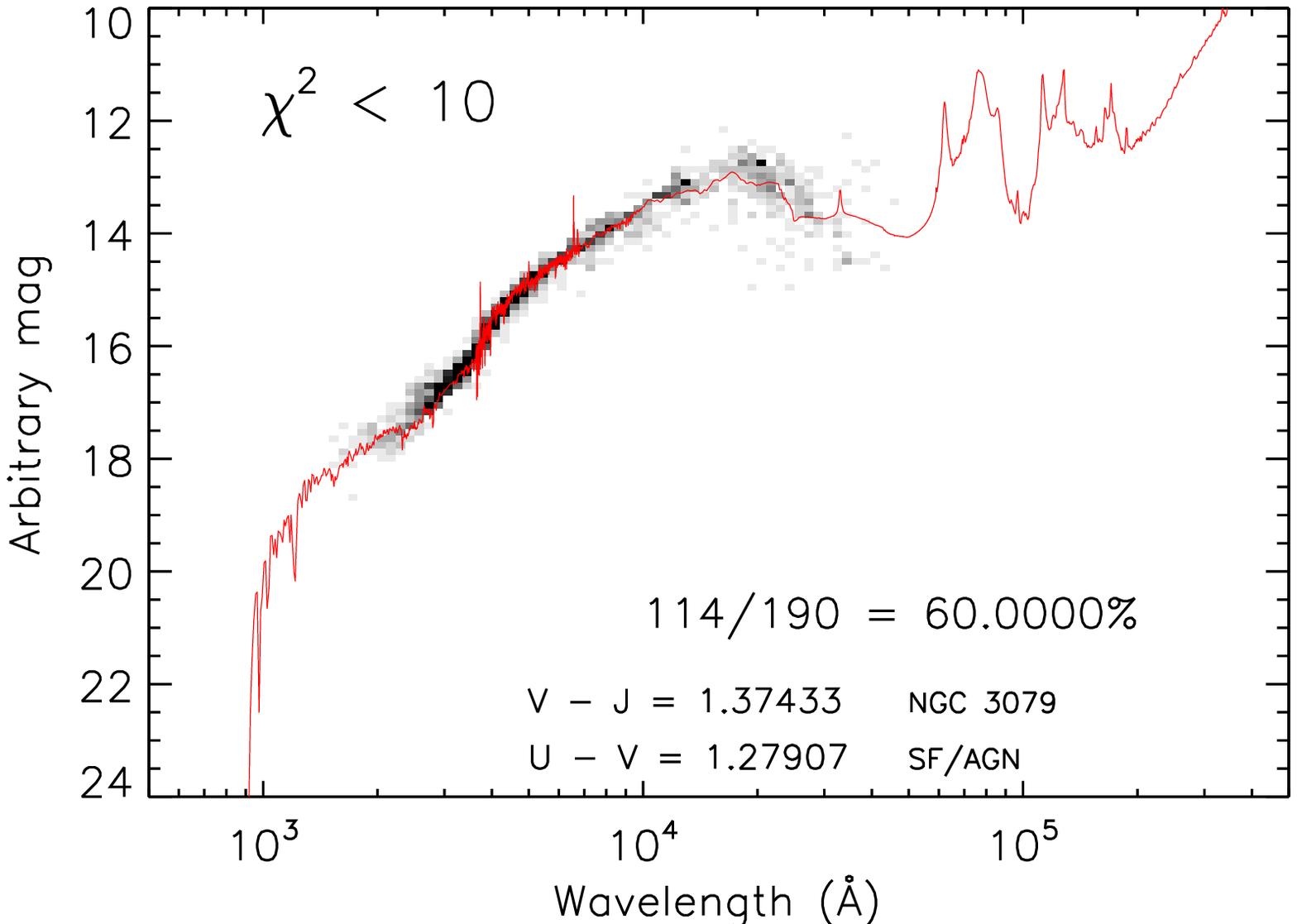}
\includegraphics[width=0.32\textwidth]{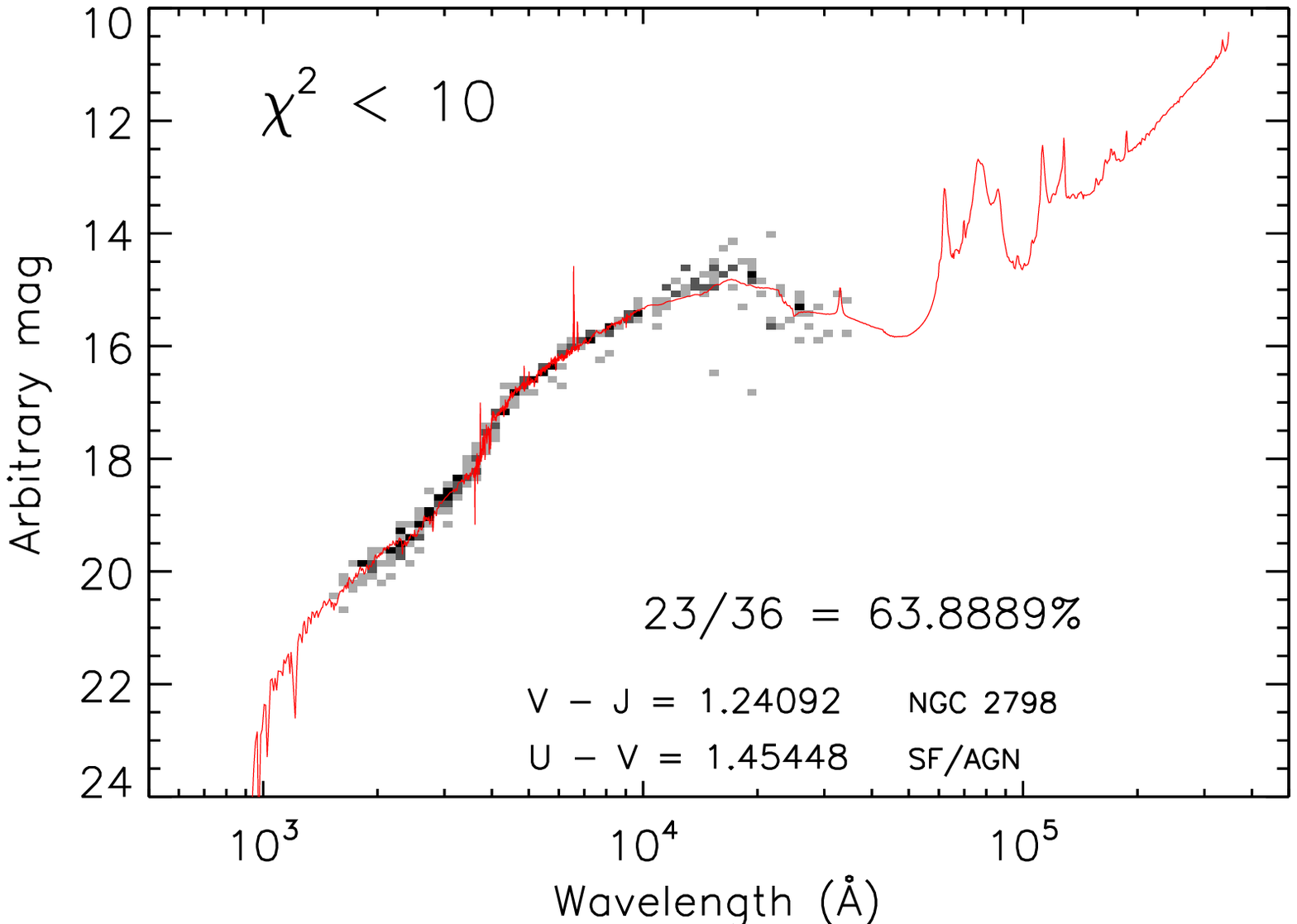}
\includegraphics[width=0.32\textwidth]{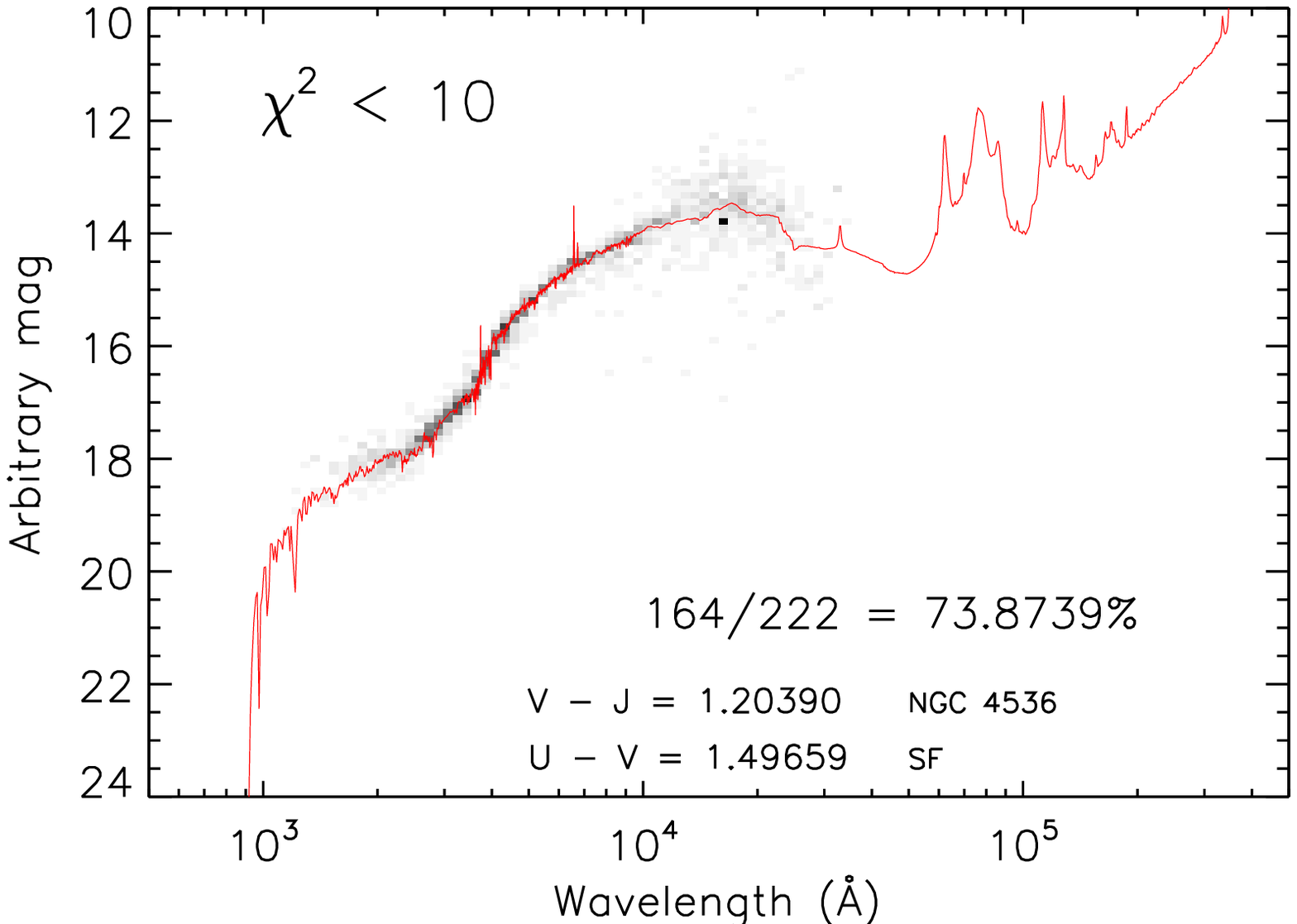}
\includegraphics[width=0.32\textwidth]{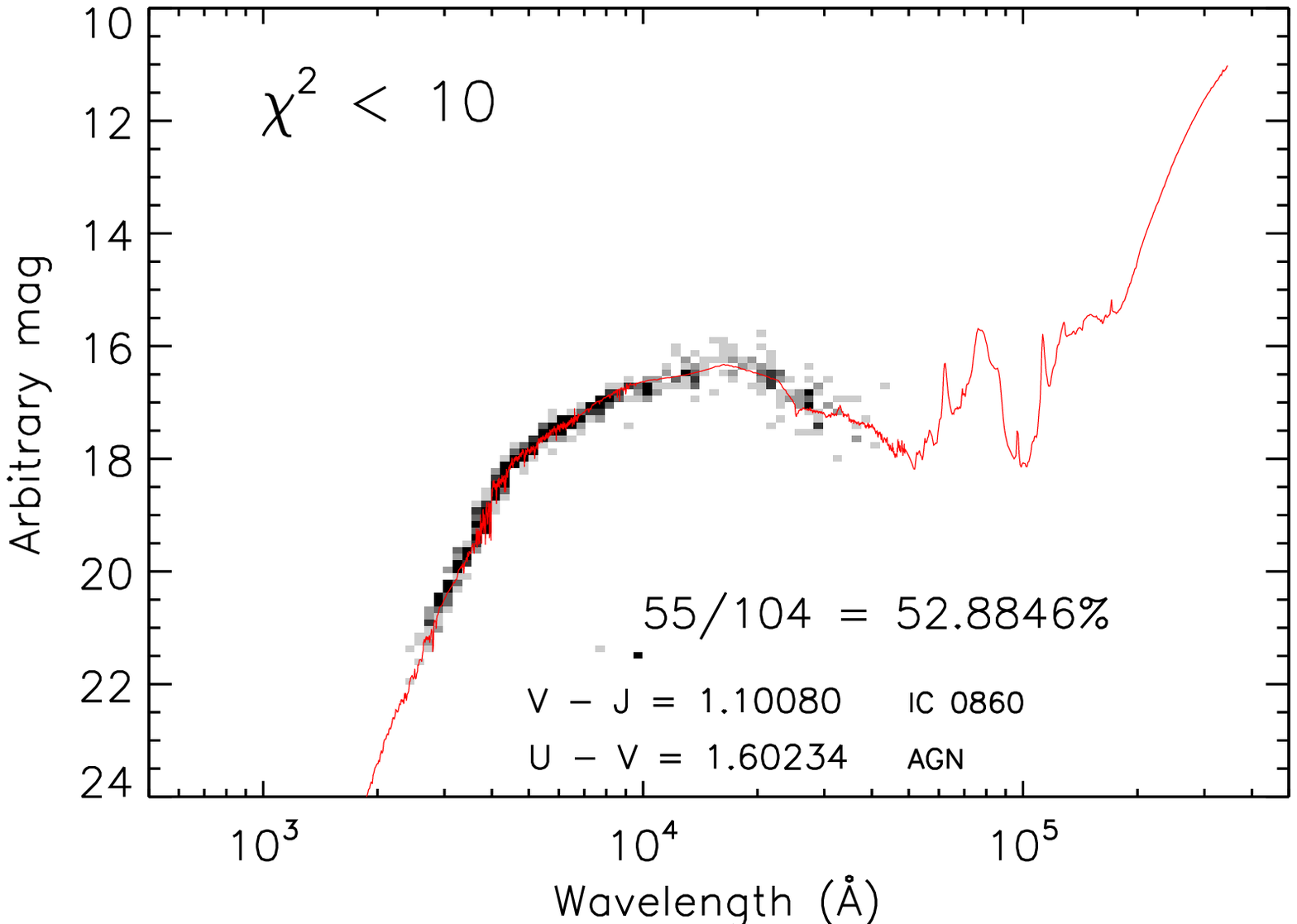}
\includegraphics[width=0.32\textwidth]{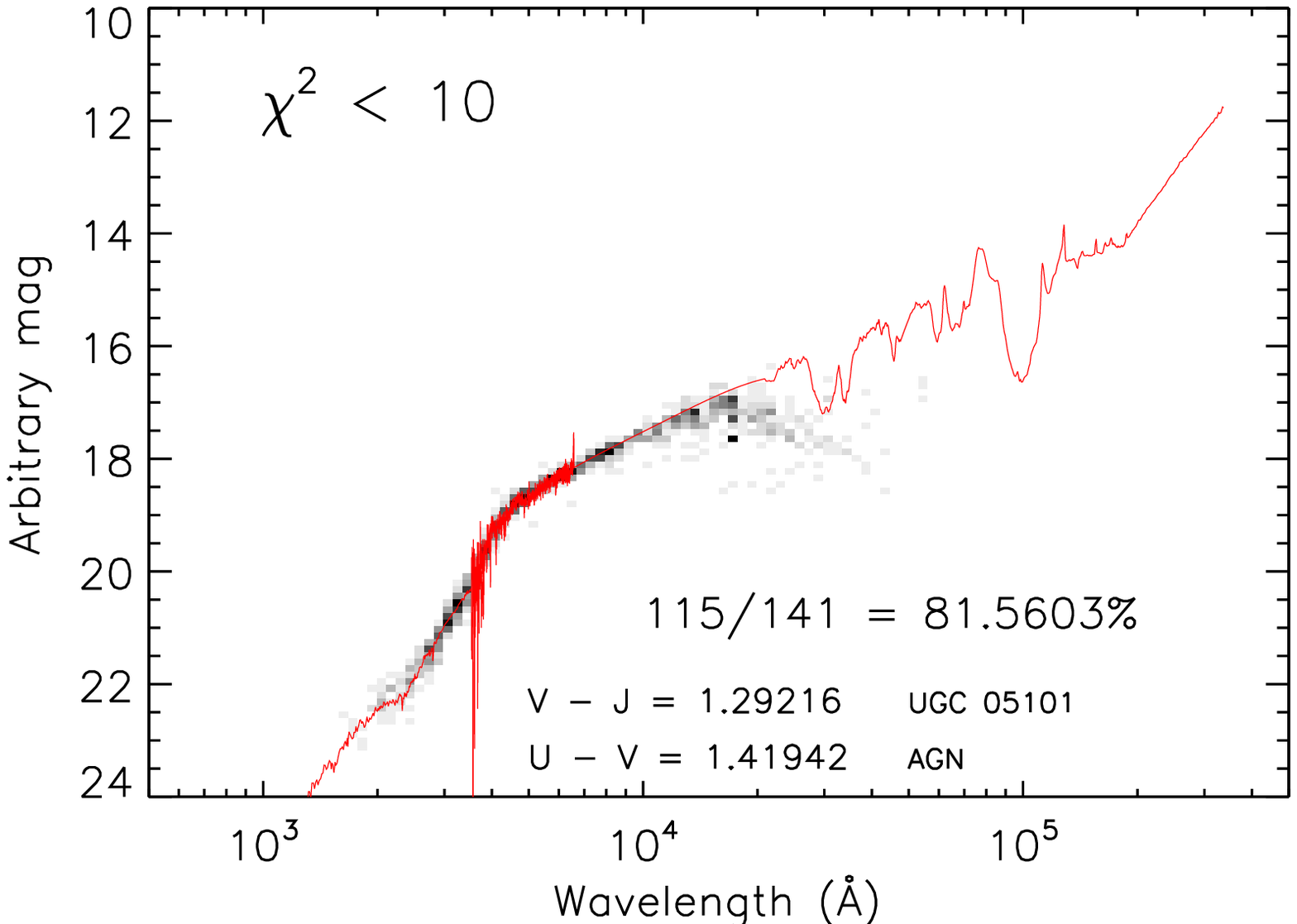}
\includegraphics[width=0.32\textwidth]{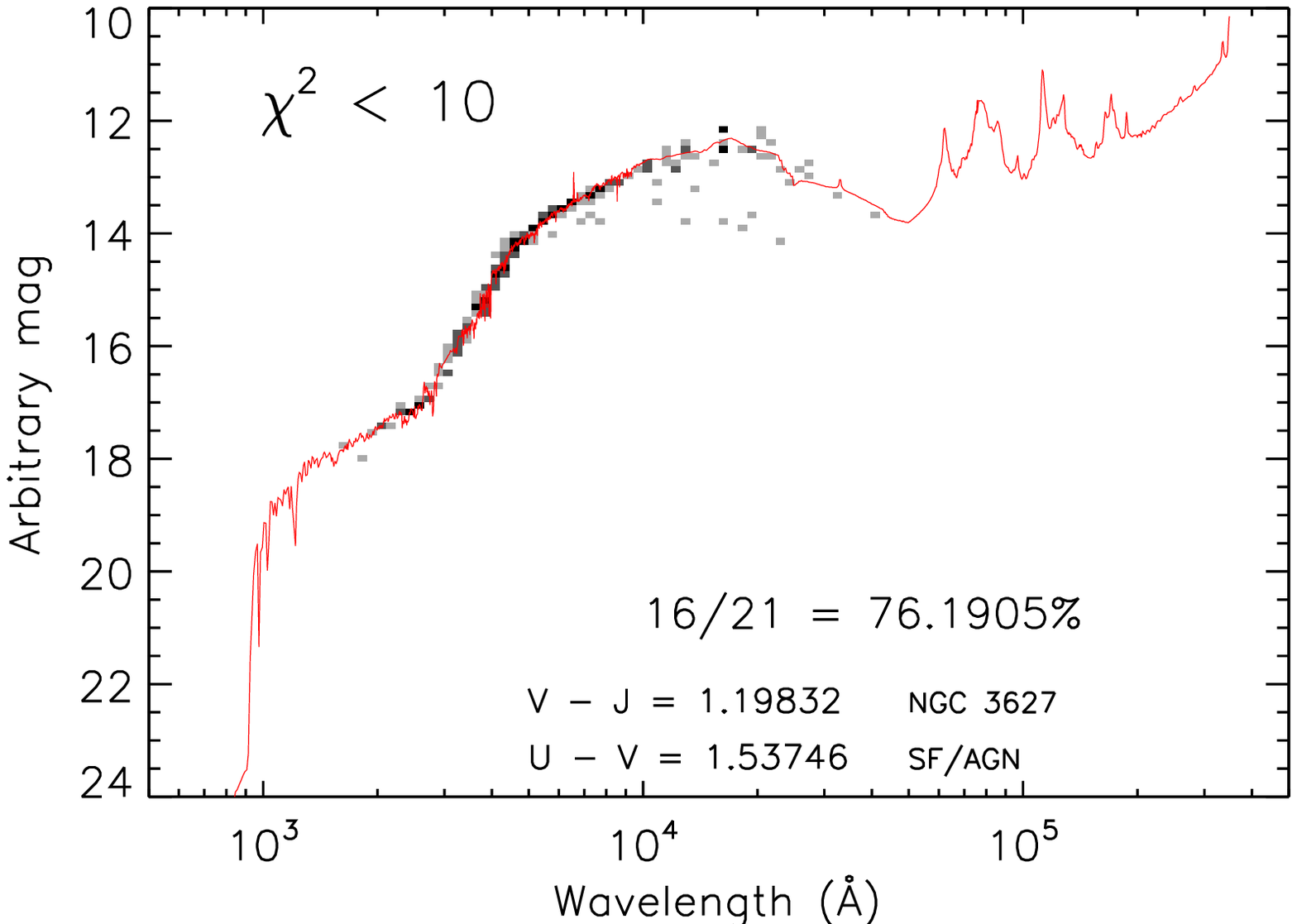}
\includegraphics[width=0.32\textwidth]{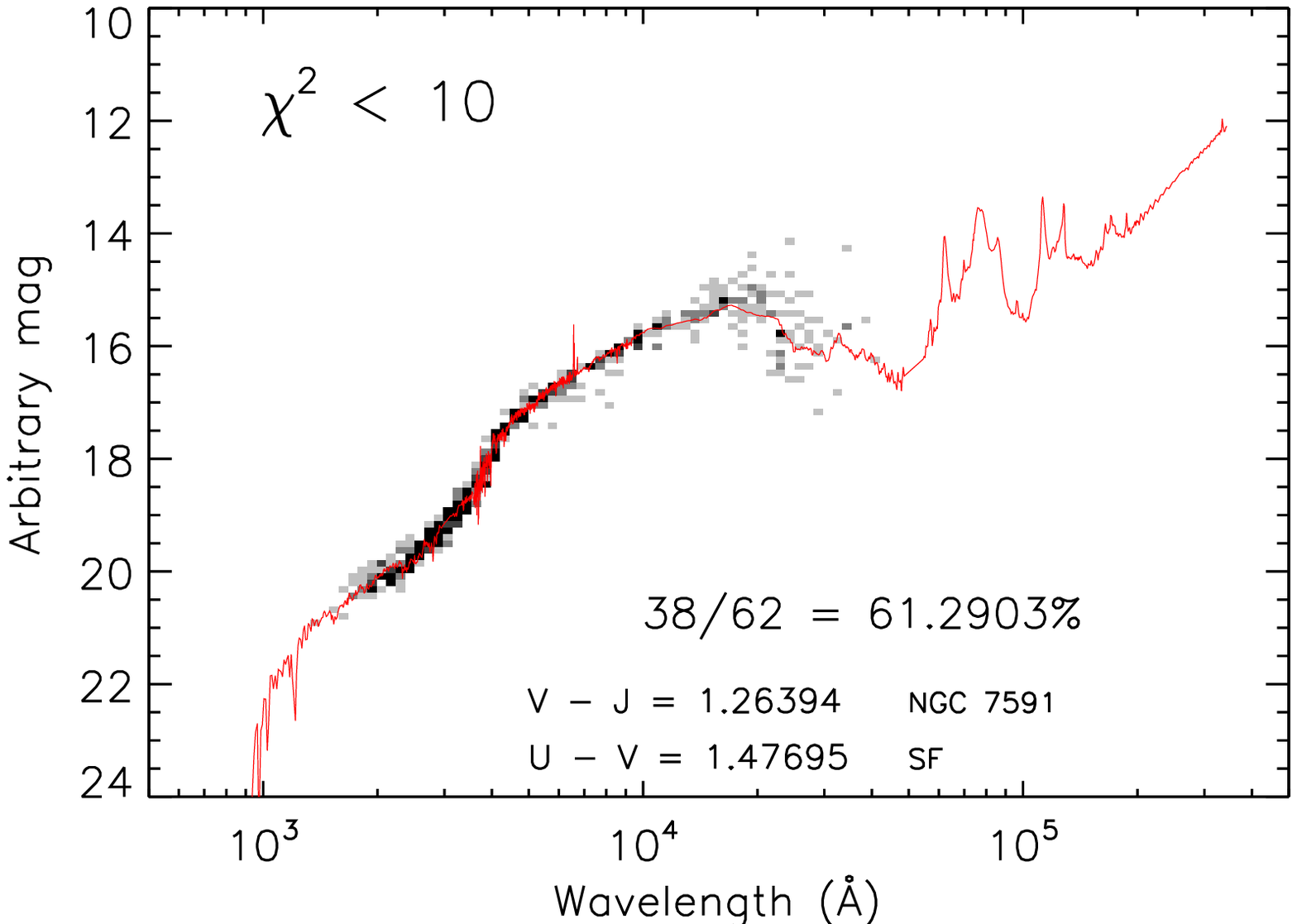}
\includegraphics[width=0.32\textwidth]{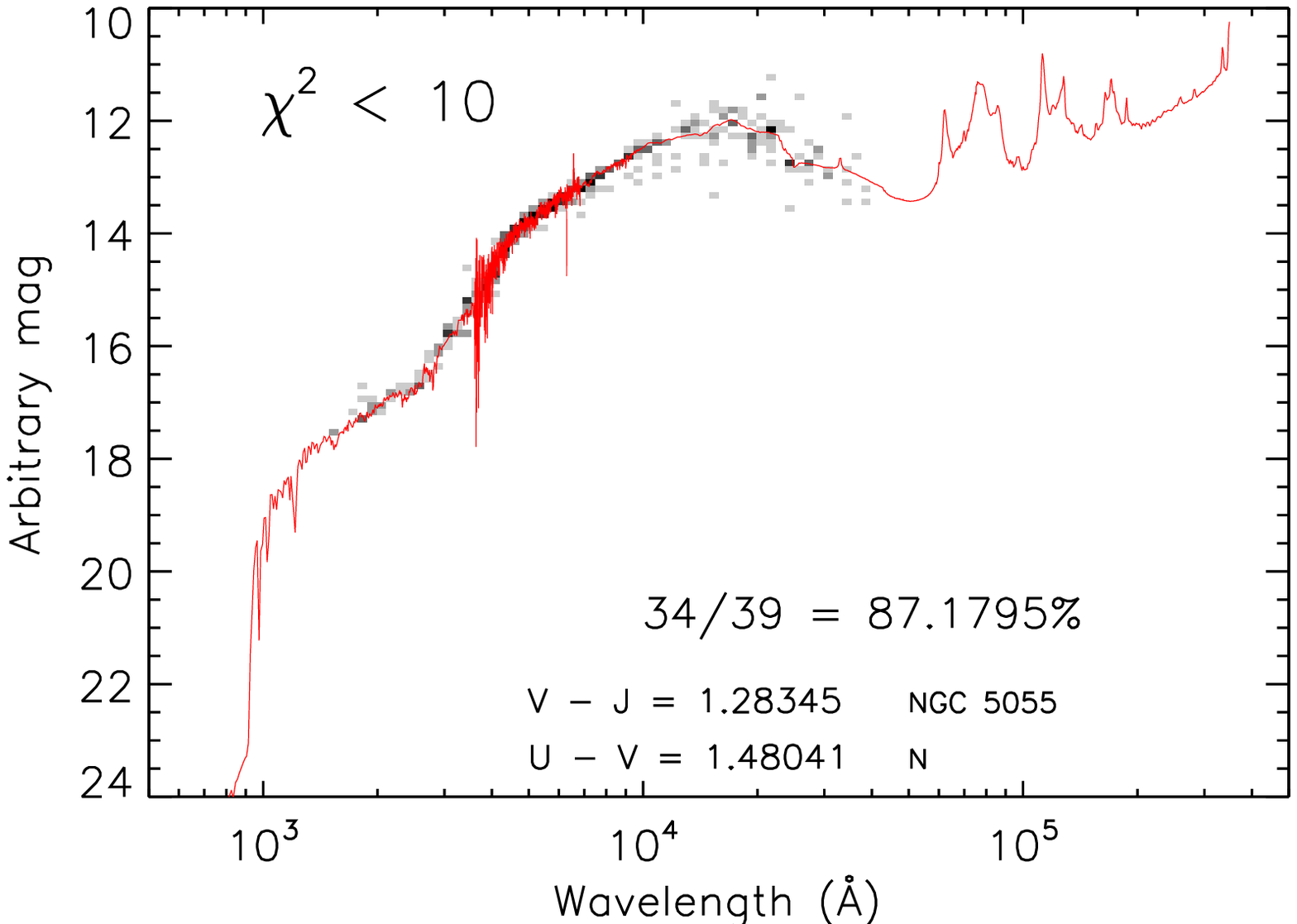}
\includegraphics[width=0.32\textwidth]{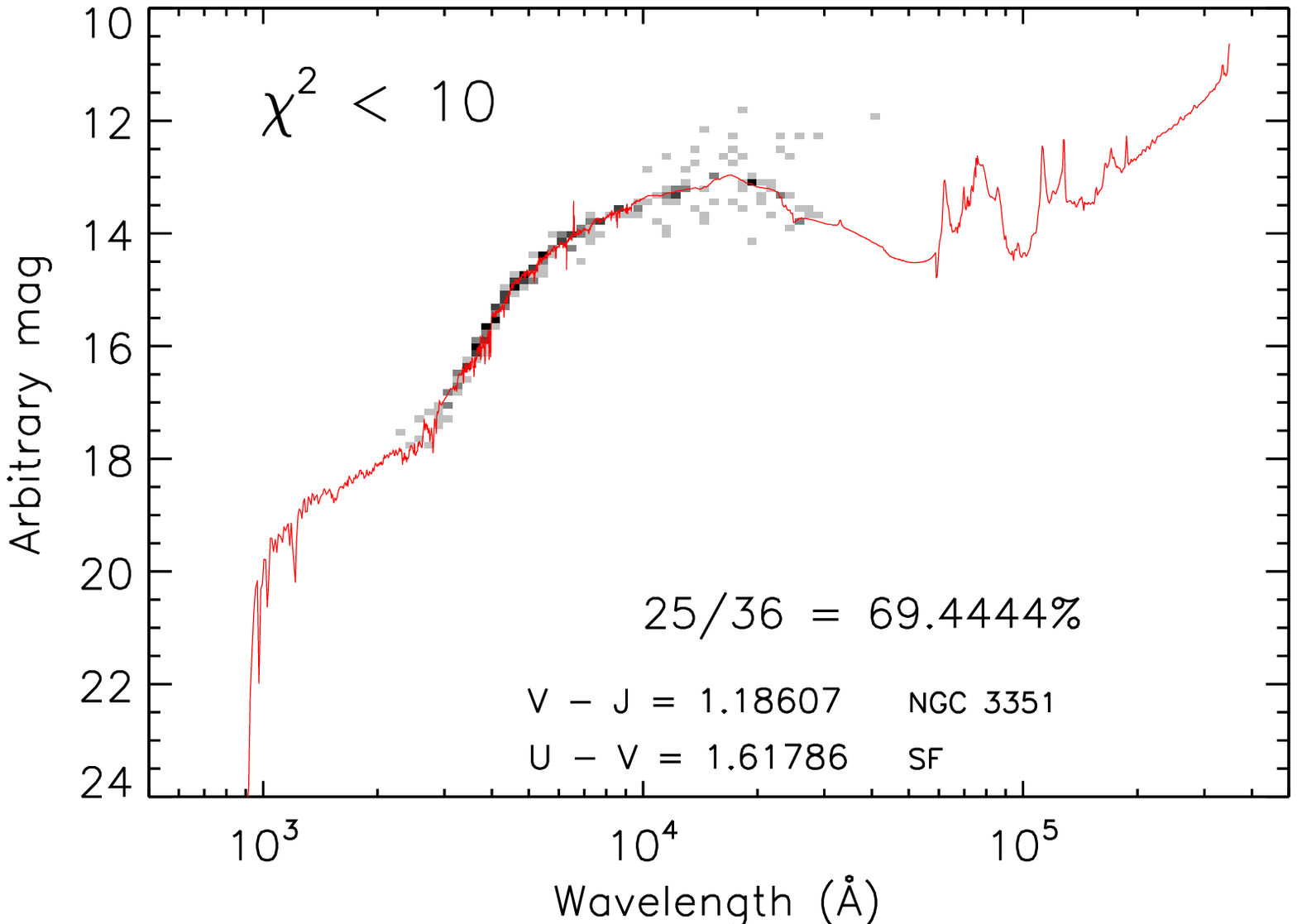}
\includegraphics[width=0.32\textwidth]{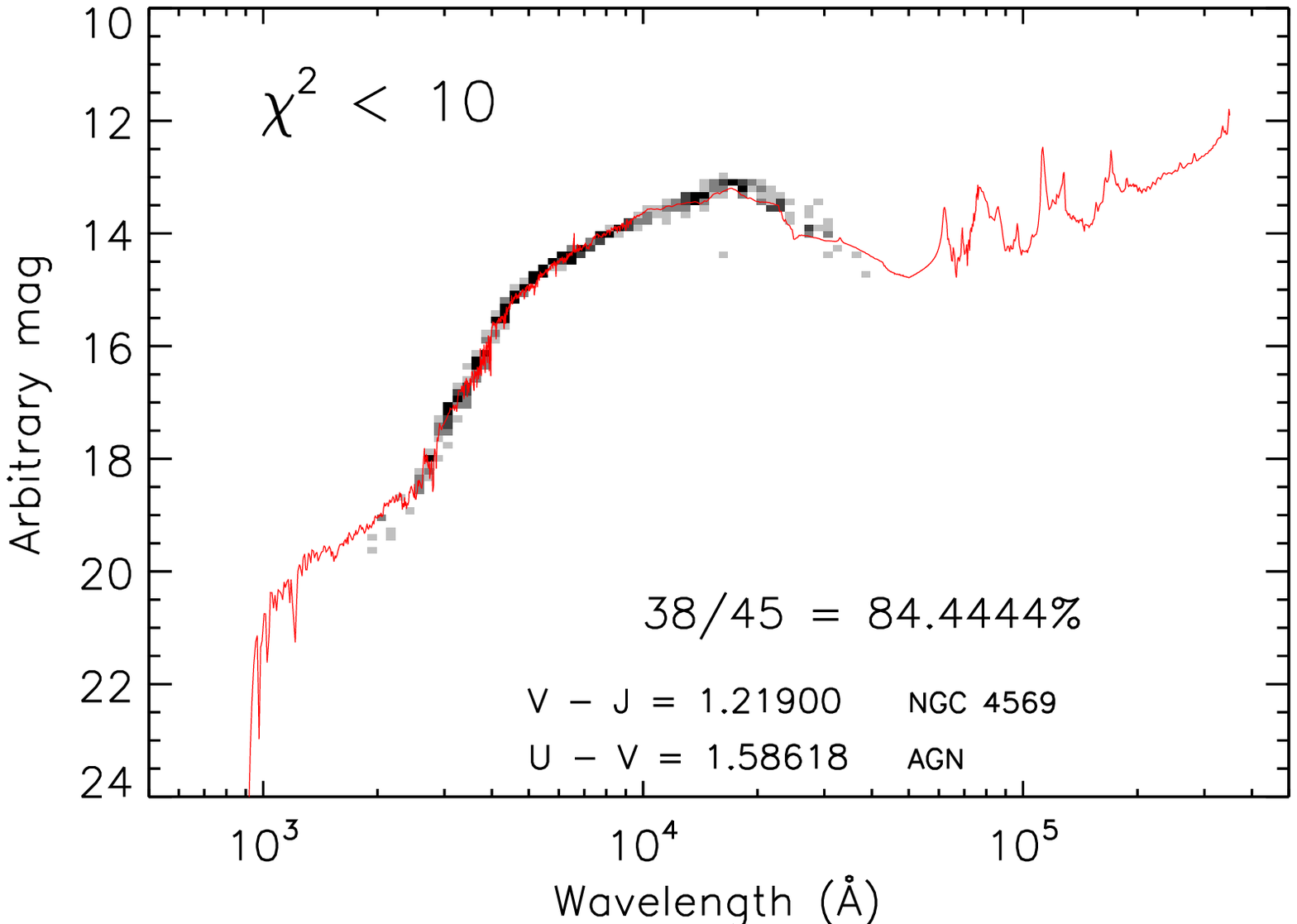}
\includegraphics[width=0.32\textwidth]{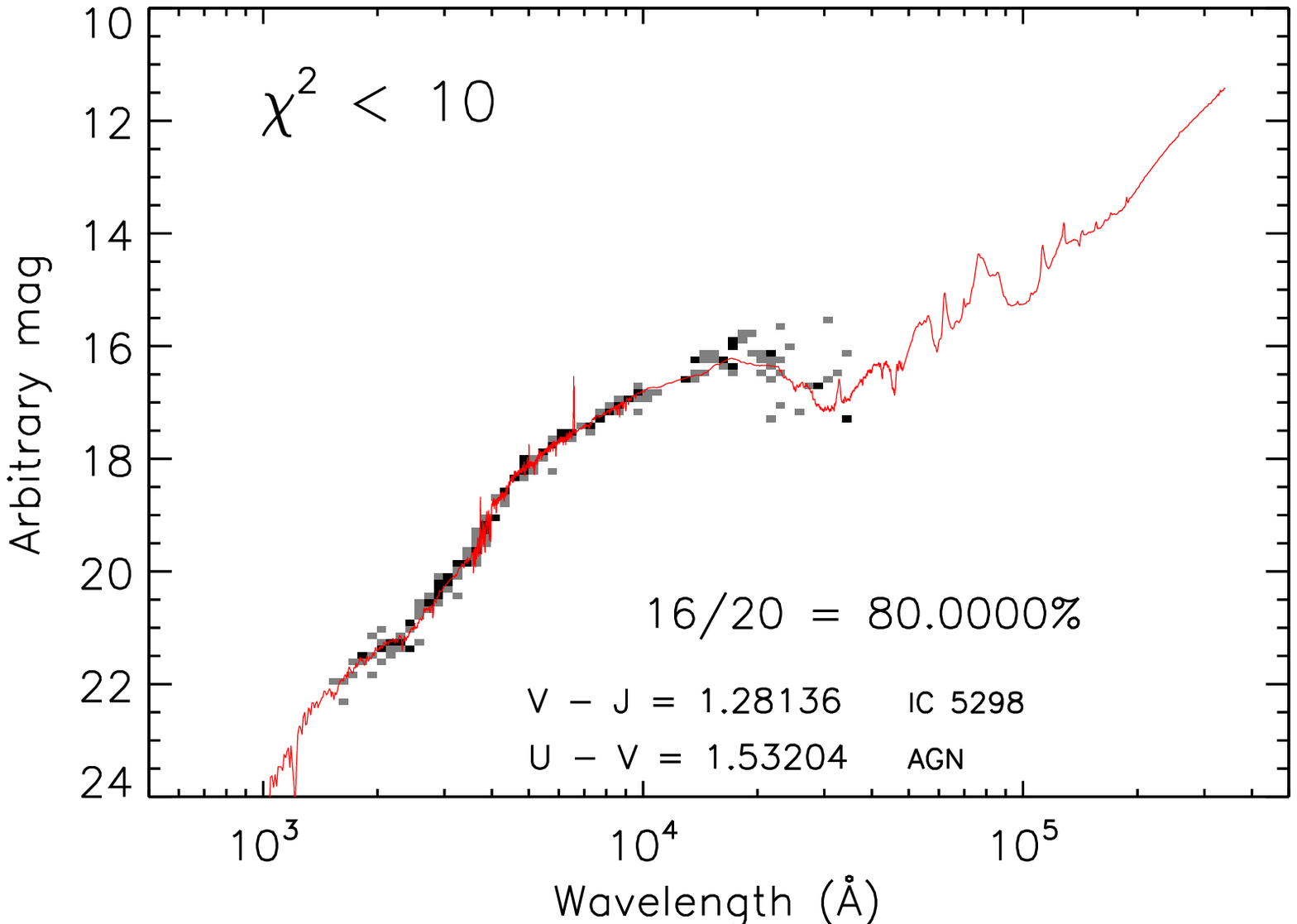}
\includegraphics[width=0.32\textwidth]{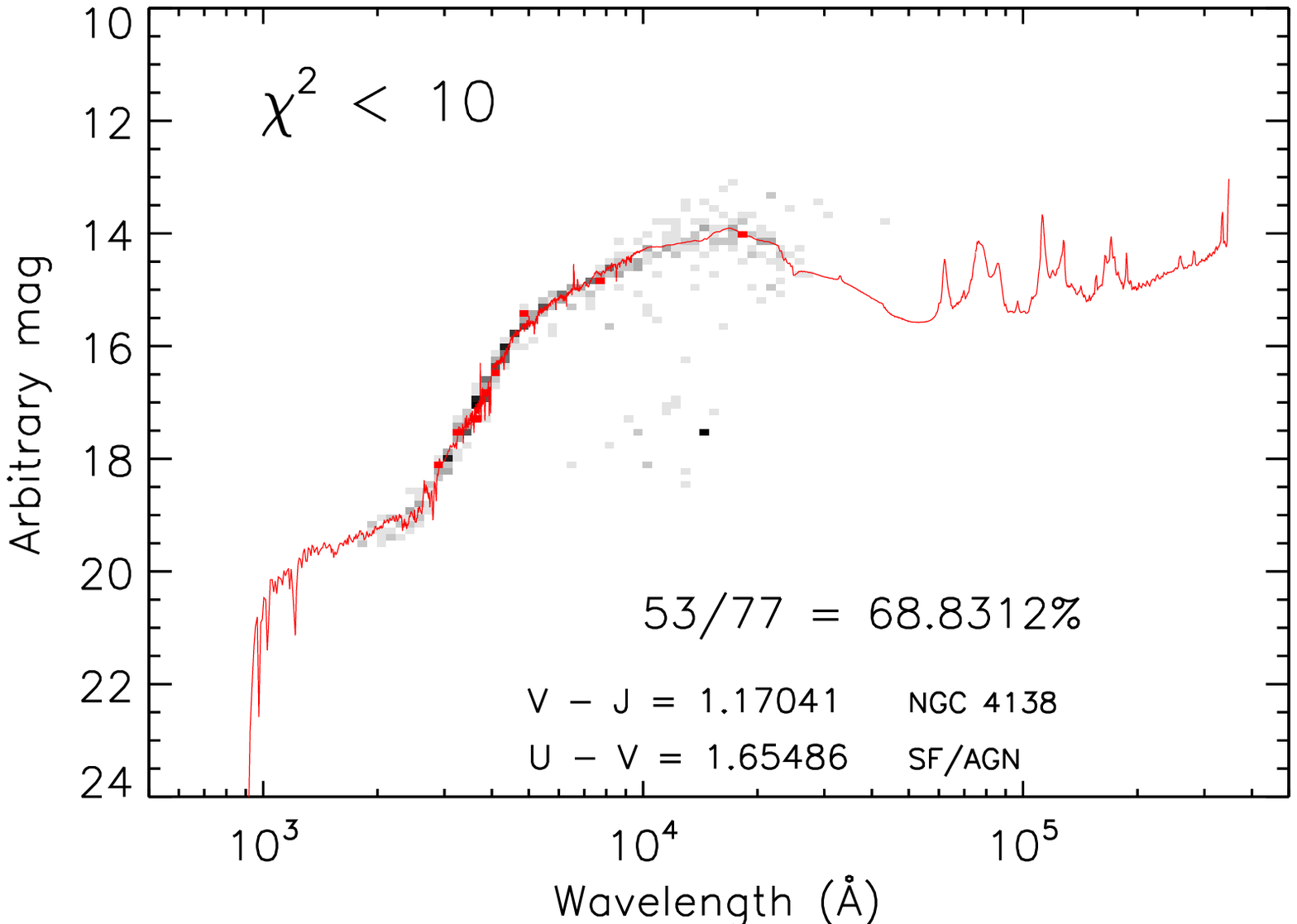}
\includegraphics[width=0.32\textwidth]{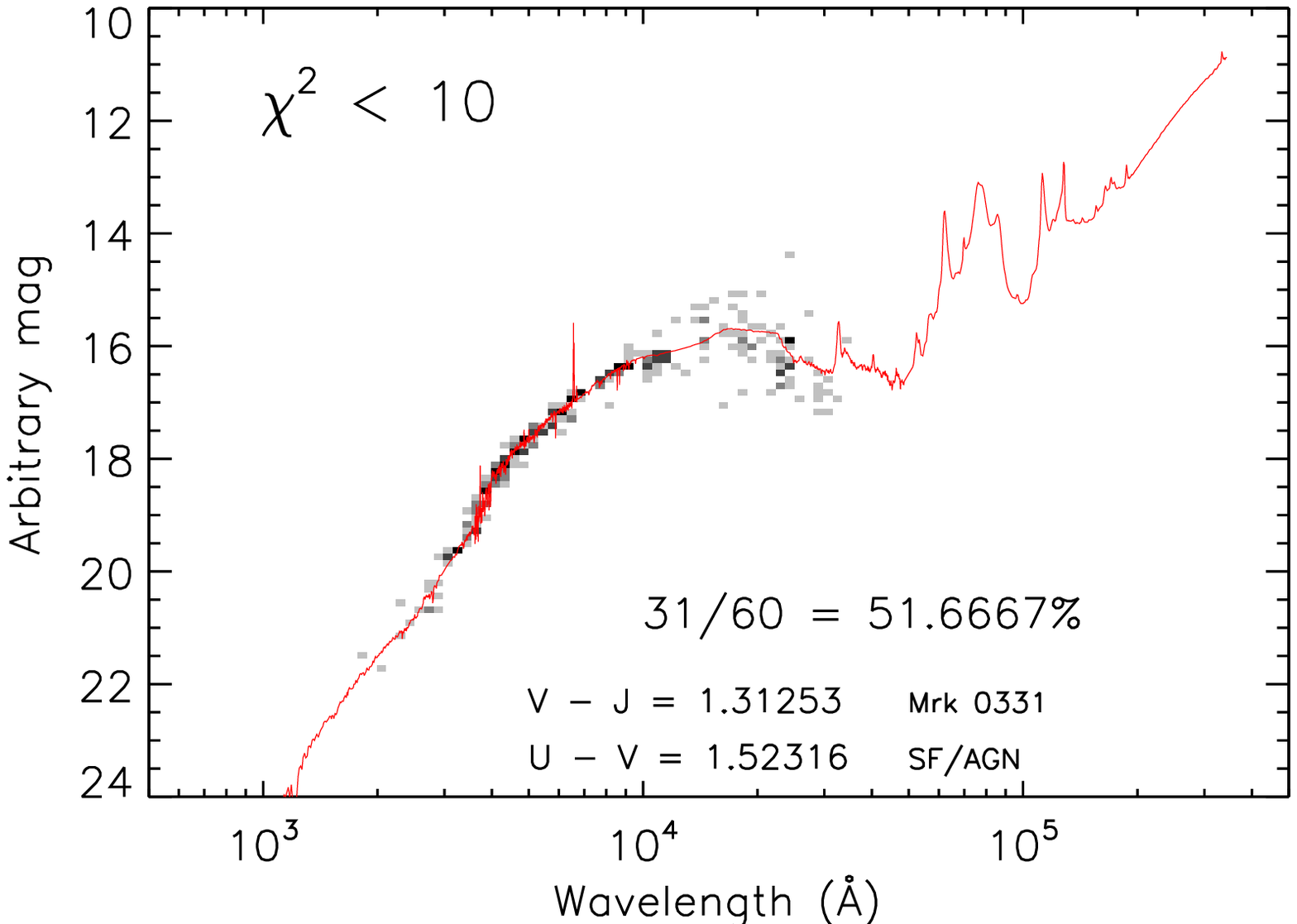}
\includegraphics[width=0.32\textwidth]{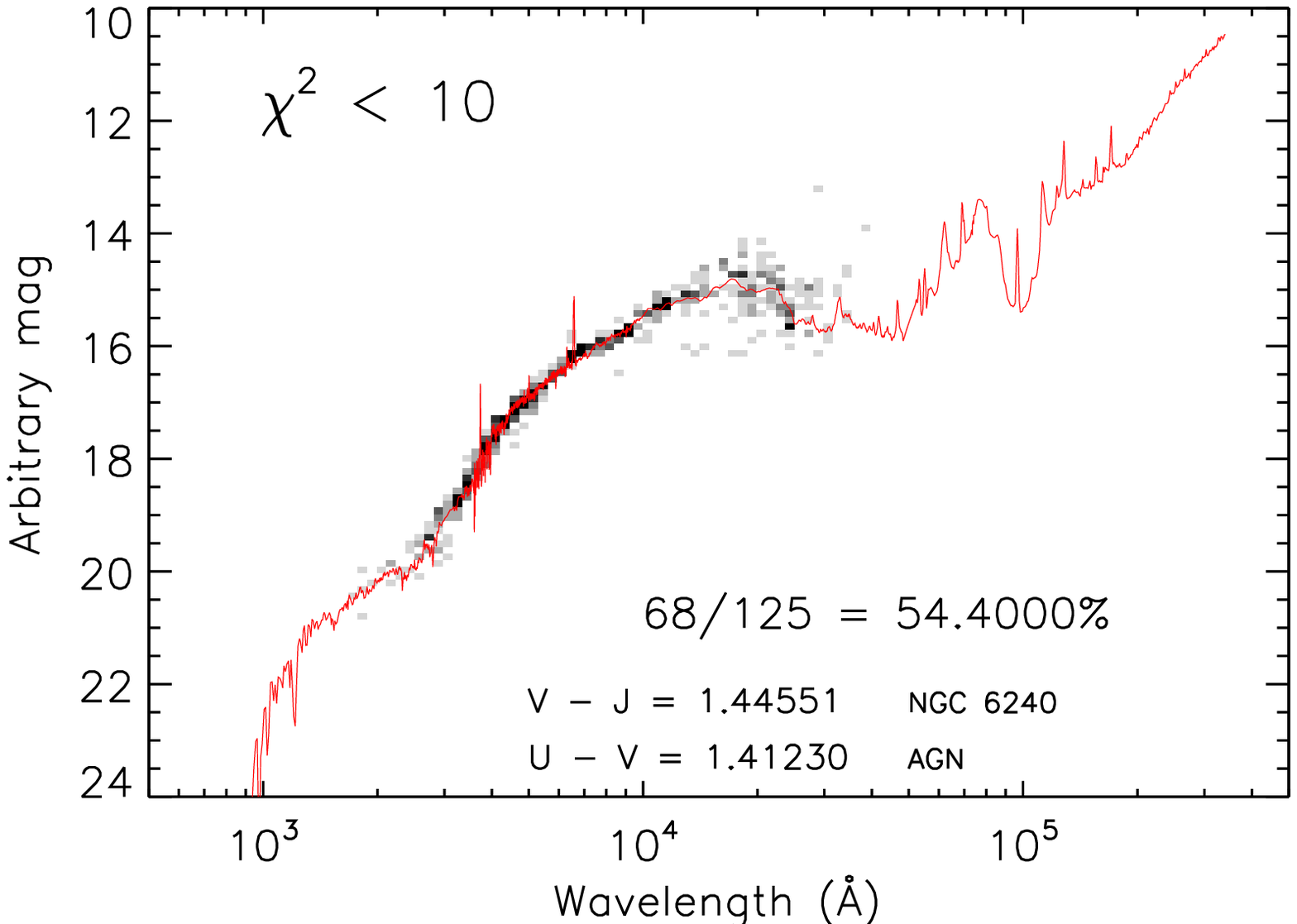}
    \caption{Continued.}
    \label{fig:my_label}
\end{figure}

\begin{figure}
    \centering
\includegraphics[width=0.32\textwidth]{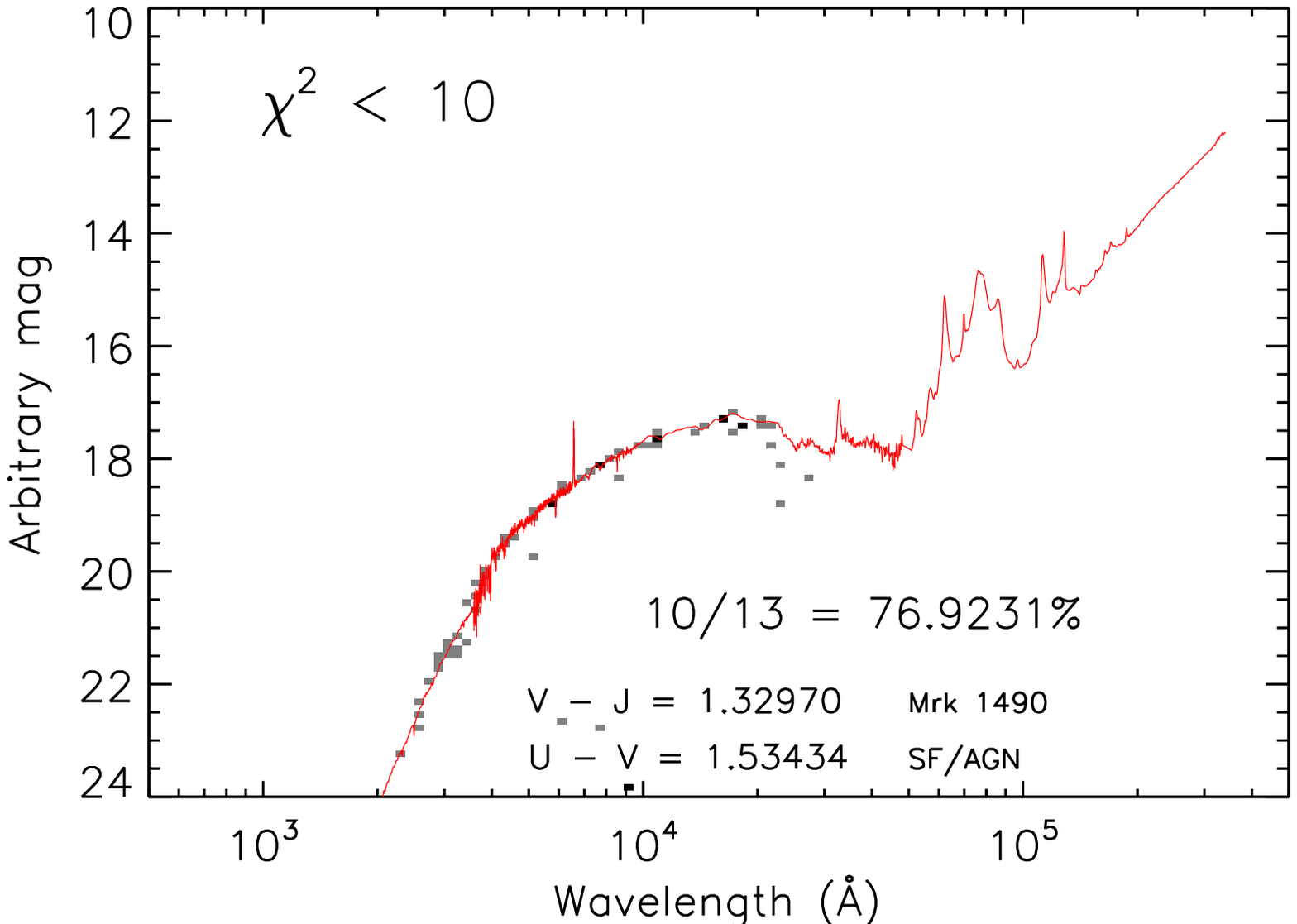}
\includegraphics[width=0.32\textwidth]{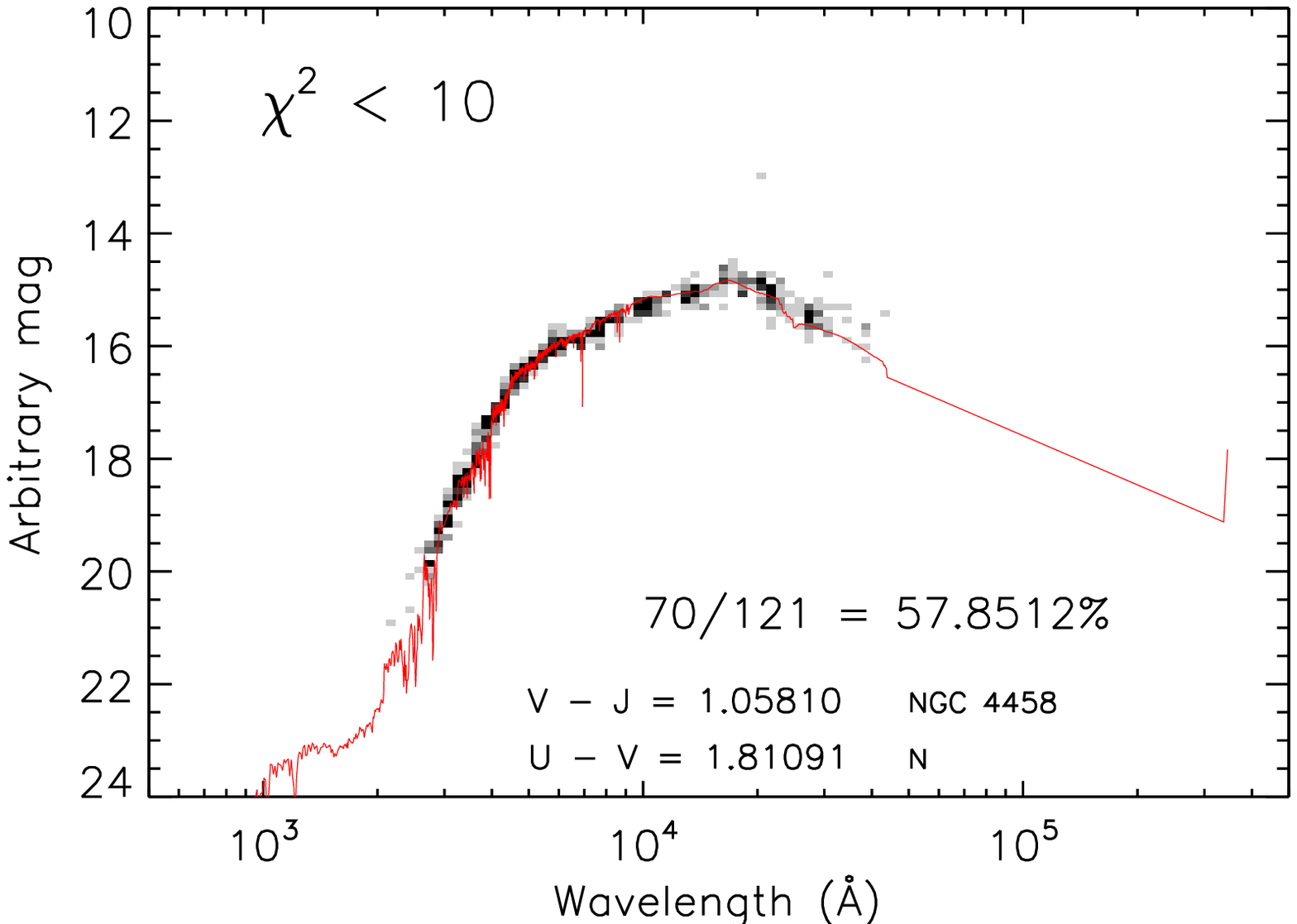}
\includegraphics[width=0.32\textwidth]{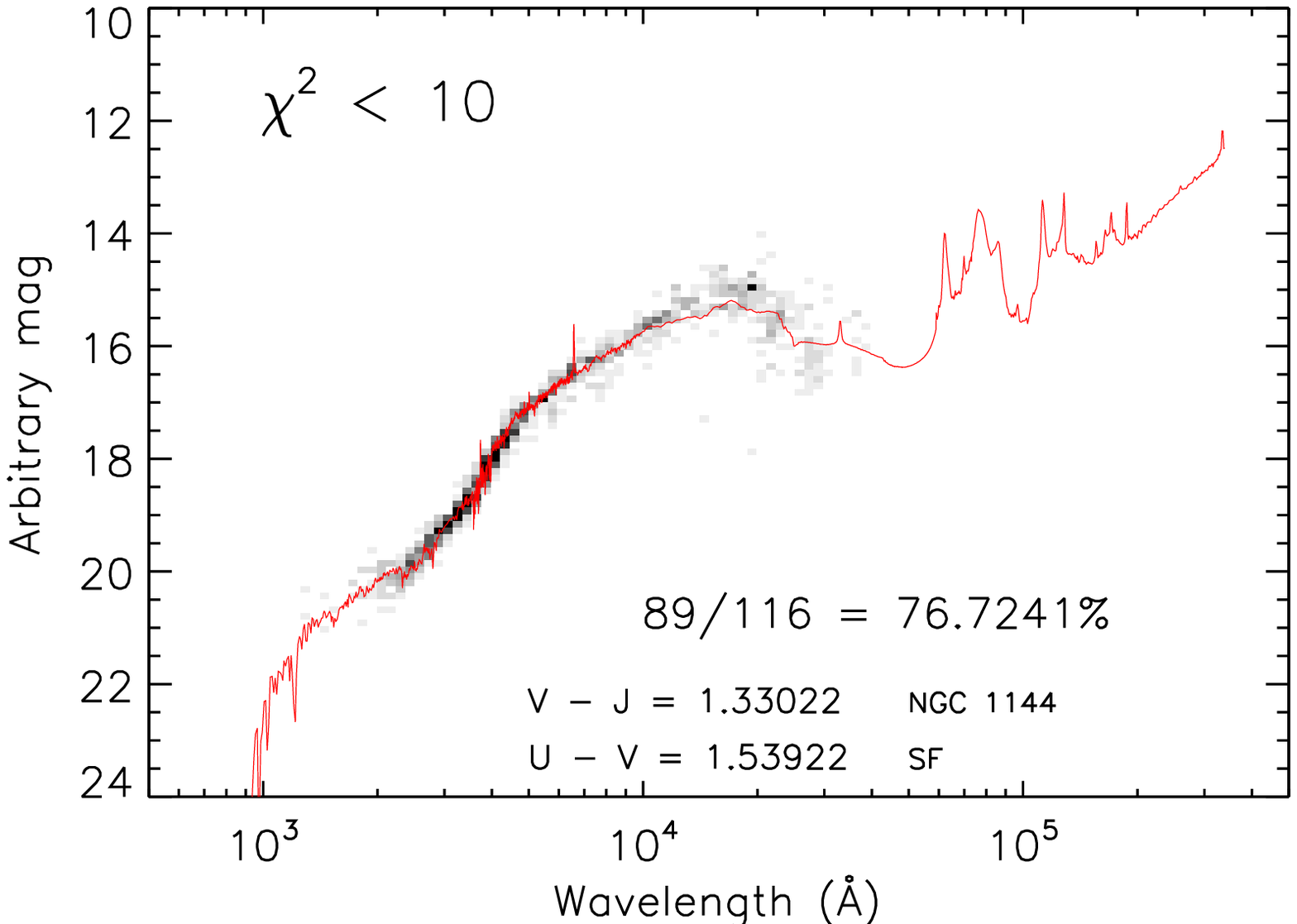}
\includegraphics[width=0.32\textwidth]{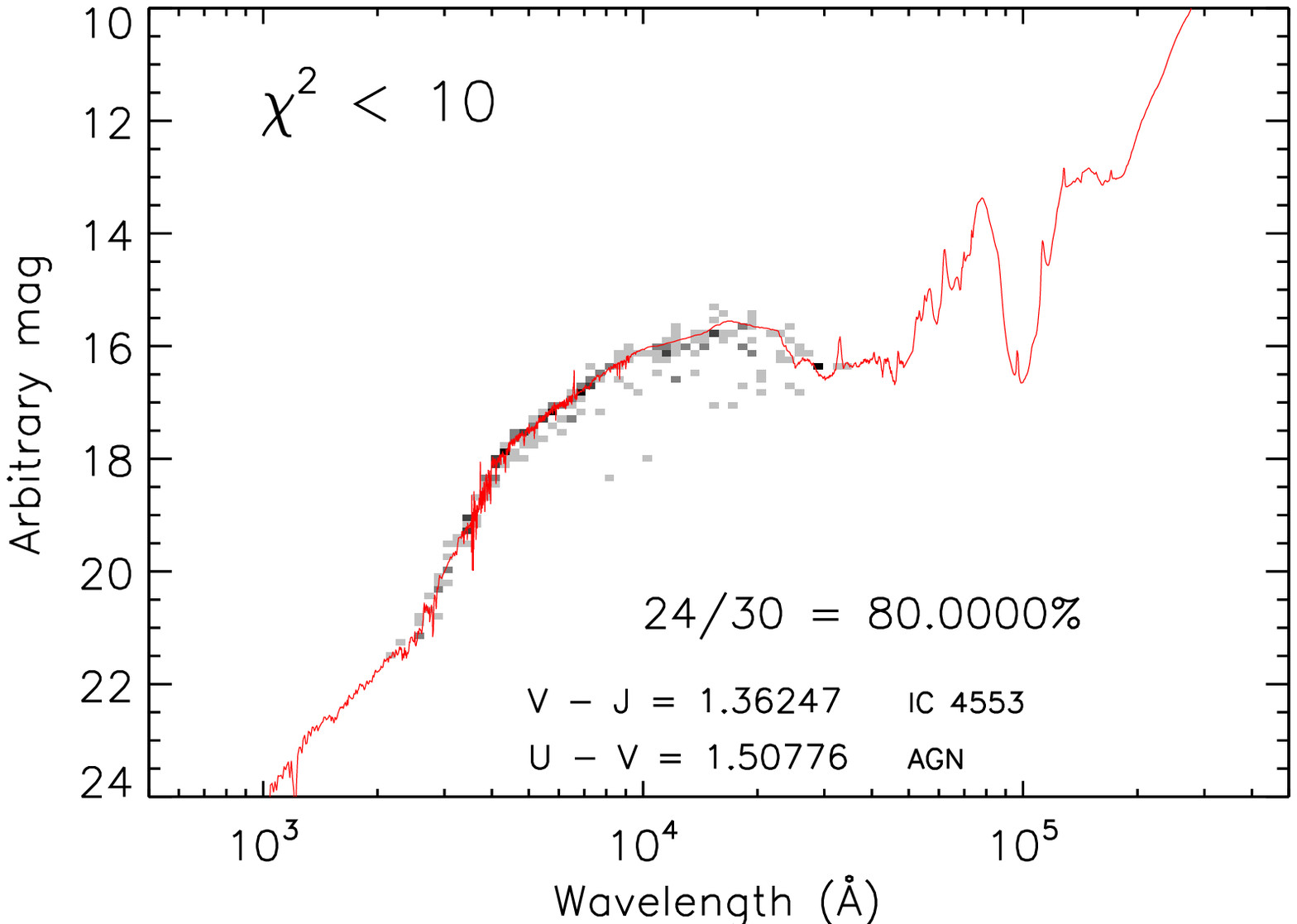}
\includegraphics[width=0.32\textwidth]{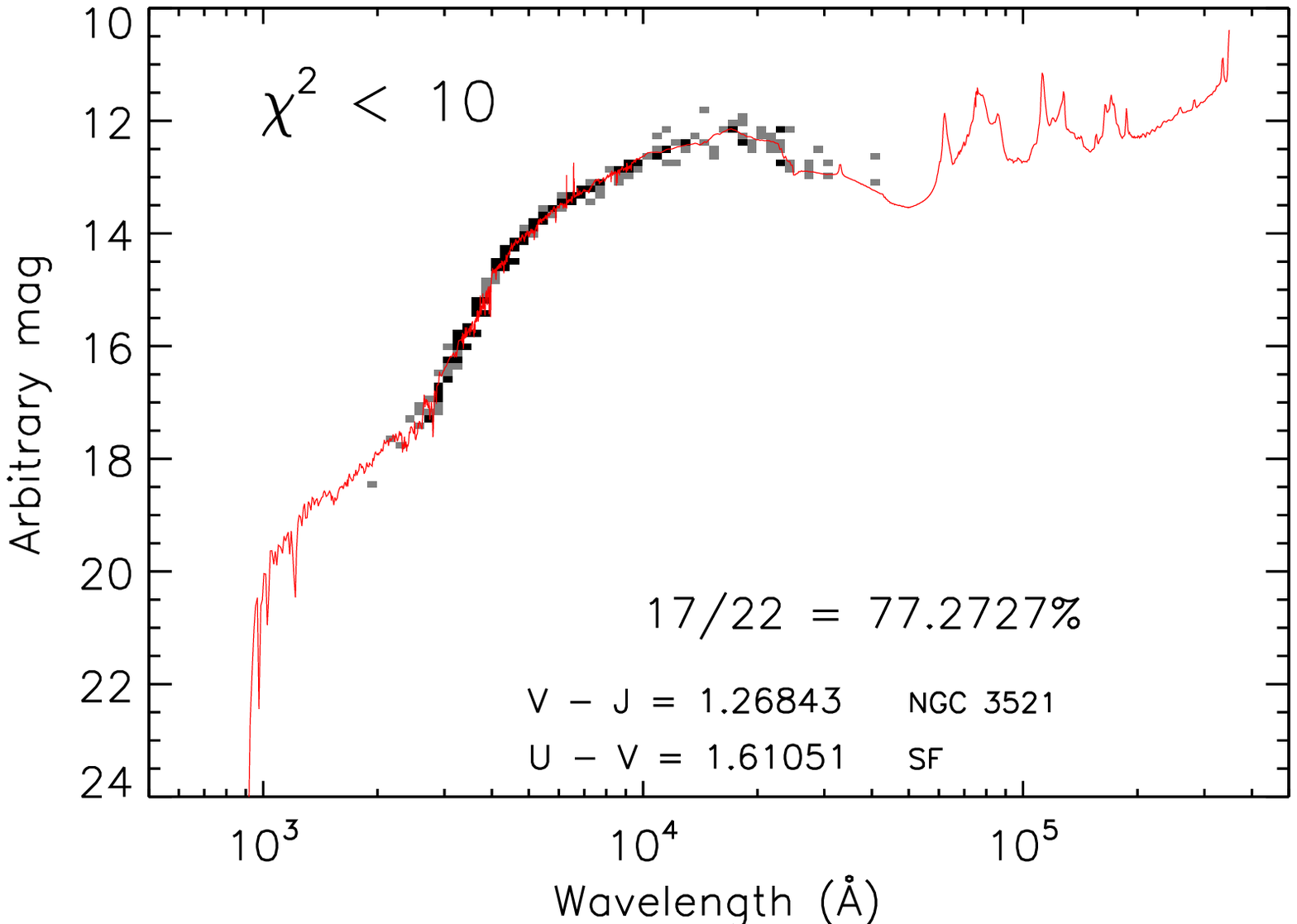}
\includegraphics[width=0.32\textwidth]{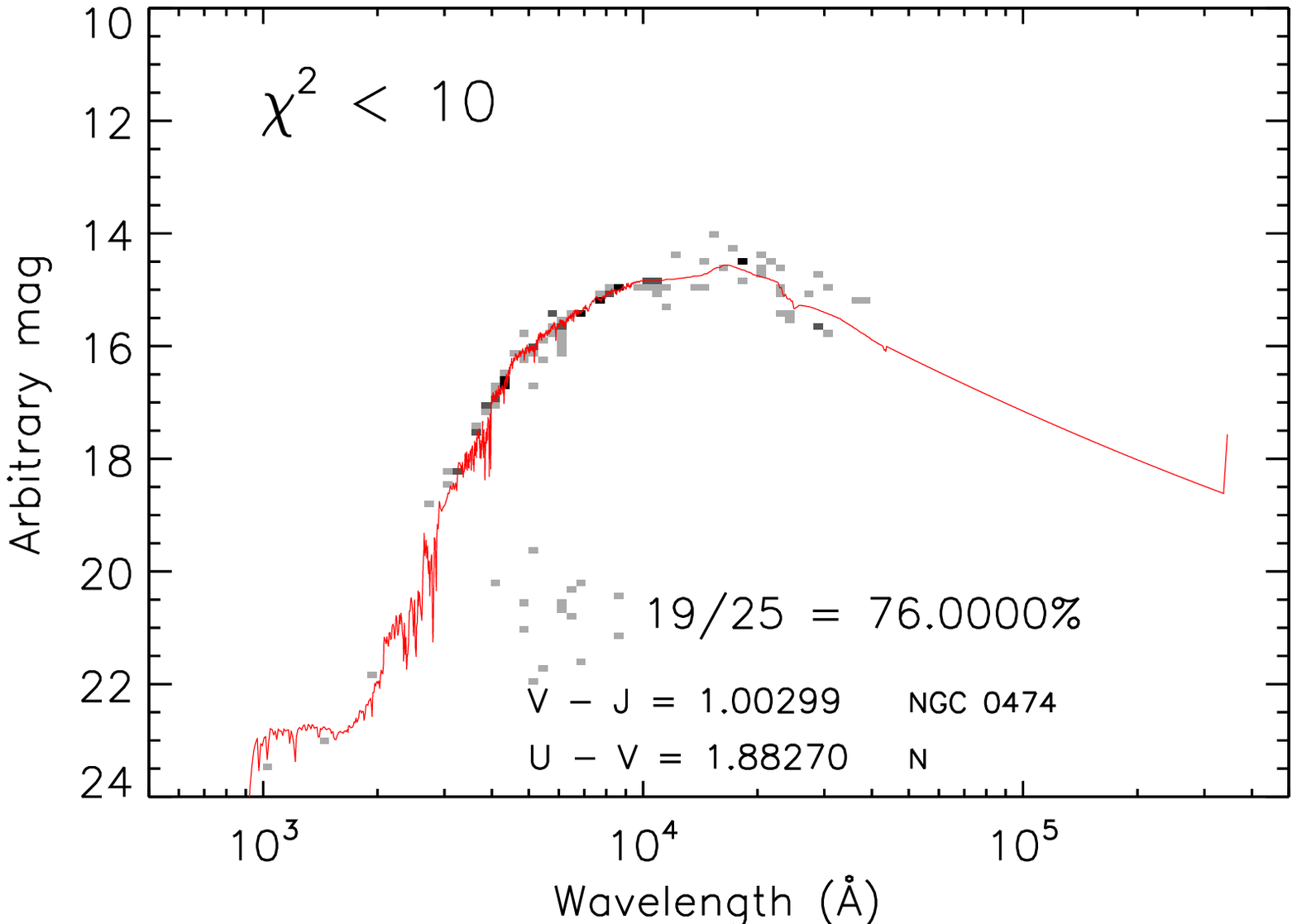}
\includegraphics[width=0.32\textwidth]{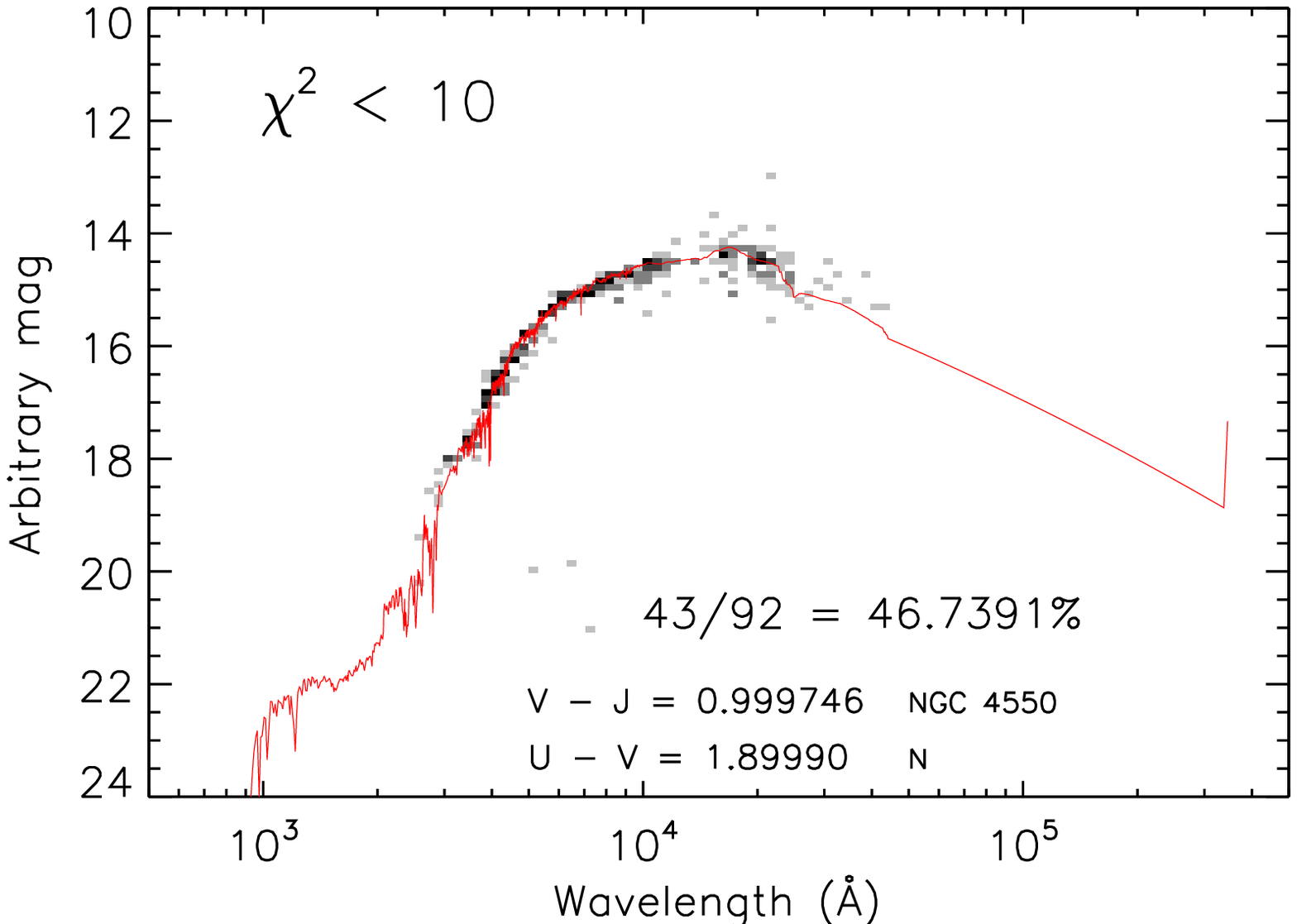}
\includegraphics[width=0.32\textwidth]{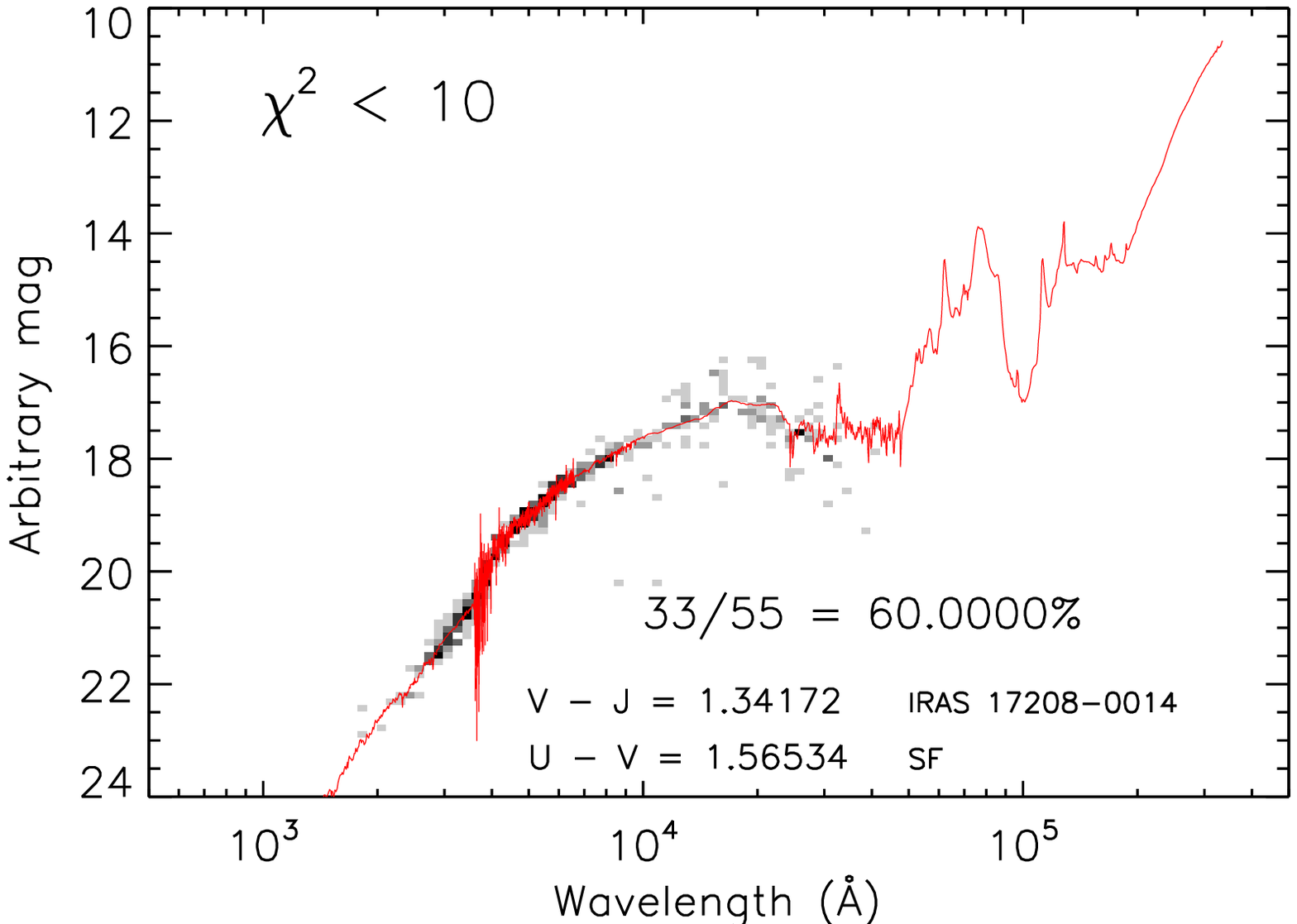}
\includegraphics[width=0.32\textwidth]{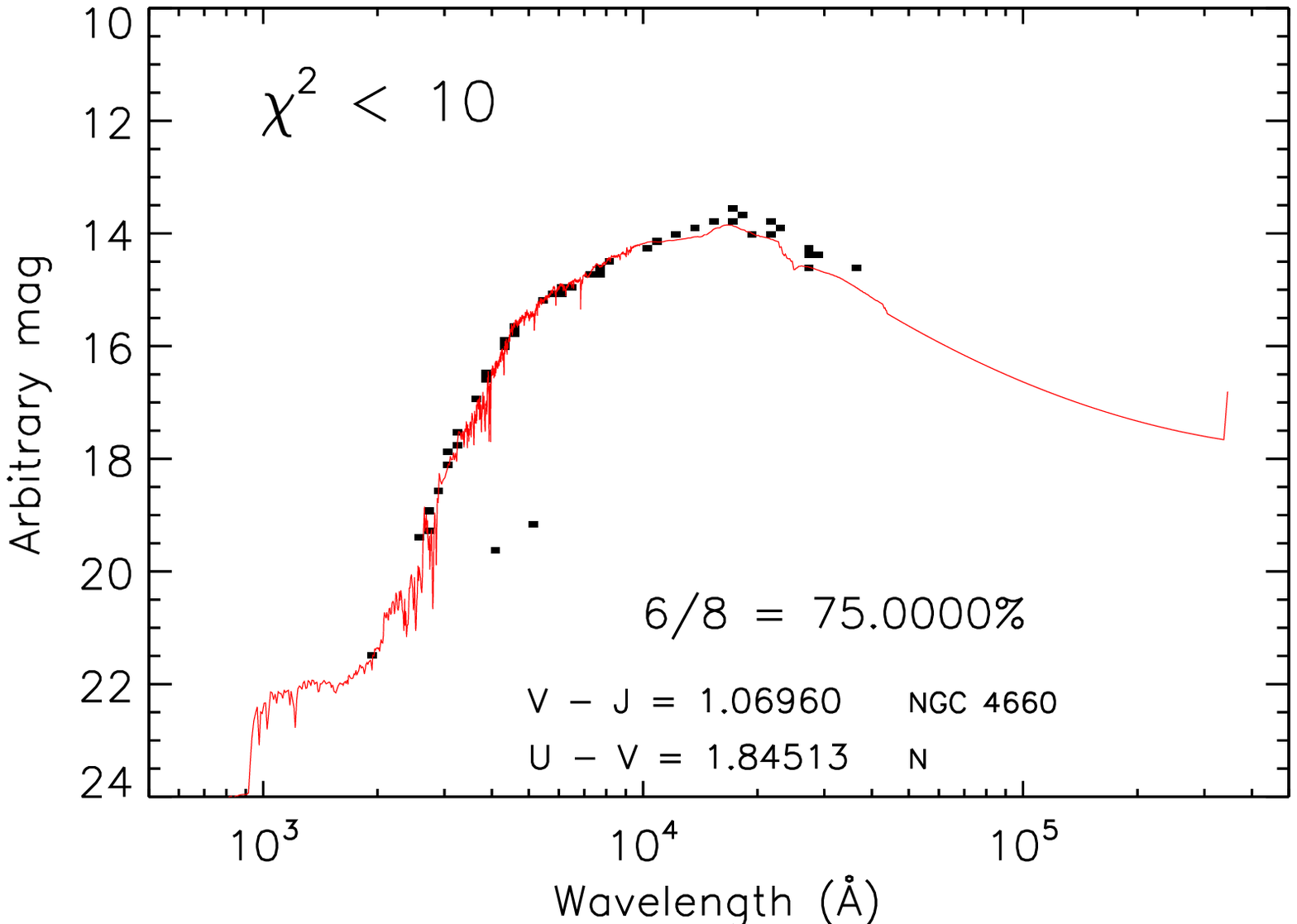}
\includegraphics[width=0.32\textwidth]{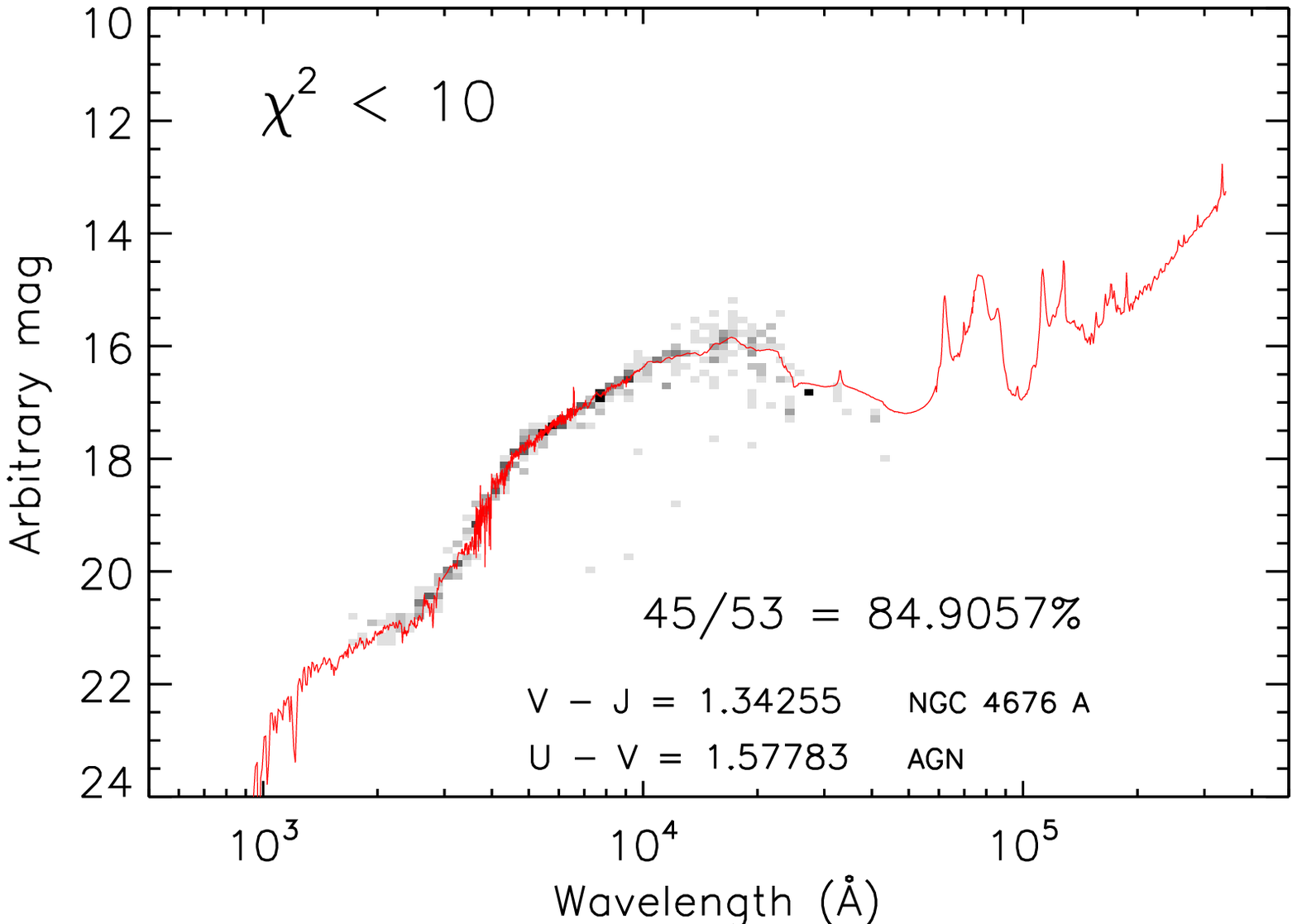}
\includegraphics[width=0.32\textwidth]{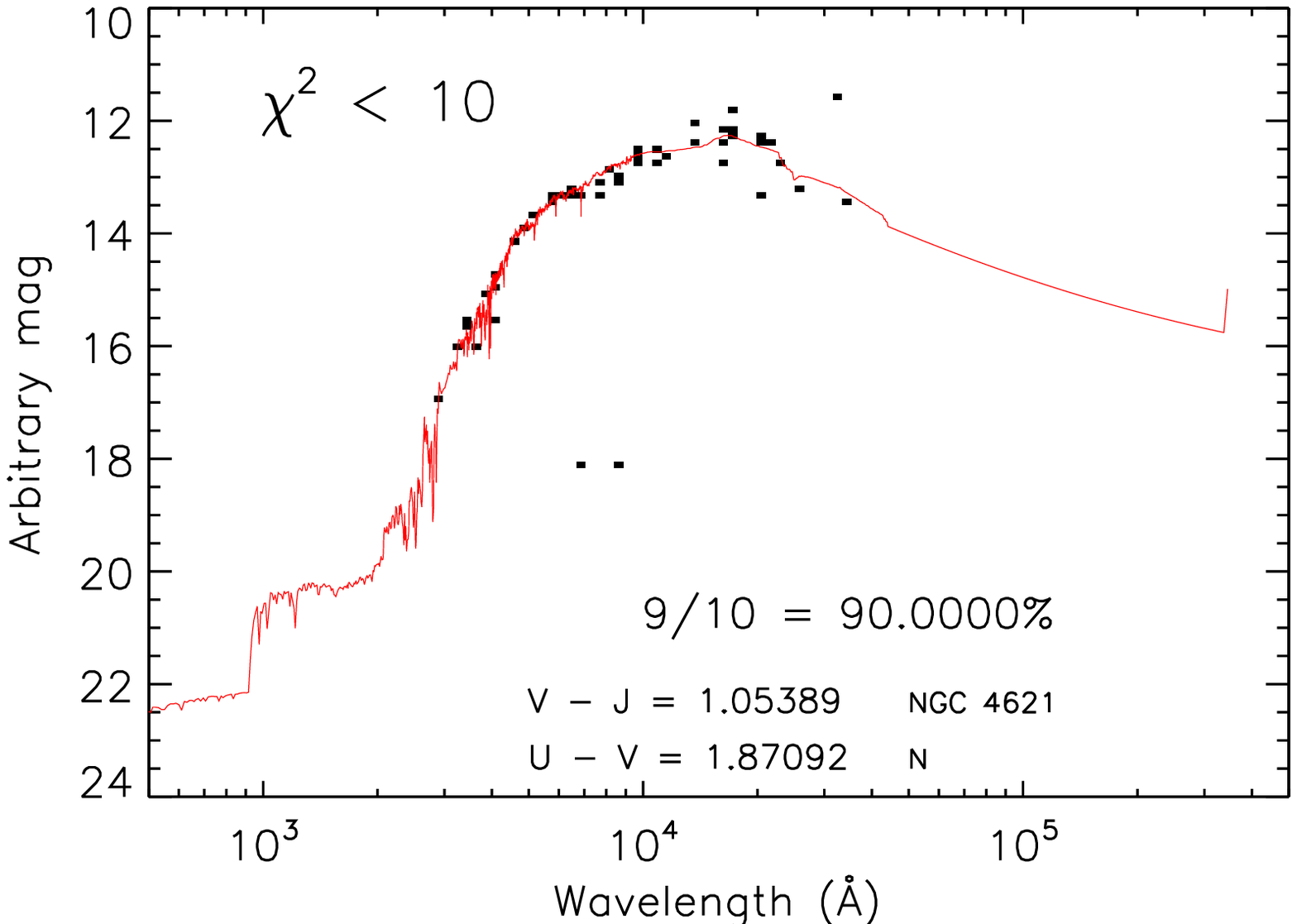}
\includegraphics[width=0.32\textwidth]{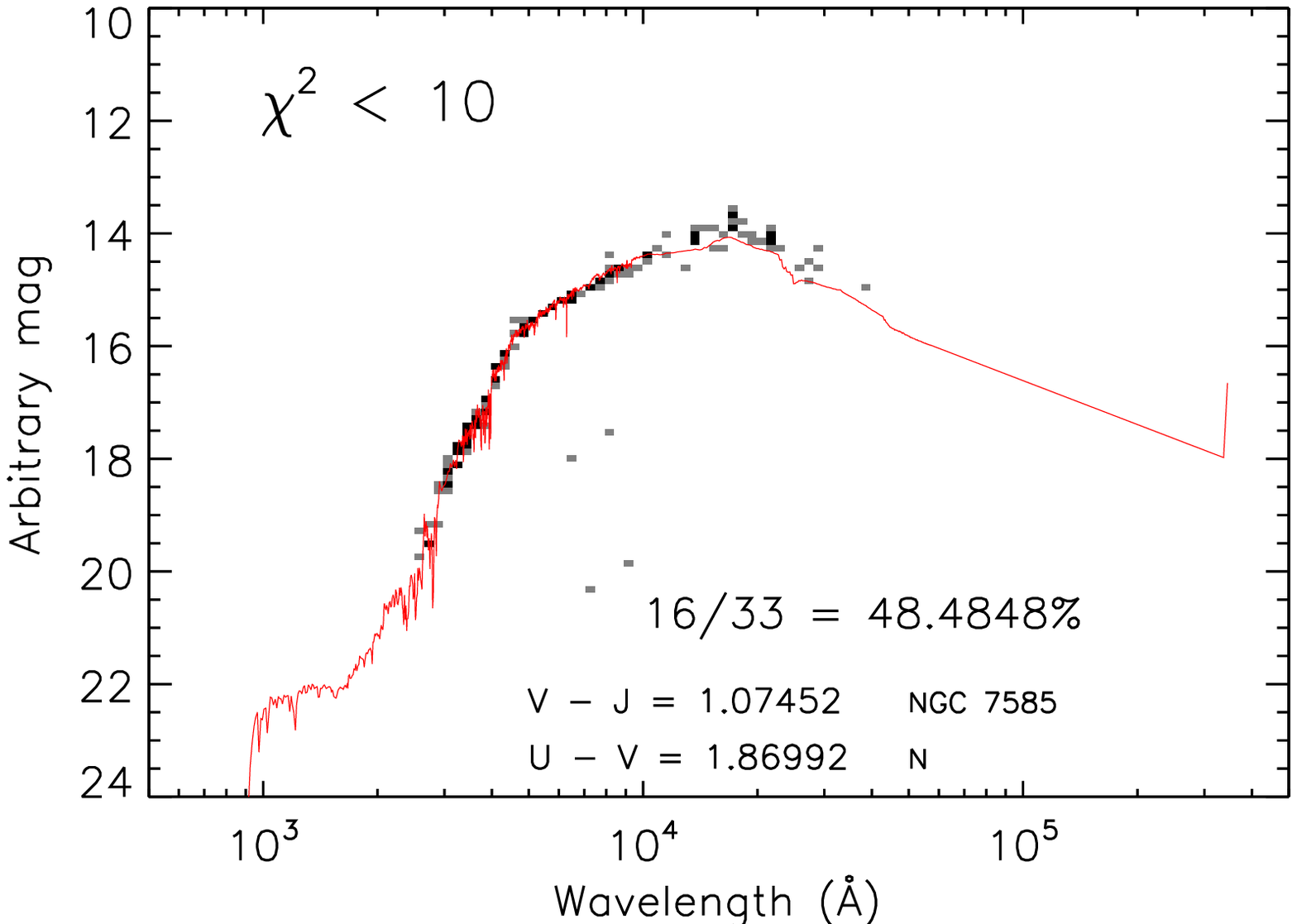}
\includegraphics[width=0.32\textwidth]{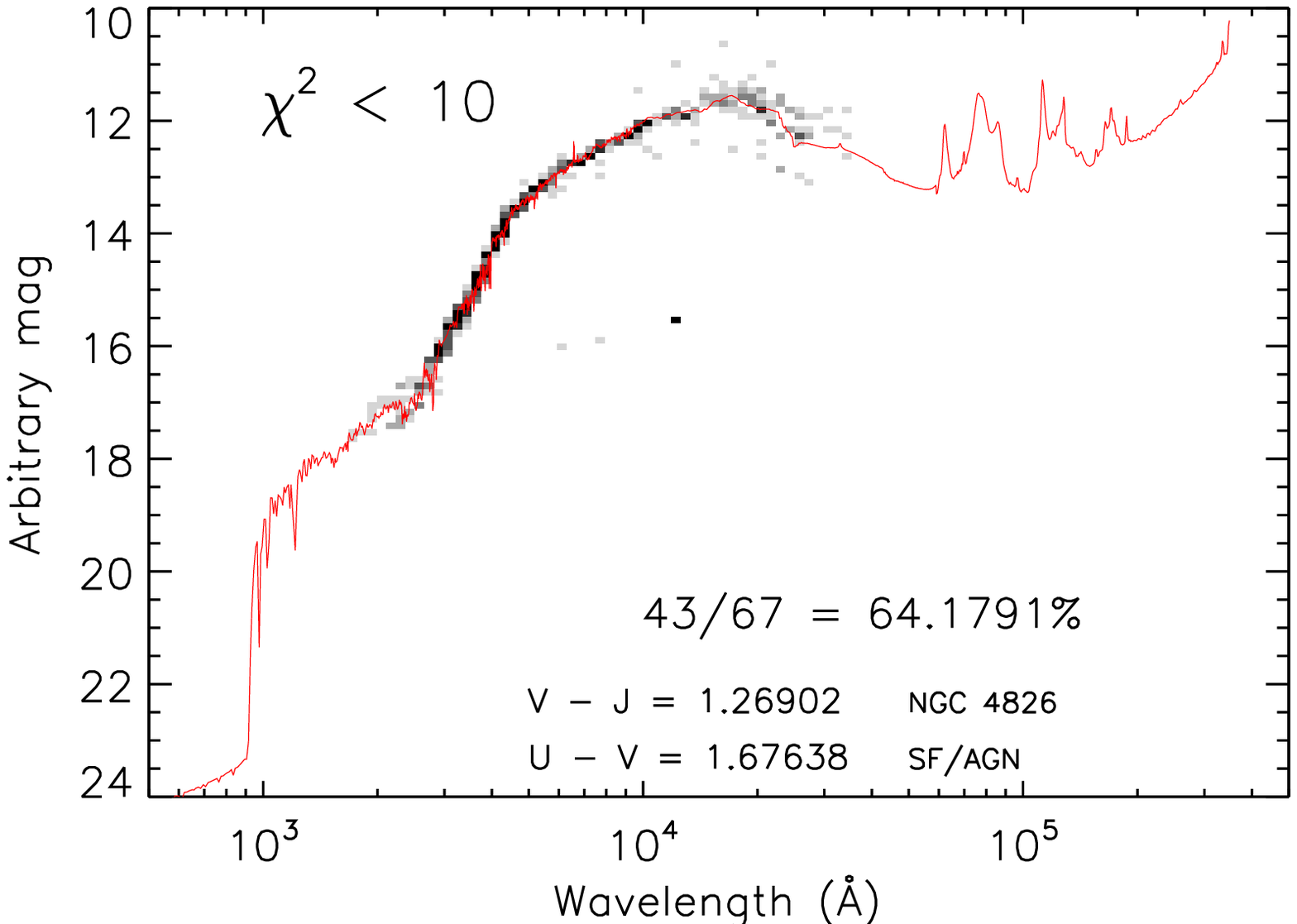}
\includegraphics[width=0.32\textwidth]{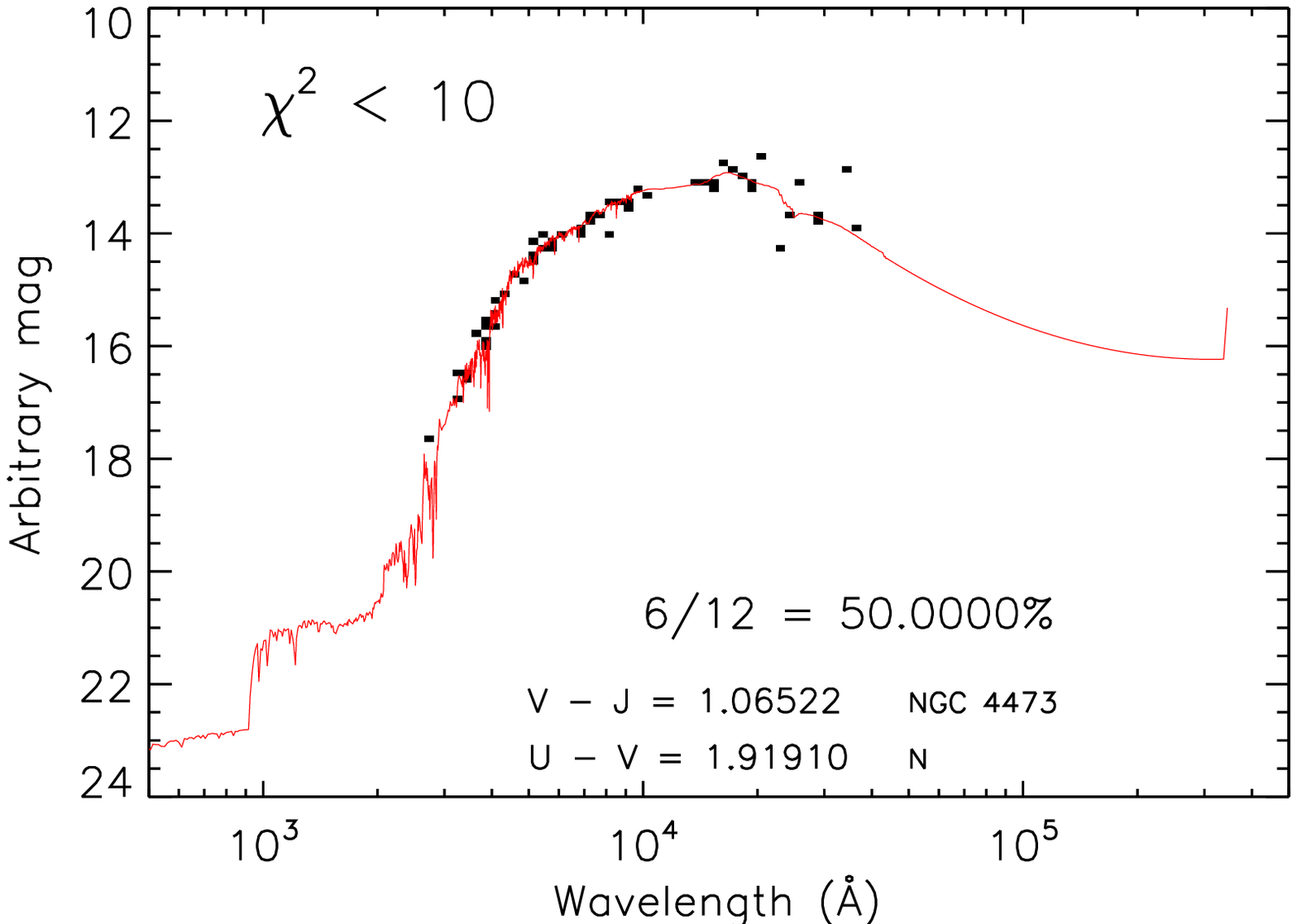}
\includegraphics[width=0.32\textwidth]{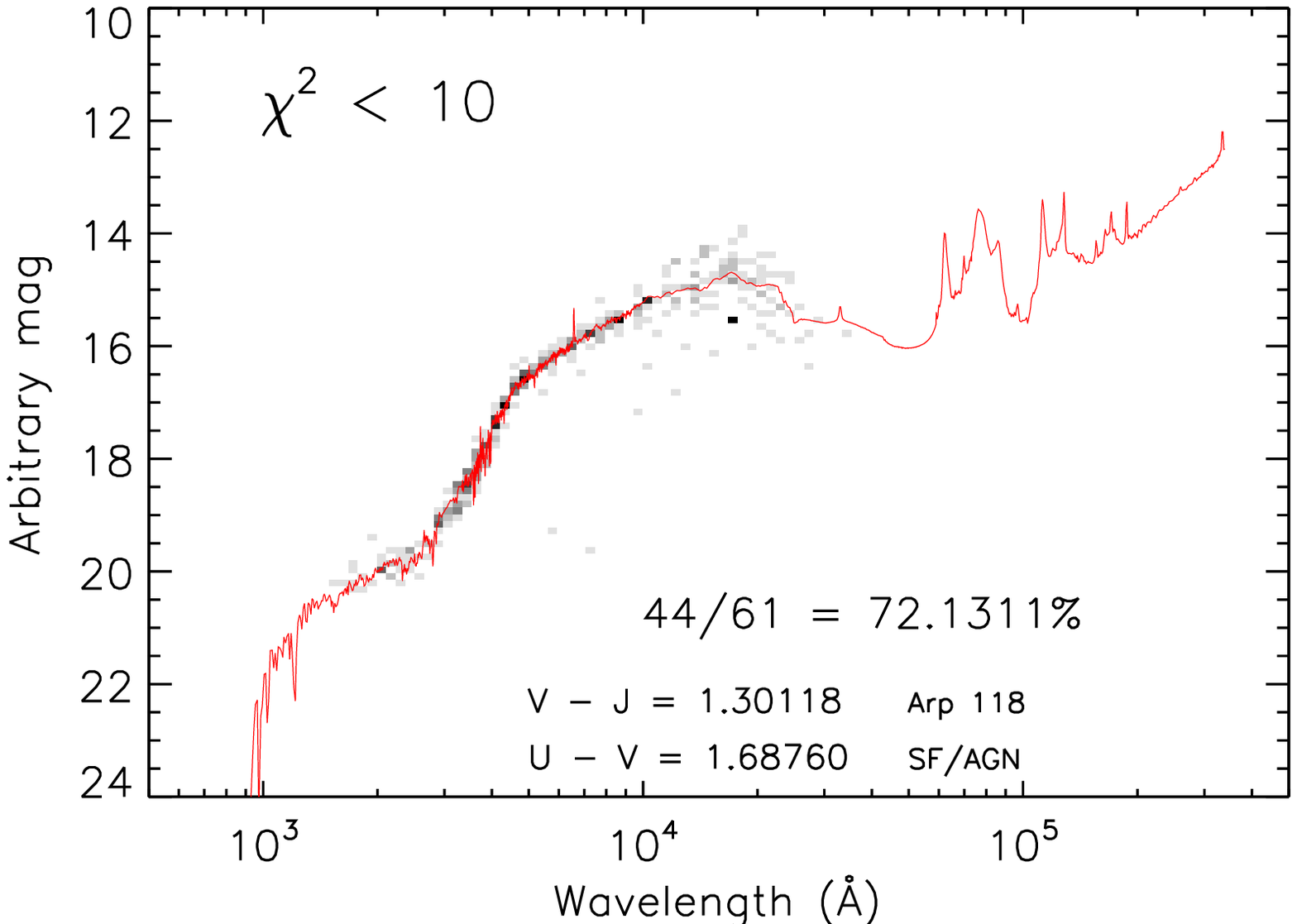}
\includegraphics[width=0.32\textwidth]{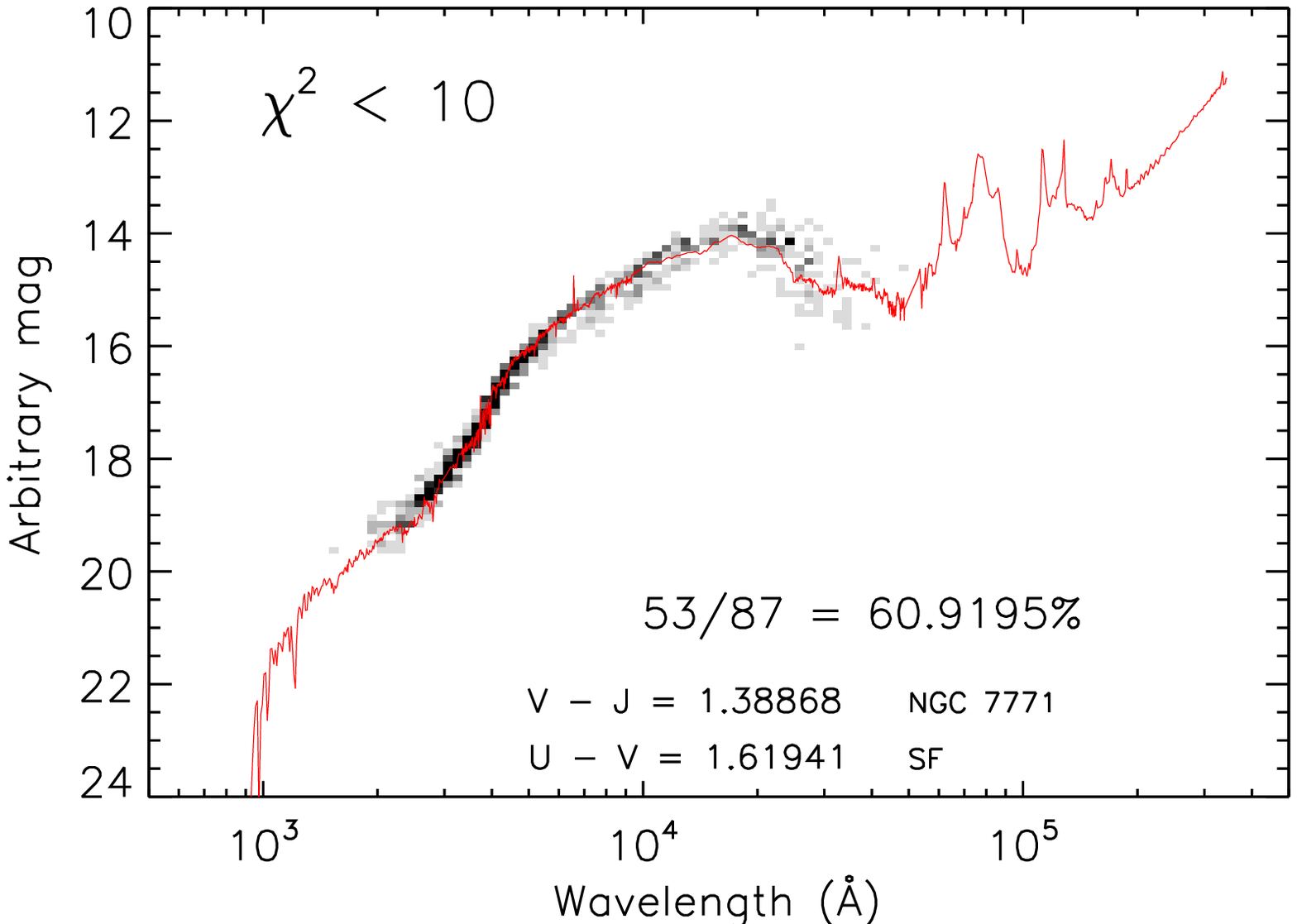}
\includegraphics[width=0.32\textwidth]{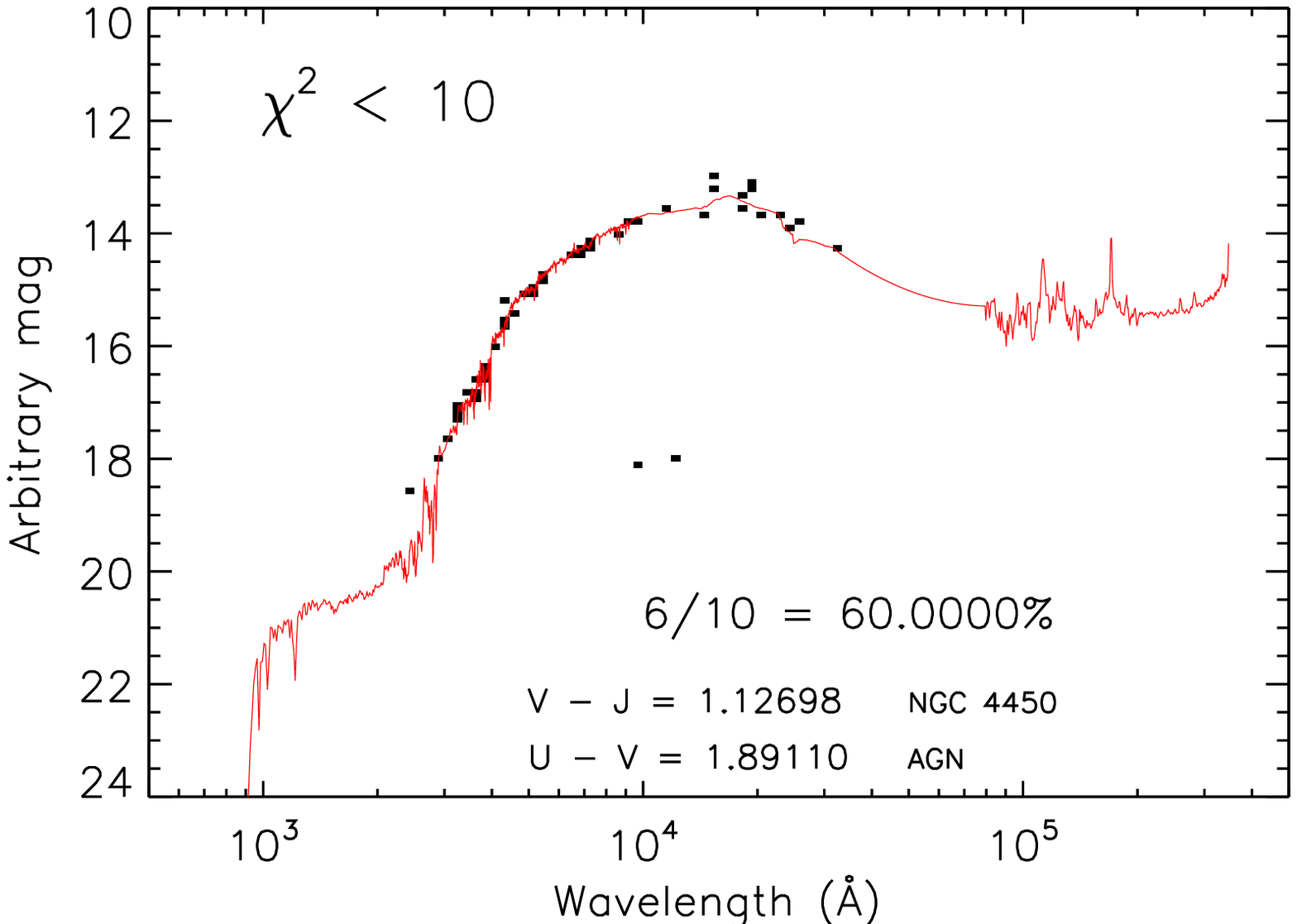}
\includegraphics[width=0.32\textwidth]{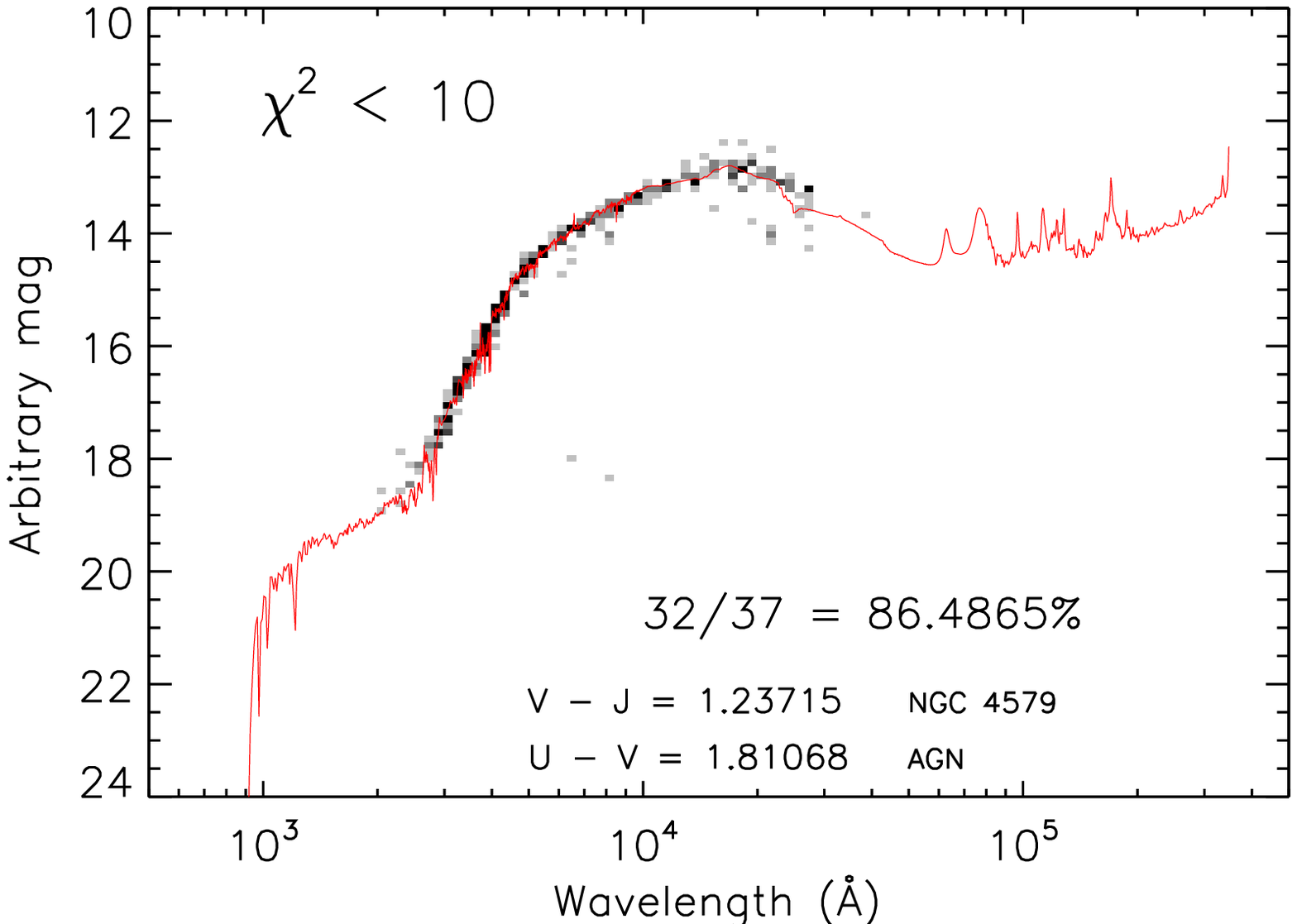}
\includegraphics[width=0.32\textwidth]{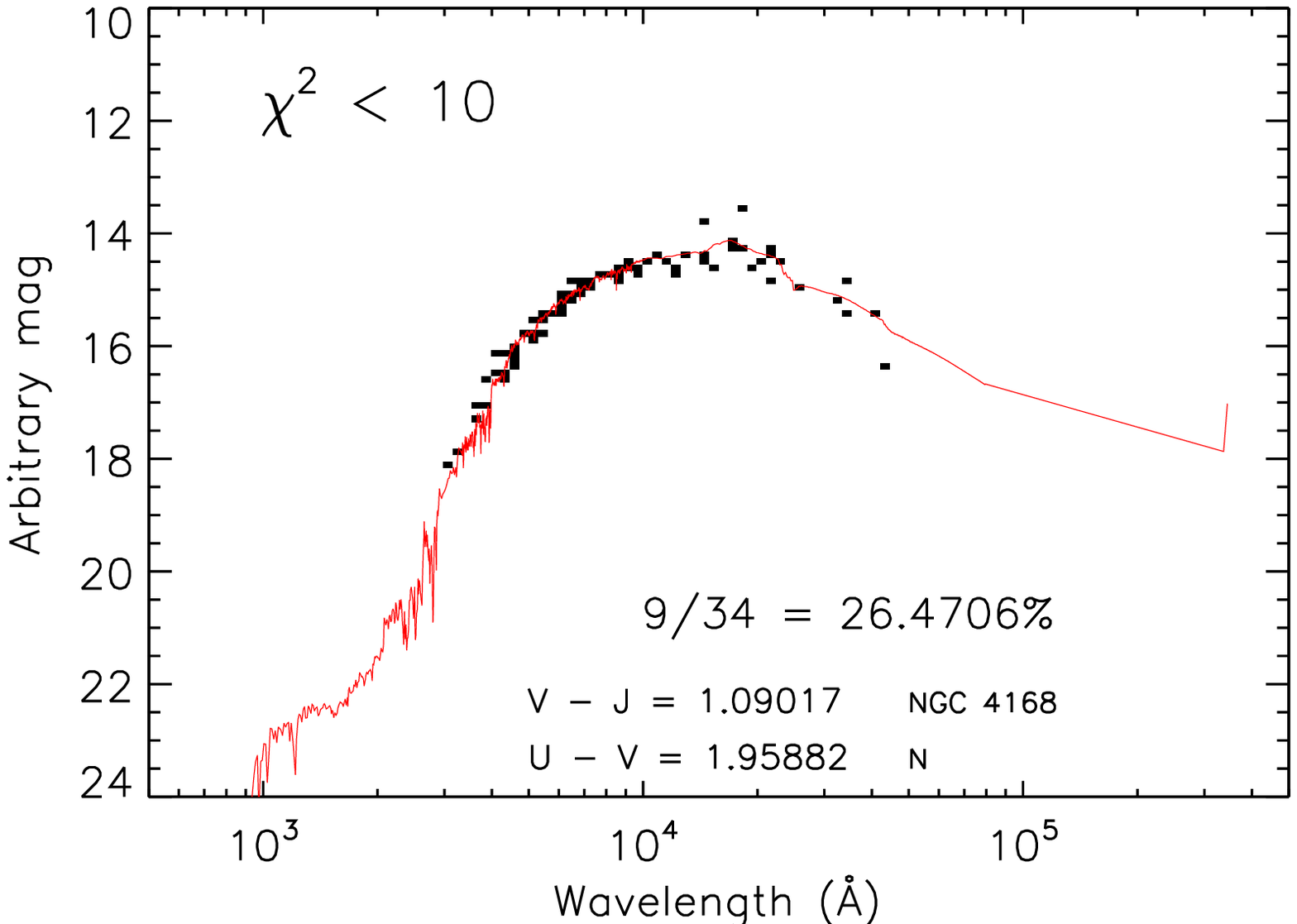}
\includegraphics[width=0.32\textwidth]{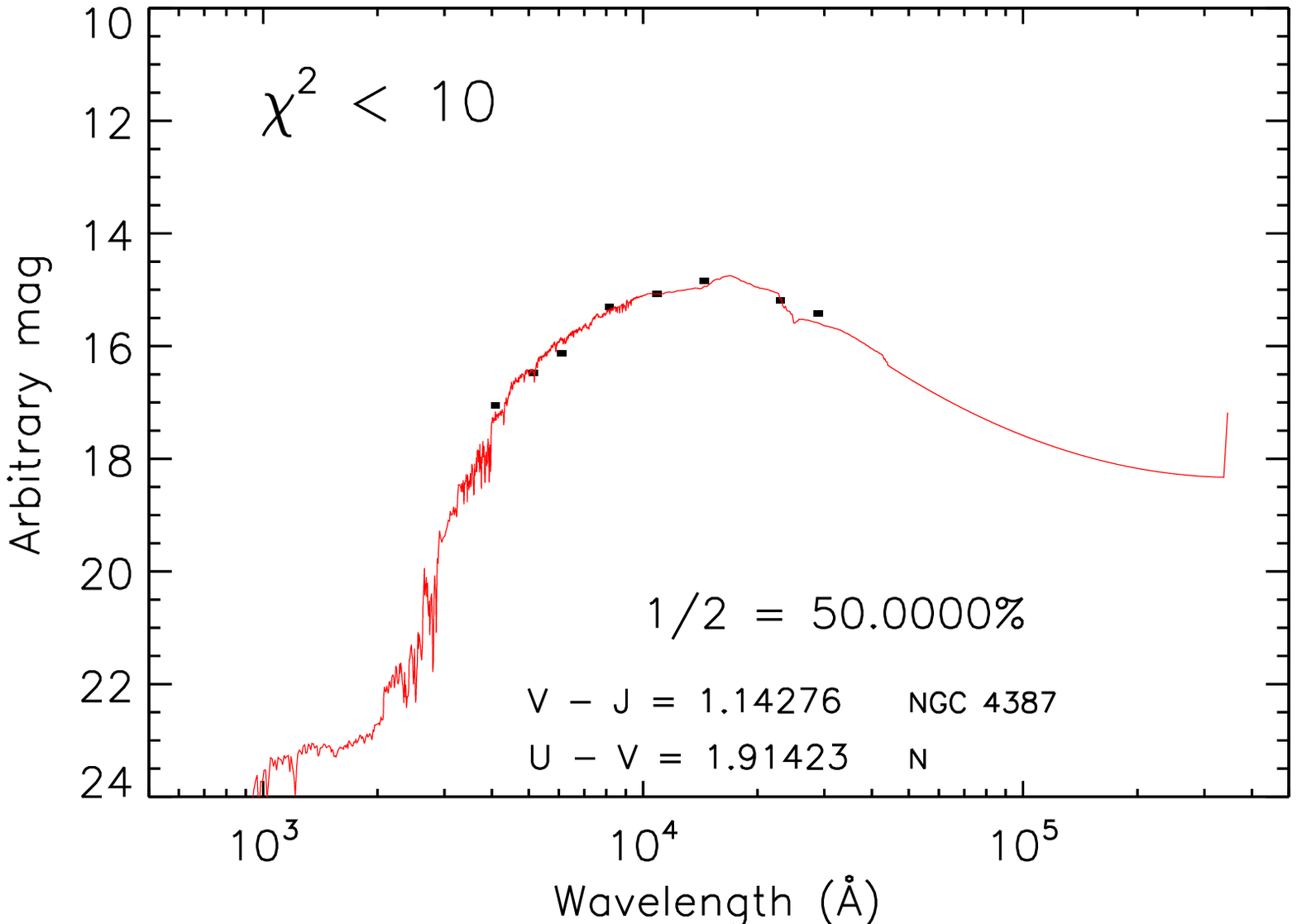}
\includegraphics[width=0.32\textwidth]{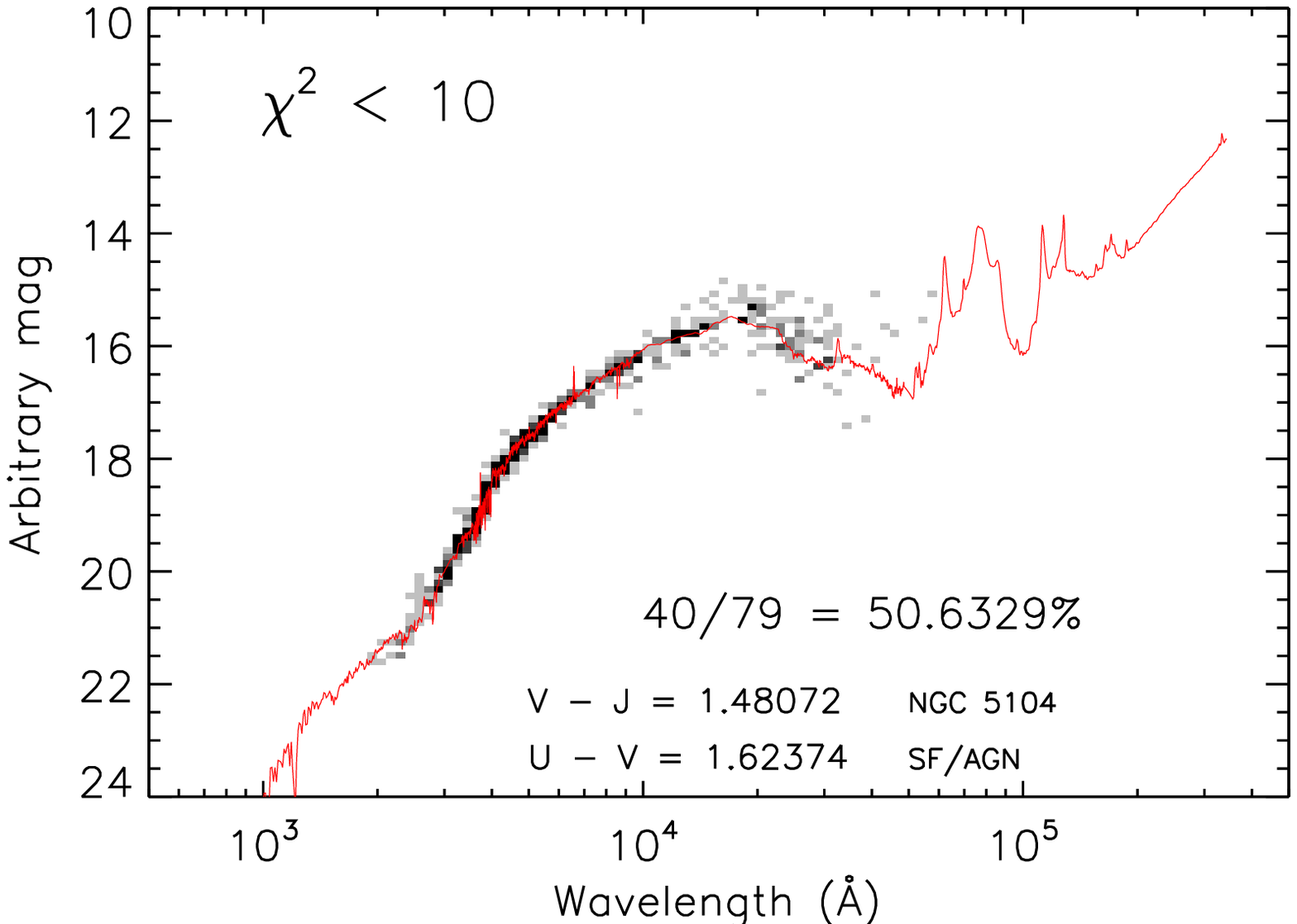}
    \caption{Continued.}
    \label{fig:my_label}
\end{figure}

\begin{figure}
    \centering
\includegraphics[width=0.32\textwidth]{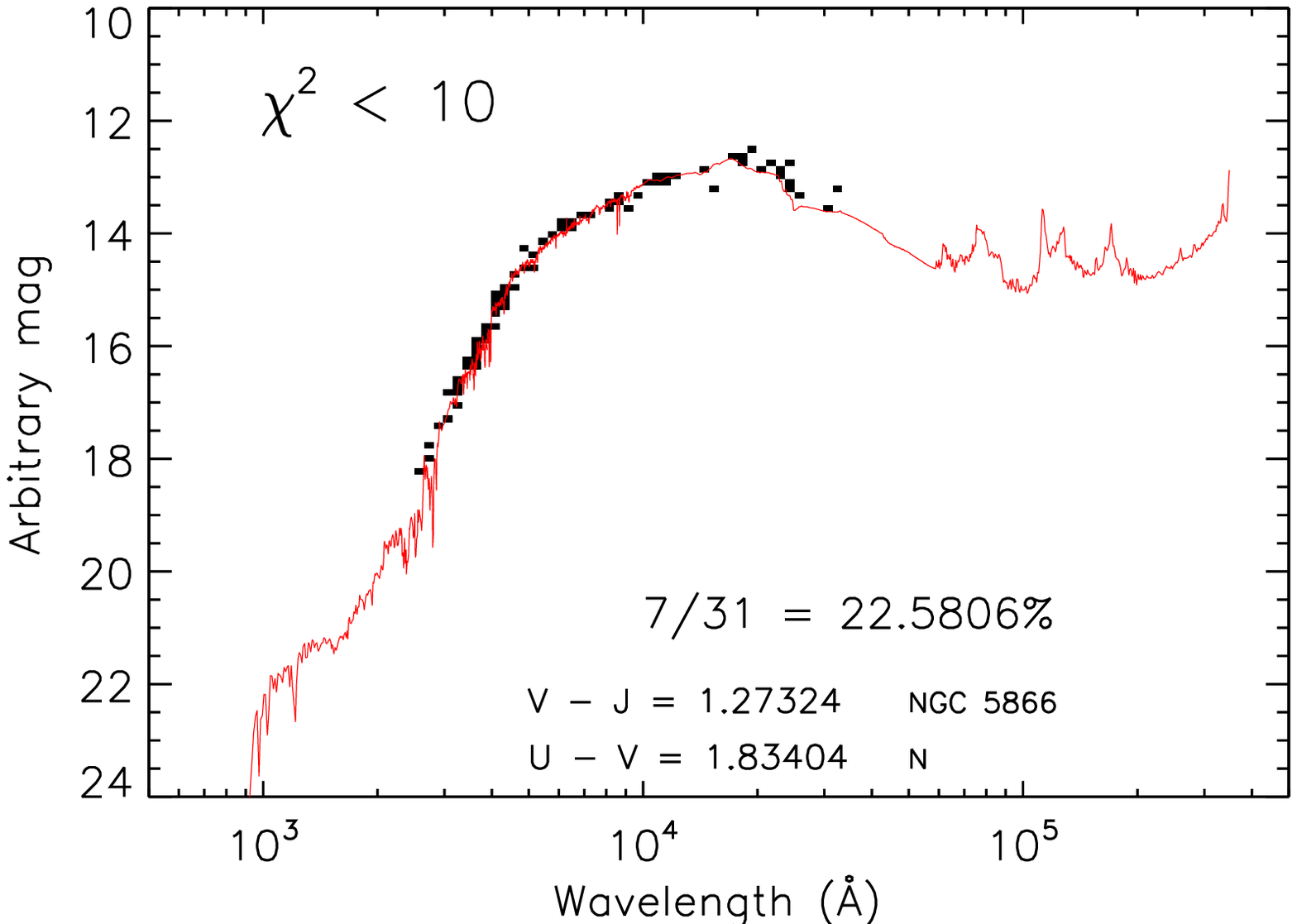}
\includegraphics[width=0.32\textwidth]{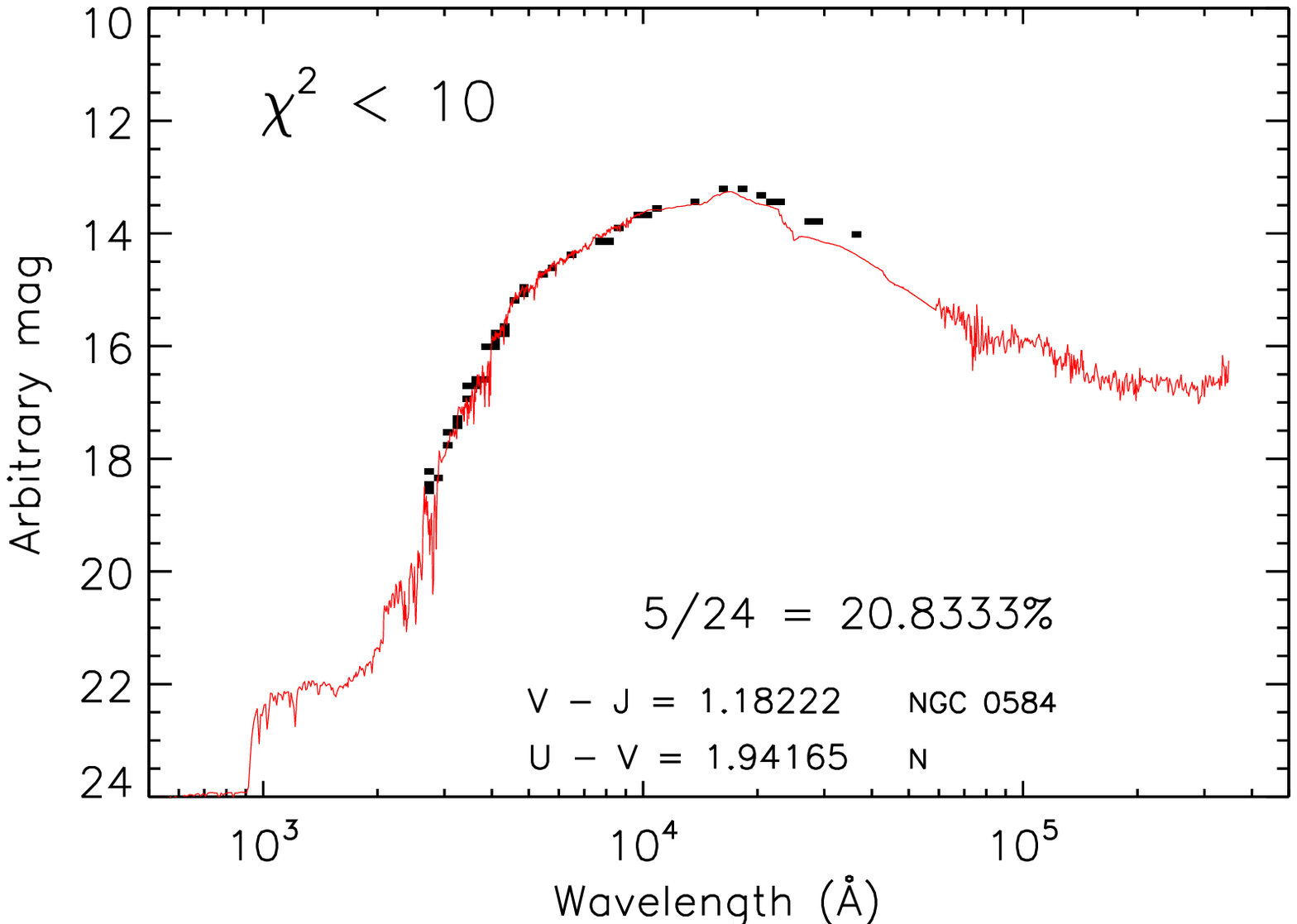}
\includegraphics[width=0.32\textwidth]{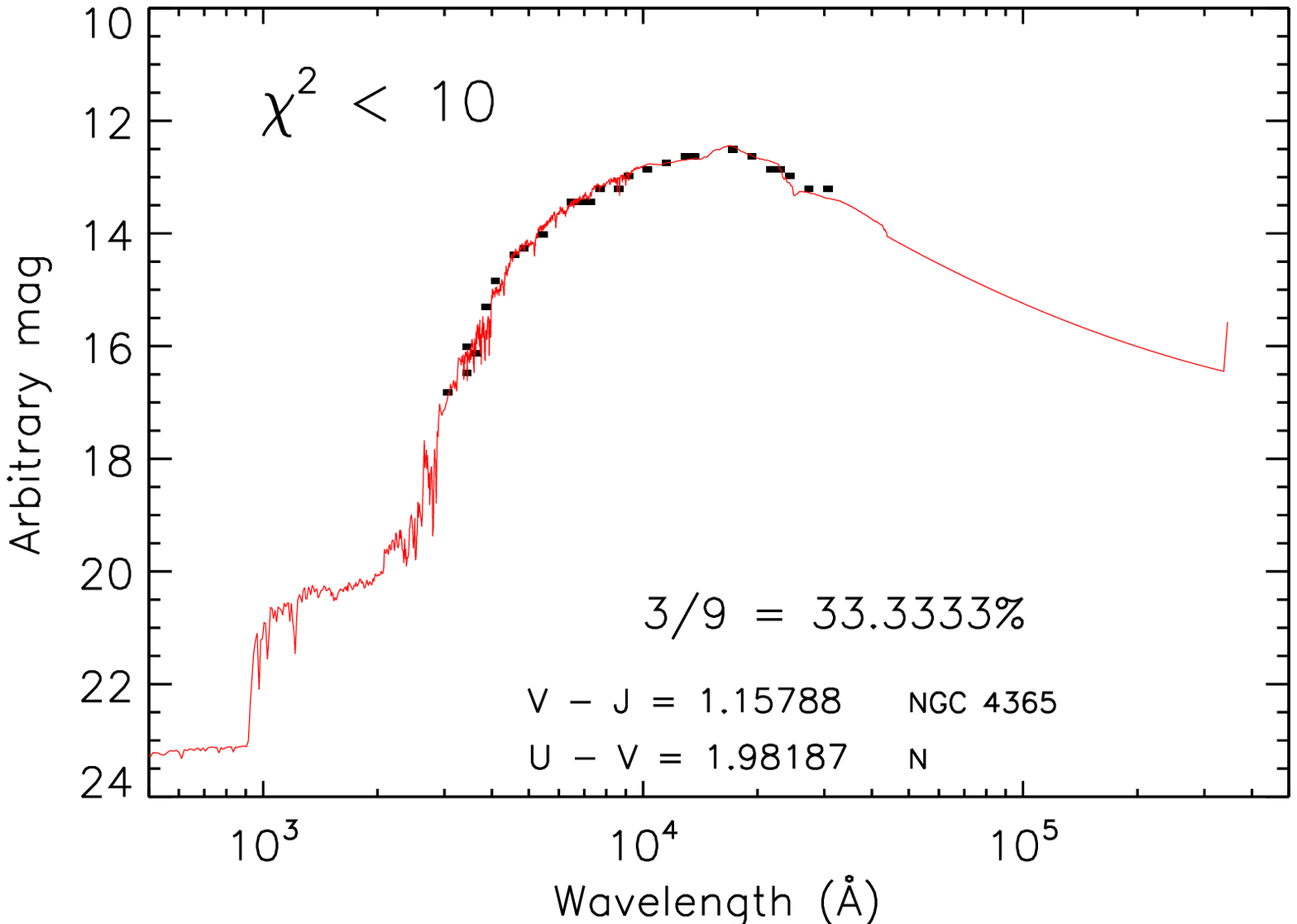}
\includegraphics[width=0.32\textwidth]{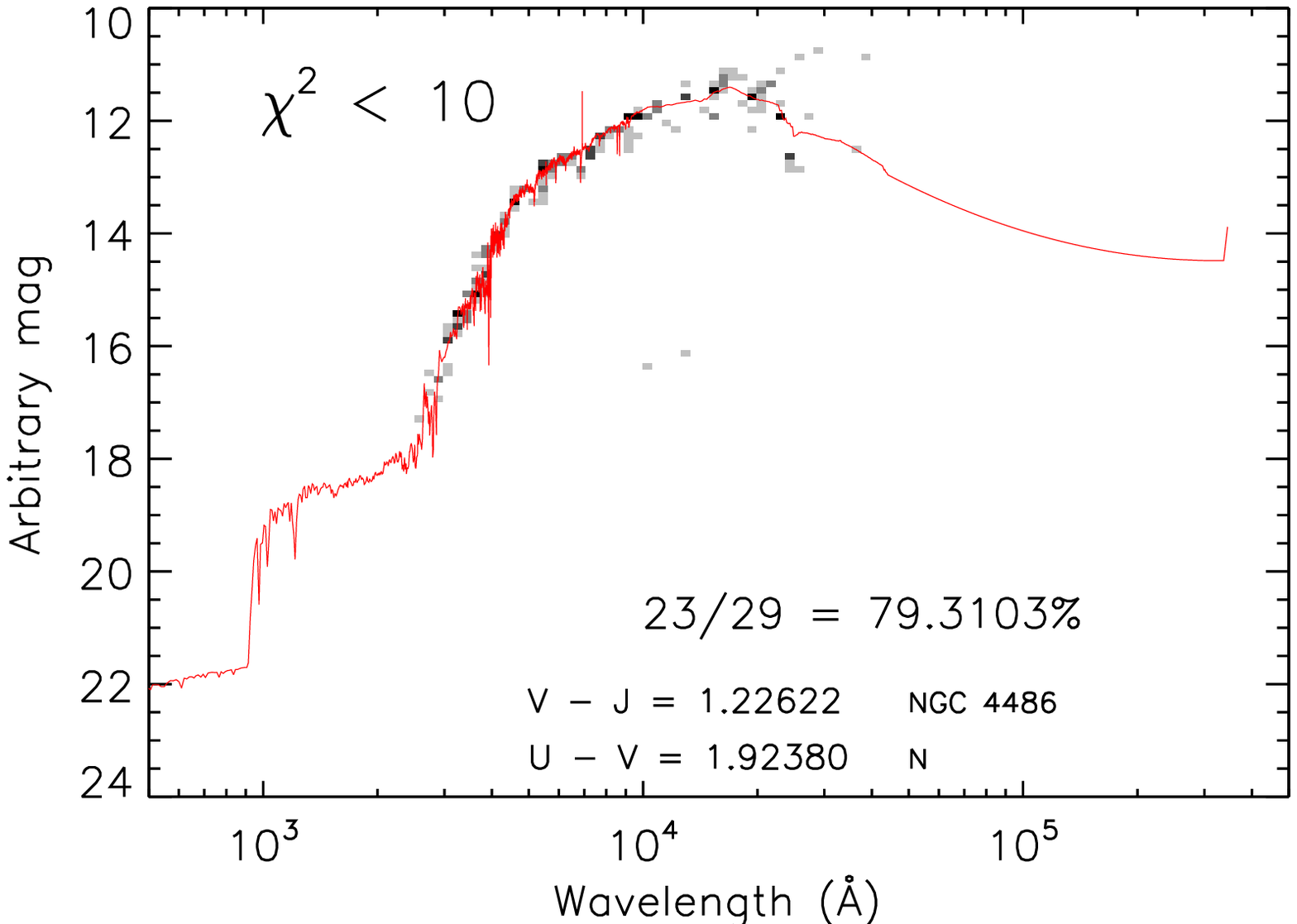}
\includegraphics[width=0.32\textwidth]{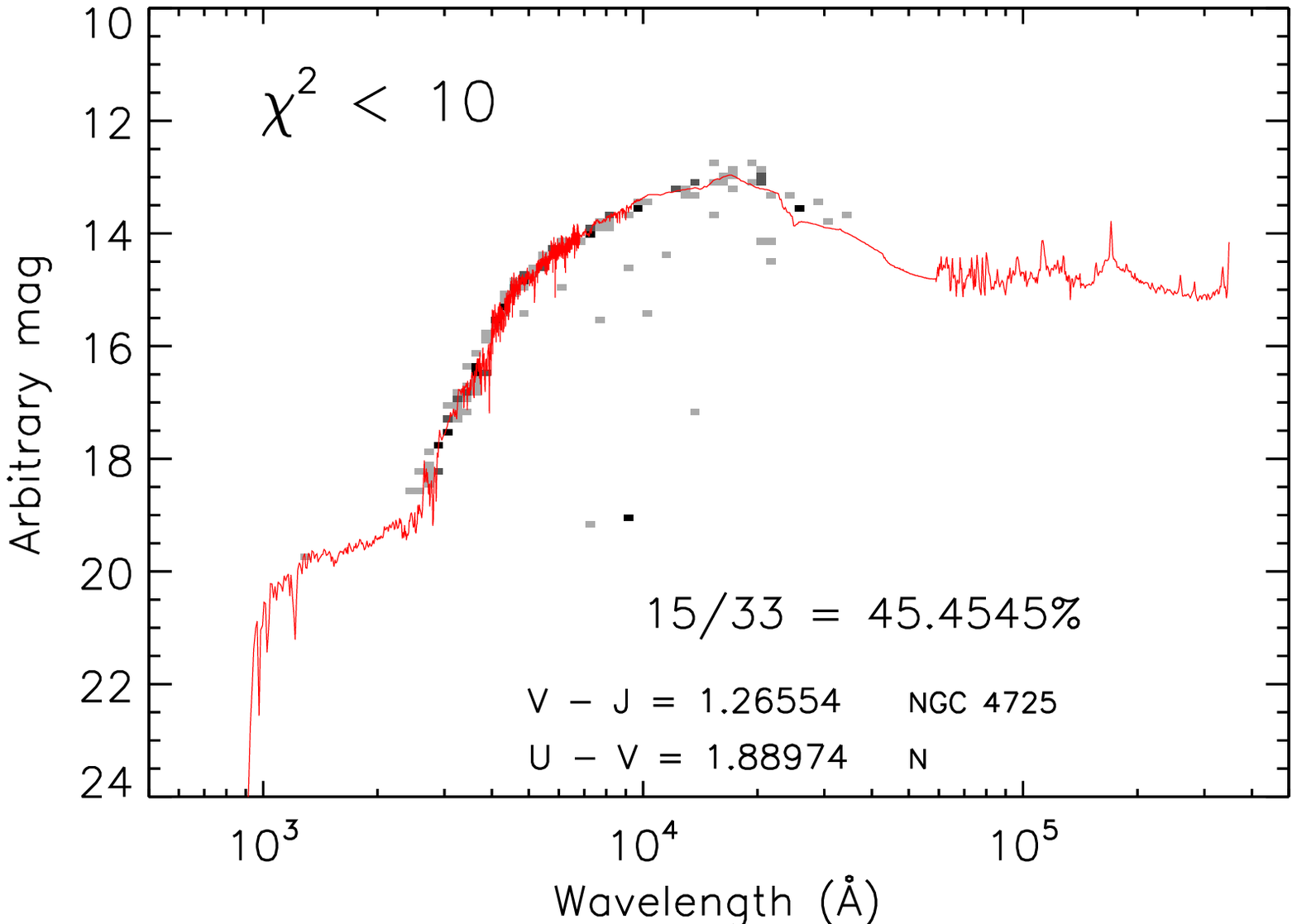}
\includegraphics[width=0.32\textwidth]{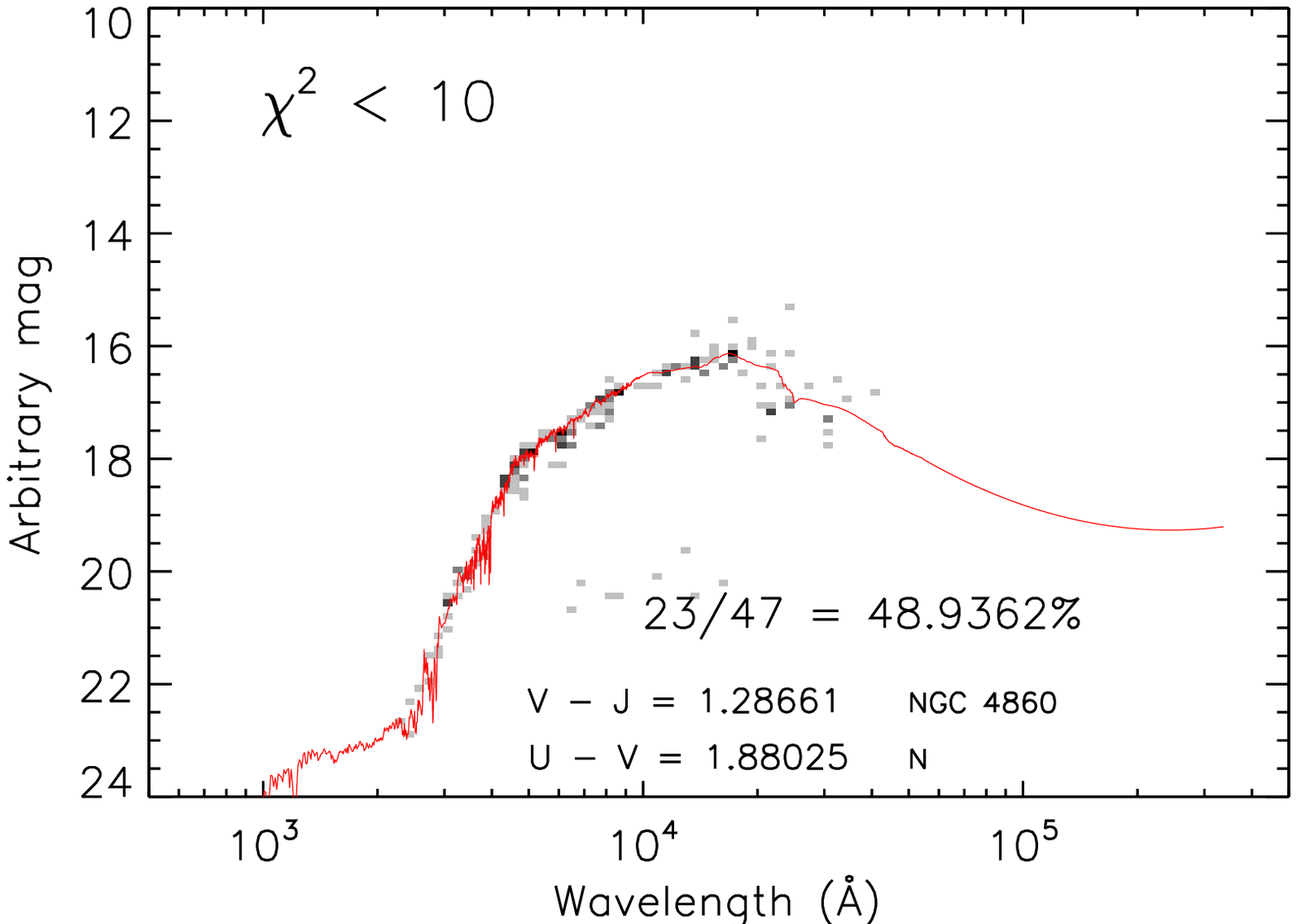}
\includegraphics[width=0.32\textwidth]{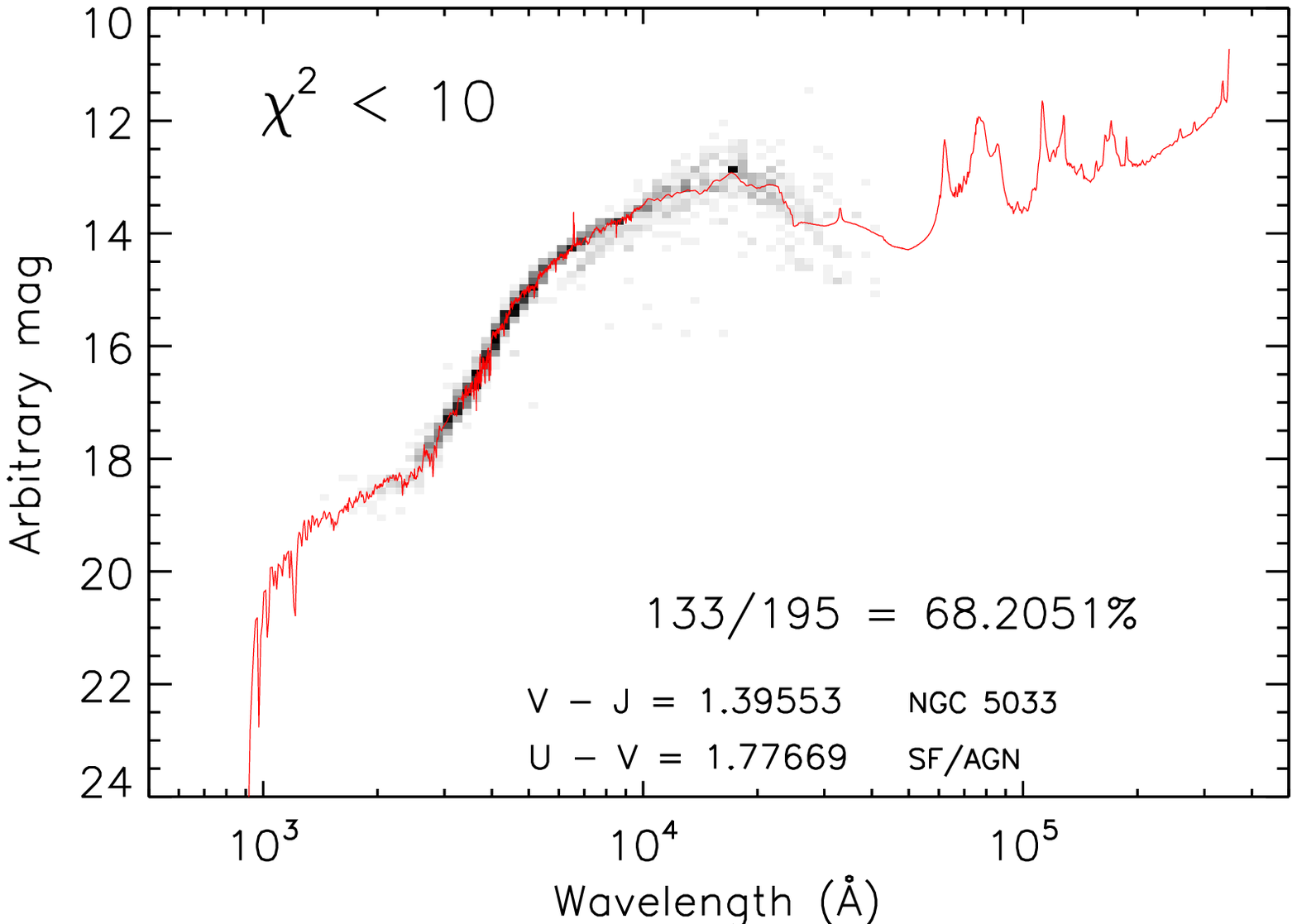}
\includegraphics[width=0.32\textwidth]{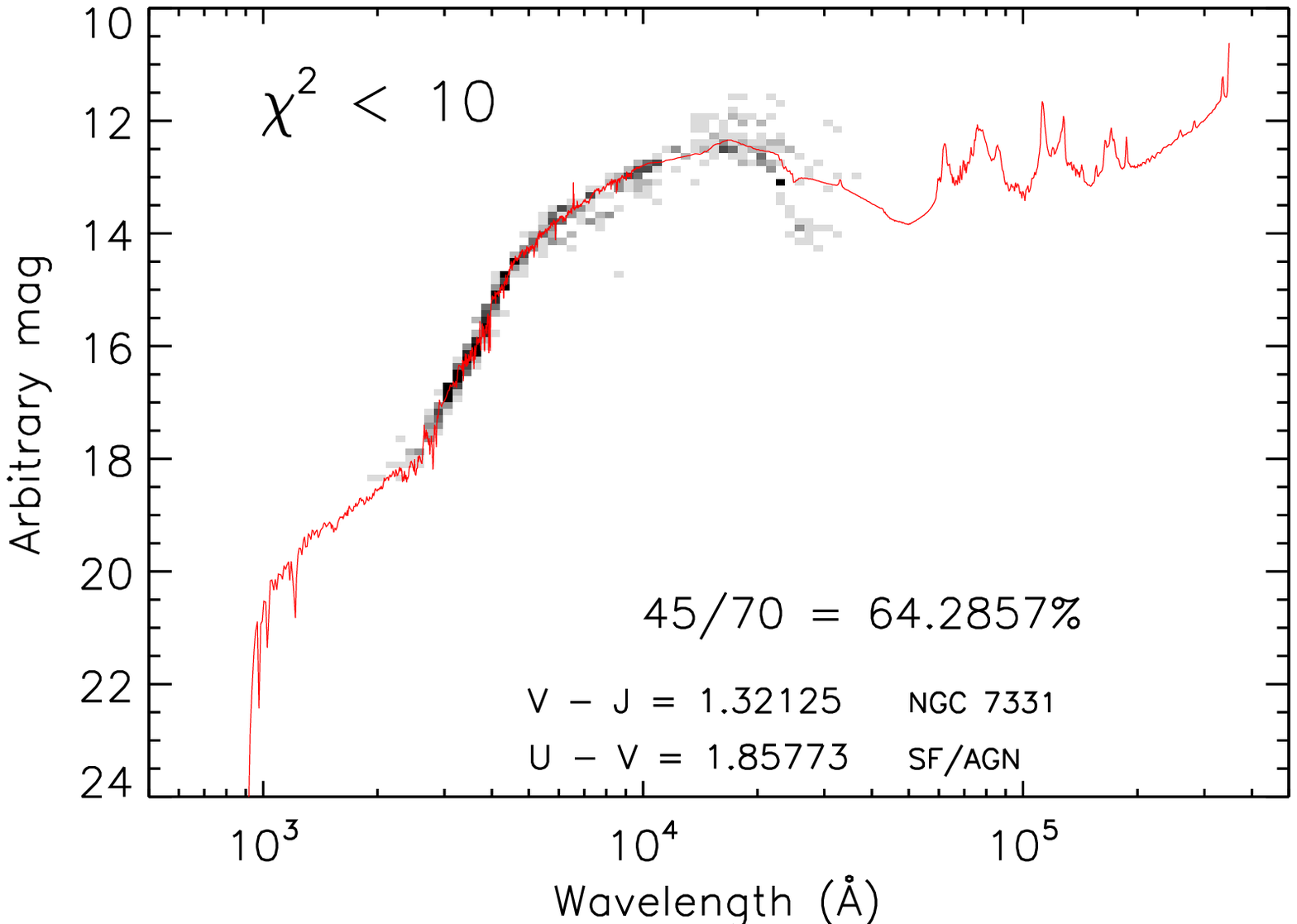}
\includegraphics[width=0.32\textwidth]{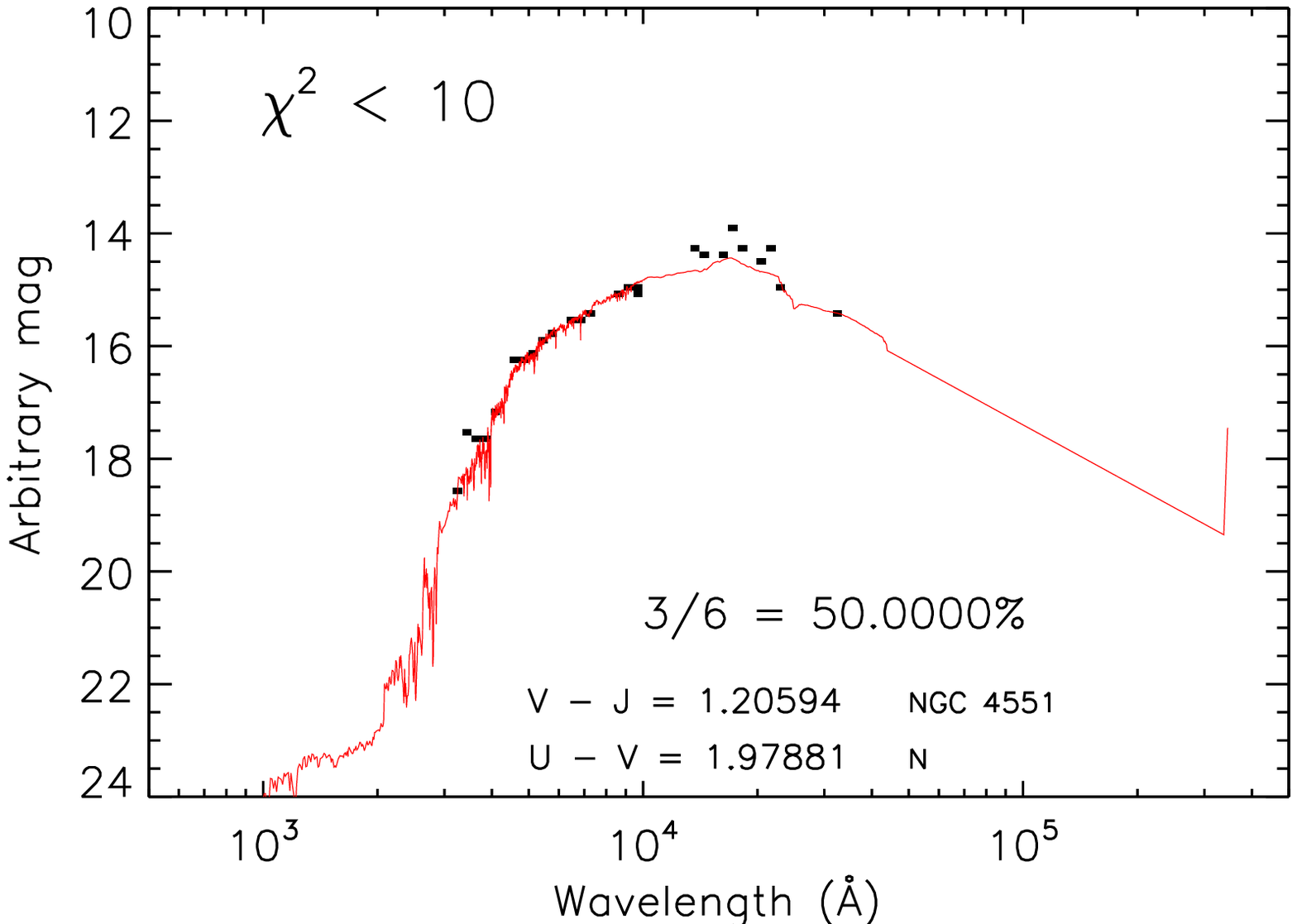}
\includegraphics[width=0.32\textwidth]{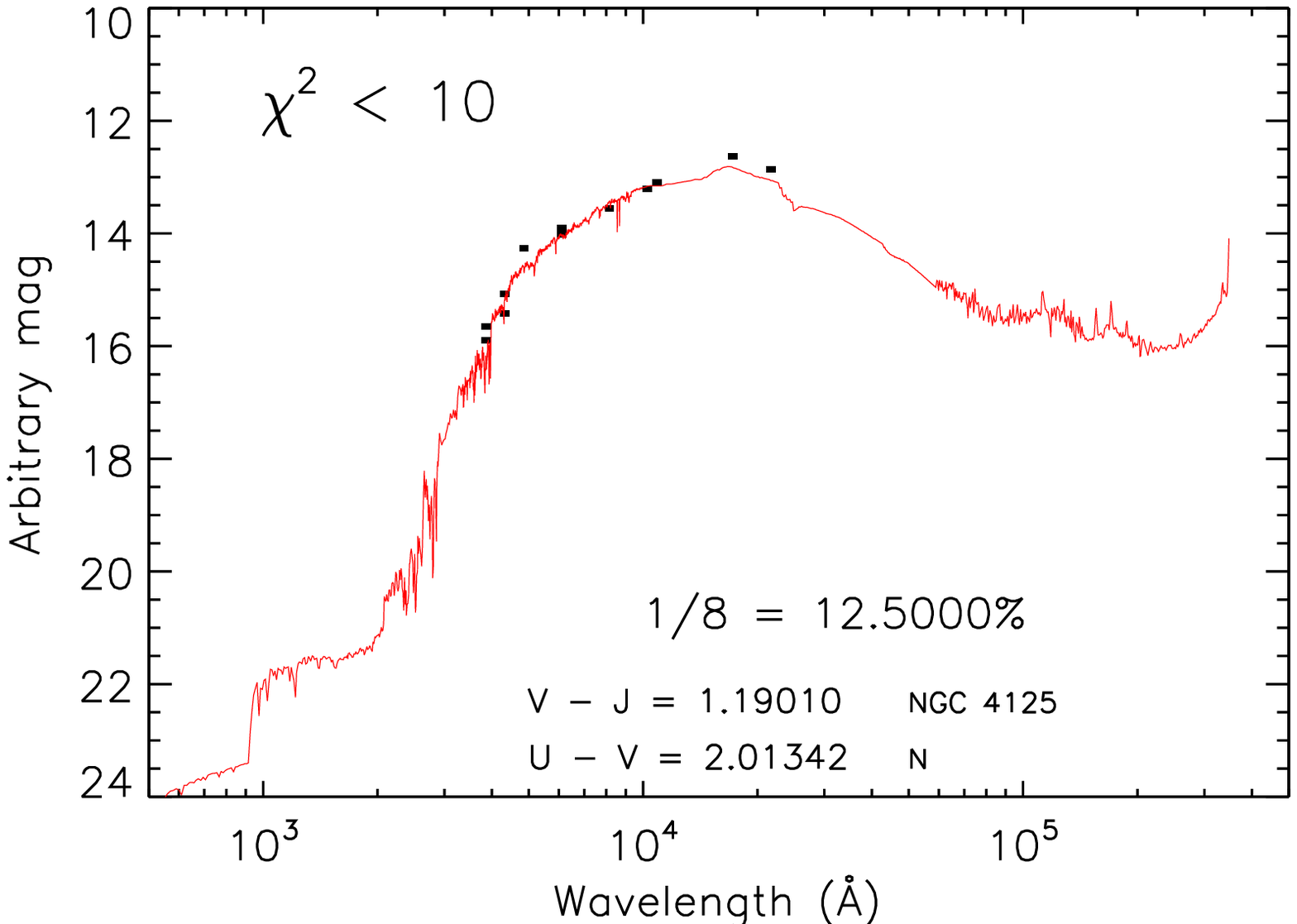}
\includegraphics[width=0.32\textwidth]{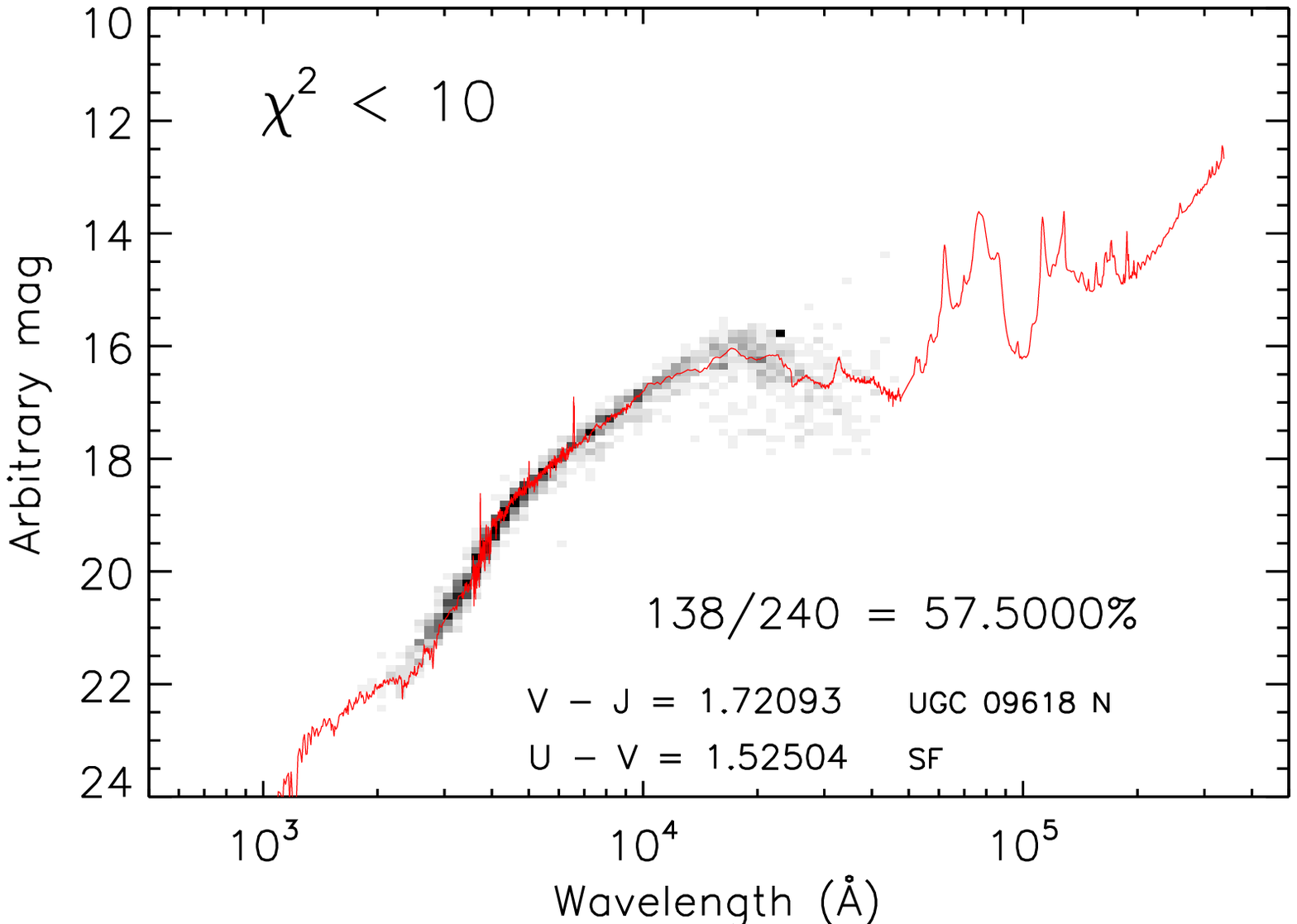}
\includegraphics[width=0.32\textwidth]{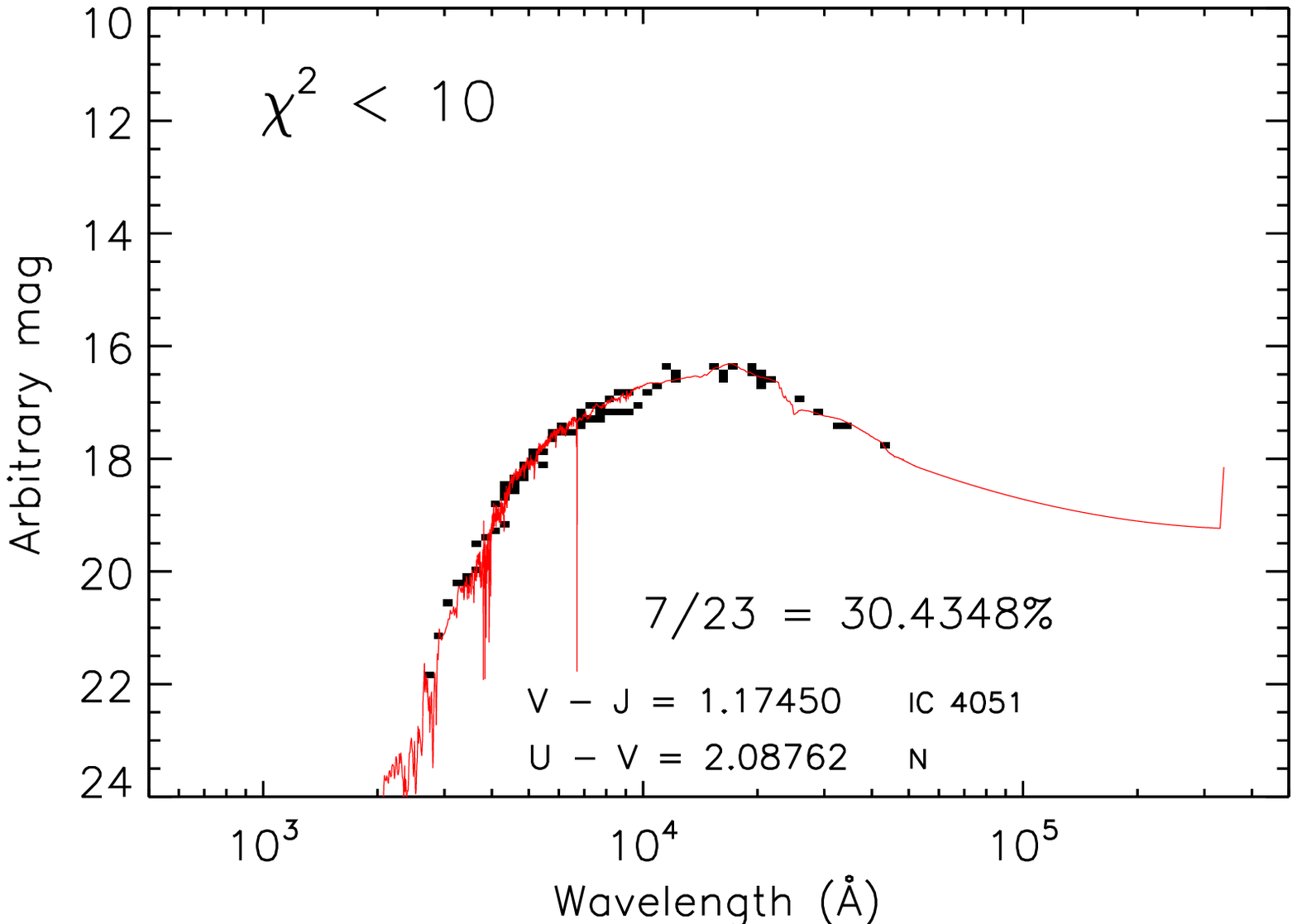}
\includegraphics[width=0.32\textwidth]{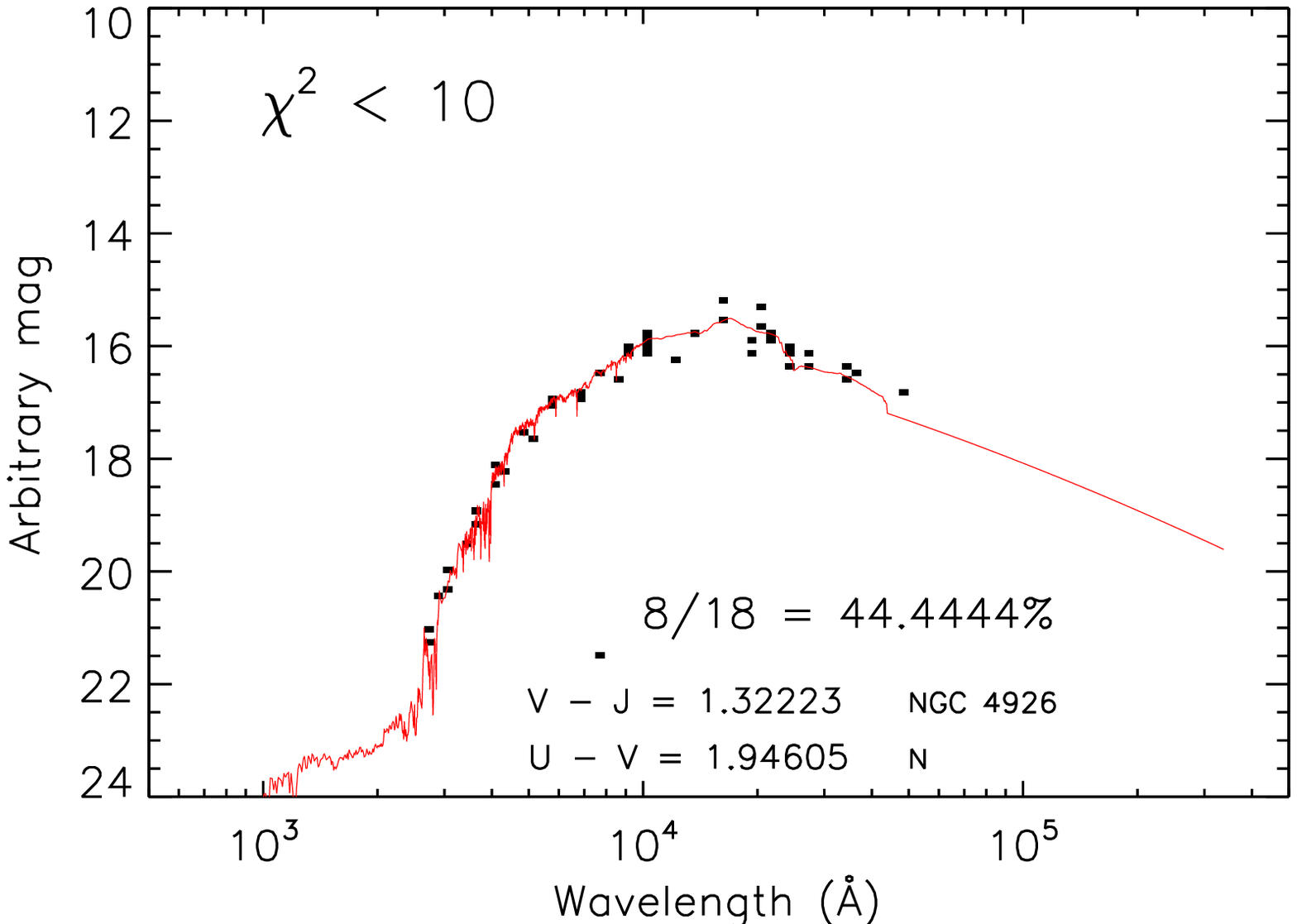}
\includegraphics[width=0.32\textwidth]{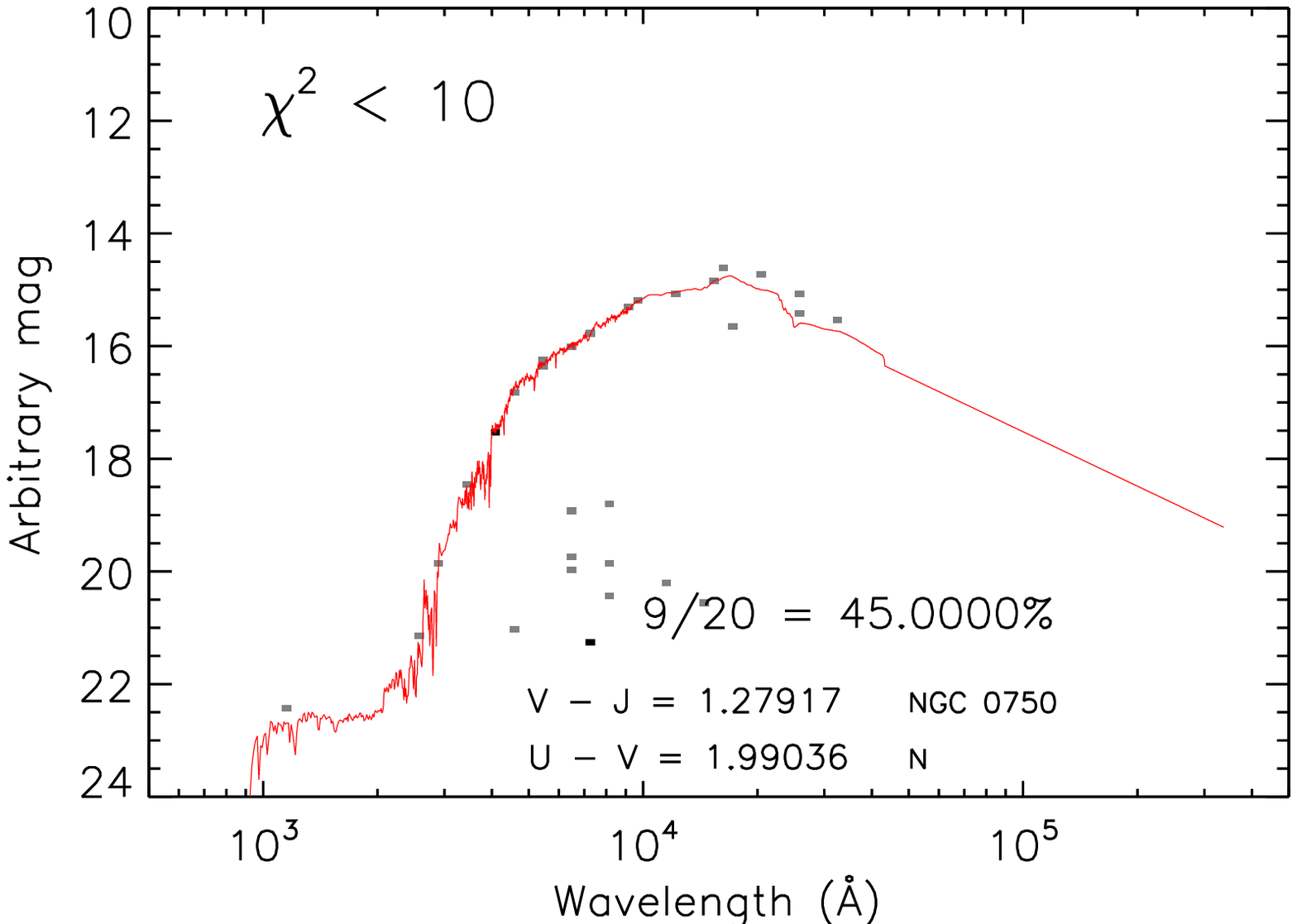}
\includegraphics[width=0.32\textwidth]{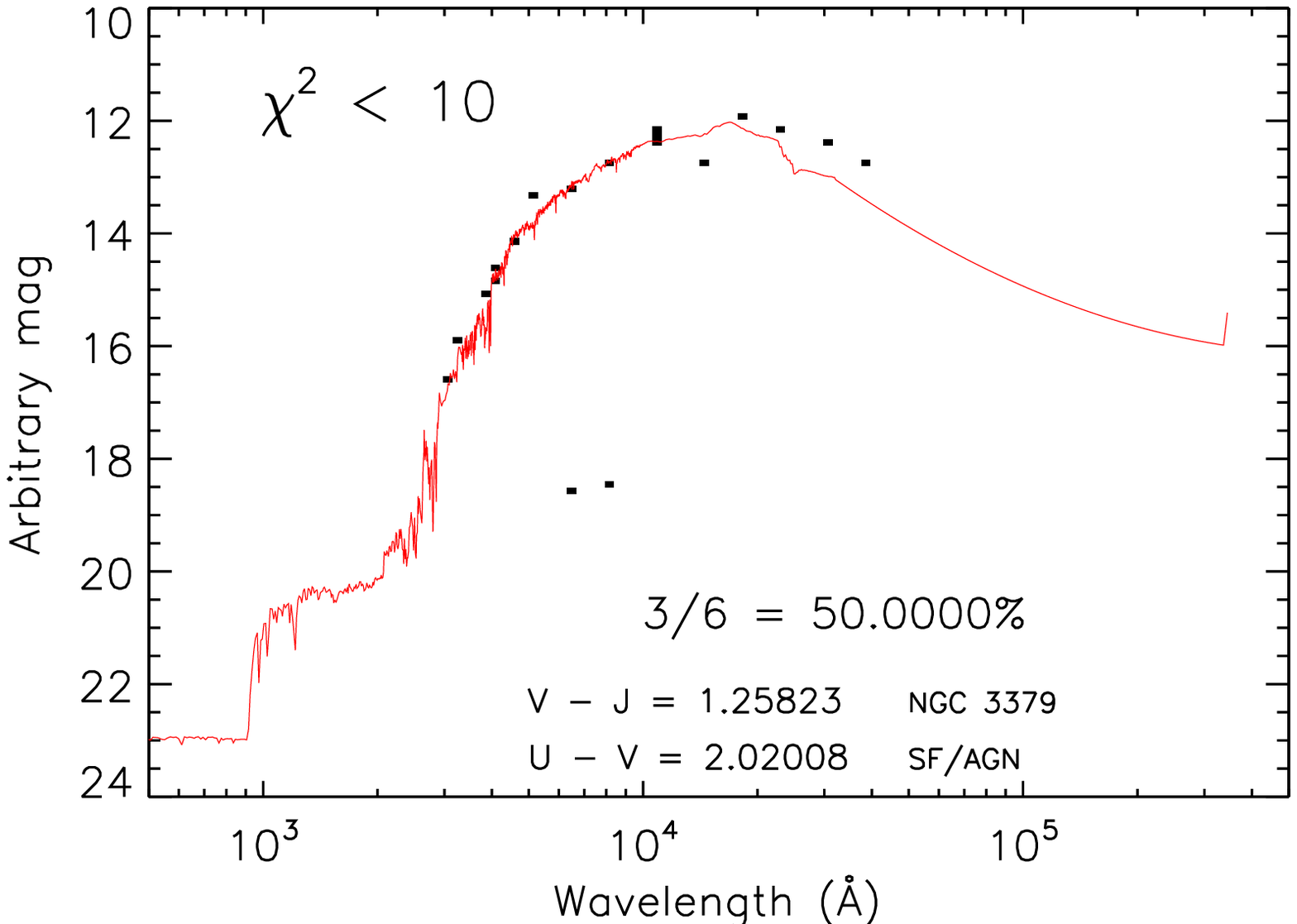}
\includegraphics[width=0.32\textwidth]{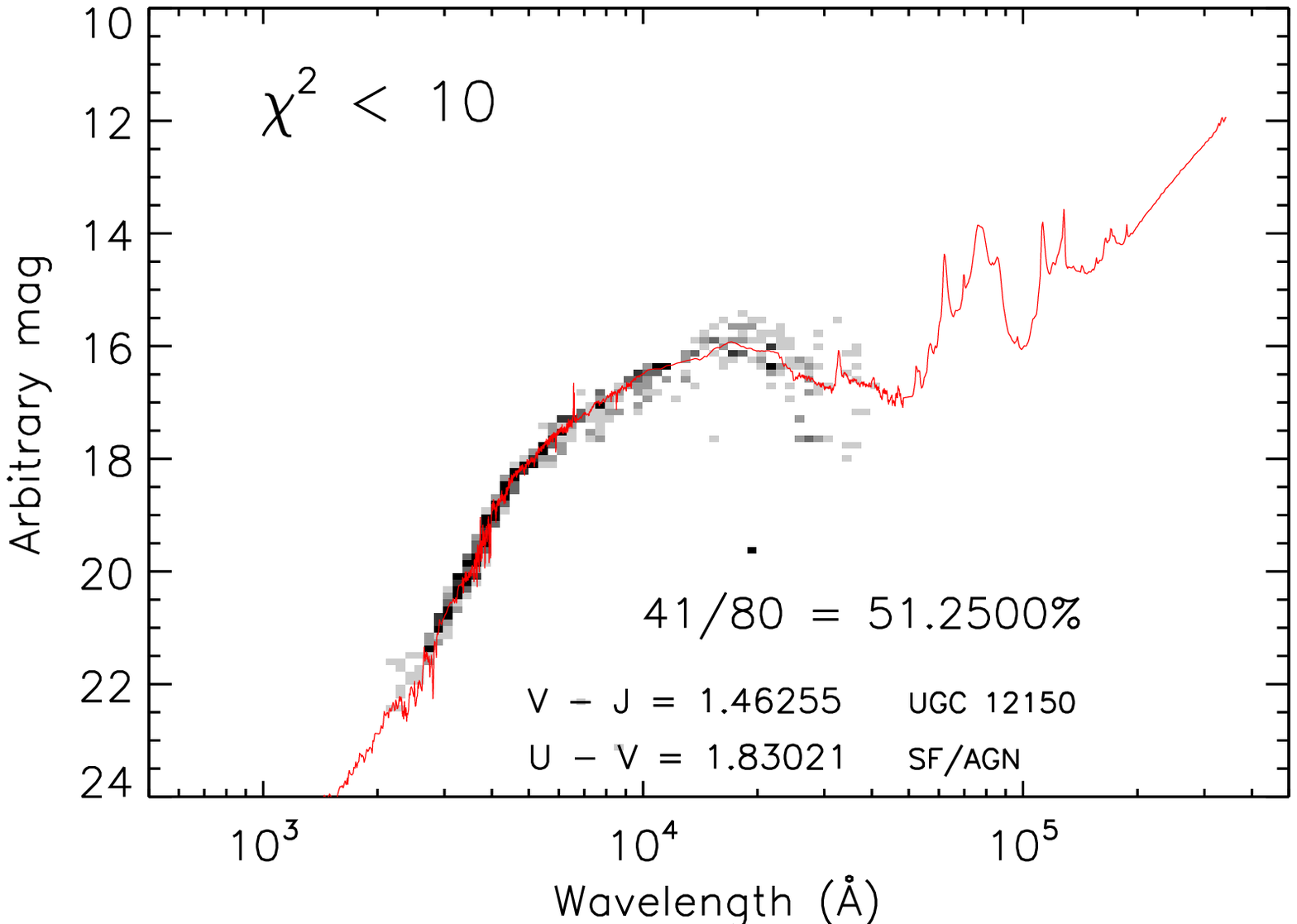}
\includegraphics[width=0.32\textwidth]{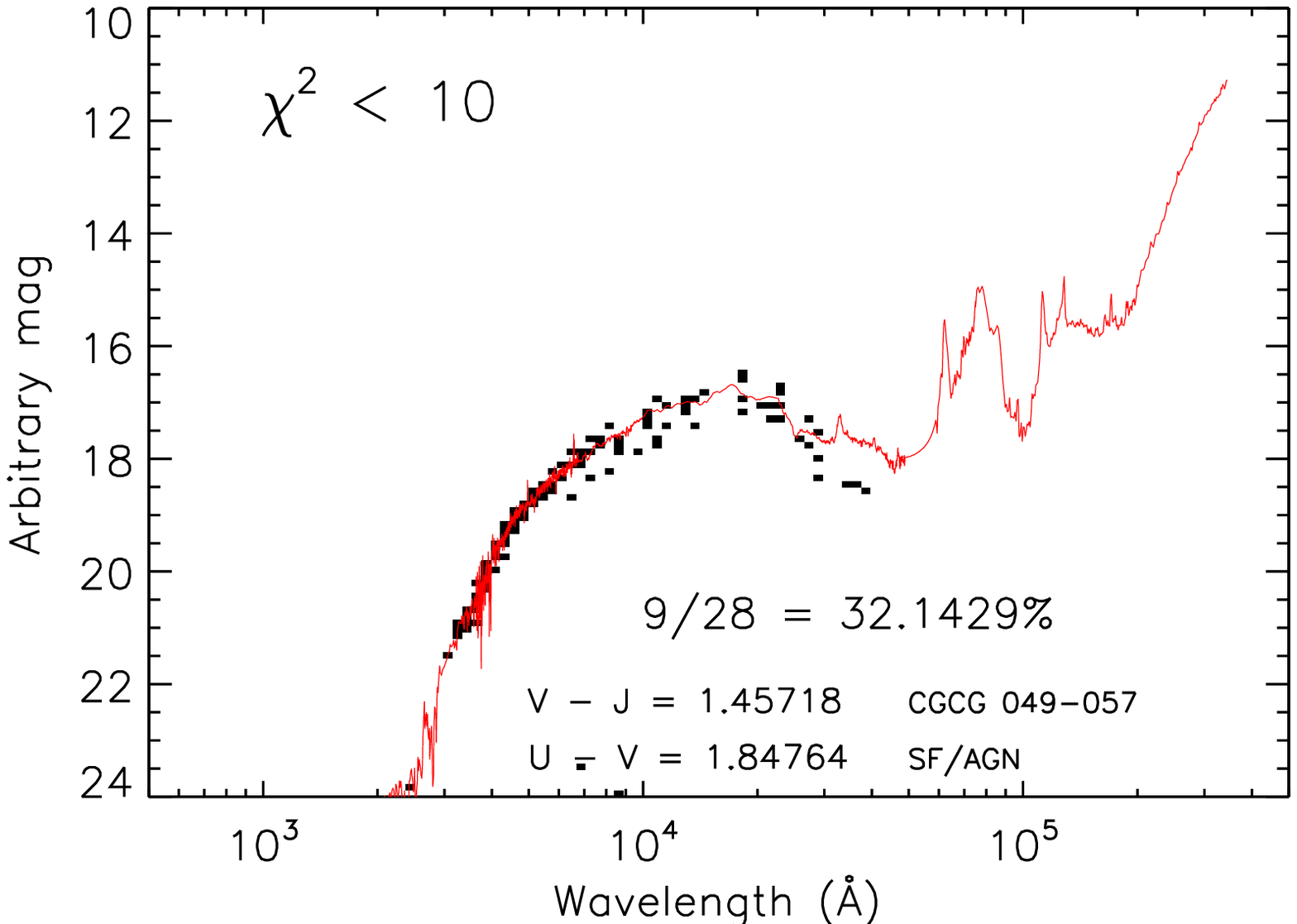}
\includegraphics[width=0.32\textwidth]{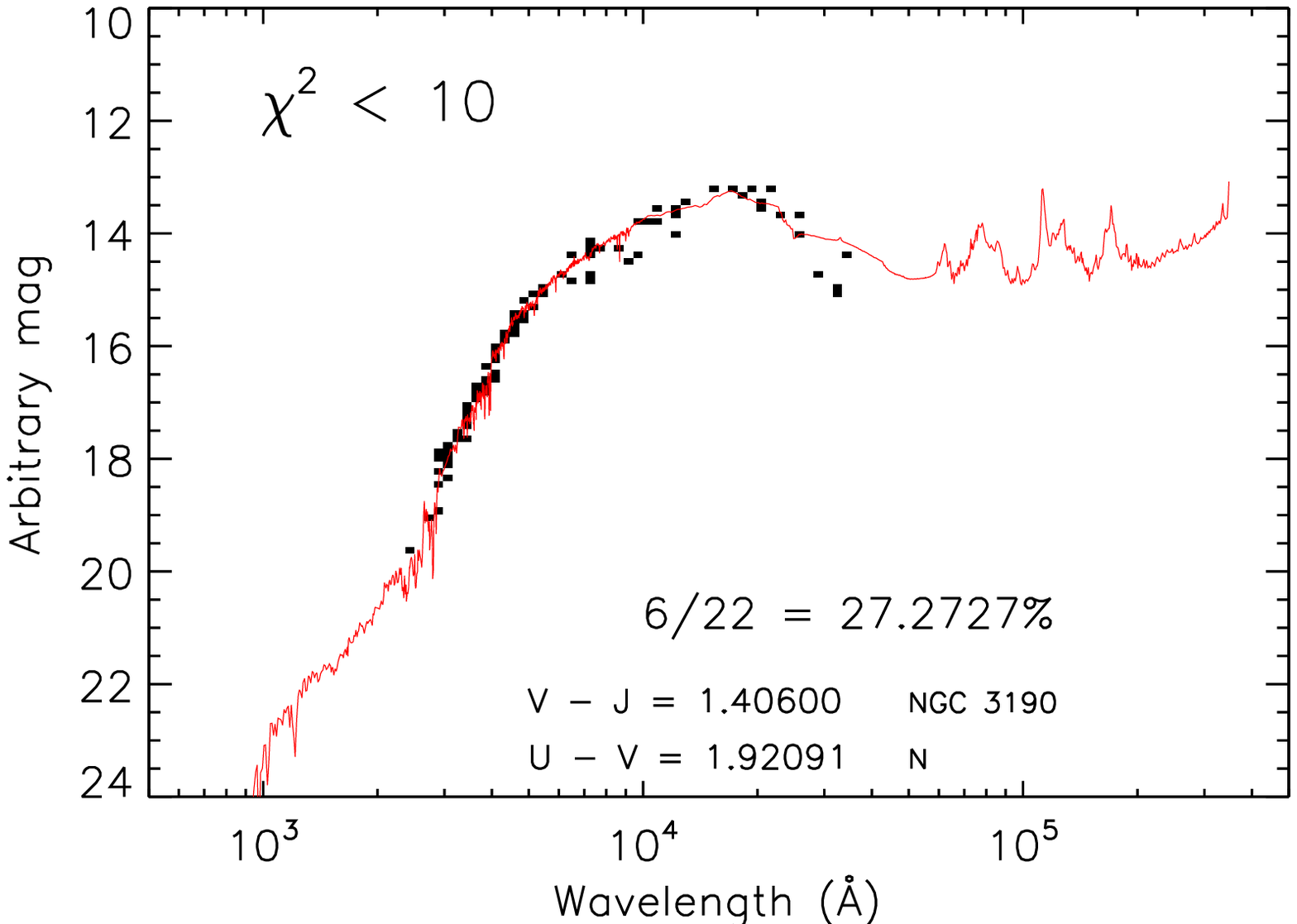}
\includegraphics[width=0.32\textwidth]{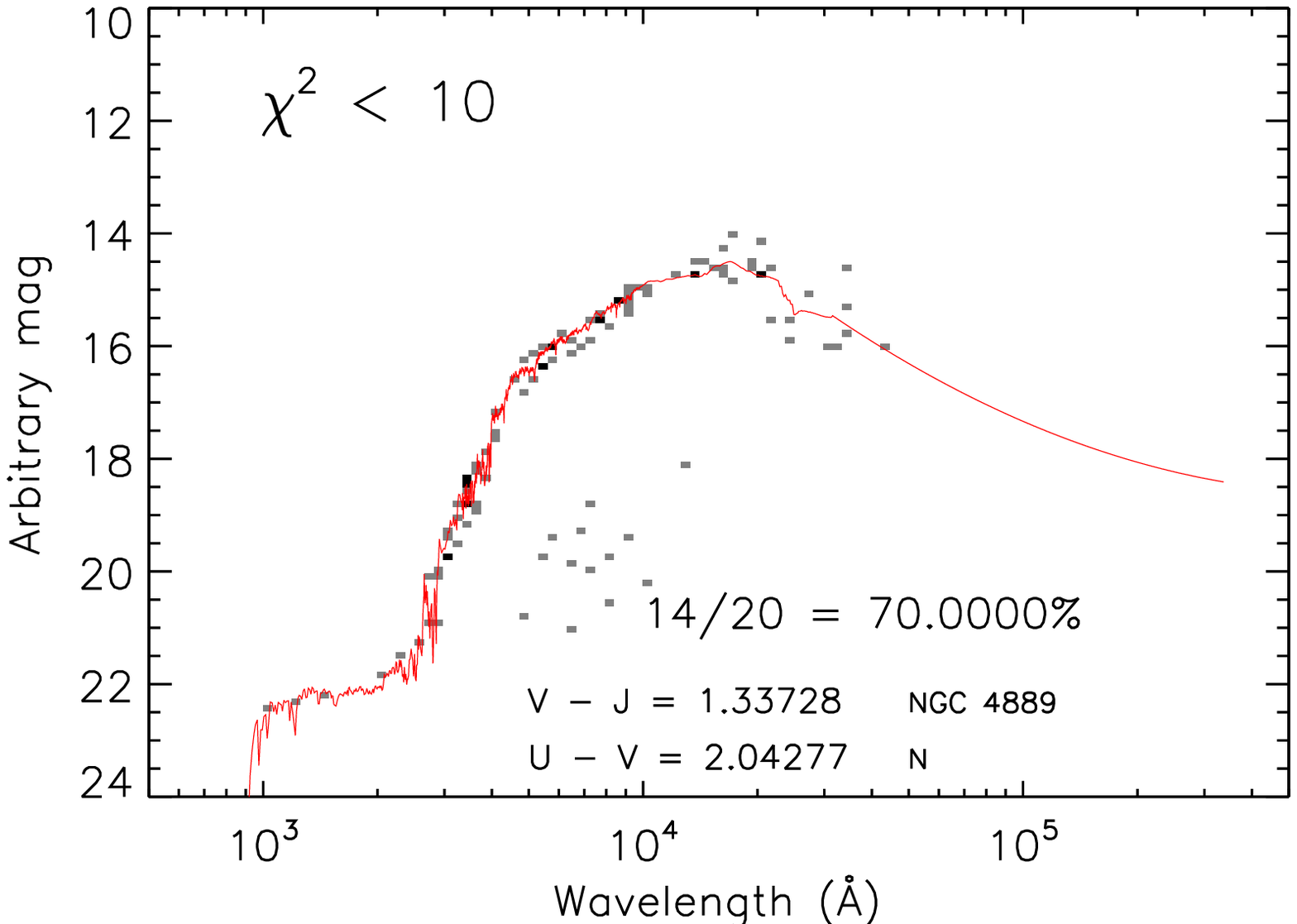}
\includegraphics[width=0.32\textwidth]{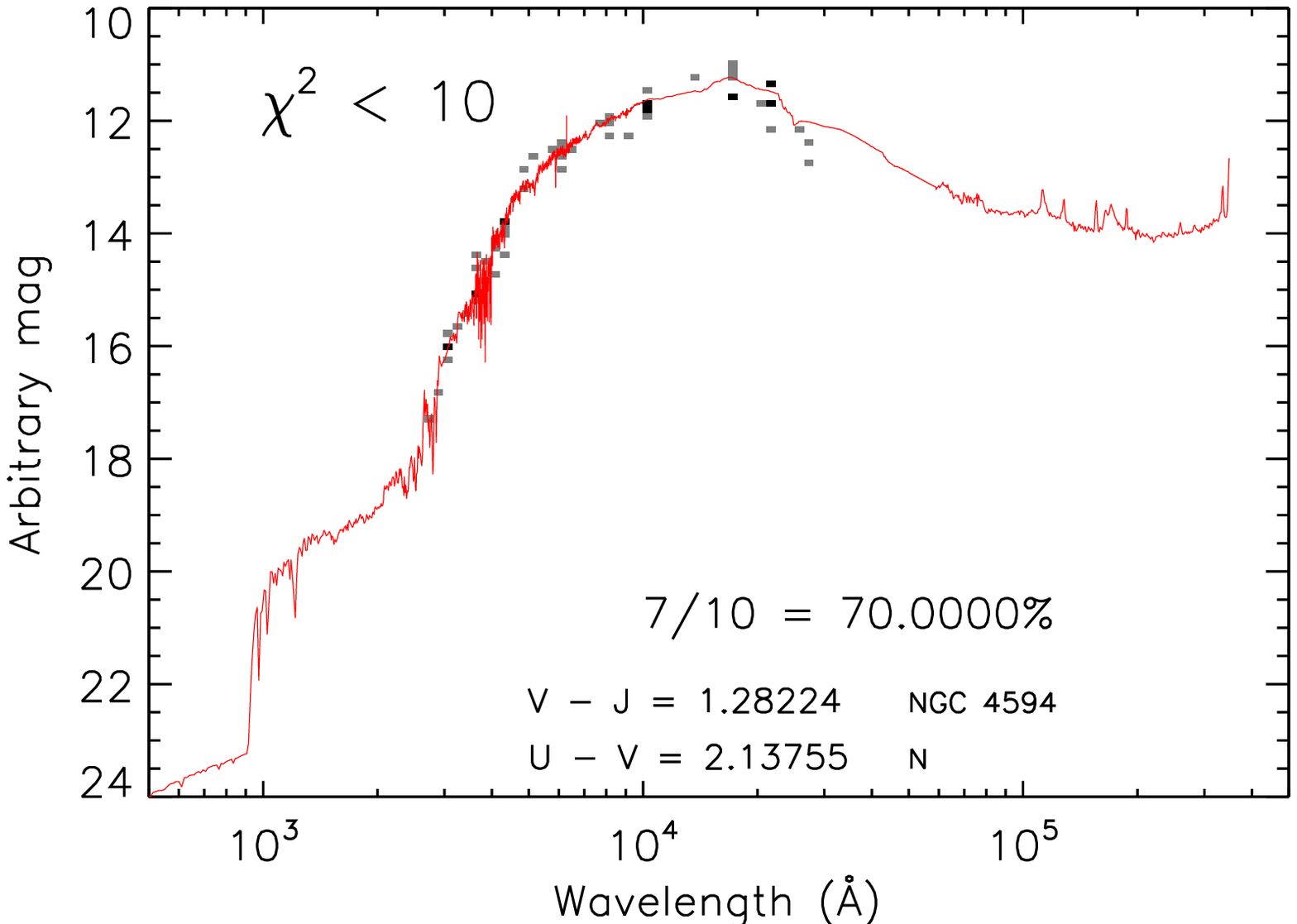}
\includegraphics[width=0.32\textwidth]{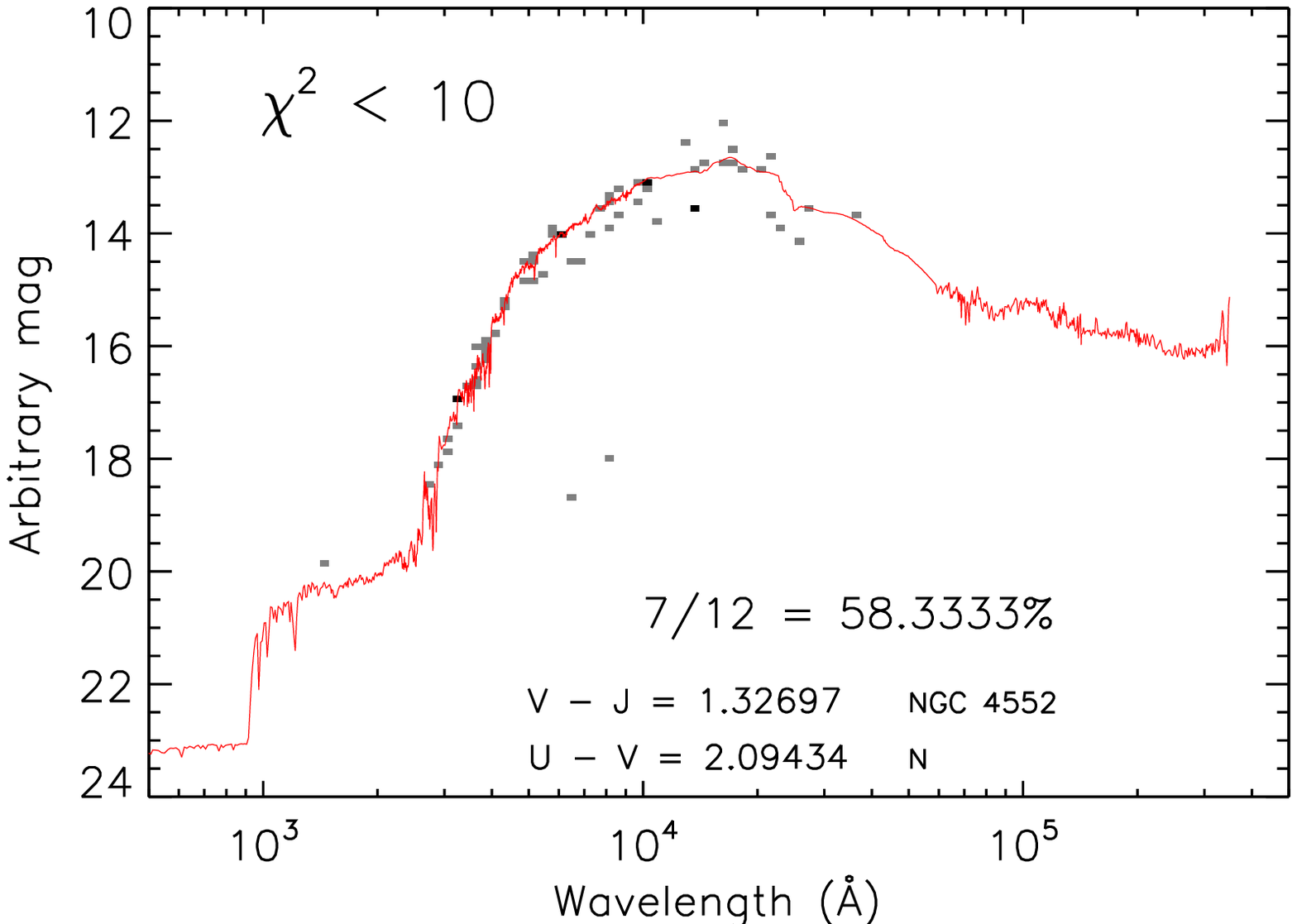}
    \caption{Continued.}
    \label{fig:my_label}
\end{figure}

\begin{figure}
    \centering
\includegraphics[width=0.32\textwidth]{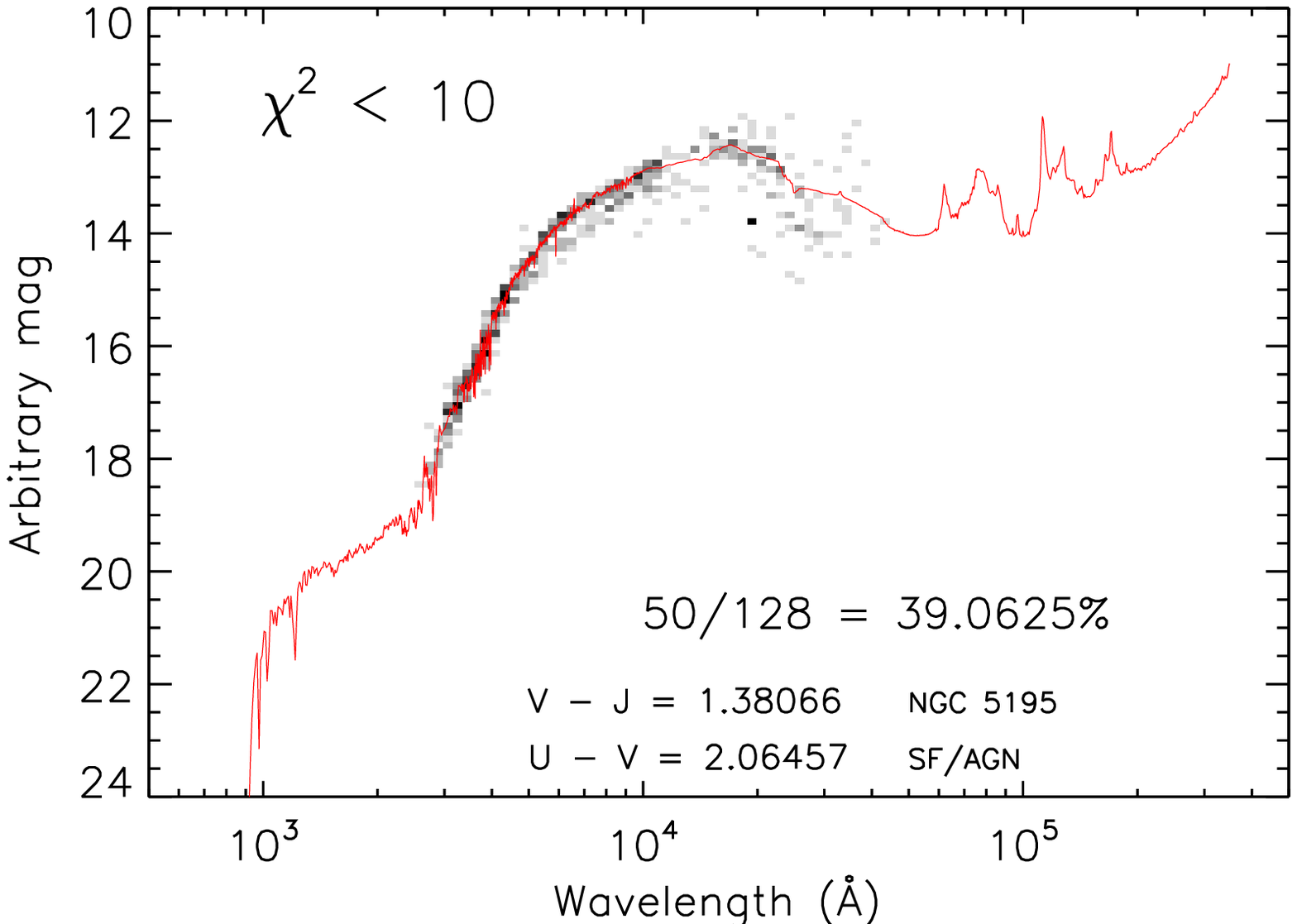}
\includegraphics[width=0.32\textwidth]{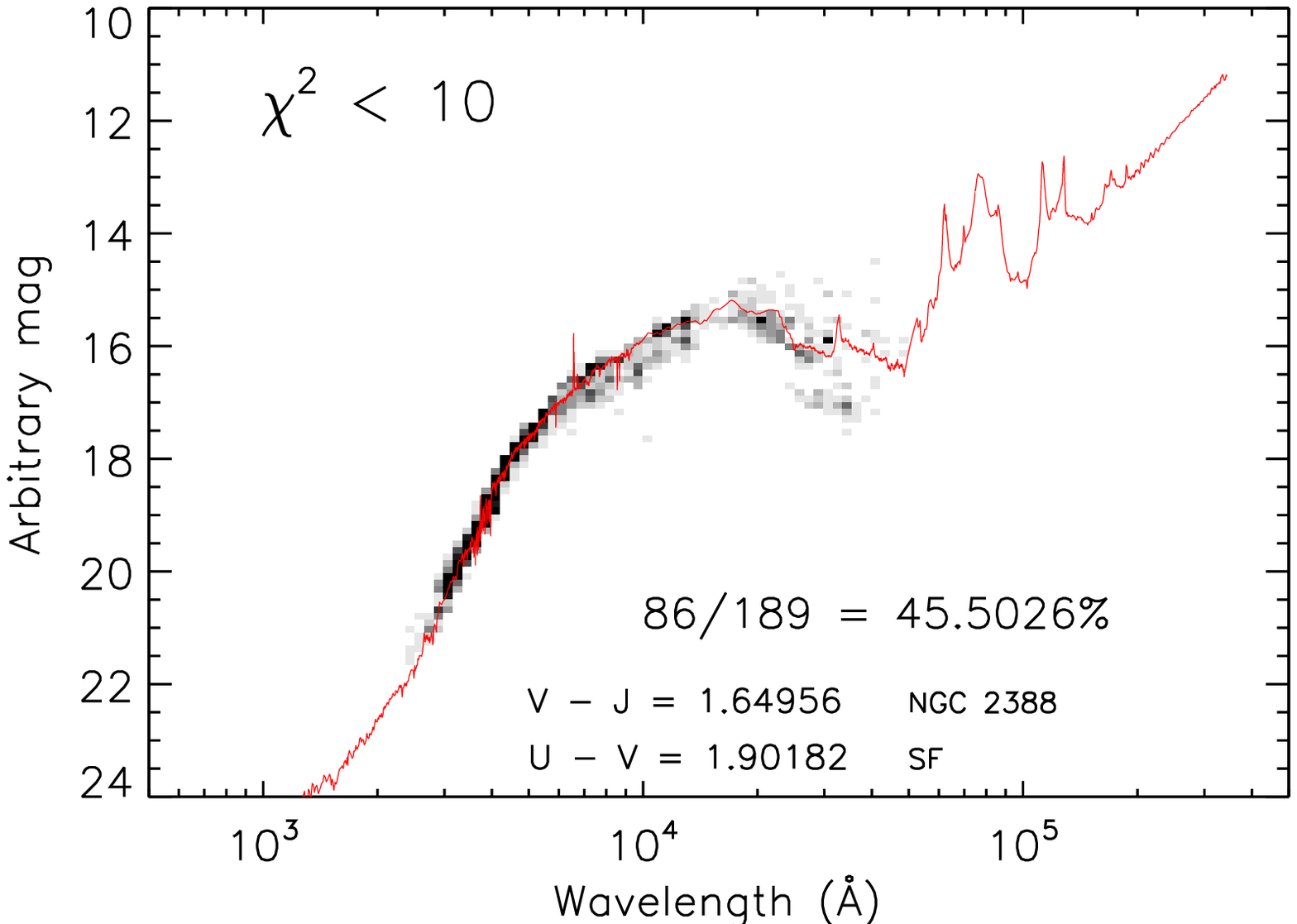}
\includegraphics[width=0.32\textwidth]{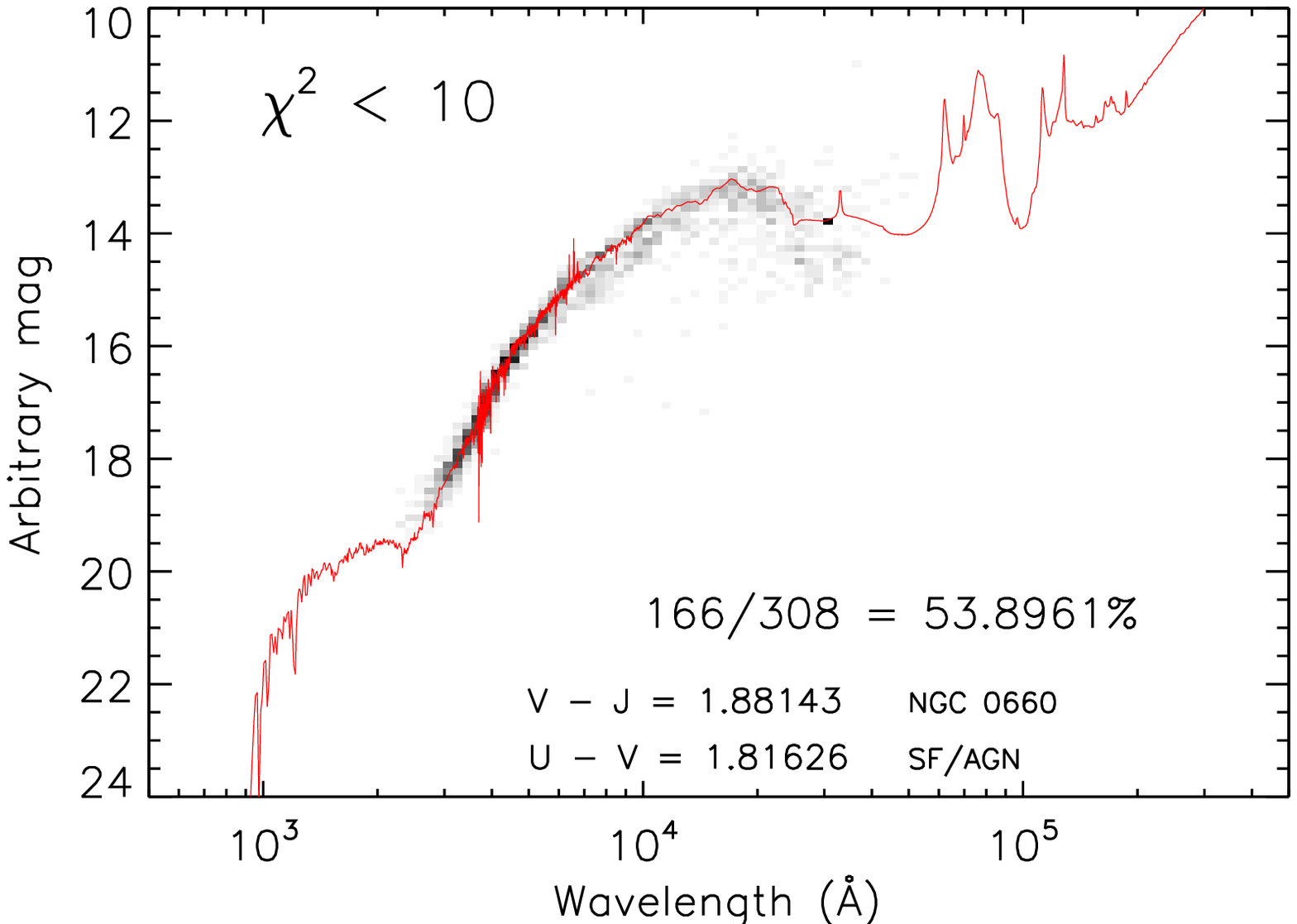}
    \caption{Continued.}
    \label{fig:my_label}
\end{figure}

\end{document}